\documentclass[11pt,a4paper,twoside]{book}

\usepackage[utf8]{inputenc}
\usepackage[english]{babel}
\usepackage{lmodern}
\usepackage{amsmath}
\usepackage{amsfonts}
\usepackage{amssymb}
\usepackage{upgreek}
\usepackage{graphicx}
\usepackage{subcaption}
\usepackage{placeins}
\usepackage{booktabs}
\usepackage{array}
\usepackage{multirow}
\usepackage{hyperref}
\usepackage{fancyhdr}
\usepackage{enumitem}
\usepackage{CJKutf8}
\usepackage{xcolor}
\usepackage{afterpage}

\author{Stefano Gramigna}

\newcommand{\tabitem}{~~\llap{\textbullet}~~}
\newcommand{\labellazzo}[2]{$#1#2$}

\newcommand\blankpage{
    \null
    \thispagestyle{empty}
    \addtocounter{page}{-1}
    \newpage
    }

\begin{document}
		\begin{titlepage}
	\begin{center}
		{\LARGE{University of Ferrara}} \\[1ex]
	   	{\large{Physics and Earth Sciences Department}} \\[1ex]
	  	{\large{Ph.D. Course in Physics}}\\[1ex]
	  	{Cycle XXXVII}\\[2ex]
	  	{\em Coordinator Prof.} {\sc Paolo Lenisa}\\
	  	\vspace{1cm}
		\includegraphics[height=6cm]{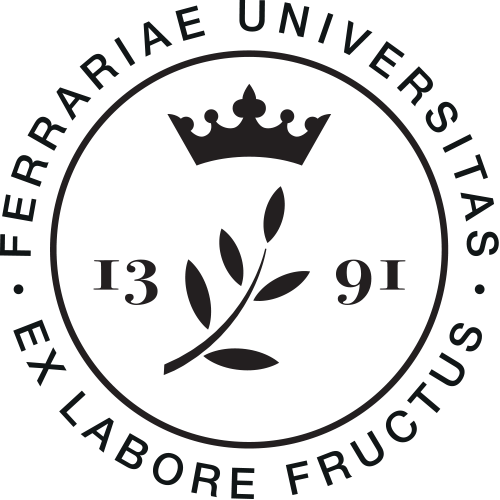}\\
		\vspace{1.5cm}
		{\Large\textbf{Construction, Commissioning, and Installation\\[0pt]
		of the Cylindrical GEM Inner Tracker\\[3pt]
		of the BESIII Experiment}}\\
		\vspace{.5cm}
		{\em Scientific/Disciplinary Sector (SDS)} {FIS/01}\\
	\end{center}
	
	\vfill
	
	\begin{flushleft}
		{\em Supervisor}  \\
		{\em Dr}. {\sc Gianluigi Cibinetto}\\[3ex]
	\end{flushleft}
	\begin{flushright}
		{\em Candidate} \\
		{\sc Stefano Gramigna}\\[3ex]
	\end{flushright}
	\begin{flushleft}
	\begin{CJK*}{UTF8}{gbsn}
		{\em Referees}\\
		{\em Dr}. {\sc Mauro Iodice}\\
		{\em Dr}. {\sc Jianbei Liu (刘建北)}\\
	\end{CJK*}
	\end{flushleft}
	\vfill
	\begin{center}
	{Year 2021/2024}
	\end{center}
\end{titlepage}
 
	\pagestyle{empty}
	\pagenumbering{roman}	
  		\tableofcontents
  		\listoffigures
  		\listoftables
  		
	\newpage
	\afterpage{\blankpage}

 	\pagenumbering{arabic}
 		\chapter*{Foreword}
\addcontentsline{toc}{chapter}{Foreword}
My research activity as a PhD candidate revolved around the development, the operation, and ultimately the installation of micropattern gaseous detectors.
My main effort was directed towards the advancement of the CGEM-IT (Cylindrical Gas electron Multiplier Inner Tracker) project, which aims to provide the BESIII (Beijing spectrometer III) experiment with an upgraded detector to replace its aging inner tracker.

I was initially tasked with developing the accelerometric measurement system used in the drop test that validated the new design of the detector's outermost layer.
Later, I joined all other phases of the test: the construction of a mock-up reinforced with PEEK spacer grids, the preparation of a setup to achieve repeatable falls, and the analysis of the CT scans.

The success of the test green-lighted the split construction of the new Layer 3, whose electrodes were built in Italy and then assembled in China.
I suggested the use of laser triangulation sensors in the search of a new method for aligning the detector's assembly machine.
The idea was approved and I was charged with the development, the test, and the commissioning of the new laser alignment system.
As a member of the construction team I also joined: the preparation of the new cleanroom, the restoration of the setup for testing the GEM foils, the construction of all electrodes, the commissioning and alignment of the assembly machine, the assembly of the layer, and its validation.

With the completion of Layer 3, the CGEM-IT entered pre-com\-mis\-sion\-ing.
The detector was deployed, alongside all its ancillary systems, on a cosmic ray telescope setup, to acquire data and asses its performance.
Aside from participating in the cabling of the detector, I designed the testing station and commissioned the new gas system.
During this time I also became the contact person for the installation of the front-end electronics and most on-detector maintenance.

In parallel to the cosmic ray data taking, the installation of the detector was being prepared.
As coordinator of the insertion team, I was responsible for: preparing the two insertion tests, redesigning the installation tooling, and writing the operations' procedures.
I also prepared the detector's cabling schemes and designed a mock-up to study and validate them.
During the installation of the CGEM-IT I coordinated the insertion of the detector, its fixing to BESIII, and the extraction of the installation trolley.
As an expert of the operations occurring within the endcaps of BESIII's MDC (Multilayer Drift Chamber), I trained my colleagues in the routing of the detector's cables and their connection to the patch cards according to the schematics.
  		\chapter*{Introduction}
\addcontentsline{toc}{chapter}{Introduction}
The Beijing Spectrometer III (BESIII) is a particle physics experiment located at the Institute of High Energy Physics (IHEP) in the capital of the People's Republic of China.
Exploiting the particle beams of the Beijing Electron Positron Collider II (BEPCII), the experiment studies a vast physics program in the $\tau$-charm energy region.
The high luminosity reached by BEPCII and the clean background of the $e^+e^-$ collisions make BESIII particularly well suited for precision measurements, allowing it to contribute some of the most accurate measurements to date of the properties of the tau lepton and to discover new exotic charmonium states.

The Cylindrical Gas Electron Multiplier Inner Tracker (CGEM-IT) is the candidate detector for the upgrade of BESIII's aging inner Multilayer Drift Chamber (MDC).
The main improvements achievable through the use of Micropattern Gaseous Detector (MPGD) technology are a factor at least 2 of improvement to the spatial resolution in the beam's direction and increased aging resistance, to better face the planned upgrades to the machine's luminosity.
The development of the new tracker is sponsored by the Italian component of the BESIII collaboration, whose members refer to universities or sections of the Italian National Institute for Nuclear Physics (INFN) located in the cities of Ferrara, Frascati, Turin, and Perugia.
The project was officially proposed in 2013 and, at the time of writing, the now complete CGEM-IT is being installed in the spectrometer.

The new tracker consists of three concentric layers, each of which is an independent cylindrical triple-GEM detector.
The largest of these, Layer 3, suffered from mechanical instabilities that initially prevented it from powering on.
This thesis, encompassing the three final years of the CGEM-IT's development, begins by describing how these issues were overcome through a review of the layer's design.
The fragile nature of CGEM detectors at these dimensions also required to rethink the construction pipeline to safeguard the delicate floating electrodes.
The differences from the previous experiences are thus highlighted, with a particular focus on the technological solutions adopted.
Since the completion of Layer 3, the CGEM-IT entered commissioning, with the purpose to acquire cosmic ray data and assess its resolution and efficiency, thus informing the internal review process.
The reassessment and the finalization of pre-existing tools and procedures to safely install the detector in BESIII was also an integral part of this review.

The thesis is structured as follows:
\begin{itemize}
\item \textbf{Chapter 1} is intended to provide the scientific context represented by the BESIII experiment: the BEPCII collider, the design of the spectrometer, and its physics program.
Here, general information about the CGEM-IT project that may help understanding the contents of the following chapters is also collected.
\item \textbf{Chapter 2} explores the review of Layer 3's design. The chapter revolves around the work that went into testing the solution to Layer 3's mechanical instability.
\item \textbf{Chapter 3} delves into the particularity of Layer 3's split construction, highlighting the differences with respect to previous experiences.  The technological solution adopted to overcome one of the main hurdles in pursuing a split construction, the alignment of the assembly machine, is given particular focus.
\item \textbf{Chapter 4} describes the commissioning of the CGEM-IT, from the assembly of the three layers to the acquisition of cosmic ray data. The chapter also contains the first values for the efficiency and resolution of the complete detector.
\item \textbf{Chapter 5} analyzes of a series of tests conducted to prepare for the installation of the detector. These prompted an overhaul of the installation tooling that is here explored in detail.
The chapter concludes with a snapshot of the progress of the installation at the time of writing.
\end{itemize}

    \pagestyle{fancy}
    \fancyhf{}
    \fancyhead[RE]{\small{\nouppercase{\rightmark}}}
    \fancyhead[LO]{\small{\nouppercase{\leftmark}}}
    \fancyhead[RO,LE]{\small{\thepage}}
    \setlength{\headheight}{14pt}
    		\chapter{BESIII at BEPCII and the CGEM-IT}
\label{intro}
\section{BEPCII}
The Beijing Electron Positron Collider II (BEPCII) is a double ring $\mathrm{e^+ e^-}$ collider located at the Institute of High Energy Physics (IHEP) in Beijing. BEPCII can operate both as a collider, to allow the Beijing Spectrometer III (BESIII) experiment to collect data, and as synchrotron radiation (SR) source. The electron and the positron rings are identical, they have a circumference of 237.53$\,$m, and they cross in two points: within the hall that hosts the radiofrequency (RF) cavities and at the center of BESIII's collision hall. The SR ring consists of the outer half of both rings and has a circumference of 241.13$\,$m. The design parameters of the machine at commissioning are reported in table$\,$\ref{bepc2params}.

\begin{table}[htbp]
\centering
\begin{tabular}{ll}
\multicolumn{2}{c}{\textbf{BEPCII Design Parameters}}\\
\midrule
Beam energy & 1$\,$-$\,$2.3$\,$GeV\\
Optimum energy & 1.89$\,$GeV\\
Luminosity & $\mathrm{10^{33}\,cm^{-2}s^{-1}\ @\ 1.89\,GeV}$\\
SR current & $\mathrm{250\,mA\ @\ 2.5\,GeV}$\\
\end{tabular}
\caption[BEPCII Parameters]{BEPCII design parameters at commissioning$\,$\cite{bepcii_2009}.}
\label{bepc2params}
\end{table}

\begin{figure}[htbp]
\centering
\includegraphics[keepaspectratio, width=.7\textwidth]{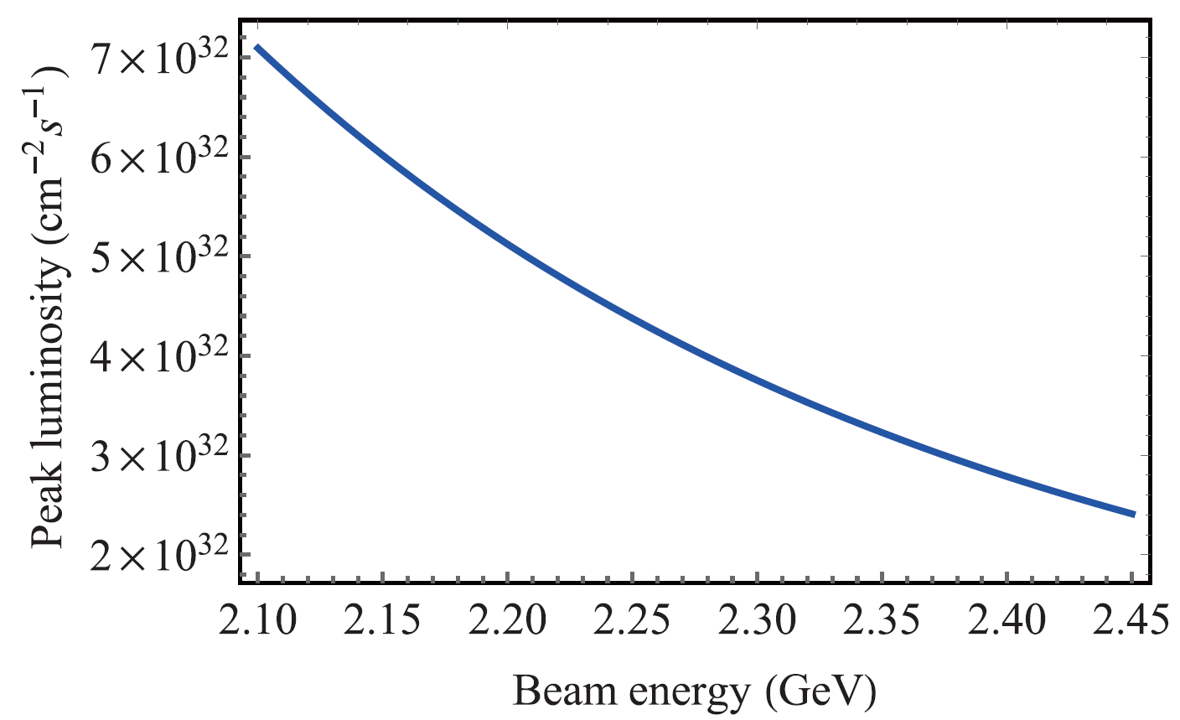}
\caption[BEPCII's luminosity decrease above optimum beam energy]{Estimated peak luminosity above 2.1$\,$GeV$\,$\cite{BESIII:2020nme}.}
\label{bepcplot}
\end{figure}

In 2019 a campaign of hardware upgrades was carried out to bring the beam energy to 2.45$\,$GeV. Above 1.89$\,$GeV, the luminosity decreases as shown in figure$\,$\ref{bepcplot} due to limitations in the power of the RF cavities and increasing difficulties in controlling the beam bunches.
Because of this, proposed machine upgrades aim not only to expand the energy range even further to 2.5$\,$GeV, and to 0.9$\,$GeV towards lower energies, but also to increase the luminosity to cope with the losses. While crab-waist schemes have been considered, increasing the beam currents by means of standard upgrades to the accelerator's components seems to be the most feasible solution to yield a factor 2 increase in peak luminosity in the near future$\,$\cite{BESIII:2020nme}.

\subsection{BEPCII's Interaction Region}
BEPCII's interaction region extends 14$\,$m from the interaction point (IP) at both sides. It houses six pairs of quadrupoles and a pair of beam bending dipoles. The closest to the IP, residing within the 1$\,$T magnetic field of BESIII's superconducting solenoid, are the compact superconducting quadrupole micro-$\mathrm{\beta}$ magnets called superconducting interaction magnets (SIMs). Each SIM consists of a vertical focusing quadrupole, a horizontal bending dipole, a vertical steering dipole, a skew quadrupole, and three anti-solenoid windings. The design of the SIMs is strictly constrained by the tight layout of BESIII, as they have to penetrate the spectrometer to be positioned as close as possible to the IP.

\begin{figure}[htbp]
\centering
\includegraphics[keepaspectratio, width=.5\textwidth]{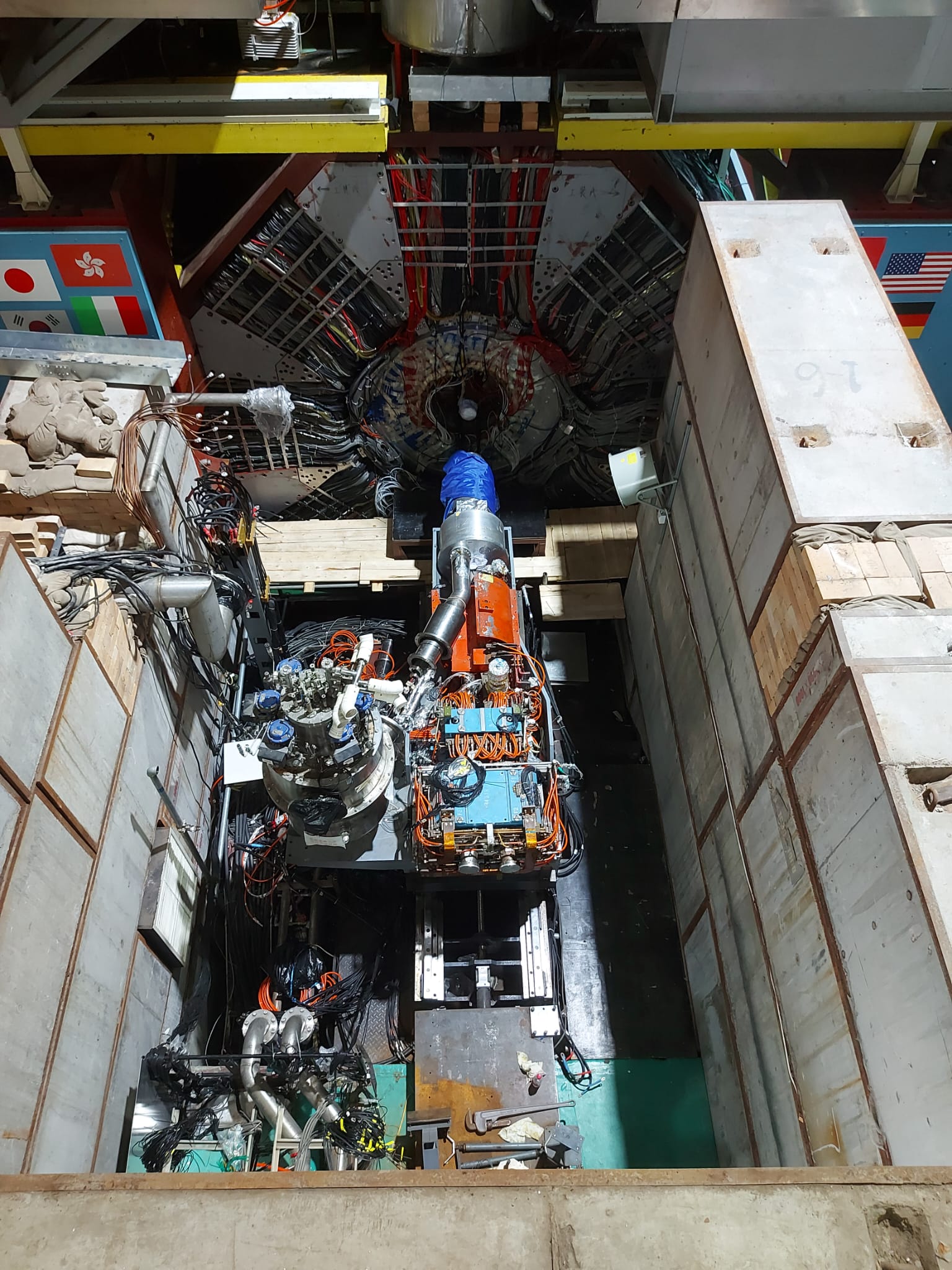}
\caption[Eastern superconducting interaction magnet]{The eastern superconducting interaction magnet, before its removal, extracted to free the access to BESIII's innermost subdetectors.}
\label{quads}
\end{figure}

The SIMs are mounted on moving trolleys, so that they can be extracted to access the innermost BESIII subdetectors, as shown in figure$\,$\ref{quads}. For the installation of the Cylindrical GEM Inner Tracker (CGEM-IT), the eastern magnet was temporarily removed, to make space for the installation tooling. The western SIM was also extracted but it was kept in place on its rails.

Many competing requirements constrain the design and material choices of the section of beam pipe that encloses the IP:
\begin{itemize}
\item The material within the detector's acceptance region should be minimized to prevent multiple scattering.
\item Good thermal and mechanical properties are required for withstanding both high thermal loads and and large differential pressure.
\item Good electric conductivity is necessary to ensure EM shielding from beam bunches.
\end{itemize}

BESIII's beam pipe consists of three sections: a central, thin-walled beryllium pipe and two copper extensions, welded to the first.
Due to the high thermal load, all three sections must be actively cooled. The central beryllium pipe is cooled by flowing high purity mineral oil in between its inner and outer walls, while the two copper sections are water cooled. The adopted cooling scheme was designed to cope with a maximum heat load of about 700$\,$W and to maintain the temperature of the outer surface of the beam pipe within $\mathrm{\pm\,1\,^{\circ}C}$ of ambient temperature. The estimated material budget at normal incidence is 1.04\% of $\mathrm{X_0}$\cite{BESIII:2009fln}.

\section{BESIII}
The Beijing Spectrometer III is a high energy physics experiment that investigates the portion of the energy region between 2.0 and 4.9$\,$GeV, also known as the $\tau$-charm region. Energies below 2.0$\,$GeV can be accessed through initial state radiation processes. The very high luminosity, the highest for experiments in this range of energies, and the clean backgrounds of $\mathrm{e^+e^-}$ collisions make BESIII particularly well suited for high precision measurements.

The experiment covers a vast physics program, spanning many topics:
\begin{itemize}
\item Studies of $\tau$ physics, having contributed one of the the most precise measurements of the $\tau$-lepton mass \cite{BESIII:2014srs}.
\item Precision tests of the Standard Model, in particular measuring the cross section of $\mathrm{e^+ e^- \rightarrow hadrons}$ processes$\,$\cite{BESIII:2015equ}\cite{PhysRevLett.128.062004}.
\item The study of charmonium and charmonium-like XYZ states. BESIII discovered the first confirmed charged charmonium-like tetraquark, the $\mathrm{Z_{c}(3900)^\pm}\,$\cite{BESIII:2013ris}.
\item Precision measurements of charmed hadrons decays, like the most precise determination of the branching ratio of the leptonic $D_s^+ \to \tau^+ \nu_{\tau}$ decay$\,$\cite{BESIII:2021bdp}.
\item Light hadron spectroscopy in charmonia decays, i.e. the observation of the exotic quantum number state $\eta_1(1855)\,$\cite{BESIII:2022riz}$\,$\cite{PhysRevLett.130.159901}.
\item The search for new physics beyond the Standard Model, especially studying rare charm decays such as $D^0\rightarrow\pi^0\nu\overline{\nu}\,$\cite{BESIII:2021slf}.
\end{itemize}
Further information about the present and future physics program of BESIII can be found in references \cite{Yuan:2019zfo} and \cite{BESIII:2020nme}, respectively.

\begin{figure}[htbp]
\centering
\includegraphics[keepaspectratio, width=\textwidth]{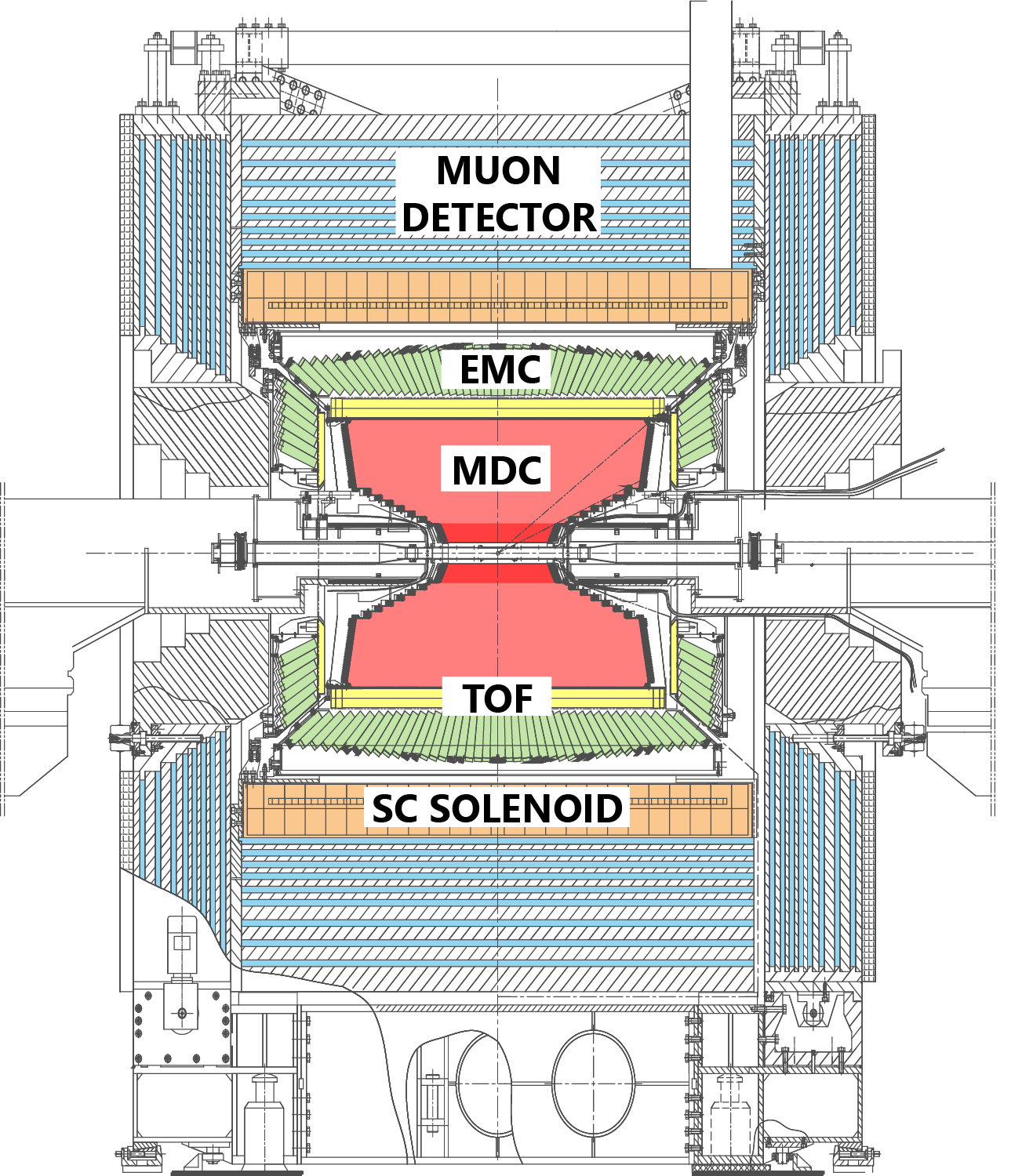}
\caption[BESIII's layout]{Cross-sectional view of the BESIII experiment.}
\label{besiii}
\end{figure}

In BESIII the collisions are symmetric and so is the layout of the detector, consisting of a barrel and two identical endcaps, which manages to achieve a 93\% coverage of the whole solid angle. The spectrometer, in figure$\,$\ref{besiii}, comprises 5 main subsystems: the multilayer drift chamber (MDC), the time of flight (TOF) system, the electromagnetic calorimeter (EMC), the superconducting solenoid magnet (SSM), and the muon counter (MUC). A detailed description of the BESIII subsystems can be found in reference$\,$\cite{BESIII:2009fln}.

The innermost subdetector and main tracker of the experiment is the multilayer drift chamber. BESIII's MDC has a spatial resolution of 130$\,\upmu$m in the r$\phi$ direction and of about 2 mm in the beam direction. It can measure charged particles' transverse momenta with a resolution $\sigma_p/p$ of 0.5\% at 1$\,$GeV/$c$ and their energy loss $dE/dx$ with a resolution $\sigma(dE/dx)/(dE/dx)$ of 6\%, for perpendicular tracks. Due to the centrality of the MDC's design for this thesis, it will be thoroughly described later in a dedicated section.

The time of flight system is divided into a barrel and two endcaps. The barrel consists of two layers of staggered plastic scintillator bars (BC-408) read by Hamamatsu fine mesh PMTs (Hamamatsu R5924), directly attached to each end of the bars. The time resolution of the barrel is about 80$\,$ps for 1$\,$GeV muons$\,$\cite{Cao:2020ibk}. For kaons and pions it is expected to be about 20\% larger, still allowing a 3$\,\sigma$ $K/\pi$ separation of approximately 700 MeV/$c$ for perpendicular tracks. The endcaps have  been upgraded from a similar configuration to arrays of multi-gap resistive plate chambers (MRPCs). Each endcap of the TOF system now consists of two layers of MRPCs, staggered to avoid dead regions due to the gaps separating the chambers. The time resolution has been improved from 140$\,$ps to 65$\,$ps, granting a 2$\,\sigma$ $K/\pi$ separation up to 1.4$\,$GeV/c.

BESIII's electromagnetic calorimeter is made of CsI(Tl) crystals read by large area photodiodes. The crystals are arrangend in rings, 44 in the barrel and 12 in the two endcaps, fanning out from the IP to maximize coverage. A 5$\,$cm gap separates the endcaps from the barrel, allowing cables and service lines to reach the inner detectors. The endcaps themselves can be extracted for extraordinary maintenance to the MDC and TOF systems. The calorimeter was designed to measure energies of electrons and photons from 20$\,$MeV up to about 2$\,$GeV, ensuring good $e/\pi$ discrimination at momenta larger than 200$\,$MeV/$c$.
The energy resolution $\mathrm{\sigma_{E}/E}$ is 2.5\%, for 1$\,$GeV particles, and the position resolution $\sigma$ is $\mathrm{6\,mm/\sqrt{E(GeV)}}$.

The MDC, the TOF and the EMC are enclosed in a 1$\,$T superconducting solenoid magnet with a mean radius of about 1.5$\,$m and a length of about 3.5$\,$m. The SSM is capable of generating a uniform 1.0$\,$T axial field for the 2.58$\,$m long MDC.

The steel structure for the flux return of the magnetic field houses the RPCs of the muon counter. In the barrel there are 9 RPC layers while in the endcaps there are only 8, for a total area of about 700$\mathrm{\,m^2}$. The RPCs have strips in both the $\theta$ and the $\phi$ direction, with a typical strip width of about 4$\,$cm. The special bakelite laminates employed in the construction of the electrodes were engineered to avoid the application of the linseed oil coating that is common for such kind of detectors. The single 2$\,$mm gap separating the electrodes is filled with an Ar:$\mathrm{C_2F_4H_2}$:$\mathrm{C_4H_{10}}$ gas mixture in 50:42:8 proportions.
In this configuration the single layer efficiency reaches $\sim96\%$ for the barrel counters and $\sim95\%$ for the ones in the endcaps. The main purpose of the muon counter is to separate muons from pions and other long-lived hadrons. The lower limit of muon momentum at which the detector starts to become effective is approximately 0.4$\,$GeV/$c$.

\subsection{The Design of BESIII's MDC}
BESIII's MDC has an outer radius of 810$\,$mm, an inner radius of 59.2$\,$mm, and a length of 2582$\,$mm. The chamber is built as two separate parts, an inner and an outer chamber, sharing the same gas volume. Such subdivision makes it possible to replace the inner chamber, more subject to radiation damage due to its proximity to the IP. The two chambers are joined at the endplates, where gas tightness is ensured by O-rings and epoxy sealing.

The MDC adopts a small-cell design, where each sense wire is surrounded by 8 field wires arranged in a square pattern. The cell size in the inner chamber is 12$\,$mm while it is on average 16.2$\,$mm in the outer chamber. The drift chamber has 43 sense layers overall. The inner chamber hosts the first 8, which are all small angle stereo layers. The other stereo layers, from layer 21 to 36, reside in the outer chamber, while all other layers are axially oriented. Sense wires have a diameter of 25$\,\upmu$m and they are made of a gold-coated tungsten and rhenium alloy. Field wires are 110$\,\upmu$m in diameter and made of gold-plated aluminum. Neighboring groups of 4 layers (3 for the last one) are arranged in superlayers, where field wires are shared between neighboring cells. Axial and stereo superlayers are separated by additional field wires to regularize the fields across the transition.

\begin{figure}[htbp]
\centering
\includegraphics[keepaspectratio, width=\textwidth]{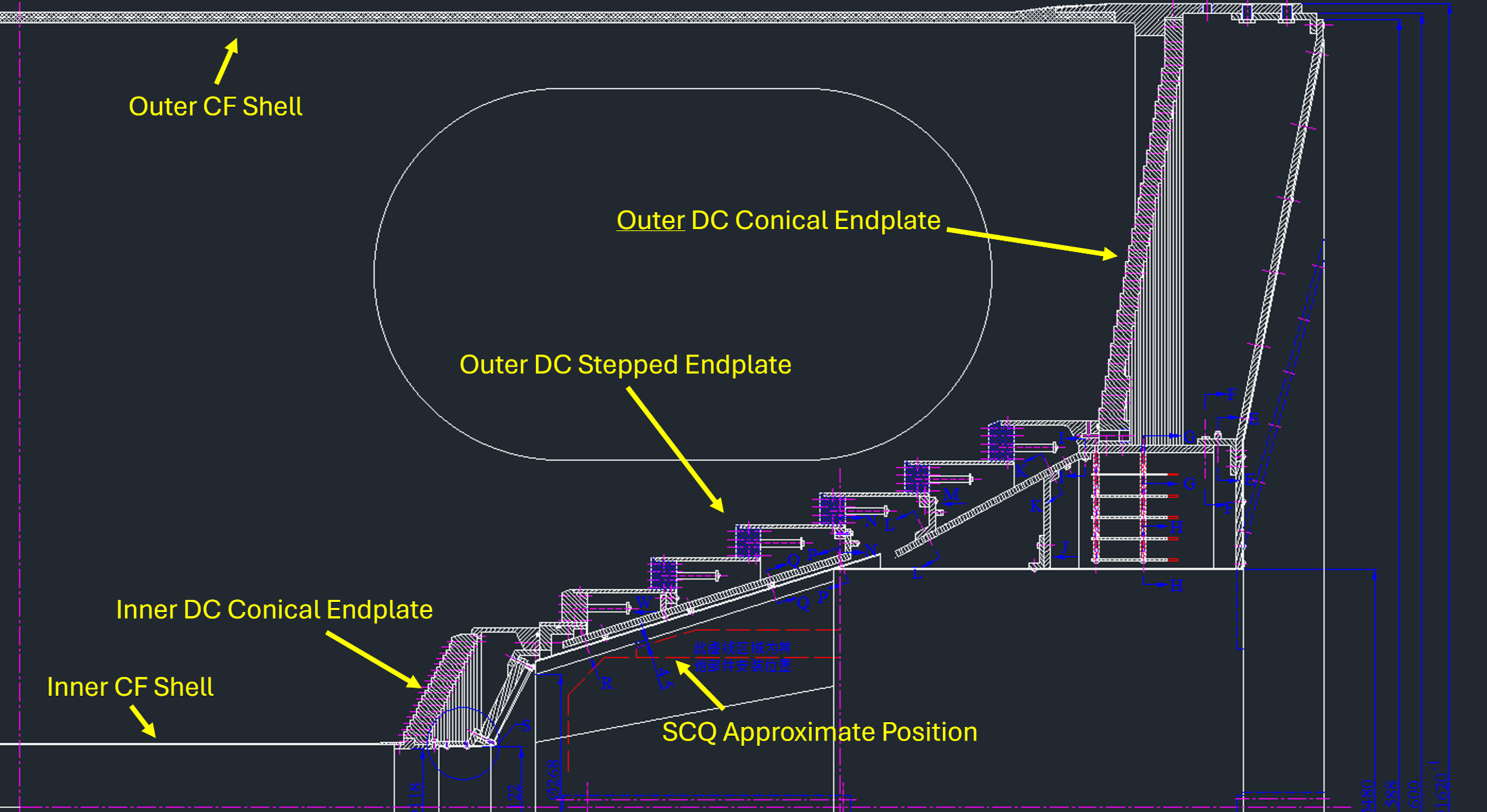}
\caption[Cross section of the MDC mechanical supports]{Cross-sectional view of the MDC's mechanical support structures.}
\label{mdc_det}
\end{figure}

The inner chamber has conical endplates, while the outer chamber has an inner stepped part and an outer conical one, with the stepped sections providing the necessary space for the insertion of the interaction quadrupoles. All endplates are machined from high strength aluminum alloy, while the cylindrical shells that separate them, bearing the axial load of the MDC's many wires, are made of carbon fiber composites. The outer shell of the MDC is 11.5$\,$mm thick, while the inner shell is 1.2$\,$mm thick. The cross-sectional view in figure$\,$\ref{mdc_det} shows how the mechanical structures of BESIII's MDC fit together and the approximate position of the quadrupole when fully inserted.

The operating gas mixture of BESIII's MDC is He:$\mathrm{C_3H_8}$ (60:40), with a gas gain of $3\times10^4$ at the reference voltage of 2200$\,$V. The helium based mixture was chosen for its long radiation length of 550$\,$m, which minimizes multiple scattering, and sufficient hydrocarbon component to ensure stable operation. The insurgence of aging phenomena in 2012 required slight alterations of the original mixture, as described in the following section.

\subsection{Aging of the Inner MDC}
Since its commissioning in 2009, the inner MDC has been suffering a substantial decrease in performance due to aging phenomena. The amount of aging experienced by the different layers is correlated to hit occupancy and therefore integrated charge, with the innermost layers being the most affected as they are more exposed to the large beam related background.
Aging has affected both anodic and cathodic wires.

For the cathodes it takes the form of Malter effect$\,$\cite{PhysRev.50.48}: a self sustaining, non-localized discharge due to the polymerization of gas pollutants onto the surface of the field wires. The insulating layer prevents ions from being neutralized, their accumulation then leading to an increase in field strength near the field wires. Electrons extracted from the field wires may drift to the anode, get multiplied, and produce more ions, thus reinforcing the effect. Malter discharges in BESIII's MDC were first observed in 2012 and initially contained by adding 5\% $\mathrm{CO_2}$ to the gas mixture, which led to a decrease of 23\% in gas gain. A more complex solution was later implemented by replacing the $\mathrm{CO_2}$ with 0.2\% water vapor. This allowed stable operation since, with only a 9\% decrease in the MDC's gas gain at reference HV settings$\,$\cite{Dong:2024mzz}.

\begin{figure}[htbp]
\centering
\includegraphics[keepaspectratio, width=.7\textwidth]{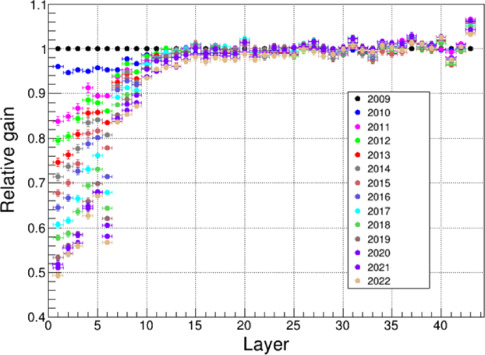}
\caption[Relative gain loss of BESIII's MDC]{Relative gain loss of the MDC layers since commissioning$\,$\cite{Dong:2024mzz}.}
\label{rel_gain}
\end{figure}

\begin{figure}[htbp]
\centering
\begin{subfigure}{.45\textwidth}
\includegraphics[keepaspectratio, width=\textwidth]{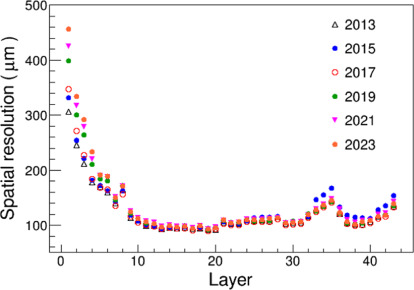}
\caption{Spatial resolution}
\end{subfigure}
\hfill
\begin{subfigure}{.45\textwidth}
\includegraphics[keepaspectratio, width=\textwidth]{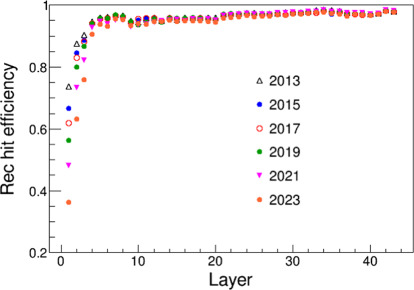}
\caption{Hit reconstruction efficiency}
\end{subfigure}
\caption[MDC performance losses]{Yearly variation of the MDC's spatial resolution and hit reconstruction efficiency per layer$\,$\cite{Dong:2024mzz}.}
\label{fom}
\end{figure}

Anodic aging can also be traced back to the deposition of insulating compounds on the wires. In this case, the deposits increase the diameter of the sense wires, lowering the electric field and therefore the effective gain of the affected cell. Figure$\,$\ref{rel_gain} shows the gain losses experienced by different layers of BESIII's MDC since commissioning. While only the first two layers of the outer chamber have been experiencing minor losses, all the 8 layers of the inner chamber have been heavily affected, with losses up to 50\% with respect to reference values at commissioning$\,$\cite{Dong:2024mzz}. These gain losses have direct and measurable effects on resolution and hit reconstruction efficiency, whose degradation in time is shown in figure$\,$\ref{fom}. While the effect can be counteracted by raising the HV settings of the affected cells, this also accelerates the aging process. Although BESIII has been collecting data successfully by tuning the MDC's operating parameters in the past years, the collaboration has been developing an innovative cylindrical GEM inner tracker, main subject of this thesis work, to replace the aging inner chamber. A new and improved drift chamber was also developed in parallel, as a reliable backup solution.

\FloatBarrier

\section{The CGEM-IT Project}
The project for the development of a Cylindrical GEM Inner Tracker for the BESIII experiment is led by the INFN sections and universities of Ferrara and Turin, together with INFN's Frascati National Laboratories (LNF). The CGEM-IT working group also includes members of: the universities of Ferrara, Mainz, Turin, and Uppsala; and IHEP. The project, officially approved in 2013 and now in its 11th year, is at the time of writing reaching its most important milestone of completing the installation of a replacement for BESIII's aging inner chamber. The expected duration of the experiment's data taking with the new tracker is at least 5 years. 

\subsection{GEM Operating Principles}
\begin{figure}[htbp]
\centering
\includegraphics[keepaspectratio, width=.6\textwidth]{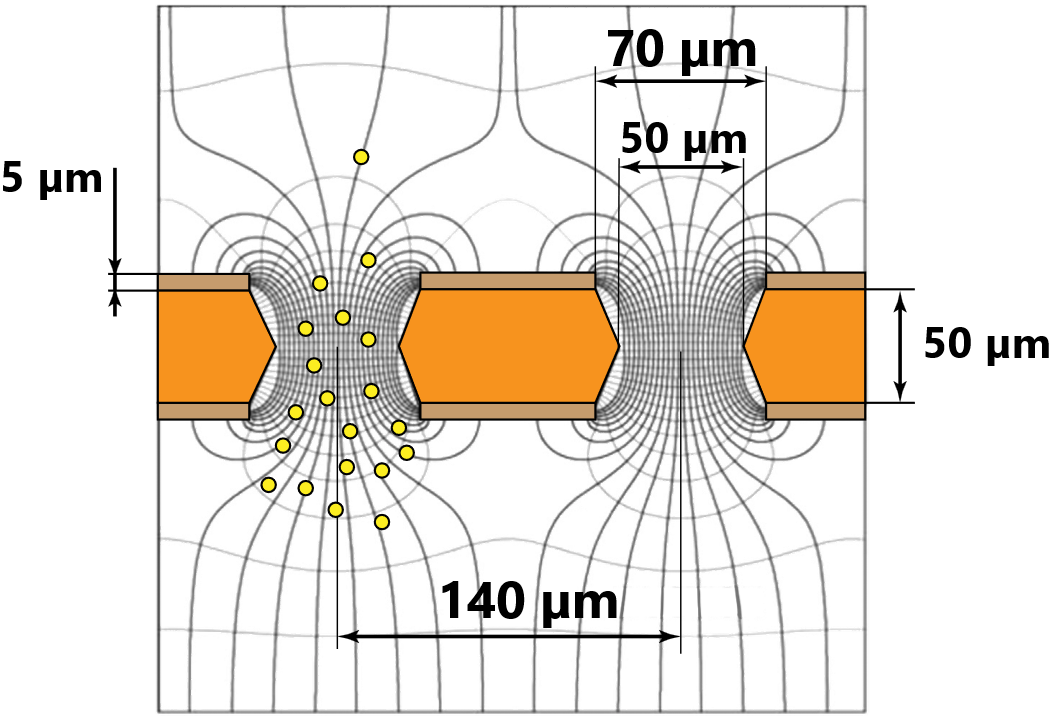}
\caption[Cross section of a GEM foil]{Cross sectional drawing of a GEM foil.}
\label{gemholes}
\end{figure}

Gas Electron Multiplier (GEM) is a Micropattern Gaseous Detector (MPGD) first introduced by Fabio Sauli in 1997 \cite{SAULI1997531}.
GEM foils are produced by etching a dense matrix of holes on a 50$\,\upmu$m thick polyimide foil plated by a 5$\,\upmu$m thin copper layer on both faces.
The holes, obtained through photolitography and chemical etching, have a biconical cross section, with a diameter of 70$\,\upmu$m at the two copper surfaces and 50$\,\upmu$m in the narrowest point of the polyimide bulk.
The pitch of the holes' matrix in the standard fabrication pattern is 140$\,\upmu$m, but foils' parameters can be optimized for specific applications.
A cross-sectional drawing of a GEM foil and its holes' structure is provided in figure$\,$\ref{gemholes}.

\begin{figure}[htbp]
\centering
\includegraphics[keepaspectratio, width=.6\textwidth]{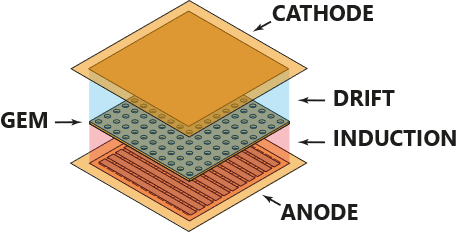}
\caption[Example of single GEM detector]{Simple example of a single GEM detector with two-dimensional strip readout.}
\label{gemdet}
\end{figure}

By applying a voltage of the order of a few hundreds of volts to the two faces of a GEM foil, an electric field of the order of 10$\,$kV/cm can be generated inside the holes. The simplest example of GEM detector consists of a cathode, a GEM foil, and a patterned readout anode, stacked and separated by gaps of a few millimeters as shown in figure$\,$\ref{gemdet}. The electrodes's stack is immersed in a gas mixture, which include an easily ionizable noble gas and one or more quenchers.
Quenchers are usually hydrocarbon-based or fluorinated gases, which mitigate the discharge by capturing the UV photons generated in the de-excitation of the noble gas. These photons might otherwise generate electrons through photoelectric effect upon reaching the copper electrodes, thus leading to a continuous discharge.

By applying appropriate cascading voltages to the electrodes, three main regions are established. The gap between the cathode and the GEM identifies the drift region, permeated by the drift field. The multiplication region is distributed within the many holes of the GEM foil and houses the strongest electric fields in the system. Finally, the gap that separates the foil from the anodic readout circuit identifies the induction region, where the induction field resides. 

A charged particle traversing the detector generates electron-ion pairs by ionizing the operating mixture. Only the electrons generated in the drift region have a chance to be multiplied and therefore to contribute significantly to signal formation. The drift field leads the ions towards the cathode, where they are neutralized, and the electrons towards the holes of the GEM foils. Upon reaching the strong electric field inside the holes, the electrons are quickly accelerated, gaining enough energy to cause secondary ionization in the gas mixture, and leading to the formation of an electron avalanche.

About a 30\% of the avalanche is lost on the other side of the GEM foil, while the rest is led by the induction field towards the anodic circuit. The avalanche moving through the induction region induces a signal on the anodic circuit, which can be picked up by suitable readout electronics.

A single GEM detector can reach gains well above $10^3$ before discharges start to arise$\,$\cite{Sauli:2016eeu}. Spreading the multiplication among multiple stacked GEM foils not only allows to reach higher gains, up to order $10^4\,$\cite{Sauli:2016eeu}, but also renders it possible to operate the foils at lower voltages, thus increasing the detector's stability and lifetime, at the cost of significantly more complicated design and assembly. The fields used to lead the electrons from one multiplication stage to the next are called transfer fields.

When compared with their wire based counterparts, GEM detectors allow to reduce the distance between the electrodes from O(1$\,$cm) to O(1$\,$mm), which facilitates the evacuation of the positive ions, thus reducing space charge build-up and increasing rate capability. The larger conductive surfaces also make MPGDs less prone to aging phenomena that involve polymerization of pollutants onto the electrodes.

The great versatility of the flexible GEM electrodes allows them to be employed in a variety of geometries based on specific application needs. The inner tracker at the center of this thesis work is an example of the application of GEM technology in a cylindrical configuration.

\subsection{CGEM-IT}
\begin{figure}[htbp]
\centering
\includegraphics[keepaspectratio, width=\textwidth]{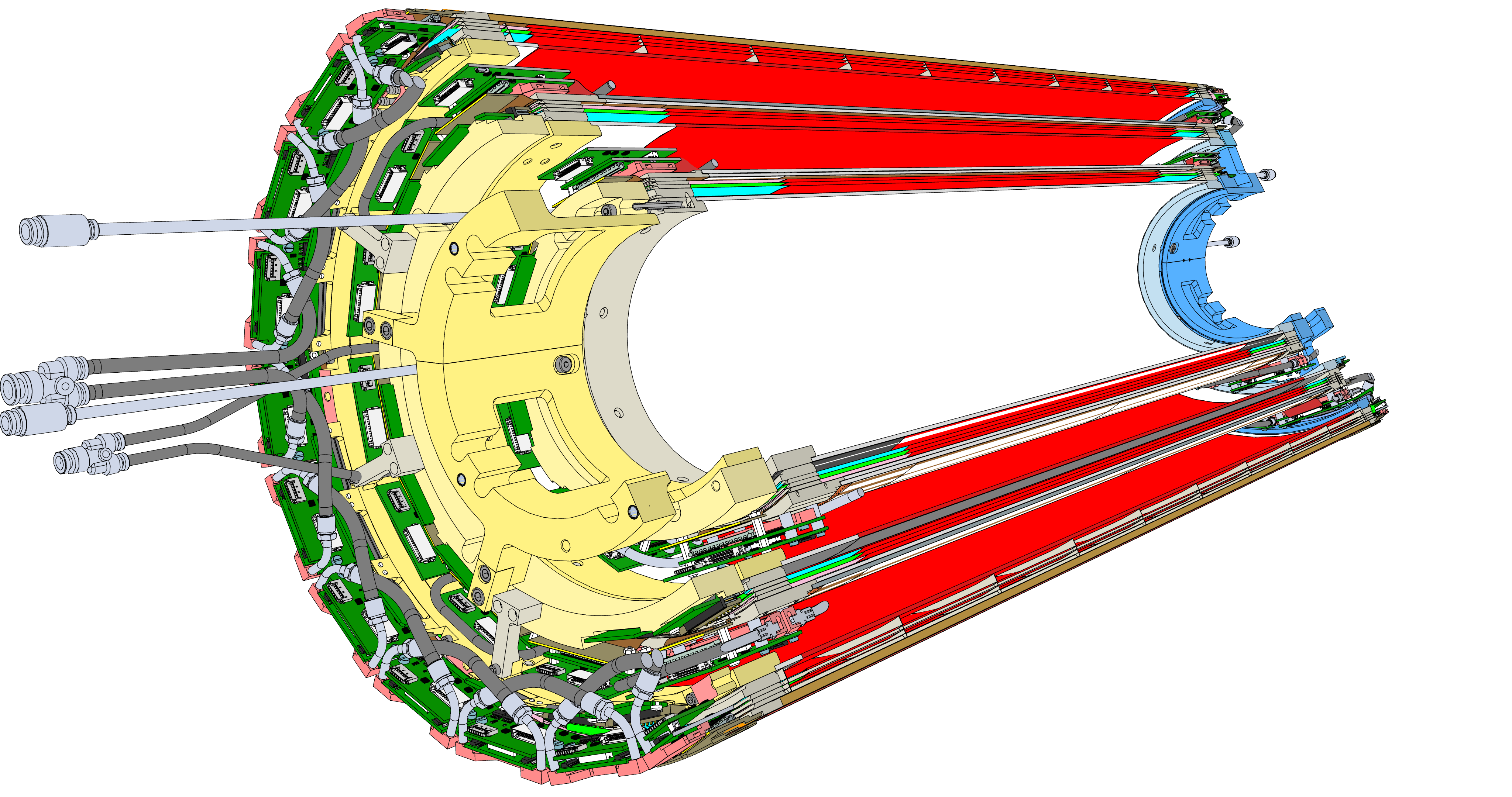}
\caption[Cross section of the CGEM-IT]{Cross-section of the CGEM-IT's CAD model.}
\label{cgemit}
\end{figure}

The Cylindrical GEM Inner Tracker (CGEM-IT) is the proposal advanced by the Italian part of the BESIII collaboration to restore and improve the performance of the experiment's aging inner drift chamber. The new tracker consists of three concentric layers, each of which is an independent cylindrical triple-GEM detector. A cross section of the CGEM-IT's CAD model, showing it's internal structure, is provided in figure$\,$\ref{cgemit}, while table$\,$\ref{cgemreq} summarizes the upgrade requirements for the project.
The upgrade is primarily meant to improve the resolution along the beam's direction while matching the performance of the inner MDC in the radial/azimuthal direction and in momentum resolution.
The new detector should also surpass the inner drift chamber in rate capability and radiation hardness, ensuring reliable operation at high luminosity for the remaining duration of BESIII's data taking.

\begin{table}[htbp]
	\centering
	\begin{tabular}{ll}
	\multicolumn{2}{c}{\textbf{CGEM-IT Upgrade Requirements}}\\\midrule
	$\mathrm{\sigma_{r\phi}}$ & $\mathrm{\leq150\,\upmu m}$\\
	$\mathrm{\sigma_{z}}$ & $\leq\,1\,$mm \\
	$dp/p$ ($\mathrm{1\,GeV}$) & 0.5\% \\ 
	Material budget & $\leq\,1.5\%\,\mathrm{X_0}$\\ 
	Angular Coverage & 93\%$\,\times\,4\pi$\\
	Rate capability & $10^4\,$Hz/cm$^2$
	\end{tabular}
	\caption[CGEM-IT upgrade requirements]{List of the CGEM-IT upgrade requirements.}
	\label{cgemreq}
\end{table}

The tracker's design is strictly constrained by the size of the cavity freed by the removal of the previous inner chamber. Outer and inner diameters are 361.4 and 131 mm respectively, whereas the overall length of the detector is 1070$\,$mm. Permaglass rings at the extremities provide structural support, maintain the spacing between the electrodes, and provide mounting points for on-detector electronics, HV connections, and gas pipes. The cylindrical structures supporting anodes and cathodes of the three layers are realized with lightweight sandwich-structured composites to minimize the material in the acceptance region. The requirement, for the entire detector, is to keep the material budget below 1.5\% of X$_0$.

A drift gap of 5$\,$mm ensures the occurrence of abundant primary ionization for MIPs in the chosen operating gas mixture, while all other electrodes in the stack are separated by 2$\,$mm gaps. The three anodes have nominal diameters of 180, 265, and 350 mm for layers 1, 2, and 3, respectively. Each anode is patterned by two views of strips, with a pitch of 650$\,\upmu$m.
X strips are 570$\,\upmu$m wide and run parallel to the beam direction, while V strips are 130$\,\upmu$m wide and form stereo angles with respect to the X strips.
The stereo angles are limited by the diagonal of the foils used in construction and differ for each layer: approximate angles of 47°, -31°, and 33° are obtained for Layer 1, Layer 2 and Layer 3 respectively. These larger stereo angles achievable through flexible circuits provide the key advantage with respect to a new drift chamber: a factor from 2 to 3 of improvement in the resolution along the beam direction. The design parameters of the CGEM-IT are summarized in table $\,$\ref{cgem_params}.

\begin{table}[htbp]
\centering
\begin{tabular}{@{}rrrrr@{}}
\textbf{Layer} & \textbf{Anode $\mathbf{\varnothing}$} & \textbf{AA Length} & \textbf{Stereo Angle} & \textbf{HV Sectors} \\
\midrule
1     & 180$\,$mm                & 532$\,$mm                        & 47$\,^\circ$                      & 4               \\
2     & 265$\,$mm                & 690$\,$mm                        & -31$\,^\circ$                     & 8               \\
3     & 350$\,$mm                & 847$\,$mm                        & 33$\,^\circ$                      & 12             
\end{tabular}
\caption[CGEM-IT design parameters]{Design parameters of the CGEM-IT.}
\label{cgem_params}
\end{table}

The GEM foils used in the construction of the CGEM-IT were produced at the CERN EST-DEM workshop. The foils adopt the standard holes' dimensions ($\mathrm{\varnothing\,70/50\,\upmu m}$) and pitch ($\mathrm{140\,\upmu m}$) but they are divided in sectors to reduce the energy released in case of discharge, which could permanently damage the electrodes. The side of the foils that faces the anode is divided in macro-sectors while each macro-sector is divided into ten micro-sectors on the side facing the cathode. The GEMs of Layer 1 have 4 macro-sectors, those of Layer 2 have 8, and the Layer 3 ones have 12. Two CAEN A1515 modules, installed into a CAEN SY4527LC power supply, generate the voltage levels necessary to power the detector. Patch panels inherited from the KLOE2 inner tracker handle the distribution of the voltage levels to macro and micro sectors. A jumper board on each patch panel allows to disconnect sectors, either individually or in groups, for debugging purposes.

\begin{figure}[htbp]
\centering
\includegraphics[keepaspectratio, width=.6\textwidth]{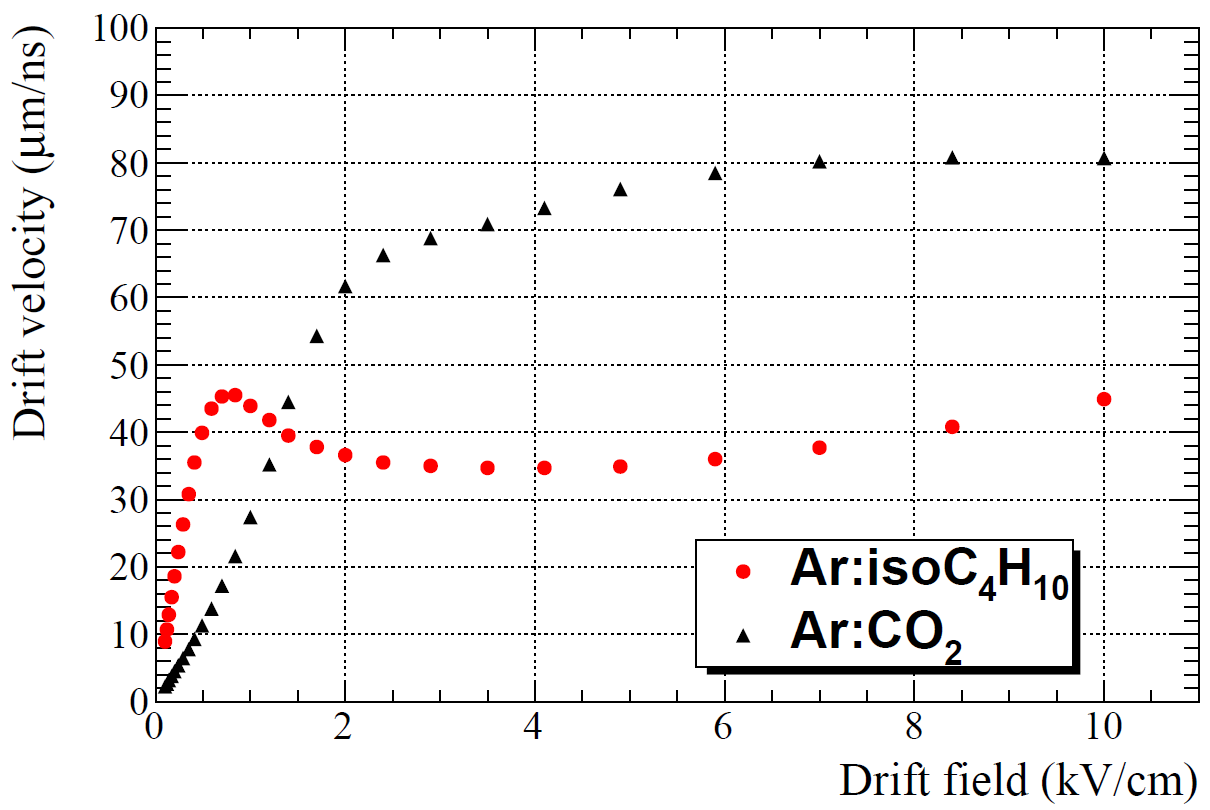}
\caption[Drift velocity VS electric field]{Drift velocity as a function of the electric field in a 1T magnetic field from Garfield simulations$\,$\cite{Alexeev:2019rng}.}
\label{vdrift}
\end{figure}

The chosen operating mixture for the CGEM-IT is Ar:iC$_4$H$_{10}$ in 90:10 proportions.
This choice of gases provides a good balance between the performance of charge centroid and $\upmu$TPC reconstruction and yields, on average, larger clusters when compared with Ar:CO$_2$ (70:30).
The relatively slow drift velocity, as shown in figure$\,$\ref{vdrift}, boosts the precision of the $\upmu$TPC algorithms used in the reconstruction of inclined tracks$\,$\cite{Garzia:2018aip}.

\begin{figure}[htbp]
\centering
\includegraphics[keepaspectratio, width=\textwidth]{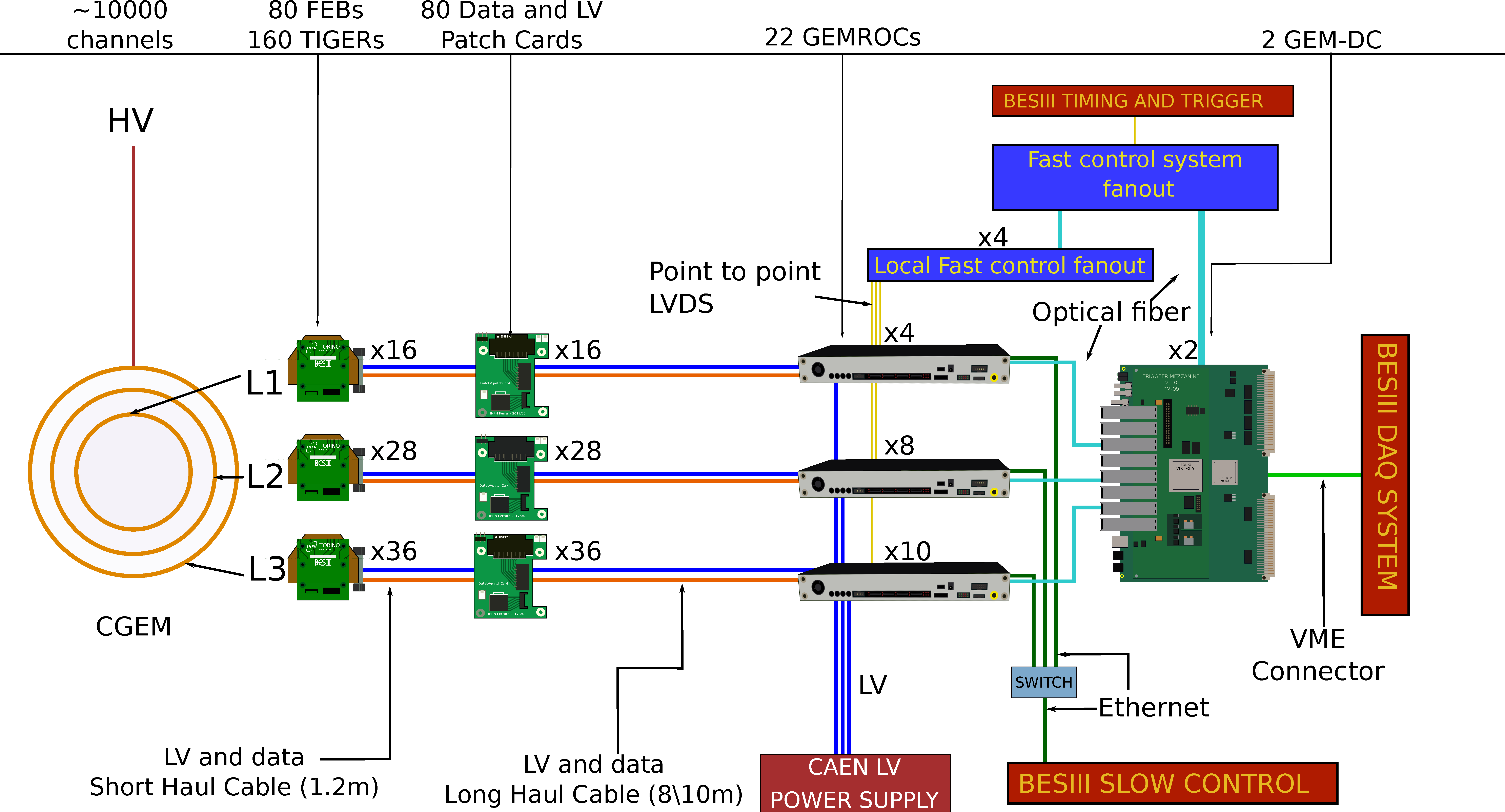}
\caption[Layout of the CGEM-IT readout chain]{Complete Layout of the CGEM-IT readout chain$\,$\cite{Amoroso:2021glb}. The schematic shows a configuration employing the minimum possible number of GEMORCs, the chosen readout scheme currently uses 22 GEMROCs overall to ensure symmetry of the two endcaps.}
\label{elec_layout}
\end{figure}

Alongside the detector, a readout chain based on the TIGER (Torino Integrated GEM Electronics for Readout) ASIC and on the GEMROC (GEM Readout Card) module has been developed. A full breakdown of the system is provided in figure$\,$\ref{elec_layout}.

\begin{figure}[htbp]
\centering
\includegraphics[keepaspectratio, width=\textwidth]{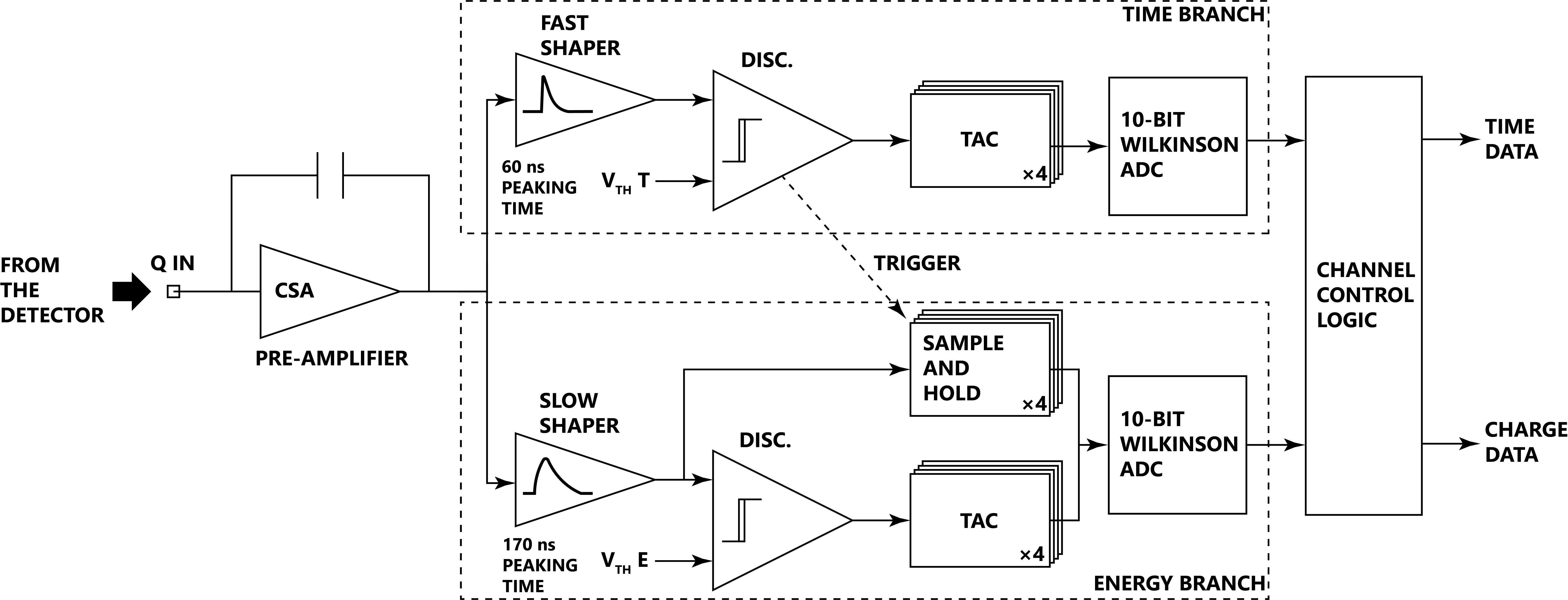}
\caption[TIGER channel architecture]{Simplified architecture of one of TIGER's channels.}
\label{charc}
\end{figure}

TIGER$\,$\cite{Cossio:2019dwg} is capable of performing simultaneous charge and time measurements on each of its 64 channels. A simplified scheme of one of these channels is provided in figure$\,$\ref{charc} while a summary of the chip's performance can be found in table$\,$\ref{tigerparams}. The signal induced on the strips is amplified by a charge sensitive amplifier (CSA), split, and fed to two branches, optimized for time (T-branch) and charge (E-branch) measurements. The T-branch has a fast shaper with a 60$\,$ns peaking time, to minimize the effects of jitter and time walk on time resolution. The E-branch's shaper has a peaking time of 170$\,$ns, more suitable for accurate charge integration.
Both branches carry hysteresis discriminators, allowing the setting of independent thresholds and trigger-less operation, and four time-to-amplitude converters (TACs) coupled with a Wilkinson ADC, which can operate time measurements at hit rates up to 100$\,$kHz. The charge can be measured in two different ways: through the sample-and-hold (S\&H) circuits present on the E-branch or through time-over-threshold (ToT) by the TDCs. S\&H is generally preferred due to its better linearity with respect to ToT. The digital part of TIGER is derived from the TOFPETv2 ASIC$\,$\cite{DIFRANCESCO2016194} and handles data transmission to the GEMROCs via low voltage differential signaling (LVDS) in 8b/10b encoding. Triple redundancy of its digital registers and manufacturing in 110$\,$nm CMOS technology allow the chip to operate in close proximity to BESIII's interaction point.

\begin{table}[htbp]
	\centering
	\begin{tabular}{ll}
	\multicolumn{2}{c}{\textbf{TIGER's Performance on Silicon}}\\\midrule
	Input dynamic range & 2-50$\,$fC\\
	Gain (E-branch) & 11.8$\,$mV$\,$fC$^{-1}$\\
	Noise (E-branch) & $<\,$1800 e$^-$ ENC (0.29$\,$fC) \\ 
	Jitter (T-branch) & $<\,$4$\,$ns \\ 
	S\&H residual non linearity & $<\,$1\% in the whole dynamic range
	\end{tabular}
	\caption[TIGER's performance]{TIGER's performance measured on silicon with $\mathrm{C_{in}\,=\,100\,pF}$ and $\mathrm{Q_{in}\,=\,10\,fC}\,$\cite{Amoroso:2021glb}.}
	\label{tigerparams}
\end{table}

\begin{figure}[htbp]
\centering
\includegraphics[keepaspectratio, width=.8\textwidth]{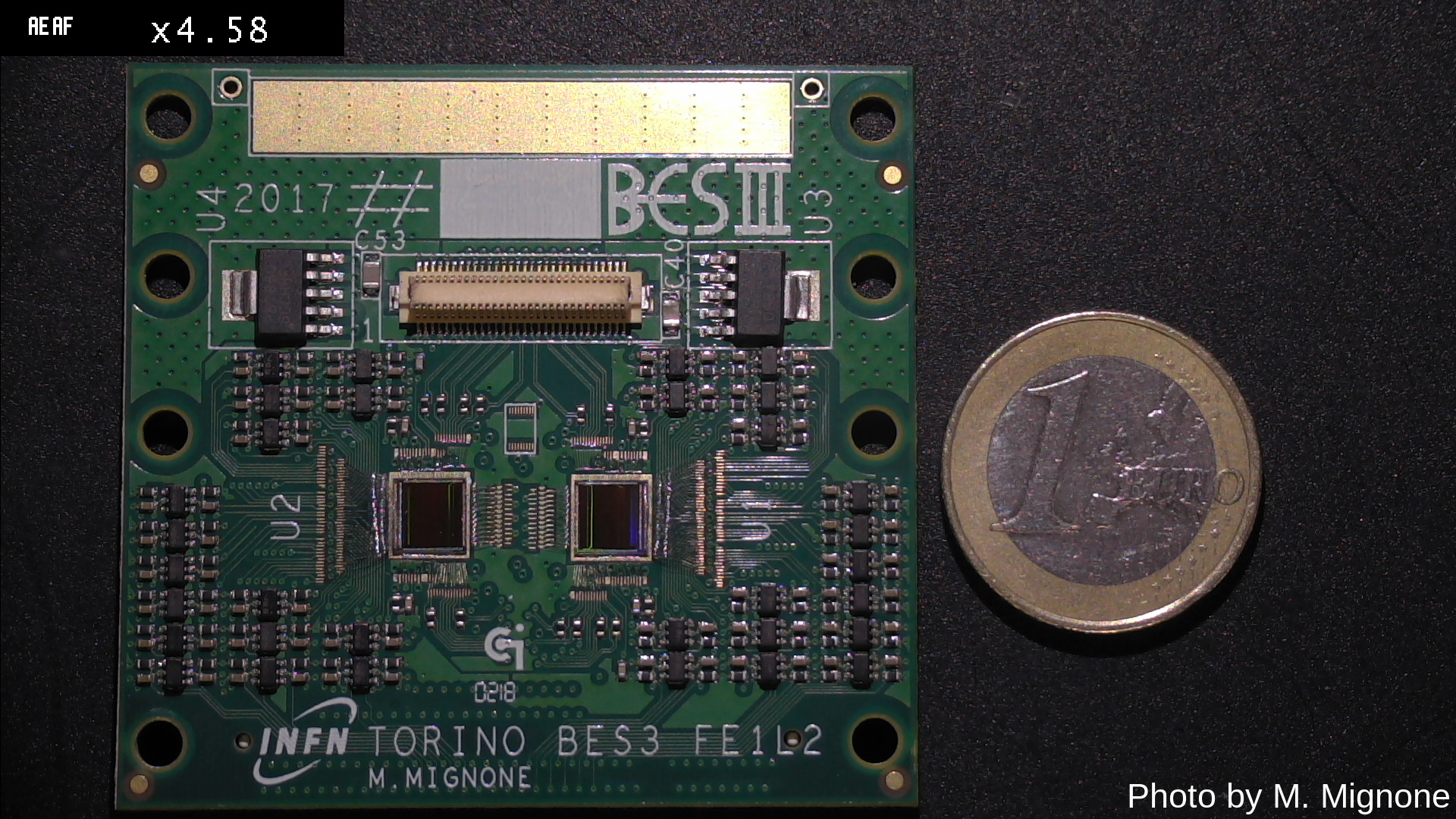}
\caption[CGEM-IT front-end board]{Photograph of one of Layer 2's front-end boards.}
\label{febphoto}
\end{figure}

TIGERs are mounted in pairs on the detector's front-end boards (FEBs), as shown in figure$\,$\ref{febphoto}. Layer 3's FEBs have a more compact design with respect to the ones of layers 1 and 2, due to the very limited clearance that separates the outer diameter of the detector and the walls of the cavity. On each FEB is mounted a copper heat sink, inside of which flows demineralized water. Active cooling is necessary to dissipate the 12.5$\,$mW of heating power generated by each TIGER channel in the confined space of BESIII's IP.

The data output from the TIGERs is handled by the GEMROCs, FPGA based back-end modules that build the events before forwarding them to the rest of the chain. GEMROCs also distribute digital and analog voltage levels, configure the TIGERs, and monitor their operating parameters during the data acquisition. Each GEMROC handles up to four FEBs and therefore up to eight TIGERs. The system's modularity ensures a high degree of versatility and scalability.

Trigger-matched data from the GEMROCs are transmitted via fiber optics links to  the GEM-DCs (GEM Data Concentrators). These FPGA based modules, borrowed from the inner tracker of the KLOE2 experiment, handle the data packets and forward them to the BESIII DAQ system$\,$\cite{gemdc}.

\subsection{Recent History of the CGEM-IT Project}
The only other existing example of cylindrical triple-GEM detector is the inner tracker of the KLOE2 experiment at DA$\Phi$NE$\,$\cite{KLOE-2:2010ana}\cite{Gauzzi:2019iks}, which concluded its data taking in 2018 and has since been decommissioned.
The design of BESIII's CGEM-IT and many of the techniques employed in its construction are derived from this experience.
The BESIII CGEM project officially began in 2013.
In early 2020 Layer 3 failed to power on despite having suffered no accident during construction.
This prompted an in depth review of Layer 3's mechanical design, aimed at understanding the issue with the detector and providing a viable solution, before the ever nearing deadline for installation in late 2024.
It is in the context of this review that the work that led to the writing of this thesis begins.

    		\chapter{Investigation on the Mechanical Issues of Layer 3}

\section{Layer 3's Electrical Failure}
\begin{figure}[htbp]
\centering
\includegraphics[keepaspectratio, width=\textwidth]{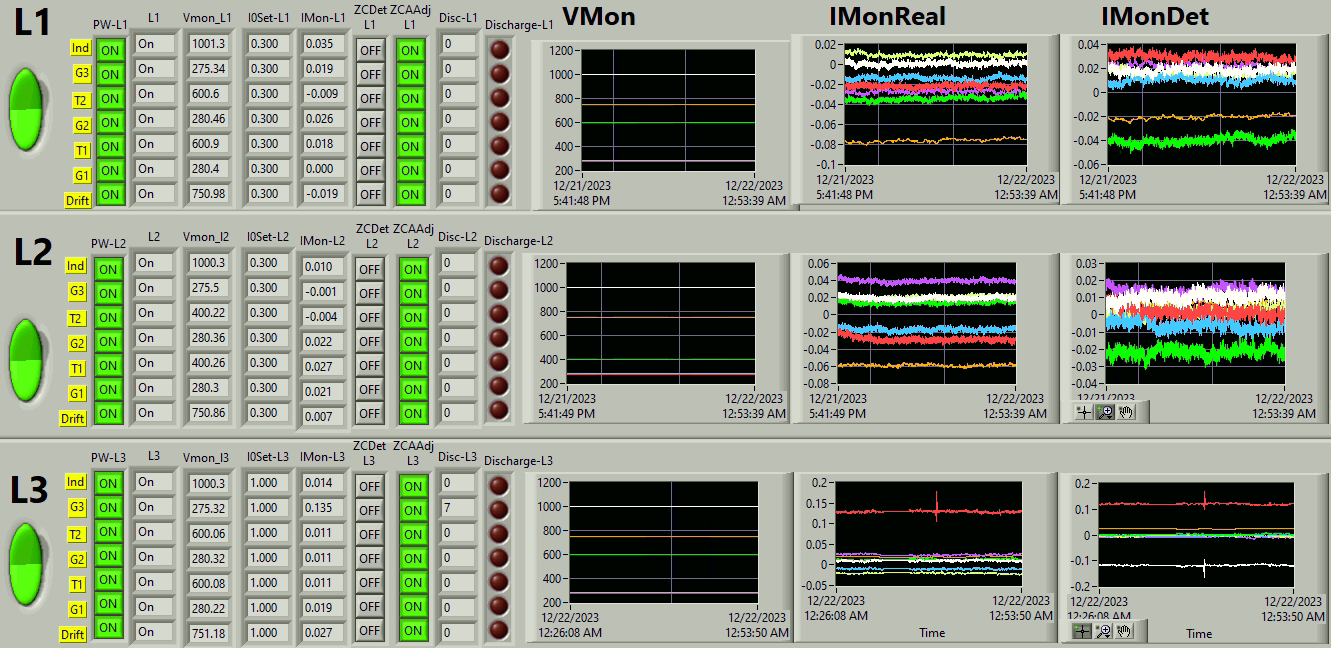}
\caption[HV control and monitoring interface]{Control and monitoring interface of the HV power supply system used for the commissioning. The IMonReal parameter indicates the factory-calibrated reading of the absolute current circulating on each channel. IMonDet instead represents the current circulating in the individual meshes formed by two adjacent electrodes.}
\label{hvmon}
\end{figure}

At the end of their construction CGEM layers undergo a series of quality control checks.
These include: a gas leakage test, measurements of resistance and capacitance for all HV sectors, electrode-by-electrode power on and validation, and finally the power on of the whole detector at both sub-nominal and nominal HV settings.
Despite suffering no accidents during its construction, Layer 3 immediately started displaying unstable electrical behavior, even when very low voltages were applied to its electrodes.
The nature of electrical issues in GEM detectors varies, with some phenomenology being indicative of physical abnormalities in the geometry or cleanliness of the electrodes' stack.
The custom interface shown in figure$\,$\ref{hvmon}, developed to control and monitor the HV distribution system during the detector's commissioning, allows to display the currents circulating in each channel of the CAEN A1515 modules\footnote{
The A1515 series of power supply modules is designed specifically for triple and quadruple-GEM detectors.
The floating HV channels are internally stacked to power all electrodes of two independent detectors.
The maximum current resolution of the module's ammeters is 100$\,$pA.
}.
When functioning correctly and in the absence of a particle beam, all channels should be pulling very little current, below 100$\,$nA, at operating HV values.
A higher current flowing between the two faces of a foil, especially when displaying ohmic behavior, is usually caused by some impurity occluding one of the holes.
This kind of issues can be fixed by electrically ``cleaning'' the GEM foil after assembly by means that will be discussed in chapter$\,$\ref{commissioning}.
Current flowing through one of the transfer gaps, that is through facing surfaces of two separate foils, may be an indication of a much more serious issue, like the deformation of an electrode, which may prevent the detector from functioning altogether.

Early in its validation, many sectors of Layer 3 presented abnormal current absorption, effectively preventing the detector from safely reaching even sub-nominal operating values (4, 2 and 1$\,$kV/cm for the induction, transfer and drift fields, respectively, during the validation).
Moreover, rotating the detector 180$^\circ$ seemed to change both the number and the position of affected sectors.
Electrical tests were performed over a period of about one month, connecting and disconnecting sectors at the patch panels, to map out the location and behavior of the pathological areas.

\section{Investigating the Internal Structure of CGEM Detectors}
\begin{figure}[htbp]
\centering
\includegraphics[keepaspectratio, width=\textwidth]{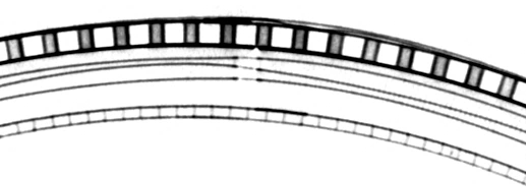}
\caption[X-ray cross-sectional picture of Layer 3]{Detail of a cross-sectional slice of Layer 3 showing extensive misplacement of the electrodes.}
\label{l3cross}
\end{figure}

The mapping of the problematic sectors was not sufficient to identify the cause of the issues affecting Layer 3.
A more direct approach was then pursued through the use of X-ray imaging.
A campaign of computed tomography (CT) scans was conducted at TEC Eurolab SRL, a company specialized in industrial testing located in Campogalliano, Italy.
The industrial CT scan machine available within the facility has a resolution of roughly 300$\,\upmu$m, allowing to collect very clear pictures of the internal configuration of the detector, an example of which can be observed in figure$\,$\ref{l3cross}.
From the raw pictures it was immediately clear that the electrodes were not cylindrical and that their deformation was not local but extended to the whole detector.

\begin{figure}[htbp]
\centering
\includegraphics[keepaspectratio, width=\textwidth]{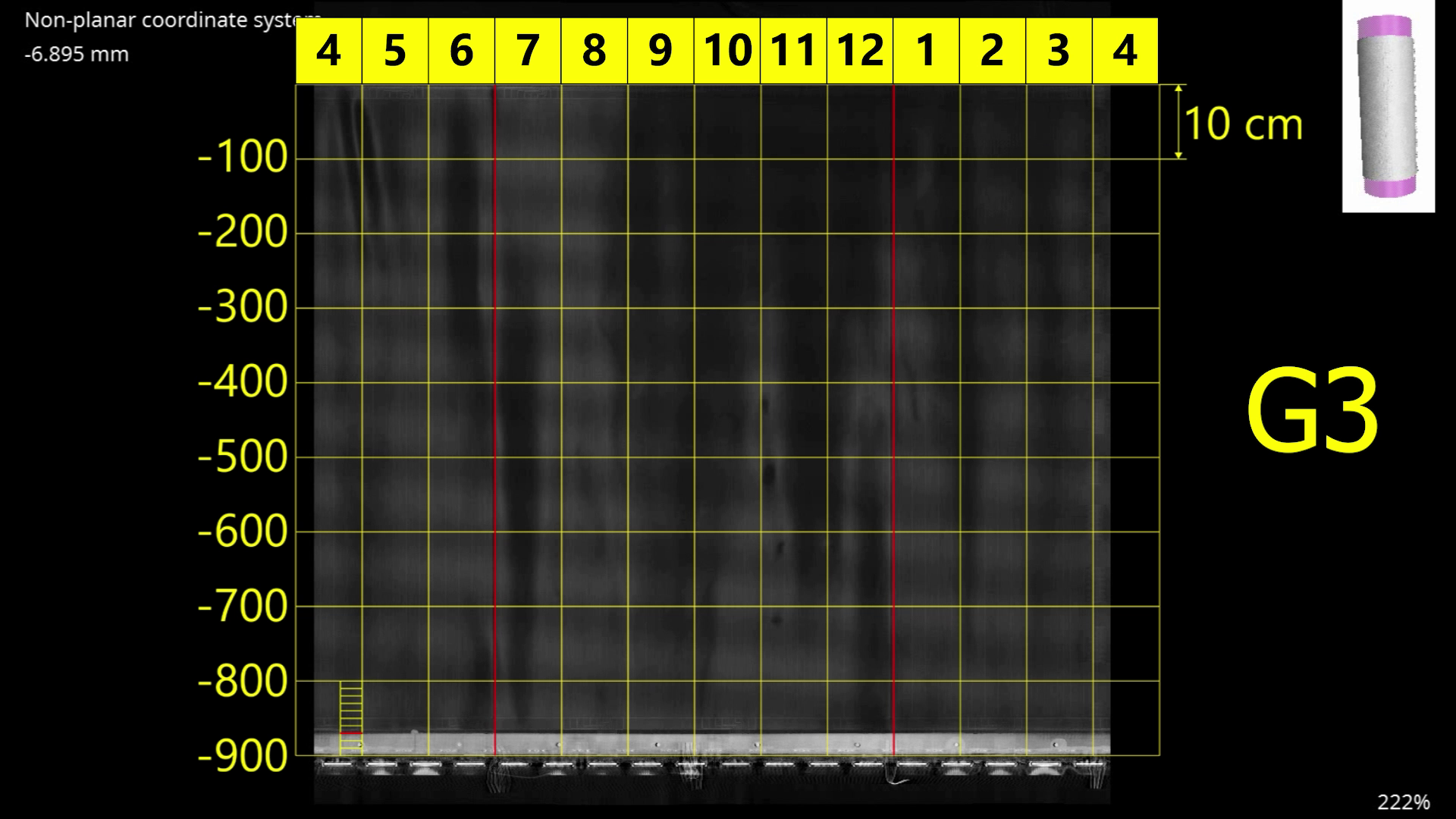}
\caption[Digital "unrolling" of layer 3 GEM 3]{X-ray image of Layer 3's GEM 3. The slices are stitched together and digitally elaborated to "unroll" the foil. A grid is superimposed to locate defects within macro-sectors. The numbers at the top correspond to the macro-sector number, while the two red lines indicate the overlaps between the two foils.}
\label{gem123}
\end{figure}

\begin{figure}[htbp]
\centering
\includegraphics[keepaspectratio, width=\textwidth]{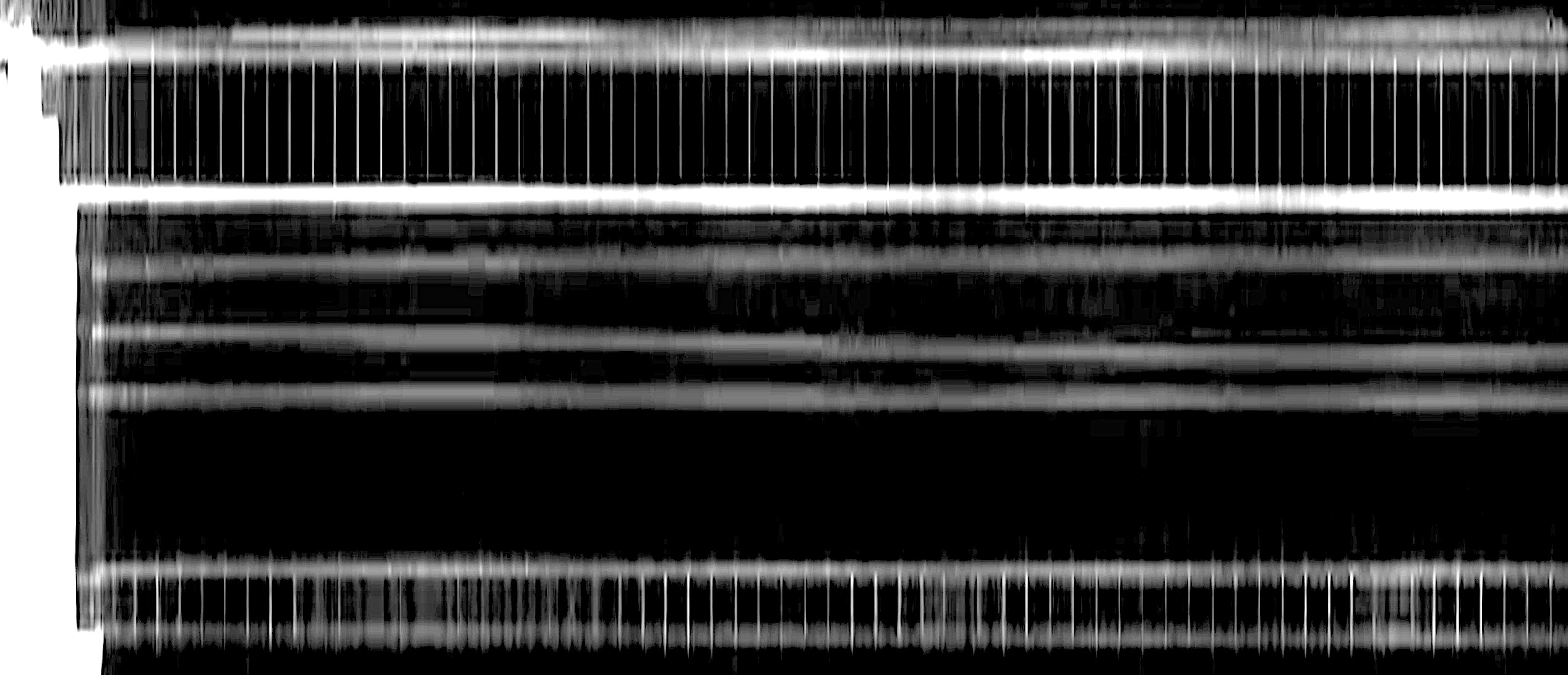}
\caption[Cross-section of layer 3's internal structure]{Lateral cross-section of Layer 3's internal structure. The image is not in 1:1 scale to accentuate the defects.}
\label{sideview}
\end{figure}

From the 3D reconstruction it is possible to obtain images such as the one shown in figure$\,$\ref{gem123}, where the GEM foils are digitally "unrolled" to precisely locate defects.
While some striping, due to imperfect alignment between the real axis of the detector and the one used for unrolling, is to be expected, some of the more evident inhomogeneities are indication of structural collapse of the electrodes and matched well with the mapped malfunctioning sectors.
Other smaller bumps and creases are also highlighted in such view.
Other views, like the one in figure$\,$\ref{sideview}, allow to roughly measure the magnitude of the deformation by superimposing a proportional grid.
The position of the anode and the cathode, together with their nominal distance of 11$\,$mm can be used as reference.

\begin{figure}[htbp]
\centering
\includegraphics[keepaspectratio, width=\textwidth]{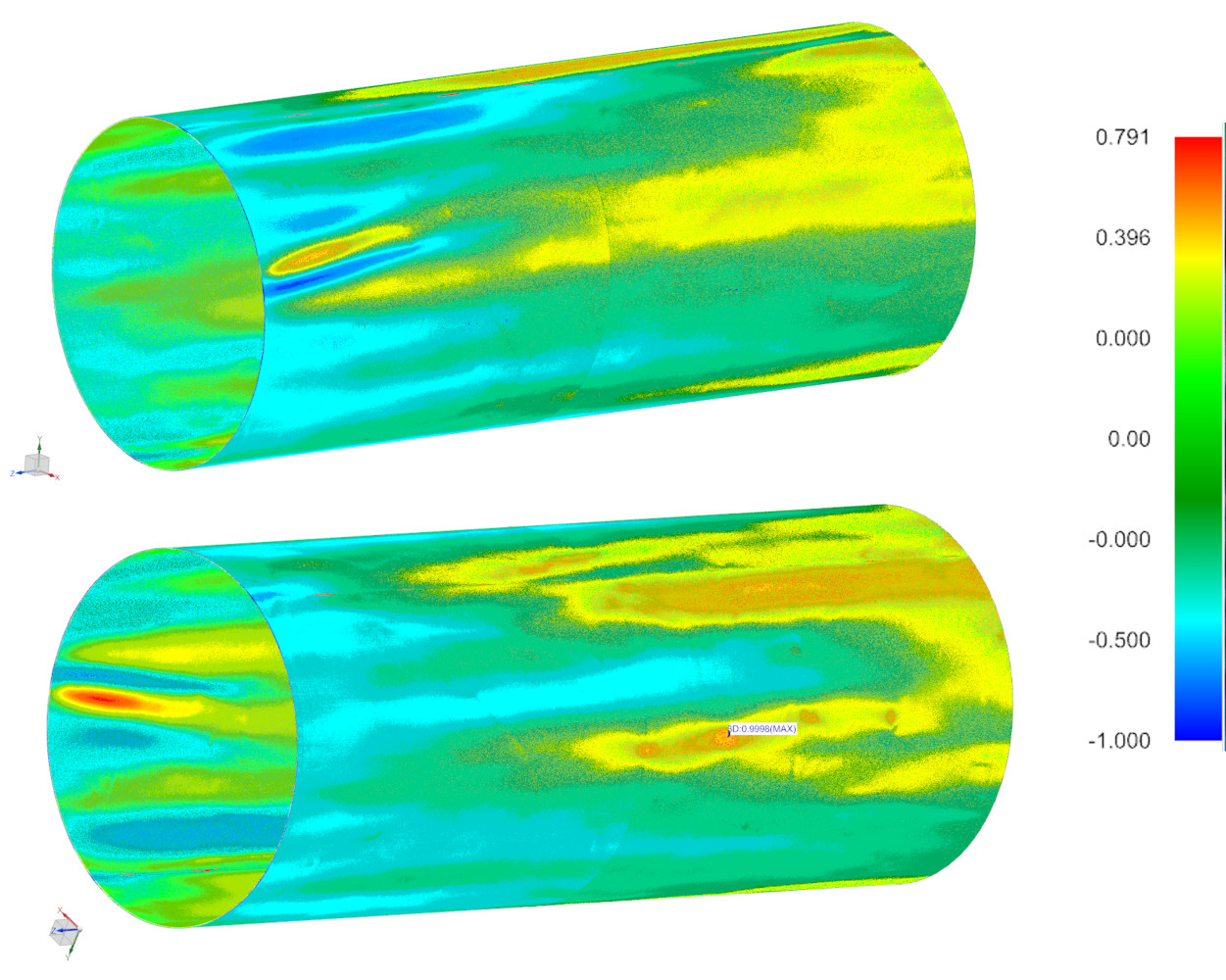}
\caption[Reverse engineering of GEM 3]{Deviation of the CT scan reconstruction of GEM 3 with respect to a nominal cylindrical surface.
The alternating pattern of ridges and valleys is compatible with buckling lobes. The unit of the color grading scale is mm.}
\label{reveng}
\end{figure}

Further elaboration of the images allows to extract a 3D model of each GEM foil, which can then be compared with a nominal cylindrical surface to measure the deviation.
This was done using the reverse engineering module of the Siemens NX suite, to obtain images as the one shown in figure$\,$\ref{reveng}.
The striping already observed in the unrolled images is now clearly visualized as an alternating pattern of peaks and valleys.
This kind of deformation is typical of a phenomenon called buckling, to which thin walled cylindrical shells, with unfavorable ratios of radius to overall length, are particularly prone.
Buckling occurs suddenly at the reaching of a critical load, which depends on the geometry and material of the structure under stress.
FEM simulations of the CGEM-IT layers were carried out, setting provisional values for their critical load: 8.7$\,$g for Layer 1, 4.0$\,$g for Layer 2, and 2.5$\,$g for Layer 3, where g is the standard acceleration of gravity.

FEM simulations of inhomogeneous materials are affected by large errors and require the support of experimental measurements.
Layer 3 suffered no accidents during construction and later tests showed that an acceleration of 2.5$\,$g is hardly reached when handling the detector normally.
This hints at the actual critical load for Layer 3 being substantially lower, if not low enough for its GEMs to collapse under their own weight.

\section{Containing Buckling-induced Deformation in CGEM Detectors}
Since there is no evidence in literature of this issue affecting the only other existing example of CGEM detector, the investigation was diverted towards the KLOE2-IT, and the design differences between the two.
Its outermost layers have PEEK spacer grids both in the transfer gaps, between the GEM foils, and in the drift gap, separating GEM 1 from the cathode.
These spacers were introduced out of concerns of thermal expansion, which is capable of producing axial loads that could similarly trigger buckling.
Due to the complexity involved in the assembly of such grids, and the fact that BESIII's beam pipe is actively cooled to maintain it close to ambient temperature, the CGEM-IT collaboration decided not to adopt spacers early on in the design of the detector.

\begin{figure}[htbp]
\centering
\includegraphics[keepaspectratio, width=\textwidth]{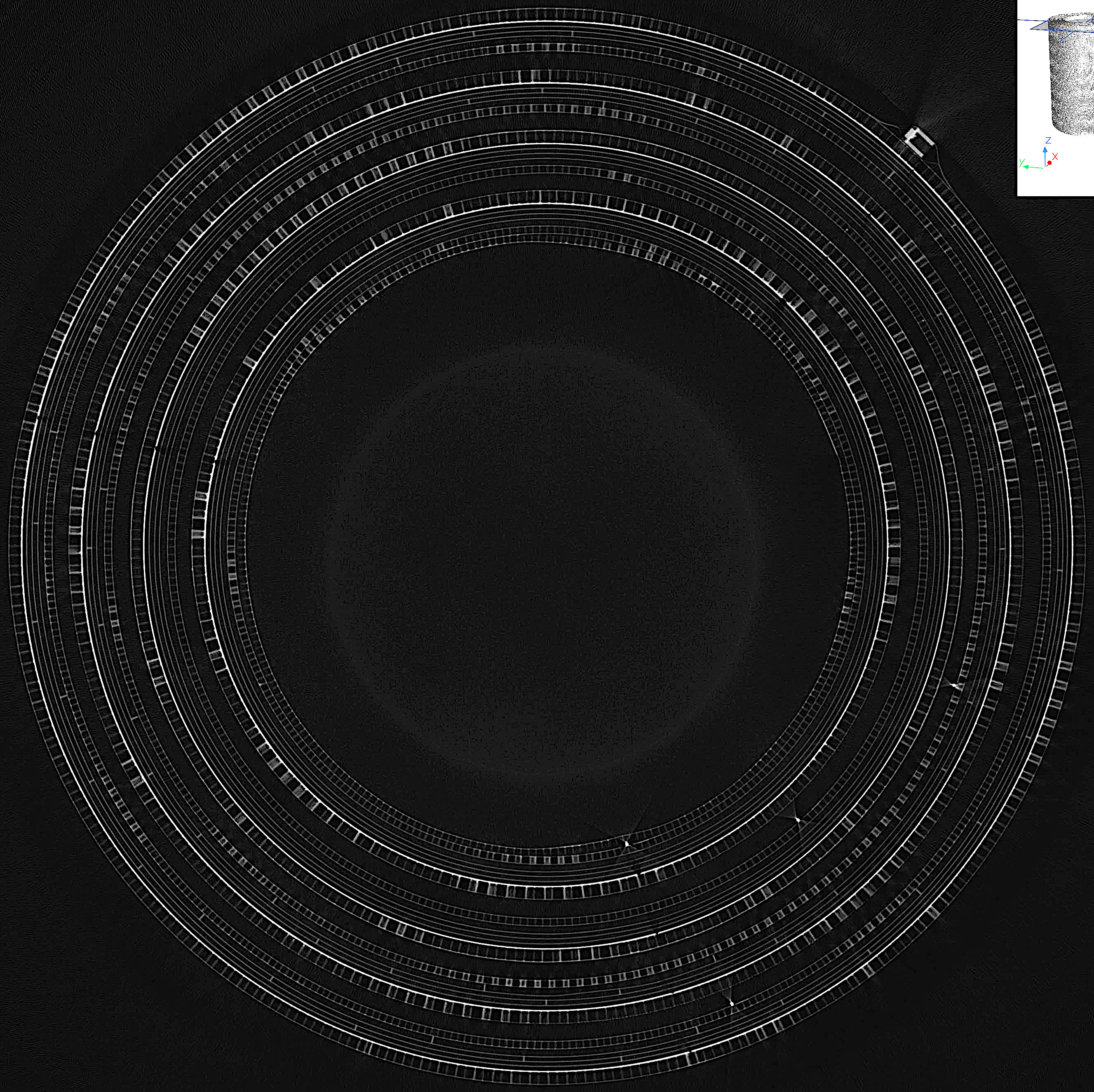}
\caption[Trasversal cross section of the KLOE2-IT]{Cross section of the KLOE2-IT. The gaps separating the electrodes appear evenly spaced in all four layers. The axial rods of the PEEK spacer grids can be observed in the two outermost layers.}
\label{kloe}
\end{figure}

The KLOE2-IT was subjected to a CT scan similar to the one performed on Layer 3, to verify the absence of buckling-induced deformation of its electrodes.
The pictures collected showed some localized defects, traceable back to accidents happened during assembly, but no signs of buckling, as shown by figure$\,$\ref{kloe}.

The grids constrain the spacing between the electrodes, not preventing buckling entirely but containing the magnitude of the large buckling lobes, to keep the gaps within tolerance.
The choice of PEEK for the fabrication of the grids was due to the polymer's high radiation resistance, good mechanical stability, and low outgassing$\,$\cite{Capeans:2002ki}.

\FloatBarrier

\section{Testing the Spacer Grid Solution}
Since FEM simulations on inhomogeneous materials and non-standard structures, as the ones represented by GEM foils, can be affected by relative errors larger than 20\%, an experimental test was deemed necessary to confirm the validity of the spacer grid solution and to validate the simulations.
The purpose of the test was dual: on the one hand, to raise the critical load and ensure safe transport and handling of Layer 3; on the other hand, to form a new construction team and have it practice the techniques necessary to build and assemble CGEM detectors.

The testing method would proceed according to the following steps:
\begin{itemize}
\item The realization of a representative mock-up of Layer 3.
\item The procurement, characterization, and positioning of accelerometric sensors on the mock-up for measuring load during test falls.
\item The arrangement of a setup for having the mock-up take repeatable, dampened falls at predictable peak acceleration values.
\item The collection of CT scan data after a number of such falls, to assess the internal condition of the mock-up.
\end{itemize}

\begin{figure}[htbp]
	\centering
	\begin{subfigure}{.49\textwidth}
		\centering
		\includegraphics[keepaspectratio, width=\linewidth]{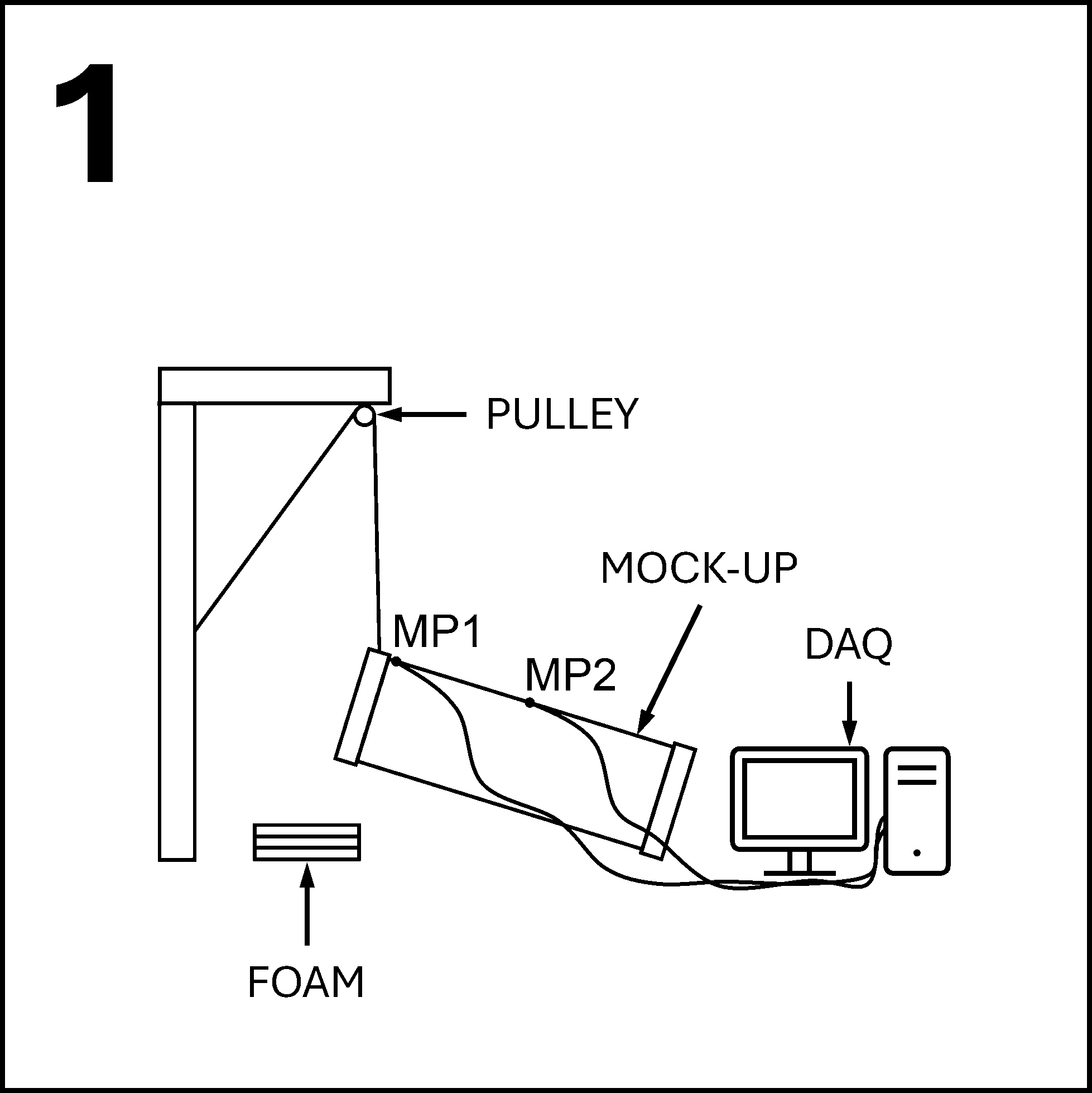}
	\end{subfigure}
\hfill
	\begin{subfigure}{.49\textwidth}
		\centering
		\includegraphics[keepaspectratio, width=\linewidth]{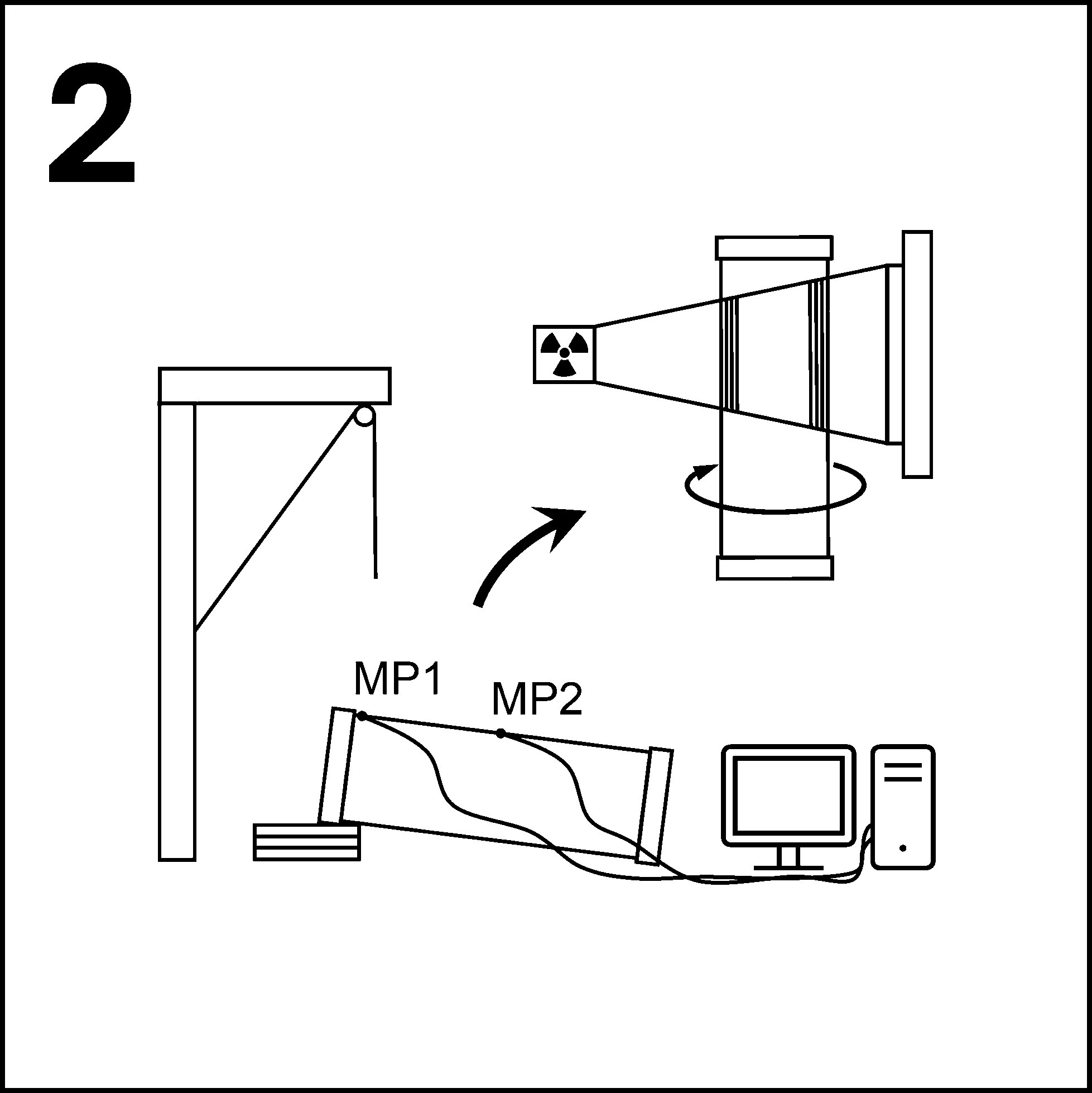}
	\end{subfigure}
	\caption[Schematic representation of the test]{Schematic representation of the test. MP1 and MP2 represent the location of the accelerometric sensors installed on the mock-up.}
	\label{drop_scheme}
\end{figure}

The test, schematically represented in figure$\,$\ref{drop_scheme}, was designed to subject a representative mock-up of Layer 3 to increasing inertial loads.
The load could be tuned by varying the height of the fall and the number of shock absorbers at the point of impact.
These absorbers also increase the duration of the impact events, making them more resembling of what the detector would experience during handling or wheeled transport with respect to a hard fall.
The mock-up was dropped while horizontal, since the detector is always maintained horizontal after leaving the assembly machine, and on a single side, to ensure the repeatability of the falls and simplify the test setup.

\subsection{Design of the Mock-up}
For the test to be significative, the mock-up had to be representative of the structural and inertial behavior of Layer 3. 
While a stronger structure would have invalidated the test, a slightly weaker build ensured some margin of safety over the resulting critical load value.
Overall cost, procurement time of the materials, and manpower were all limiting factors in the design and realization of the mock-up.

The precisely machined fiberglass rings were substituted with 3D print\-ed analogs in ASA (Acrylonitrile Styrene Acrylate).
The plastic rings do not match the elastic behavior of the composite ones but they allowed to avoid months of procurement time and the high costs of machining.
A GEM 2, built out of spare GEM foils left from a previous construction, was the representative electrode under test.
The total number of floating electrodes in the stack was reduced from 3 to 2, with the other floating electrode, an analog of GEM 1, being fashioned out of copper clad Kapton foils without holes.
The glass fiber reinforced anodic circuit was also replaced with copper-clad Kapton.
The structural sandwiches in honeycomb and carbon fiber were kept unchanged with respect to Layer 3's ones, if not for a slight reduction in the thickness of the Kapton skins.
A full stratigraphy of the Layer 3's mock-up is provided in table$\,$\ref{strat}.

\begin{table}[htbp]
\centering
\begin{tabular}{lrl}
\multicolumn{3}{c}{\textbf{Layer 3 Mock-up Stratigraphy}}                \\ \midrule
\textbf{Material}                 & \textbf{Thickness ($\upmu$m)} & \textbf{Mandrel}     \\ \midrule
Laminated   carbon fiber & 60                   & \multirow{4}{*}{GEM 3} \\
Nomex Honeycomb          & 4000                 &                        \\
Kapton                   & 125                  &                        \\
Copper clad Kapton       & 50+5                 &                        \\ \midrule
Gap                      & 2000                 & \multirow{2}{*}{GEM 2} \\
GEM 2                    & 5+50+5               &                        \\ \midrule
Clearance                & 200                  & \multirow{3}{*}{GEM 1} \\
PEEK grid                & 1800                 &                        \\
Copper clad Kapton       & 50+5                 &                        \\ \midrule
Clearance                & 200                  & \multirow{7}{*}{Cathode}   \\
PEEK grid                & 4800                 &                        \\
Copper clad Kapton       & 50+5                 &                        \\
Kapton                   & 50                   &                        \\
Nomex Honeycomb          & 2000                 &                        \\
Laminated carbon fiber   & 60                   &                        \\
Kapton                   & 25                   &                       
\\
\end{tabular}
\caption[Layer 3 mock-up stratigraphy]{Stratigraphy of the mock-up of Layer 3 used in the drop test.}
\label{strat}
\end{table}

The design of the grids is derivative of those adopted for the KLOE-2 IT, with rods running down the length of the detector and evenly spaced rings.
The rings have openings, which are aligned so that the grid can be forced open and positioned onto the electrode once it has been rolled on its mandrel.
The thickness of the PEEK films used for the production of the grids is 0.3$\,$mm.
Small 0.5$\,$mm wide slits, machined into both rods and rings, form joints that fix the position of the ones with respect to the others.
The number of rods was set to 12, to match the number of macro-sectors, while the number of rings was set to 10, which is a good trade off between grid size and assembly complexity.

\subsection{Mock-up and Grid Construction}
A complete description of the fabrication of CGEM electrodes and their support structures falls outside the scope of this paragraph and of this thesis, but exhaustive information on the subject can be found in the author's master's degree thesis$\,$\cite{Gramigna:2022ozf} or, more concisely, in reference$\,$\cite{Balossino:2022ywn}.

While the procedures adopted in the realization of the mock-up were the same used for the real detector, many extra steps were rendered necessary to accomodate for the different materials.
All electrode analogs, aside for GEM 2, had to be manually cut to shape. Positioning holes and slots were also die-cut according to hand-drawn references.
The 3D printed rings had to undergo several adjustments as well: from reaming the reference holes to reducing their thickness by filing and sanding them.  
The construction of the mock-up did not take place inside a cleanroom, but in dedicated workspaces within the workshop of INFN Ferrara and Frascati National Laboratories (LNF).
Particular care had to be put into cleaning mandrels and electrodes, to avoid trapped debris to cause bumps on their surfaces during the vacuum-assisted gluings.
All pre-existing defects on the foils were mapped and noted down, so that they could be ignored when analyzing the CT scan images. 

\begin{figure}[htbp]
\centering
\includegraphics[keepaspectratio, width=.85\textwidth]{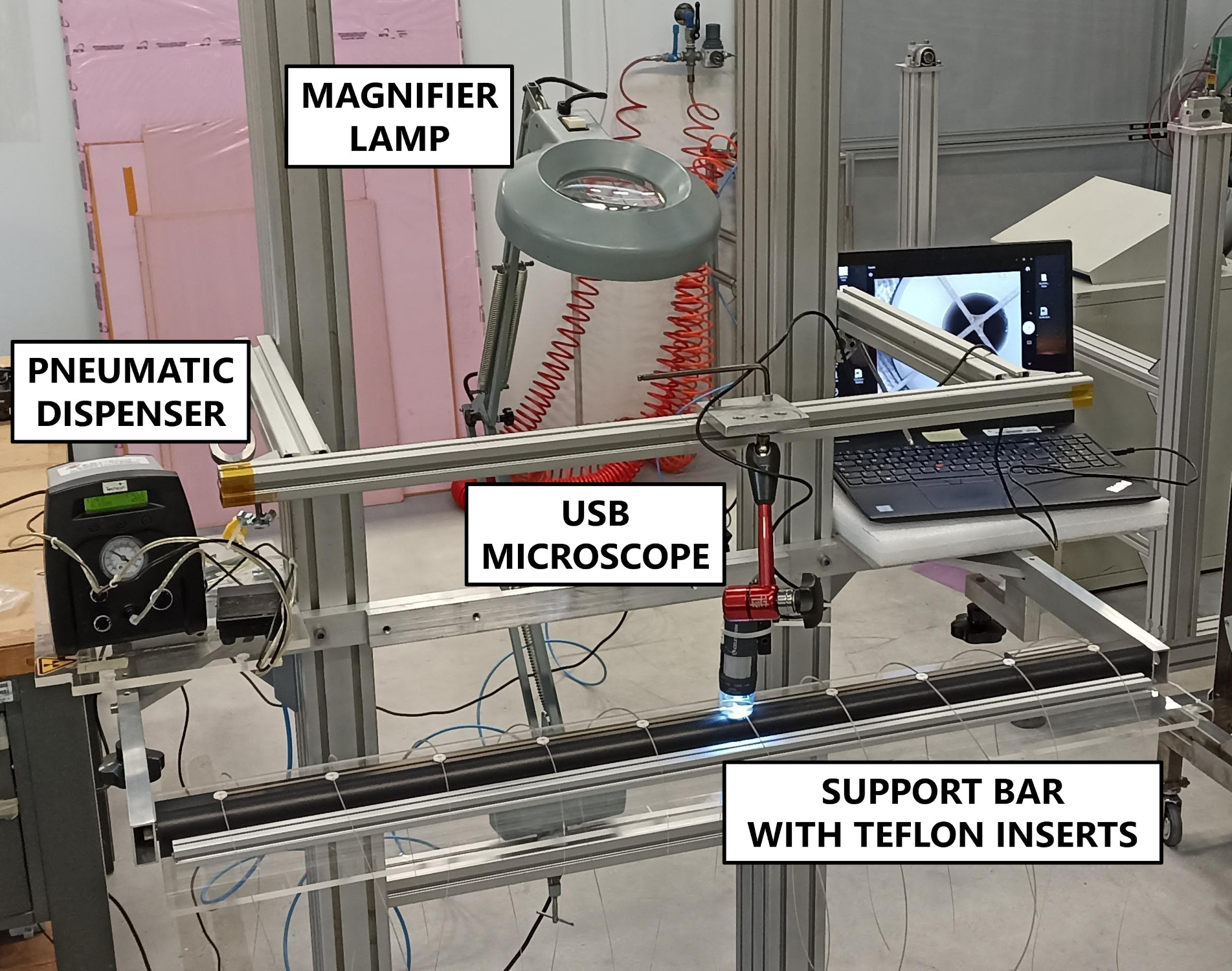}
\caption[Grid assembly machine]{Machine used for the assembly of the PEEK grids for the mock-up of Layer 3.}
\label{gridmachine}
\end{figure}

\begin{figure}[htbp]
\centering
\includegraphics[keepaspectratio, width=.5\textwidth]{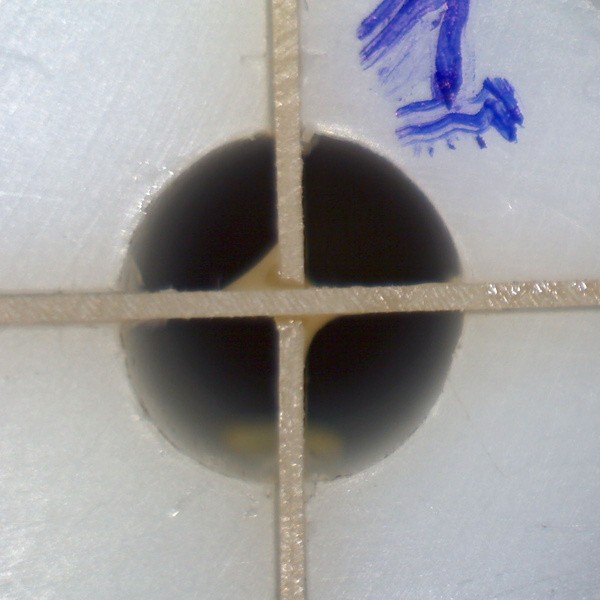}
\caption[Glued grid joint]{Glued grid joint photographed through the microscope.}
\label{joint}
\end{figure}

Rings and rods of all grids were machined at CERN EST-DEM workshop from PEEK APTIV\textsuperscript{\textregistered} 1000 film.
The tooling for the assembly of the grids, shown in figure$\,$\ref{gridmachine}, was again inspired by the one devised by the KLOE2 collaboration.
An anodized aluminum bar and Teflon inserts with cross shaped grooves hold a single rod and the ten rings in position.
A pneumatic fluid dispenser helps control the amount of glue deposited in each corner to fix the intersections.
Huntsman Araldite\textsuperscript{\textregistered} 2012 was chosen to glue the crossings, due to its faster curing time when compared with similar two-component resins.
Figure$\,$\ref{joint} shows a glued crossing, seen trough the USB microscope used for the quality control.
Fully assembled grids are glued to the electrodes at the first and last ring, which remain outside of the active area.
%

The machine used to assemble the mock-up was the same used to assemble all CGEM layers, temporarily commissioned for the occasion at LNF.
The Vertical Insertion Machine (VIM) is a custom CNC machine, which was originally built for the assembly of the KLOE2-IT and later refurbished for the construction of the CGEM-IT.
The VIM will be described thoroughly in chapter$\,$\ref{laser}, since its upgrade played an important role in making the split construction of Layer 3 possible.
During the assembly of the mock-up, defects present on the electrodes, either pre-existing or arisen during the construction process, were once again mapped and noted down to inform the analysis of the CT scan images.
Poor alignment of the VIM and the lack of dimensional fidelity of the 3D printed rings, leading to almost no clearance during the insertion of the electrodes, greatly complicated the mock-up assembly.
From its removal from the assembly machine to the end of the drop test the mock-up was kept under constant acceleration monitoring by means of a custom accelerometric measurement system developed for the test.

\subsection{Characterization of the Accelerometers and Drop Test Setup Development}
The choice of the accelerometers to be used in the test was mainly driven by: amplitude and frequency range, overall cost, and ease of implementation in a custom-made data acquisition system.
The most common accelerometric sensors on the market are either analog or digital.
Typical features and limits of both categories are summarized in table$\,$\ref{acc}

\begin{table}[htbp]
\begin{tabular}{lll}
\multicolumn{3}{c}{\textbf{Pros and Cons of Common Accelerometer Types}}\\\midrule
\multirow{2}{*}[-1cm]{Analog} & Piezoresistive       & \begin{tabular}[c]{@{}l@{}} \tabitem Widest amplitude and frequency range\\ \tabitem Low sensitivity\\ \tabitem Requires complex readout circuit\\ \tabitem Can measure stationary acceleration\\ \tabitem Continuous response\end{tabular}                                                                                                                                     \\ \cmidrule(l){2-3} 
                        & Piezoelectric  & \begin{tabular}[c]{@{}l@{}} \tabitem Narrow amplitude and frequency range\\ \tabitem Highest sensitivity \\ \tabitem Easier readout (especially IEPE variants)\\ \tabitem Cannot measure stationary acceleration \\ \tabitem Continuous response\end{tabular}                                                                                                                     \\ \midrule
Digital                 & MEMS                 & \begin{tabular}[c]{@{}l@{}} \tabitem Narrow amplitude range\\ \tabitem Bandwith limited by ODR sampling\\ \tabitem Lowest sensitivity\\ \tabitem Easiest readout (no DAQ HW required)\\ \tabitem Can measure stationary acceleration\\ \tabitem Discrete response\\ \tabitem Cheap \\ \tabitem lightweight\end{tabular}\\
\end{tabular}
\caption[Features of commercially available accelerometric sensors]{Features and limits of the most common types of accelerometric sensors available on the market.}
\label{acc}
\end{table}
%
%

Two types of accelerometers were purchased for the test: TE Connectivity Model 8201 analog IEPE (internal electronic piezoelectric) and STMicroelectronics IIS2DH digital MEMS (micro electro mechanical system).
The model 8201 has a range of $\pm$25$\,$g, good linearity between 6 and 6000$\,$Hz, and comes factory calibrated.
The IIS2DH sensors have a settable amplitude range up to $\pm$16$\,$g and output data rate up to 5.3$\,$kHz.
The choice of the sensors was informed by the many and often competing needs of the test:
\begin{itemize}
\item Capability to accurately measure up to 15$\,$g at frequencies ranging from a few hertzs to a few thousands.
\item Reduced form factor and low mass sensors for direct application on the GEM foils.
\item Possibility to install via adhesive or double-sided tape.
\item Avoiding the purchase of expensive SCADA (Supervisory Control and Data Acquisition) or similar proprietary DAQ systems.
\end{itemize}

The analog sensors were read by a digital oscilloscope with the capability of saving high resolution waveforms up to a few seconds in duration.
The digital sensors were installed on custom lightweight shuttle boards manufactured by the electronics division of INFN Ferrara.
The simultaneous acquisition of data from up to 4 sensors was handled by a Raspberry Pi, equipped with an internally designed custom shield.
Thin enamelled wires connected the shuttle boards to the shield, so to minimize the weight contribution of the cables.

\begin{figure}[htbp]
\centering
\includegraphics[keepaspectratio, width=\textwidth]{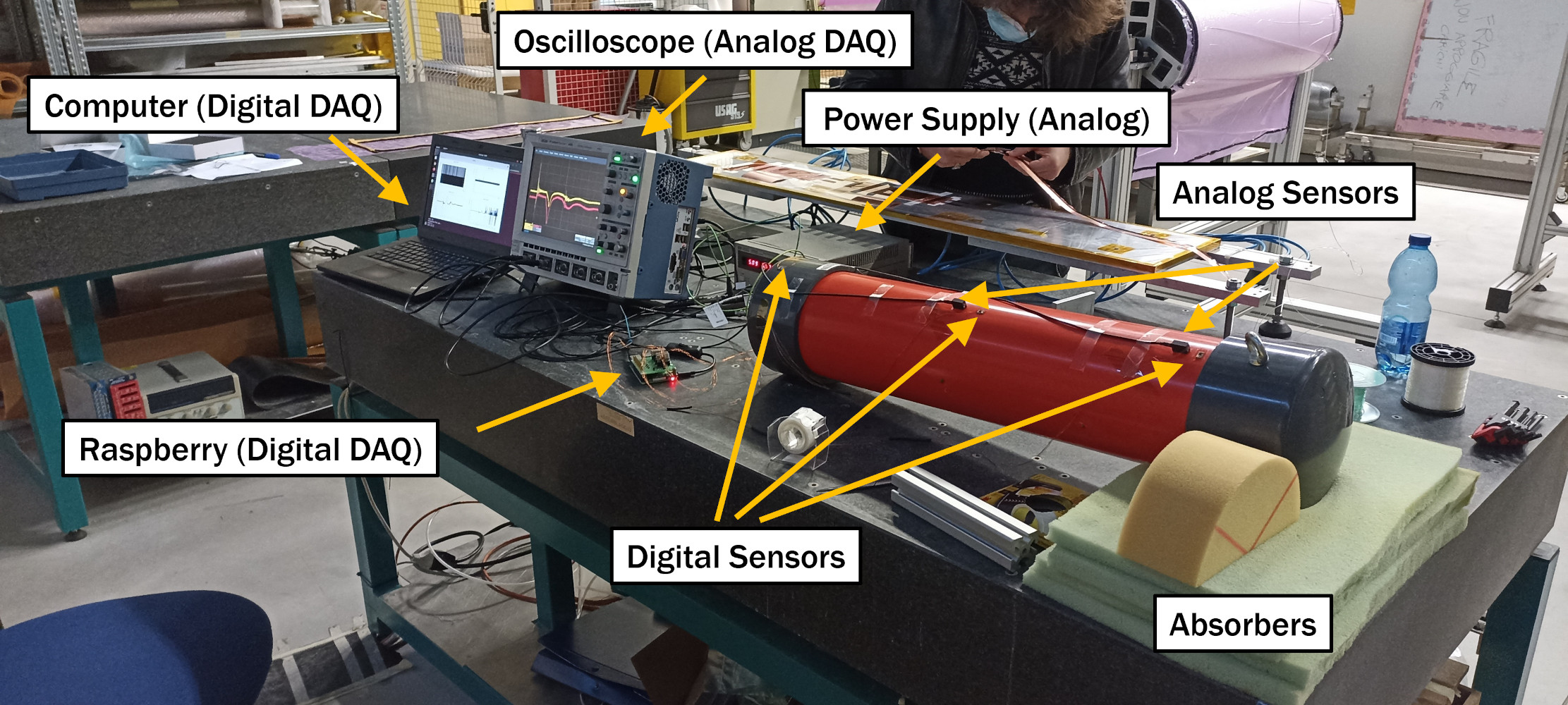}
\caption[Accelerometers' DAQ test setup]{Simple setup for the preliminary characterization of the accelerometers and the test of their DAQ system.}
\label{pipe}
\end{figure}

The simple setup in figure$\,$\ref{pipe} was prepared to test the acquisition system and begin the characterization of the accelerometers.
In this early phase, a section of plastic pipe simulated the detector.
The absence of internal structures produces simple fall spectra, allowing to focus solely on the sensors' response.
Accelerometers of both varieties were positioned in the middle of the pipe and near the point of impact.

\begin{figure}[htbp]
\centering
\begin{subfigure}{.49\textwidth}
\centering
\includegraphics[keepaspectratio, width=\linewidth]{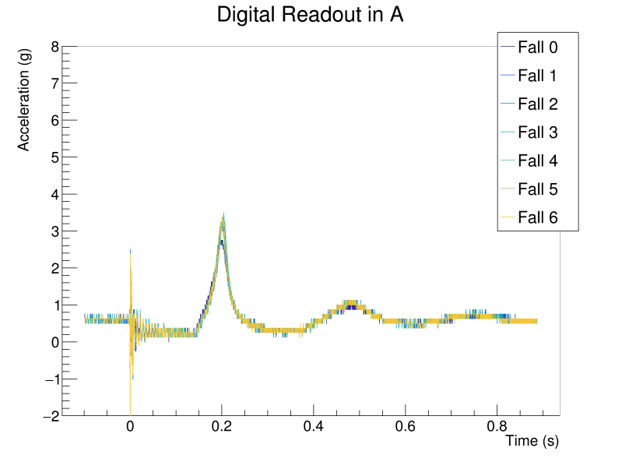}
\caption{Readings from the digital sensor positioned in the middle of the pipe.}
\label{pipe_rep_dig}
\end{subfigure}
\hfill
\begin{subfigure}{.49\textwidth}
\centering
\includegraphics[keepaspectratio, width=\linewidth]{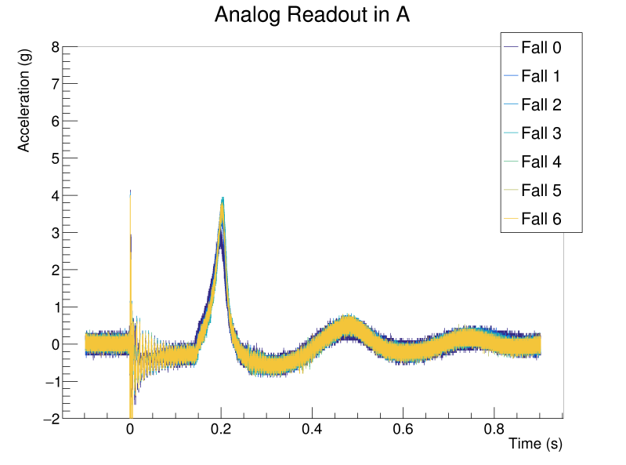}
\caption{Readings from the analog sensor positioned in the middle of the pipe.}
\label{}
\end{subfigure}
\caption[Repeatability test of dampened falls]{Repeatability test conducted by having the pipe fall several times from the same height on a pile of absorbing foam.}
\label{pipe_rep}
\end{figure}

A few examples of the spectra generated in the test falls can be found in figure$\,$\ref{pipe_rep}.
The release of the nylon wire holding the pipe suspended produces a quick acceleration spike.
The echo of the release can still be observed in the form of a high frequency oscillation until the pipe contacts the absorbers.
The first impact with the pile of absorbing foam generates a wide impact peak, where most of the energy accrued in the fall is dissipated and experienced as load by the falling object.
The impact peak is followed by low frequency and gradually dampening bumps, due to the object bouncing on the absorbers.
The analog sensors measure 0$\,$g at rest, since they cannot measure the stationary gravitational load, while the digital ones registered a value different from 1$\,$g due to a calibration error.
A scaling factor was later applied so to fix this issue but the correction still did not fully resolve the discrepancy between the digital and the analog sensors, with the latter always registering slightly larger impacts.
The more conservative measurements provided by the analog sensors were chosen to determine the load limit for Layer 3.
Digital sensors were still employed as back-up in case of failure of the analog DAQ during the final test, when falls could no longer be repeated at leisure.

\begin{figure}[htbp]
\centering
\begin{subfigure}[t]{.49\textwidth}
\centering
\includegraphics[keepaspectratio, width=\linewidth]{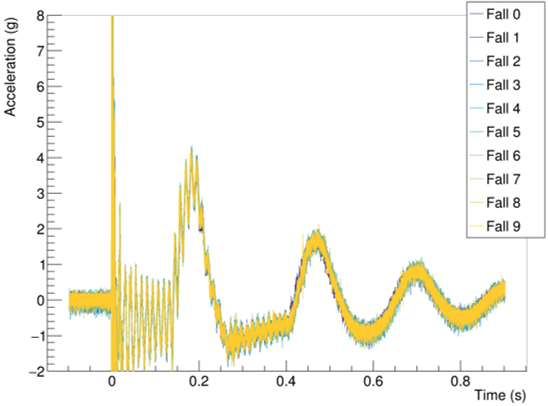}
\caption{Repeatability study performed by having the assembly take several falls at the same height.}
\label{gem_rep_ana}
\end{subfigure}
\hfill
\begin{subfigure}[t]{.49\textwidth}
\centering
\includegraphics[keepaspectratio, width=\linewidth]{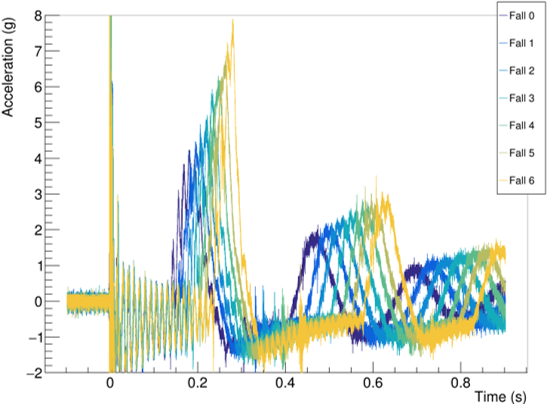}
\caption{Waveforms obtained in falls from different heights. The height of the fall was defined as the distance between the topmost absorber and the closest point of impact on the assembly. This parameter was varied between 0 and 18$\,$cm, in 3$\,$cm steps.}
\label{gem_prog_ana}
\end{subfigure}
\caption[Fall studies conducted with a CGEM-like object ]{Preliminary studies conducted on a CGEM-like assembly made out of an anode and a GEM 3 discarded in the previous construction of Layer 3.}
\label{gem_studies}
\end{figure}

The two plots of figure$\,$\ref{pipe_rep} show that this preliminary setup already granted a very good degree of repeatability to the test falls.
The plastic pipe was later substituted with an anode and GEM 3 assembly, discarded in the previous construction of layer 3.
This assembly is much more representative of the detector's elastic and inertial behavior, and provides an exposed GEM electrode to relate internal and external stresses.
Two of the light digital accelerometers were applied directly to the floating GEM, in the middle of the foil and near the point of impact, while the outer composite structure was instrumented in the same way the plastic pipe was.
The suspension structure was later upgraded to provide evenly spaced positions for the quick release mechanism, so to study the amplitude of falls from different heights.
Figure$\,$\ref{gem_rep_ana} shows the measurement's repeatability holding true despite the change of falling object, while figure$\,$\ref{gem_prog_ana} hints at an almost linear dependency between the force of the impact and the height of the falls.
The plots show a secondary high-frequency oscillation superimposed to the main one previously described. This is likely due to the presence of a resonating internal structure, since it appeared when switching from the pipe to the anode and GEM 3 assembly.
Measurements collected by the sensors positioned on the GEM and on the outer shell near the impact point show little difference.
In the middle, the acceleration recorded on the GEM was always slightly larger than the one measured on the outer shell.
The small difference of about 0.2$\,$g could either be due to the weight of the accelerometer altering the inertial behavior of the foil or to a small amplifying effect due to the material's elasticity.
Either way, the effect is too small to be relevant for the purposes of the test, and the more conservative measurements collected on the outer shell were adopted in the determination of the critical load.

Once familiarized with the analysis of fall spectra and the behavior of CGEM-like objects in dampened falls, the purpose of the studies shifted to building a model that could predict the fall parameters to use for reaching desired load values. These studies focused on the height of the falls, the number of layers of absorbing foam, and mimicking the inertial behavior of the mock-up by ballasting the anode and GEM 3 assembly to match its weight.

\subsection{Test and Analysis of the CT Scan Images}
\begin{figure}[htbp]
\centering
\includegraphics[keepaspectratio, width=\textwidth]{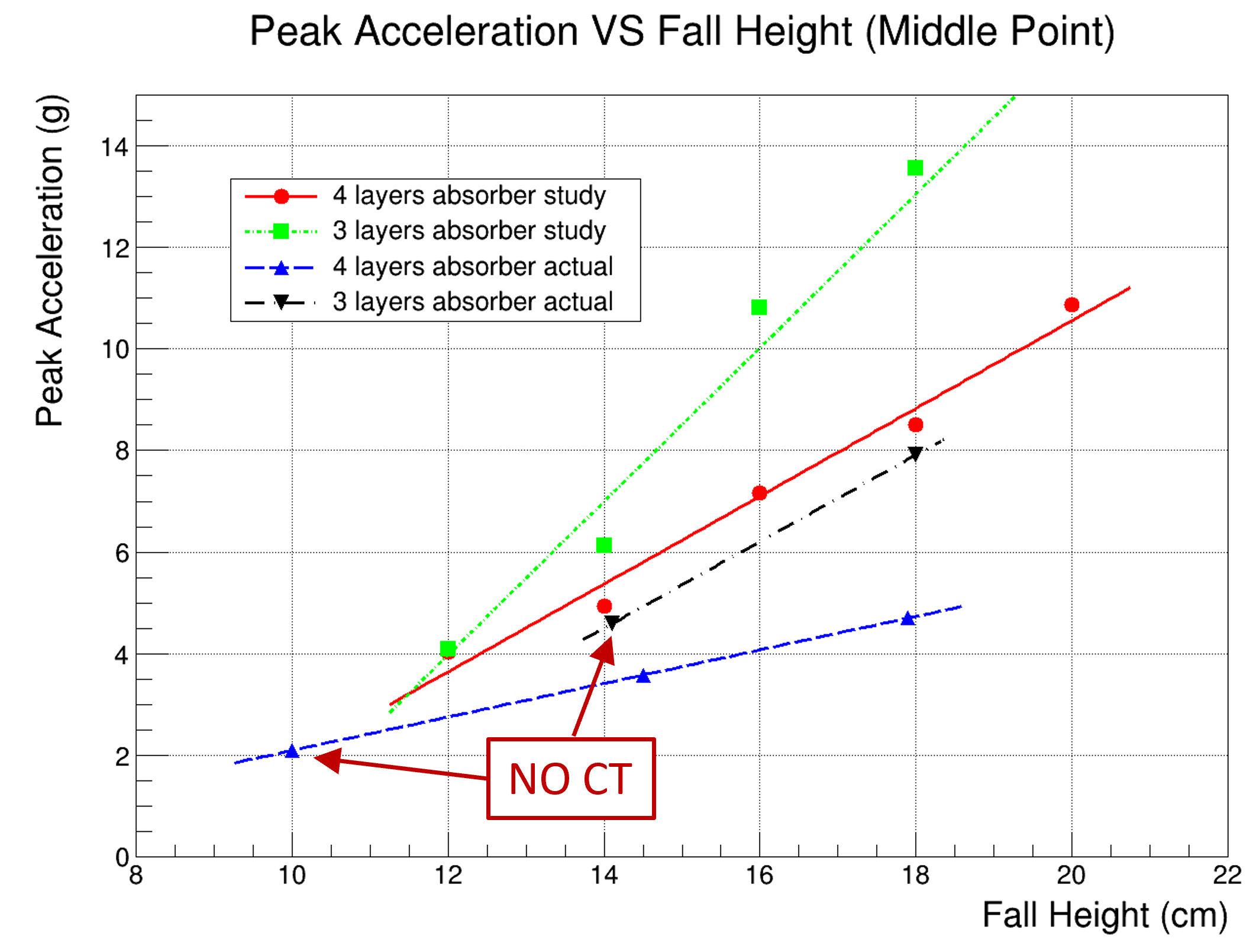}
\caption[Predictive model and test falls]{Comparison between the preliminary studies conducted on the anode and GEM 3 assembly (study) and the mock-up falls (actual). When switching between 4 and 3 layers of absorber material, to increase the maximum load achievable, an equally strong fall in the two configurations was taken, without performing a CT scan, to ensure the validity of the model and the scaling factor held true when switching between the two curves.}
\label{predmodel}
\end{figure}

The actual drop test was performed in the span of two days in a dedicated space provided within the facilities of TEC Eurolab SRL, which was selected as industrial partner to provide the CT scans.
The mock-up, assembled in Frascati, was transported to TEC Eurolab's test site.
One of the digital accelerometers continuously recorded acceleration data during transport, registering a maximum value of 3.3$\,$g at a frequency of a few hundred Hz.
Four complete CT scans of the mock-up were performed, one of them being the control scan to be taken upon arrival.
Each fall could either be too weak, and therefore uninformative, or too strong, and damage the mock-up before a relevant upper limit for the critical load had been established.
Hence, the importance of having some predictive model for the fall parameters to gauge the strength of the impacts beforehand.
Figure$\,$\ref{predmodel} shows a comparison between the studies performed in preparation to the test and the falls the mock-up was subjected to.
Since there was no way to relate the elastic behavior of the mock-up with the assembly used in the preliminary tests, the first two falls, taken below and approximately at the maximum transport load, were used as a gauge.
Once extracted the angular coefficient from these falls, and consequently the scaling factor separating the mock-up's behavior from the model, fall parameters for all values of interest could be derived.

\begin{figure}[htbp]
\centering
\begin{subfigure}{.49\textwidth}
\centering
\includegraphics[keepaspectratio, width=\linewidth]{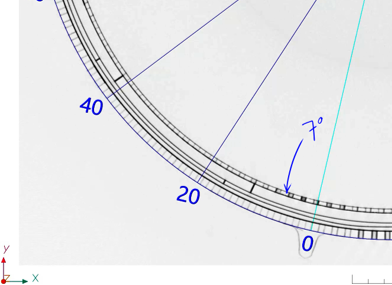}
\caption{Mock-up's internal structure upon arrival at the testing site.}
\end{subfigure}
\hfill
\begin{subfigure}{.49\textwidth}
\centering
\includegraphics[keepaspectratio, width=\linewidth]{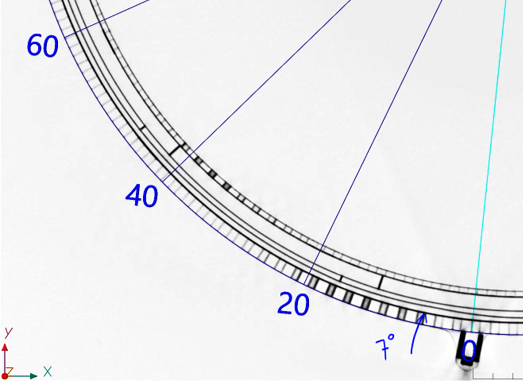}
\caption{Mock-up's internal structure after a 7.5$\,$g fall.}
\end{subfigure}
\caption[Drop test results]{Comparison between X-ray CT scan images of the mock-up collected upon arrival at the testing site and after suffering a 7.5$\,$g fall.}
\label{comparison}
\end{figure}

\begin{figure}[htbp]
\centering
\includegraphics[keepaspectratio, width=.45\textwidth]{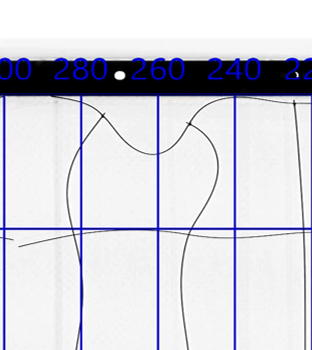}
\caption[Grid damaged during assembly]{Portion of the grid that came loose and went out of position during assembly}
\label{griddamage}
\end{figure}

The strongest load value reached, upon further analysis of the fall data, was about 7.5$\,$g.
This value refers to readings collected in the middle of the mock-up, where the internal floating electrodes are constrained the least; values recorded closer to the point of impact were even higher.
Figure$\,$\ref{comparison} shows a comparison between CT scan images collected upon arrival of the mock-up at the testing site and after the 7.5$\,$g fall.
No relevant change is observed in the internal structure between the two scans.
The electrodes remain evenly spaced and show no sign of damage, aside for the known preexisting defects.
Moreover, despite the grid detaching and going out of position due to friction during the assembly, as shown in figure $\,$\ref{griddamage}, the damage was contained, localized, and unaffected by the falls.

The test ultimately confirmed the effectiveness of the spacer grids in containing buckling induced deformation of CGEM electrodes at the dimensions of Layer 3.
The experience gained in building the mock-up not only led to the formation of a new construction team, but also helped refine the procedures and identify sources of risk.
Despite having raised the critical load sustainable by the detector to 7.5$\,$g, in the absence of reliable data on the stresses generated in the landing of cargo planes, the collaboration decided not to ship the assembled detector.
A split construction was deemed preferable, as it would have allowed preserving the delicate electrodes by shipping them while still on their mandrels to be later assembled in China.
This kind of procedure had never been attempted before and required a revamping of the assembly equipment.
    		\chapter{The Split Construction of Layer 3}
\label{laser}
The techniques adopted in the realization of CGEM detectors have already been thoroughly documented in the author's master's degree thesis$\,$\cite{Gramigna:2022ozf} and in references \cite{Gramigna:2022frx} and \cite{Balossino:2022ywn}.
This chapter aims to highlight what distinguished the construction of Layer 3 from prior experiences, with a particular focus on improvements to both tooling and procedures.
Because of this, only a concise overview of the construction process will be provided; more detailed descriptions are going to be focused on the operations that are unique to the making of Layer 3 or those that underwent major review.

\section{Construction Process of a CGEM Layer}
The construction of a CGEM layer can be divided in three main phases: an initial preparatory phase, the construction of the cylindrical electrodes, and the vertical assembly.
The preparatory phase includes all operations needed to ready the materials and the tools for the next two phases.
It includes: the procurement of all construction materials and the refurbishing of spent consumables, the dimensional checks on the fiberglass support rings, the validation and restoration of the mandrels and of their vacuum systems, the precision trimming of all flexible circuits, the high voltage testing of the GEM foils, and the alignment of the assembly machine.
This last point is fundamental to ensure the success of the detector's assembly with the narrow clearances involved, which require all five mandrels to be aligned within 100$\,\upmu$m/m on the machine.

\begin{figure}[tbp]
	\centering
	\includegraphics[keepaspectratio, width=\textwidth]{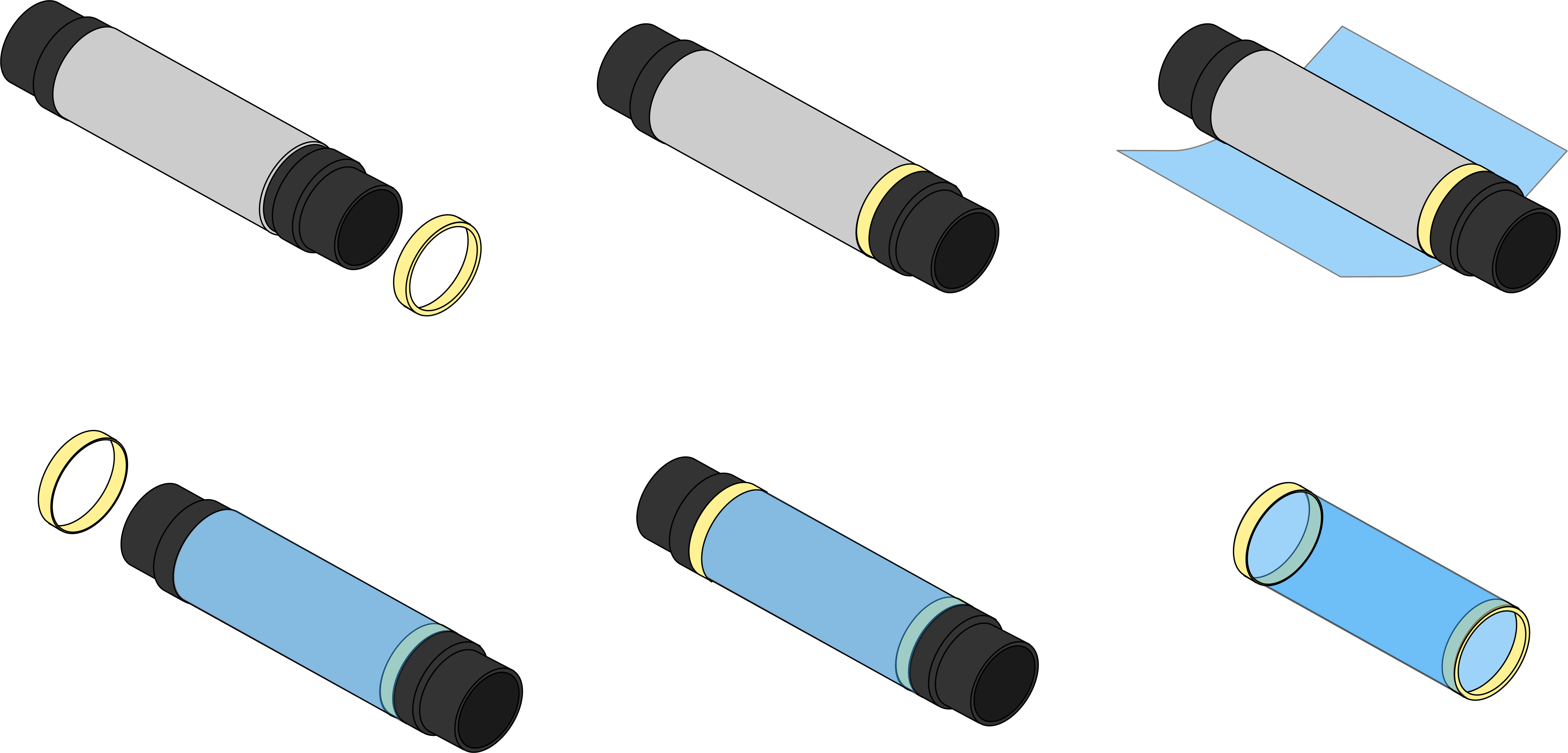}
	\caption[Electrode construction procedure]{The general idea at the basis of the construction of cylindrical electrodes: foils are wrapped around a mandrel and glued to structural rings at the two ends.}
	\label{procedure}
\end{figure}

The construction of all five electrodes follows a common principle: foils are rolled around a Teflon-coated aluminum mandrel and glued to fiberglass rings at their extremities, as shown in figure$\,$\ref{procedure}.
Cathodes and anodes have more involved construction processes, as they are supported by cylindrical sandwich-structured composites, similarly built around the mandrels.

\begin{figure}[htbp]
	\centering
	\includegraphics[keepaspectratio, width=.7\textwidth]{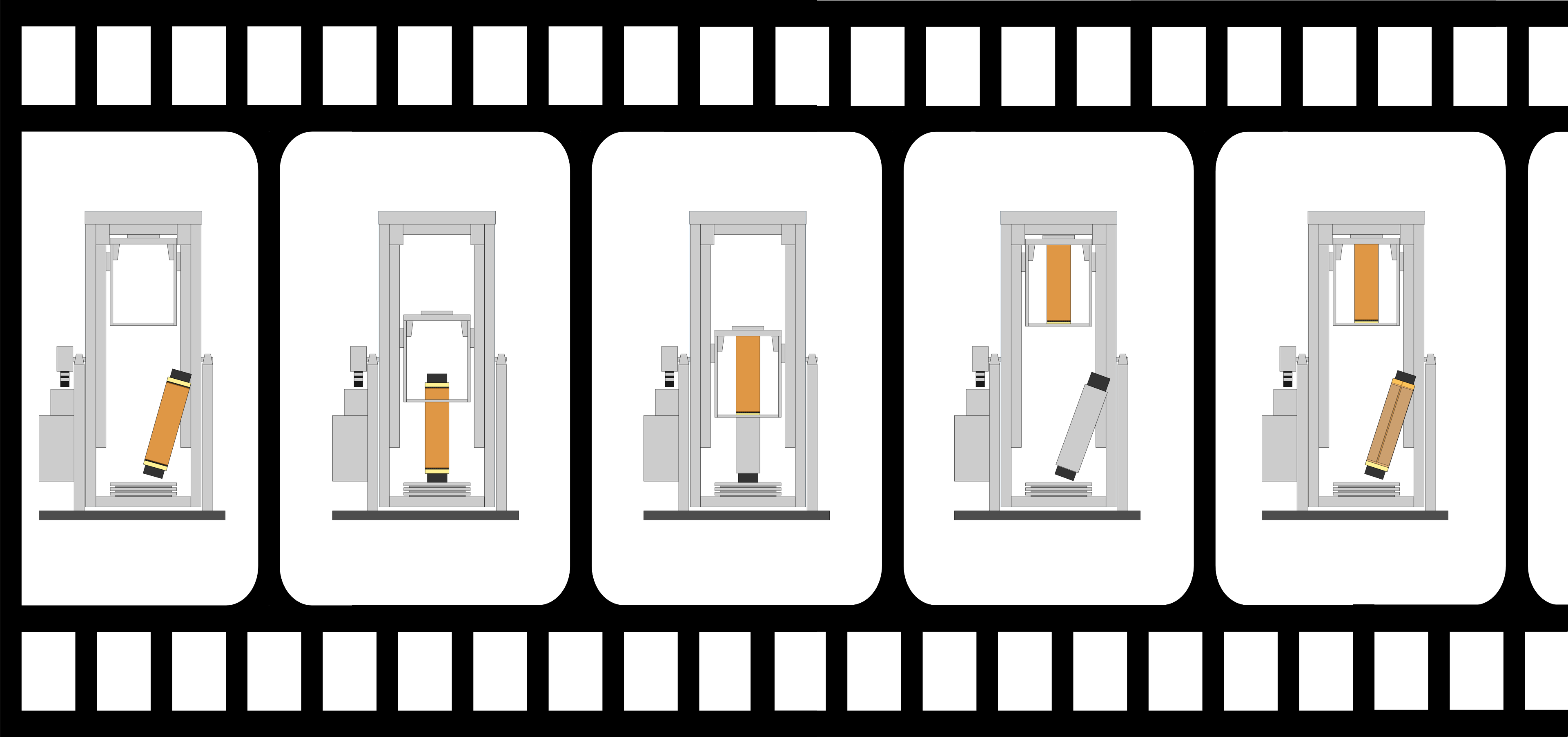}

	\vspace{3 mm}
		\includegraphics[keepaspectratio, width=.7\textwidth]{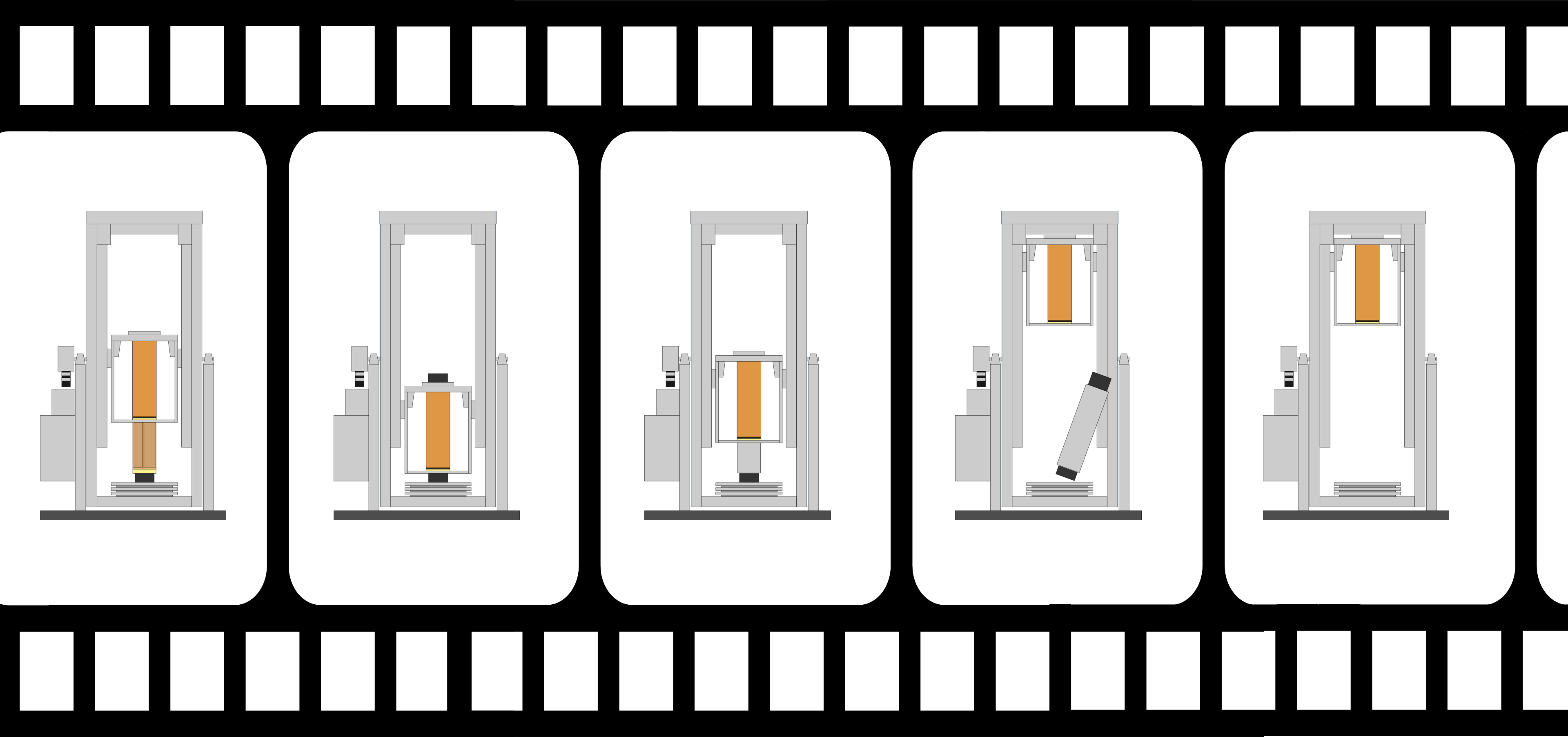}

	\vspace{3 mm}	
	\includegraphics[keepaspectratio, width=.7\textwidth]{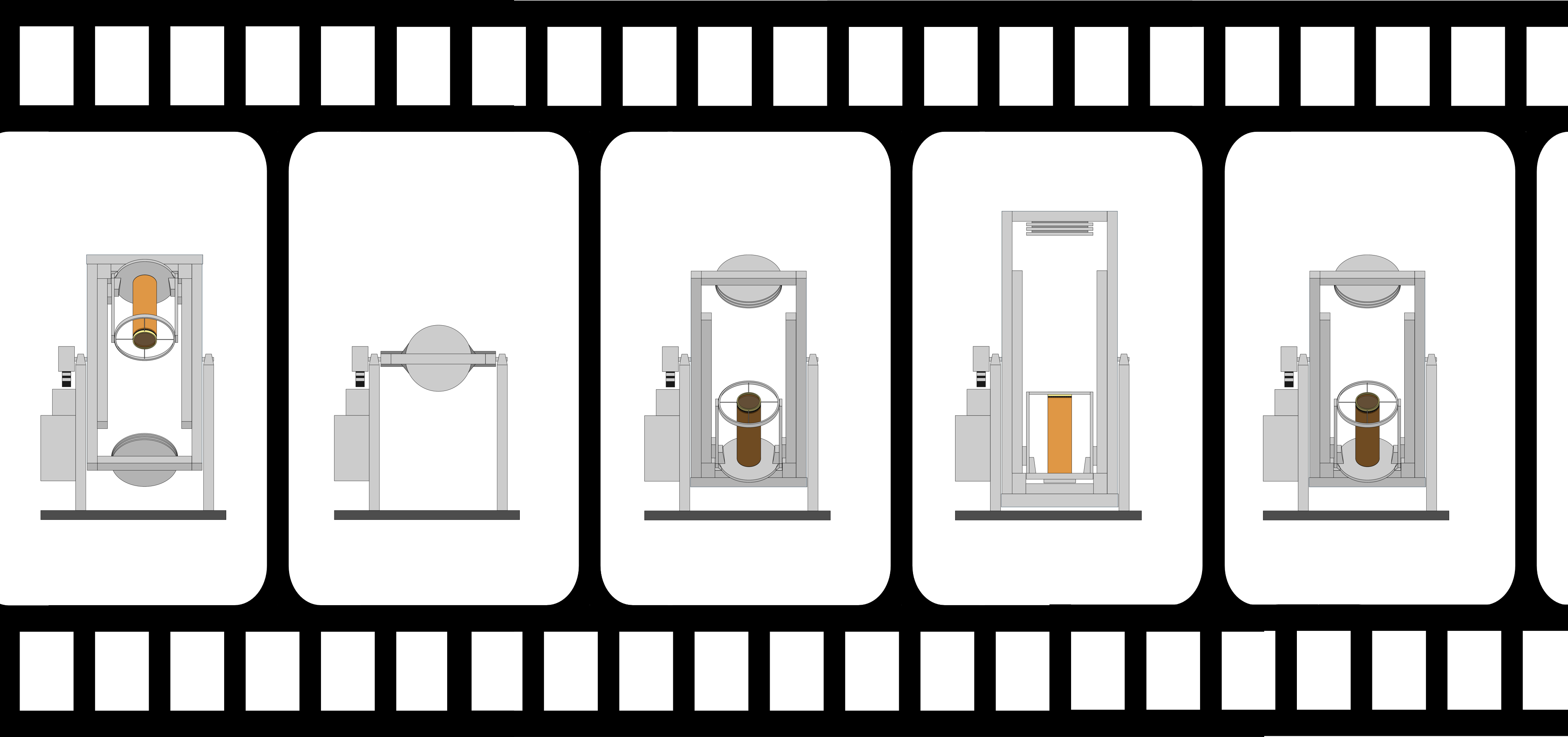}
	
	\vspace{3 mm}	
	\includegraphics[keepaspectratio, width=.7\textwidth]{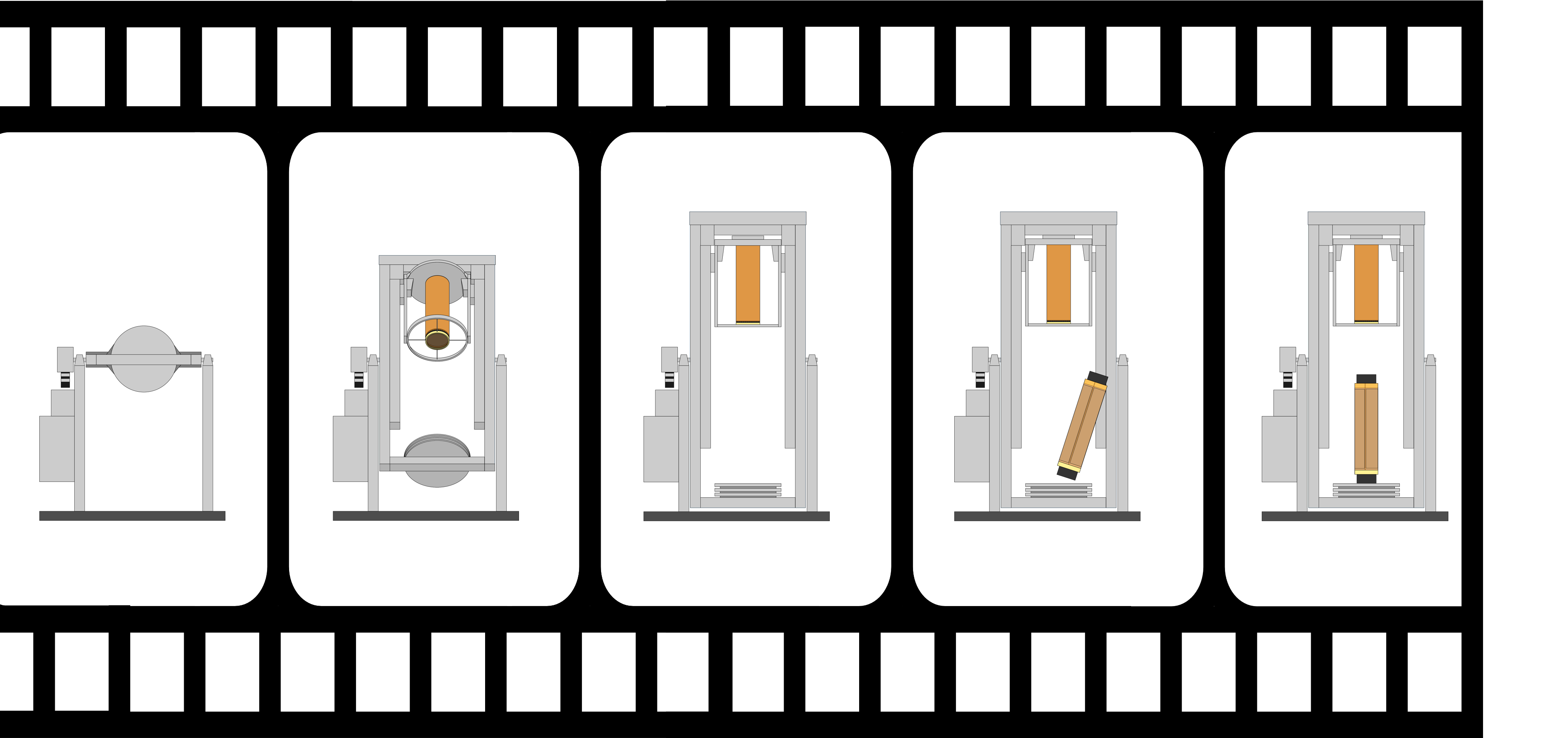}
	\caption[Assembly of CGEM electrodes]{Schematic representation of the assembly procedure of CGEM electrodes. This sequence only shows the begininning of the assembly, stopping right before the insertion of the second GEM electrode. All remaining steps necessary to fully assemble the detector follow a similar concept.}
	\label{slideshow}
\end{figure}

The assembly of the electrodes, schematically represented in the sequence of figure$\,$\ref{slideshow}, is performed vertically using a custom CNC machine called VIM.
The first step in this process is the extraction of the largest electrode, the anode, from its mandrel (frames 1 to 4).
Then, the next electrode, GEM 3, replaces the empty mandrel and is inserted into the anode (frames 5 to 7).
The top rings of the two electrodes are glued together and, once the glue has cured, GEM 3 is also extracted from its mandrel (frames 8 to 10).
The machine is then rotated 180$^\circ$ to glue the rings at the opposite end (frames 11 to 14).
After the machine is brought back to the original position (frames 15-18), the entire process is repeated until all five electrodes are glued together.
Both endcaps are finally sealed with epoxy resin to ensure gas tightness, marking the end of the CGEM layer's construction.

\section{The Logistics of a Split Construction}
To safeguard the delicate GEM electrodes of Layer 3, the collaboration decided to pursue a split construction.
The plan was to build the electrodes in Italy, ship them together with the VIM while protected by their mandrels, and finally assemble them in Beijing.
At the time Layer 3 was to enter production many logistical complications had to be overcome: the ISO 6 class cleanroom at LNF, where all previous layers of the CGEM-IT had been built, was being decommissioned; the cleanroom available at IHEP was not tall enough to accomodate the assembly machine; and finally the alignment of the VIM required the mandrels to be in China, before the electrodes could be built on top of them, so that dial gauges could be used on their surfaces.

A disused cleanroom within the locals of the University of Ferrara was refurbished, upgraded to ISO 6, and equipped to be the new site for the construction of Layer 3's electrodes.
Most of the pre-existing construction equipment was transferred from LNF to Ferrara, but had to be modified to fit the limited space available.
IHEP's cleanroom was entirely rebuilt.
The new cleanroom was larger, the height of the ceiling was almost doubled, and the air treatment unit was upgraded to provide a large ISO 6 workspace with a smaller ISO 5 enclosure.

The issue of the alignment of the VIM was a matter of both time and costs.
The construction of Layer 3 lasting too long would have shortened the cosmic ray data taking and prevented detector stability and performance studies to be completed before the planned installation date in 2024.
The air shipment of the mandrels back and forth, to align the VIM before the electrodes' construction, not only would have been very expensive but it would also have required a long time due to the delays incurred for clearing customs.
The solution to the alignment issue was ultimately technological and it is going to be one of the main subjects of this chapter.

\section{Changes to the Quality Control Protocol}
All materials involved in the construction of a CGEM layer undergo rigorous quality control checks.
The diameters of the fiberglass rings were inspected using a coordinate measuring machine at the manufacturer's production site, while all other relevant dimensions and geometrical features were checked with conventional measuring tools.
Defective rings were either adjusted or rejected and produced anew.
The Teflon-coated aluminum mandrels were also thoroughly measured by LNF's metrology division using a laser tracker system.
All mandrels were found to be within tolerance, and scratches resulting from previous extractions were evened out using 1000 grit sandpaper.

GEM foils are the most delicate materials and therefore subjected to the most rigorous quality control protocol.
Most of it was directly derived from past experiences of CGEM construction, but it was reviewed and expanded for the making of the new Layer 3.
Each of Layer 3's GEM electrodes is fashioned out of two GEM foils.
At the dimensions of Layer 3's foils, which are close to the maximum dimensions that can be achieved at CERN EST-DEM workshop, the yield is not 100\%, so a number of foils is ordered as spares.
This allows both to cope with defective foils, without incurring in delays due to the long leading times, and to select the best performing candidates.
The HV testing procedure, reviewed for the test of Layer 3's GEM foils with the compacted setup in figure$\,$\ref{hvtable}, is provided in full in appendix$\,$\ref{hvprocedure}.
Since the CT scans highlighted the presence of some localized defects that could not be traced back to buckling, a thorough visual inspection was added to the protocol.
A scoring system, accounting for both electrical behavior and mechanical integrity was implemented to provide objective criteria to select the better candidates.

\begin{figure}[tbp]
	\centering
	\includegraphics[keepaspectratio, width=\textwidth]{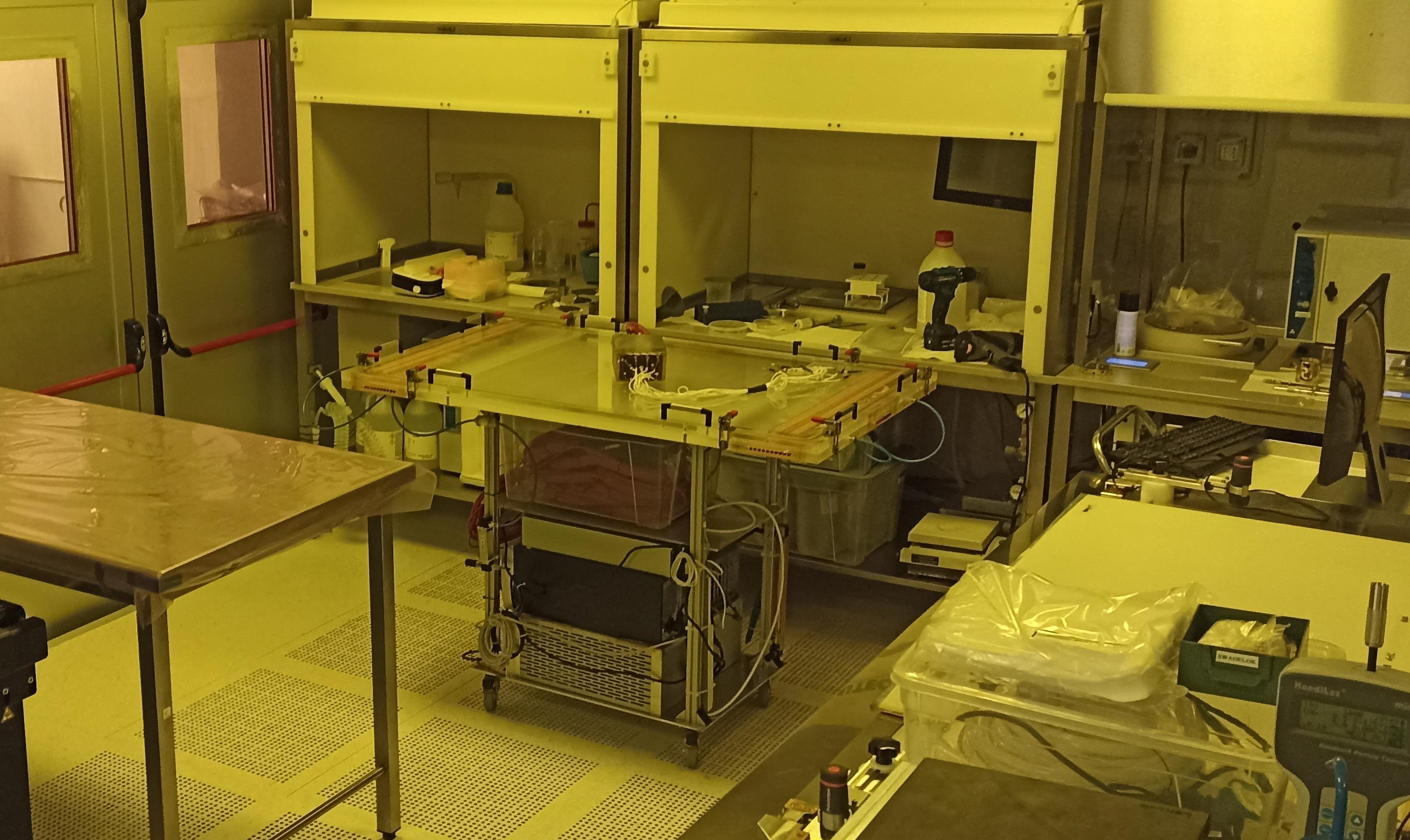}
	\caption[GEM HV test table]{The HV test setup used in the quality control of Layer 3's GEM foils. The setup was compacted on a single wheeled table to avoid crowding the smaller cleanroom space available in Ferrara.}
	\label{hvtable}
\end{figure}

\section{A New Technique for Gluing Narrow GEM Overlaps}
To make a CGEM electrode, two foils are first glued together on a vacuum table, then rolled around the mandrel to be glued both to each other one more time and to the fiberglass rings.
Gluing GEM foils together requires to deposit a thin layer of Huntsman Araldite\textsuperscript{\textregistered} 2011 epoxy adhesive so that it does not overflow from a 3$\,$mm overlap.
For the construction of all the previous CGEM layers, the glue had been deposited by hand using a spatula.
Since a new team was formed and none of the people involved were familiar with that technique, a new method was developed to perform this delicate operation.

\begin{figure}[tbp]
	\centering
	\includegraphics[keepaspectratio, width=\textwidth]{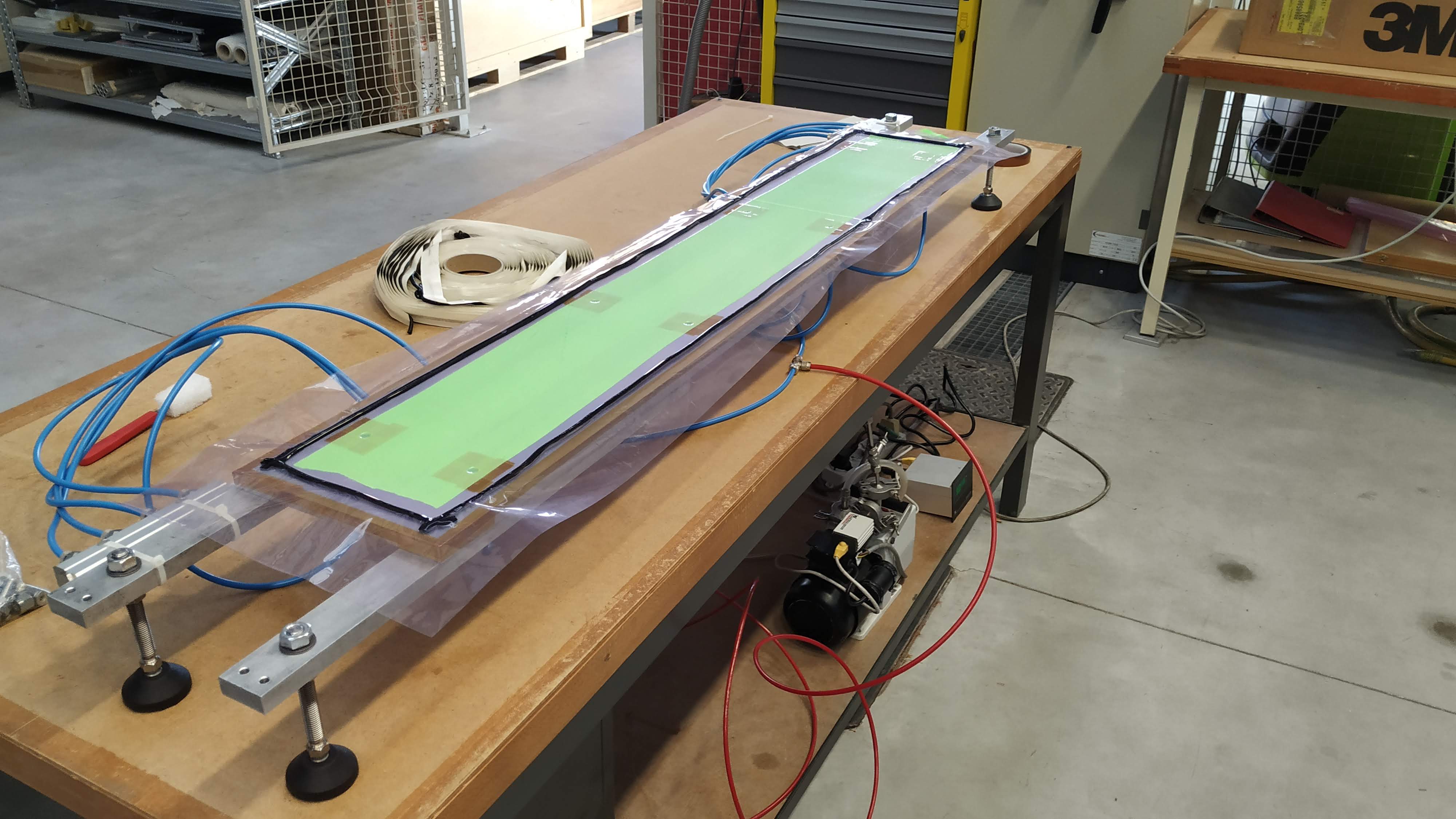}
	\caption[Practice vacuum table]{The small vacuum table used to practice the gluing of the narrow GEM overlaps.}
	\label{vactable}
\end{figure}

A small vacuum table, in figure$\,$\ref{vactable}, was designed and built to practice the gluing of GEM foils using long Kapton strips as samples.
The technique developed is a variation of the glue transfer technique used in the realization of the composite structures.
A 1.5$\,$mm wide strip is cut out of a roll of 125$\,\upmu$m thick Mylar, and serves as a disposable support to transfer the glue. The strip is fully covered in epoxy and pressed on the appropriately masked and marked margin of the GEM foil.
Wider Nylon strips can be later pressed onto the glue and peeled off to reduce the thickness of the adhesive layer.
On average this has to be done two to three times to ensure a sufficiently strong bond between the foils and prevent overflow that could occlude GEM holes and thus widen the dead region.

The new technique is faster and does not require as much expertise when compared with glue deposition by spatula.
The size and thickness of the adhesive layer are determined by the width of the Mylar strip and the number of times the glue is removed, ensuring more repeatable outcomes.

\section{VIM Alignment Overhaul}
A complete rethinking of the VIM alignment was necessary to overcome one of the main logistical obstacles involved in a split construction.
The budget for developing a viable solution had to be inferior to the cost of a double shipment of the mandrels to and from China.
The time allotted for the development and test of the system was determined by the schedule of the upgrade of the cleanroom that was to host the construction of the electrodes.
A relatively high budget and a tight schedule for development pointed towards a solution based on the integration of high-end commercially available technology.

\subsection{The Vertical Insertion Machine}
\begin{figure}[htbp]
	\centering
	\begin{subfigure}[t]{.29\textwidth}
		\centering
		\includegraphics[keepaspectratio, height=5.5cm]{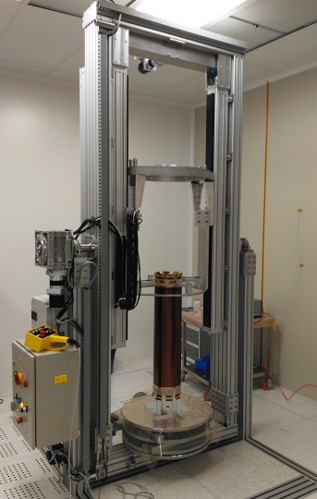}
		\caption[]{The VIM in its standard, upright position}
		\label{upright}
	\end{subfigure}
\hfill
	\begin{subfigure}[t]{.69\textwidth}
		\centering
		\includegraphics[keepaspectratio, height=5.5cm]{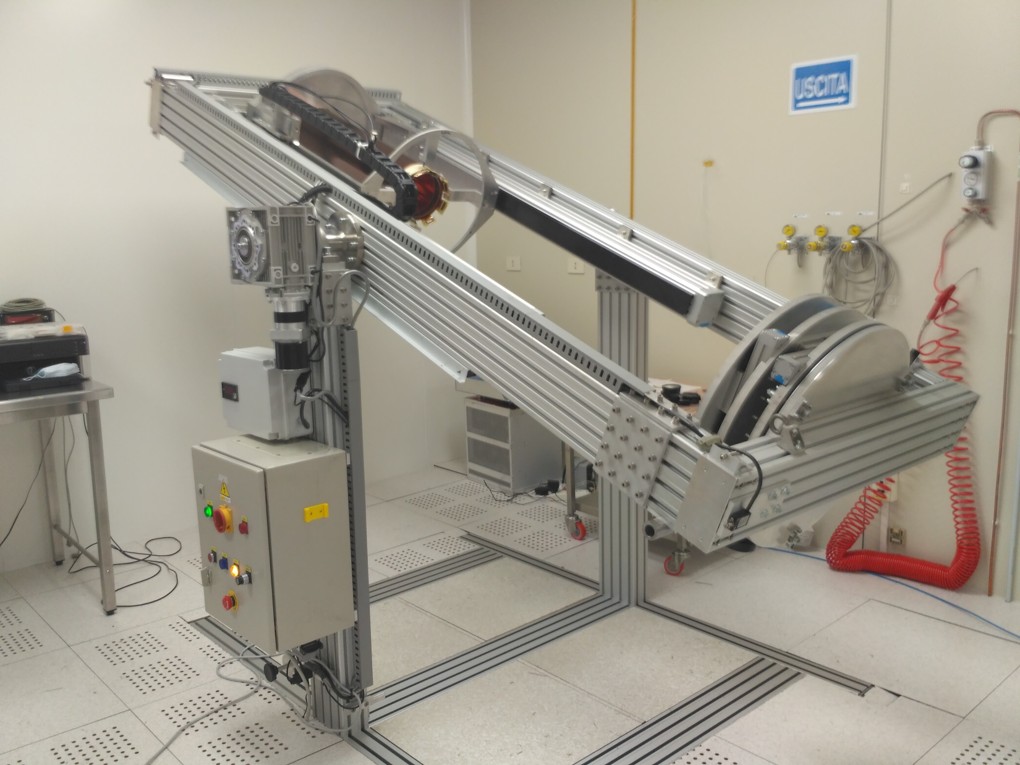}
		\caption[]{The VIM, mid rotation}
		\label{midrot}
	\end{subfigure}
	\caption[VIM's main degrees of freedom]{Two photographs of the VIM, showing its two main degrees of freedom, the vertical movement of the trolley and the rotation of the main frame.}
	\label{hourglasses}
\end{figure}

The VIM was designed and built for the construction of the KLOE2-IT, and later modified to adapt to the slightly longer CGEM-IT.
Bosch profiles were adopted for the framing, commercial electric motors and motorized linear guides control the movements, and custom machined flanges are used to interface with the detector.
Figure$\,$\ref{hourglasses} shows the VIM in its standard upright position and mid rotation.
The two principal degrees of freedom of the machine are showcased in the photographs: the vertical movement of its rail-mounted trolley and the rotation of the main rectangular frame.
The base of the machine consists of a rectangular frame, bolted to the floor, and two vertical struts, which support the main rotating structure.
This large rectangular frame has an adjustable base at the bottom, for housing the mandrels, and a moving trolley at the top, used for inserting and extracting the electrodes.

\begin{figure}[htbp]
	\centering
	\includegraphics[keepaspectratio, width=\textwidth]{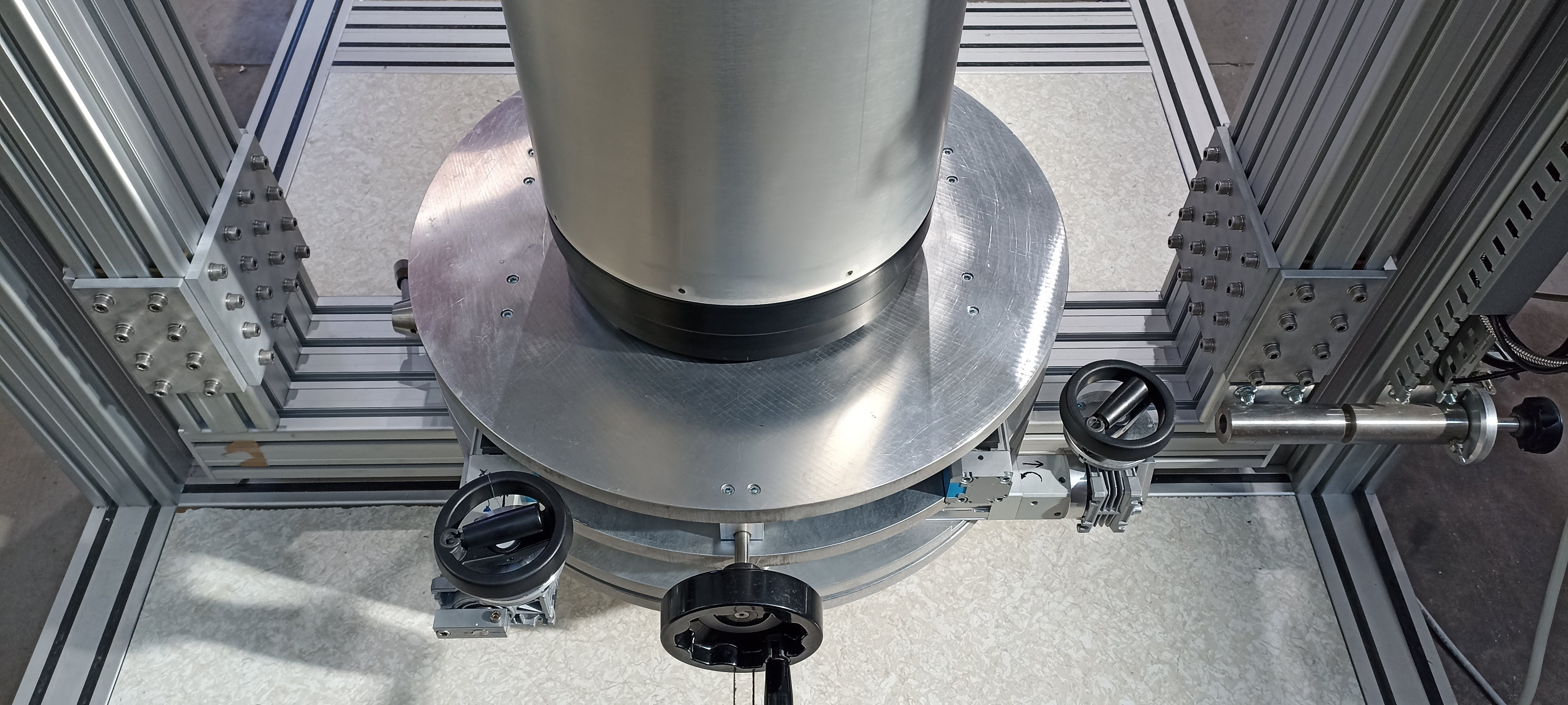}
	\caption[VIM base]{The base of the VIM showing the stacked plates it is made of and the adjustment wheels.}
	\label{base}
\end{figure}

The base of the machine, in figure$\,$\ref{base}, consists of four circular aluminum plates, stacked and connected by adjustable elements that grant the base its degrees of freedom: movement in the horizontal plane and tilt with respect to the main frame.
At the center of the base lies a self-centering chuck, which is used to keep the mandrel steady by holding onto a pin affixed to its bottom-facing three-spoke flange.

The trolley has a flange at the top, used to grab the electrodes with radial pins, and a ring at the bottom, where pushers can be installed to stabilize the detector during the rotation.
The trolley is connected to the rails near the top flange.
One of these connection points is fixed while the other is made compliant through small linear guides.
This prevents the trolley from being compressed in case of imperfect parallelism of the rails.

The main frame can rotate 180$^\circ$, to grant access to the bottom endcap of the detector.
The rotation is controlled manually through a remote, while vertical motion of the trolley can be controlled either manually, through the same remote, or by the machine's control software.
The LabVIEW-based control interface allows to move the trolley in steps as fine as 10$\,\upmu$m.

\subsection{Traditional Alignment Technique}
\begin{figure}[htbp]
	\centering
	\includegraphics[keepaspectratio, width=\textwidth]{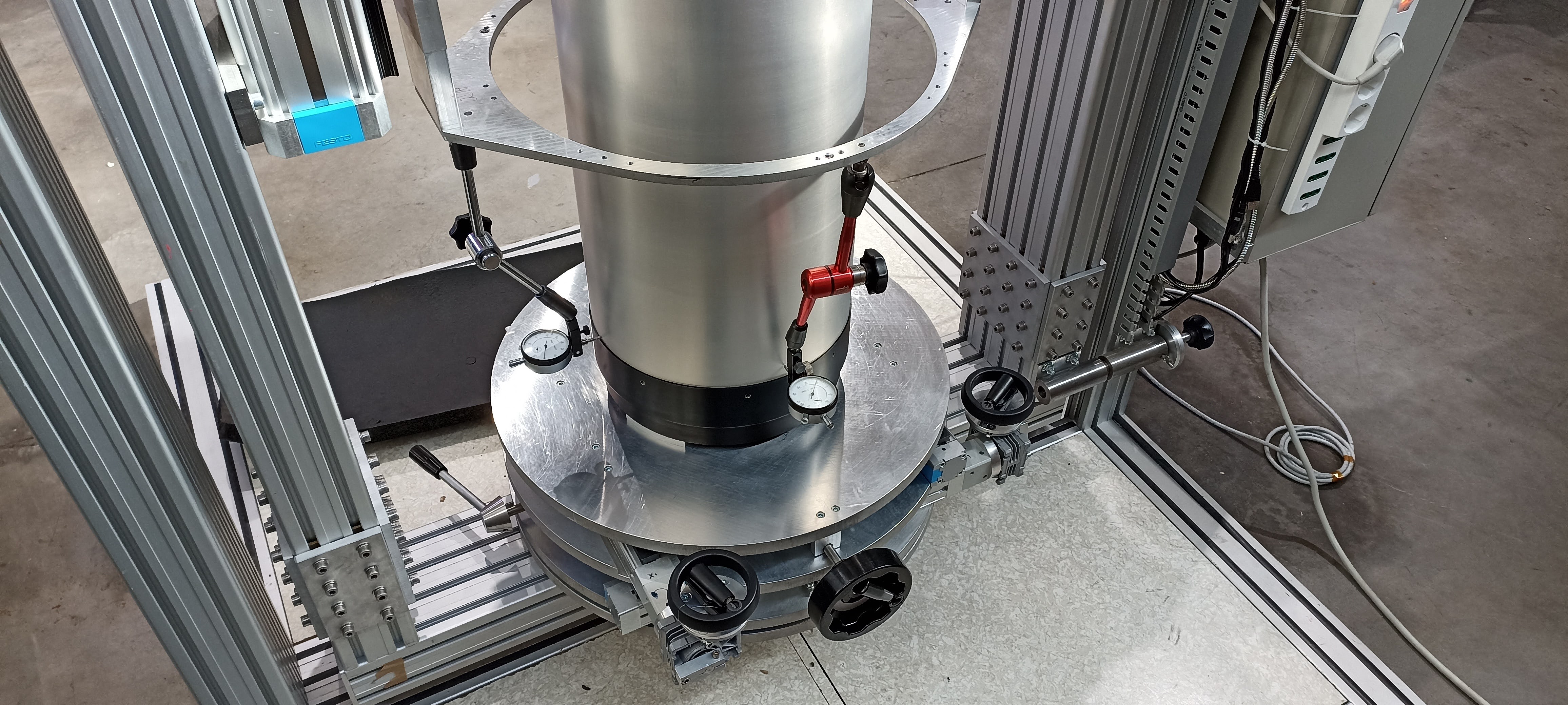}
	\caption[VIM alignment with dial gauges.]{Positioning of the dial gauges for the alignment of the VIM.}
	\label{dial}
\end{figure}

The alignment of the VIM has to satisfy two requirements:
\begin{itemize}
\item Bring each of the five mandrel's tilt with respect to the axis of the trolley's vertical movement within 100$\,\upmu$m/m  in both directions.
\item Center the top end of each mandrel with respect to the trolley's top flange within 100$\,\upmu$m in both directions.
\end{itemize}
For the construction of all CGEM electrodes prior to Layer 3 the tilt alignment was informed by the readings of dial gauges, affixed to the trolley's bottom ring and made to slide directly on the mandrels' Teflon-coated surface, as shown in figure$\,$\ref{dial}.
Moving the trolley so that the tips of the gauges travel the whole length of the Teflon surface provides a measurement of the mandrel's tilt. When the gauges are stopped near the top of the mandrel, they also provide direct feedback on the changes made by adjusting the screws at the base of the machine.
Only a single reference mandrel can be aligned using the adjustment screws, all the others, if necessary, must be aligned by placing shims between the bottom three-spoke flange and the main cylindrical body.

To asses the centering, the trolley had to be lowered until its top flange was level with the top end of the mandrel, so that the clearance separating the two could be measured with a caliper.
The clearance in the two horizontal directions could then be equalized by acting on the control wheels at the base to move the mandrel with respect to the flange.
This second part of the alignment procedure could be affected by large measurement errors, due to improper caliper placement, and was complicated by the absence of a direct feedback on the adjustments performed through the control wheels.

The traditional alignment technique relies heavily on the use of the precisely machined surface of the mandrels as reference.
Having the tips of the dial gauges slide on the delicate electrodes would damage them irreparably, so an alternative method had to be devised instead.

\subsection{A Contactless Alignment Solution}
The solution proposed to solve the alignment issue was to replace the dial gauges with contactless distance sensors. Commercially available solutions are numerous: inductive, laser, ultrasound, infrared, radar and lidar devices, each with their own strengths and weaknesses.

\begin{figure}[tbp]
	\centering	\includegraphics[keepaspectratio, width=.6\textwidth]{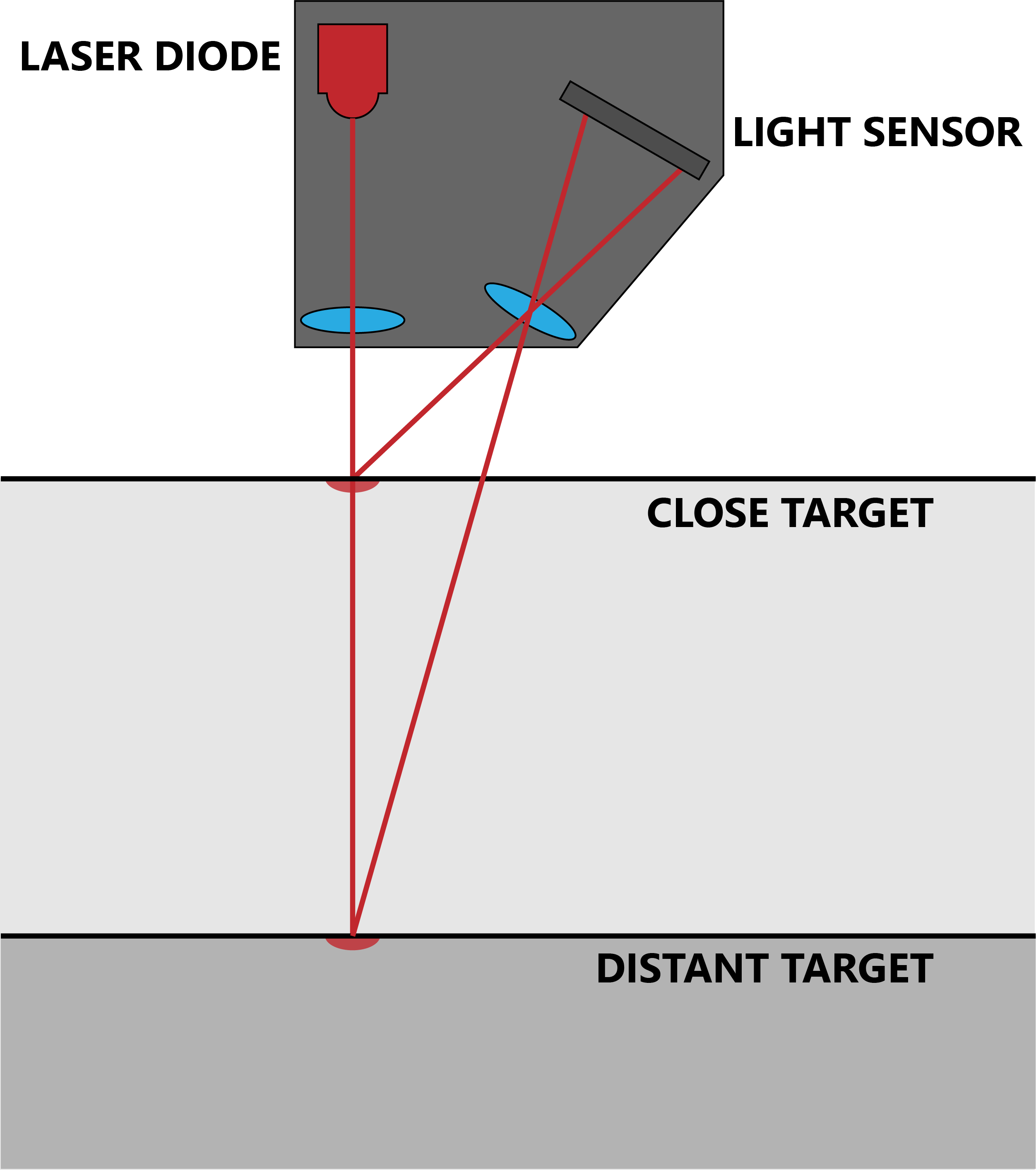}
	\caption[Laser triangulation sensor's functioning principle]{Functioning principle of a laser triangulation sensor. The light sensor is angled with respect to the emitter so that the distance can be determined by the position of the reflected light impinging on the sensor.}
	\label{lts}
\end{figure}

The choice finally veered towards laser triangulation sensors, which offered the best combination of accuracy, range, and ease of implementation for this application.
These consist of a laser diode, a light sensor, and various optics, arranged as of figure$\,$\ref{lts}.
The diode projects a laser spot onto the target object.
Shape and size of the spot are chosen according to the sensor's intended performance, and controlled by a first set of optics.
Some of the light diffused by the target's surface is conveyed onto the light sensor by a another set of optics and the distance of the object is triangulated according to its position.
The signal response of laser triangulation sensors can be affected by ambient light, temperature, reflectivity of the target, and angle of incidence of the laser projection.
The most advanced models have internal electronics which compensate for most of these effects, provided that the device is used within the operating parameters.

\begin{figure}[p]
	\centering
	\includegraphics[keepaspectratio, angle=90, width=.75\textwidth]{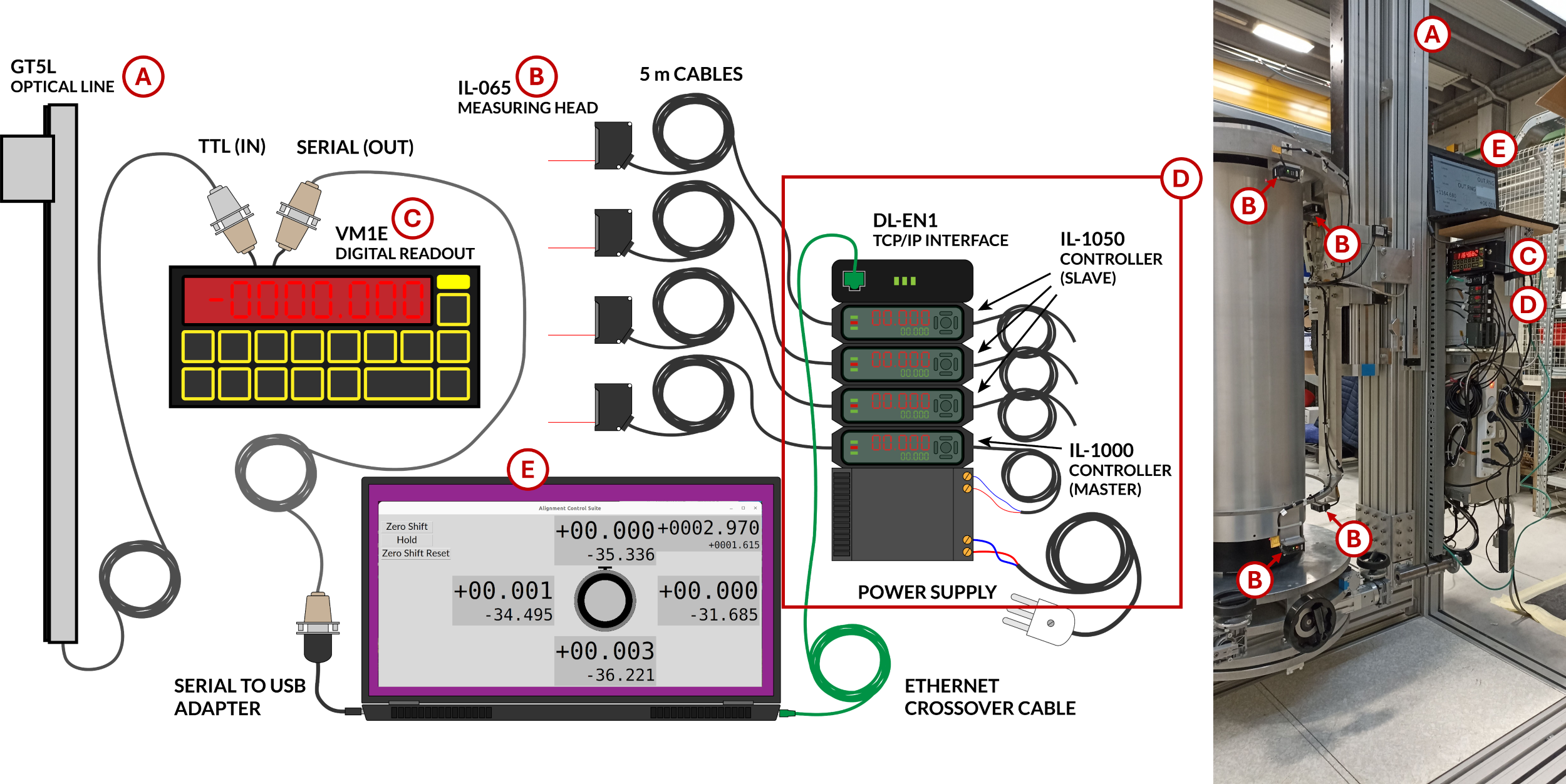}
	\caption[Laser alignment system diagram]{Diagram of the laser alignment system and position of its main components on the VIM}
	\label{lassys}
\end{figure}

A diagram of the laser alignment system, in its final configuration, can be found in figure$\,$\ref{lassys}.
Four Keyence IL-065 laser triangulation sensors were purchased, two were installed underneath the top flange of the trolley and two were mounted below its bottom ring.
The sensors were oriented so to be aligned with the translational adjustments at the base of the machine, which define the X and Y coordinates.
The sensor heads are managed by proprietary controllers and communicate with the a computer trough a TCP/IP protocol. An ELBO GT5L optical line, mounted onto one of the rails, measures the vertical position of the trolley, which is serialized and sent to the computer.

\FloatBarrier

\subsubsection{System Development and Test}
To develop and test the system, the VIM was moved and commissioned within the MPGD construction lab of the University and INFN of Ferrara, which needed to be expanded for the occasion.
The five mandrels necessary for the construction of Layer 3's electrodes, together with all the tooling necessary for their installation onto the VIM, were also moved to Ferrara and rendered available for the tests while awaiting the completion of the cleanroom upgrade.

\begin{figure}[htbp]
	\centering	\includegraphics[keepaspectratio, width=\textwidth]{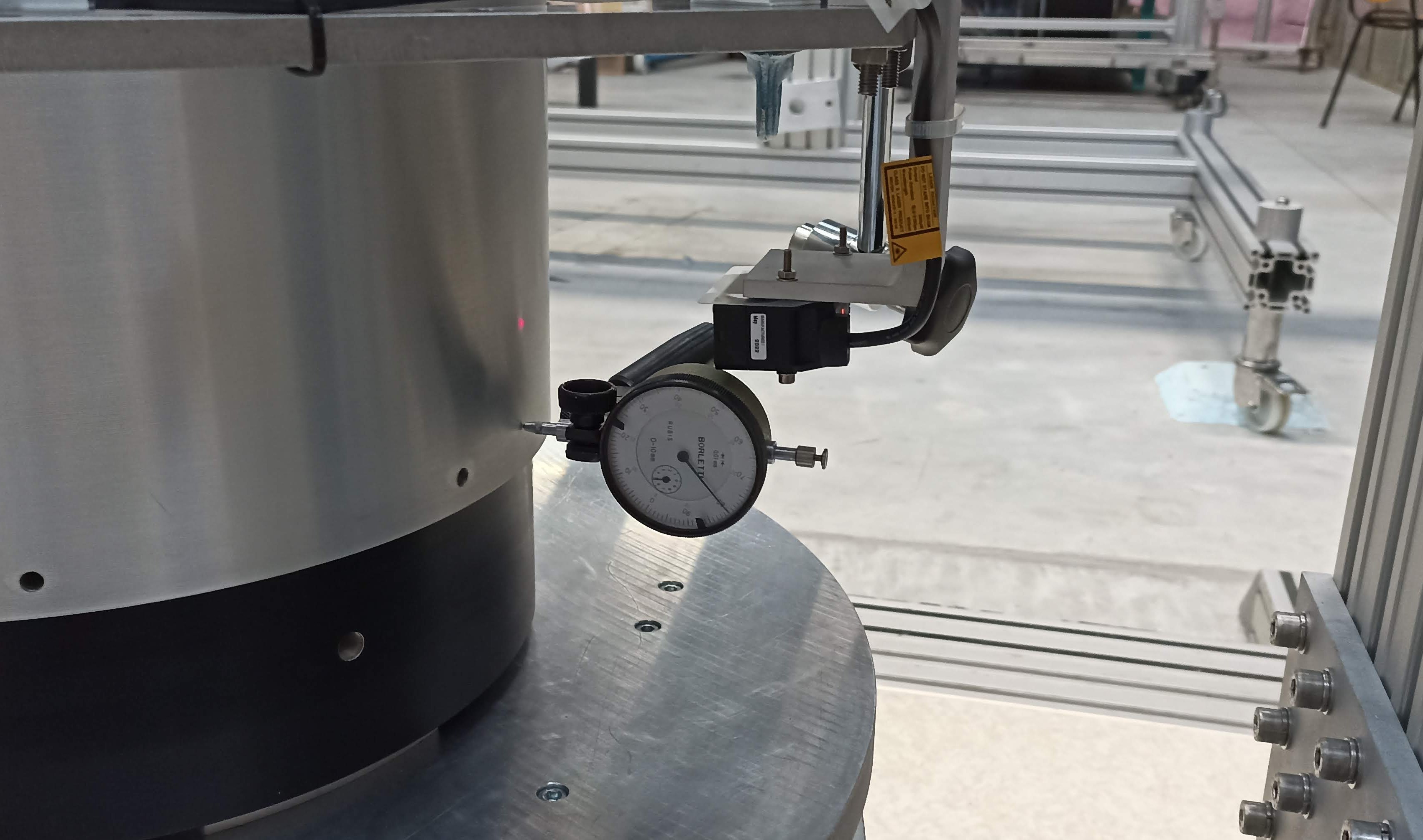}
	\caption[Dial gauge mounting for comparative studies]{Mounting of the dial gauges for their use in the preliminary comparative studies}
	\label{lasdial}
\end{figure}

All initial tests were performed comparing the results of the new laser sensors with the readings of the traditionally used dial gauges.
The tips of the gauges were pointed directly below the laser spots, as shown in figure$\,$\ref{lasdial}, and the trolley was moved up and down so that both instruments could scan the entire length of the mandrel's Teflon surface.
These preliminary measurements, an example of which is provided in table$\,$\ref{comptab}, were performed by taking a single reading at the bottom and then at the top of the mandrel. Although there appeared to be good consistency between the two instruments, within 0.1$\,$mm, the measurements collected by the laser sensors were affected by larger variations. In particular, it was observed that the choice of the initial and final point of the scans had a larger impact on the tilt measured by the laser sensors with respect to the traditional technique.

\begin{table}[htbp]
\centering
\begin{tabular}{lrr}
\multicolumn{3}{c}{\textbf{Going Downward}}               \\\midrule
            & \textbf{Dial Gauges} & \textbf{Laser Sensors} \\\cmidrule(l){2-3}
\textbf{dx (mm)} & -0.303               & -0.275                 \\
\textbf{dy (mm)} & -0.200               & -0.152                 \\

\multicolumn{3}{c}{}\\

\multicolumn{3}{c}{\textbf{Going Upward}}                   \\\midrule
            & \textbf{Dial gauges} & \textbf{Laser sensors} \\\cmidrule(l){2-3}
\textbf{dx (mm)} & 0.306                & 0.262                  \\
\textbf{dy (mm)} & 0.190                & 0.138                 \\
\end{tabular}
\caption[Comparison between laser sensors and dial gauges]{Comparison between tilt measurements based on a single reading collected by the dial gauges and the laser triangulation sensors.}
\label{comptab}
\end{table}

\begin{figure}[htbp]
	\centering	\includegraphics[keepaspectratio, width=\textwidth]{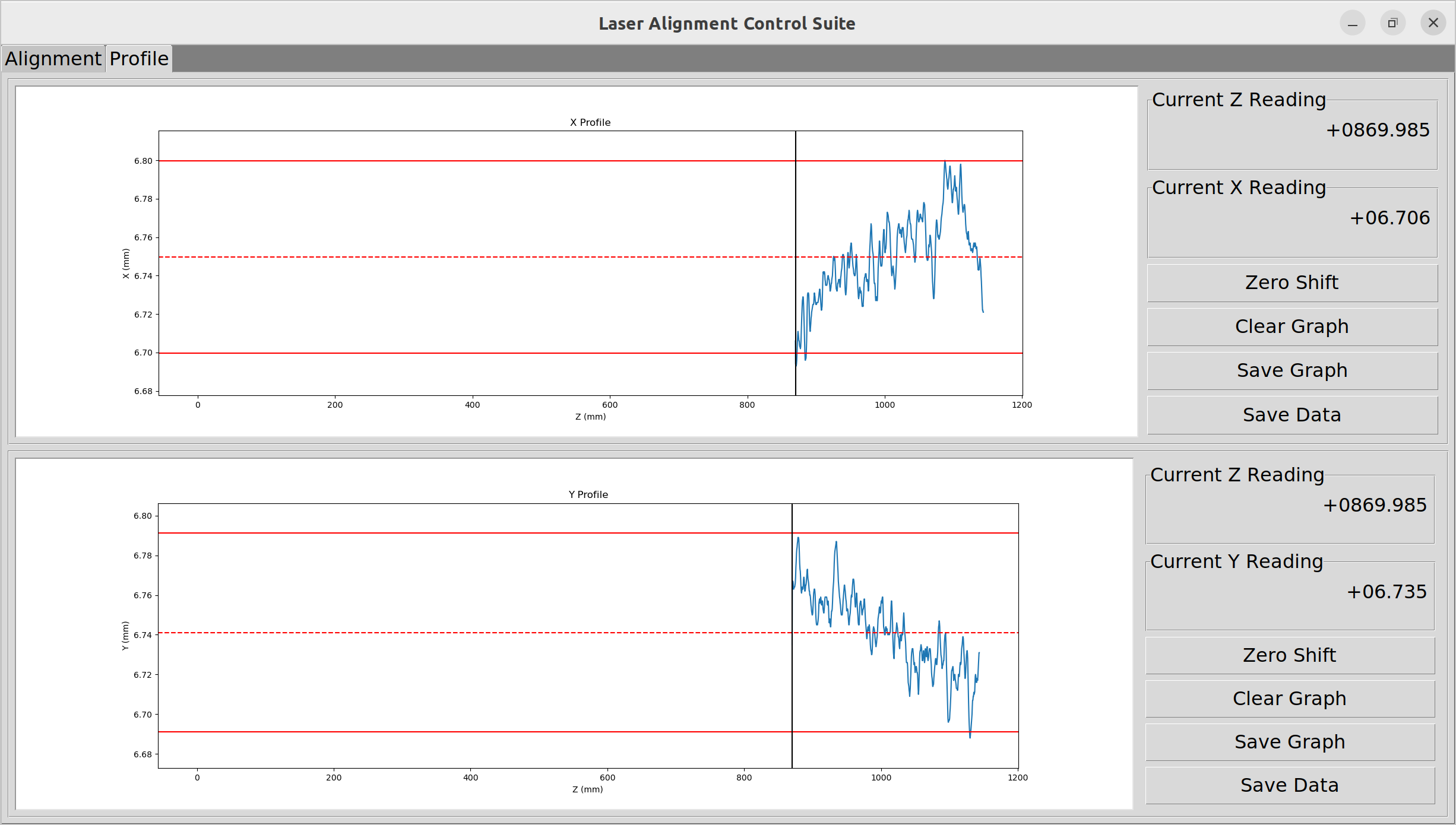}
	\caption[Early version of alignment interface]{Screenshot of an early version of the alignment interface, displaying unexpected fluctuations of the readings when scanning the Teflon surface. In this build, the monitor always displayed the full range of the profile and populated it according to the movement of the trolley, the position of which is represented by the black vertical line in the plots.}
	\label{screen}
\end{figure}

A python-based interface was being developed alongside the system, so not to rely solely on the controllers' digital readout and to help inform the operator on the adjustments to make.
A screenshot of an early version of this interface, in figure$\,$\ref{screen}, shows a concerning behavior of the sensors' readings during the scan.
Signal fluctuations were higher than expected, leading to poor scanning resolution.
The measurements were repeated on the mandrel's black anodized extremities and on a GEM sample, both of which led to even worse results, with fluctuations as high as 100$\,\upmu$m and therefore comparable with the alignment's requirements.
Possible causes for the observed fluctuations were identified in: vibrations due to the movement of the trolley, noise generated by the laboratory's neon lighting, ground disturbances, or some intrinsic limitation of the instruments.
Changing the sensors' sampling frequency, number of internal averages, light conditions, speed of the trolley's vertical movement, and grounding scheme had no effect on the dispersion of the measurements and no clear cause could be identified and addressed.
The unreliability of individual readings ultimately required the adoption of a different approach.

\begin{figure}[htbp]
	\centering	\includegraphics[keepaspectratio, width=\textwidth]{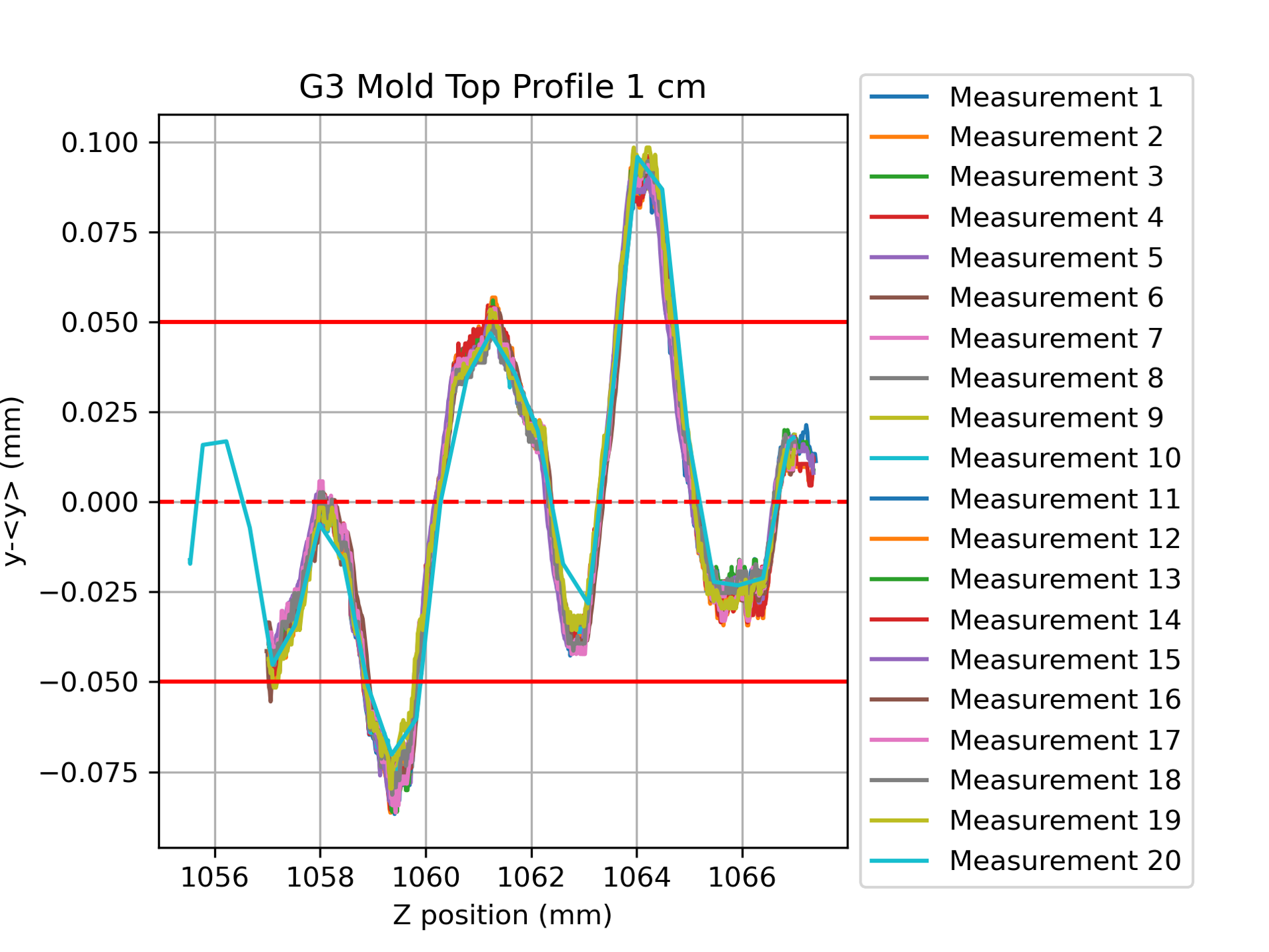}
	\caption[Repeatability of the laser sensor readings]{Repeatability study of the laser sensors' readings.}
	\label{laserrep}
\end{figure}

Despite their internal variance, the scans as a whole remain remarkably repeatable, as shown in figure $\,$\ref{laserrep}.
A new approach to the alignment procedure, based on averaging values over small scans, was therefore devised.
Since the Teflon surface of the mandrels would have been covered by the electrodes and the scanning resolution was even worse when targeting the GEM foils, the top and bottom anodized aluminum surfaces of the mandrels were chosen as targets to determine their tilt.
The two bottom sensors are first zeroed on the average values of a 20$\,$mm scan of the mandrel's bottom black end.
The trolley is raised until its bottom ring is level with the top end of the mandrel, where a second 20$\,$mm scan is collected and used to derive the tilt in both directions.
Tilt can only be corrected on a single direction per measurement, as adjusting the mandrel's position affects both. The direction in which the tilt is greater is usually corrected first.
The trolley is moved again to a position where the chosen sensor's reading matches the average of the second scan, so that the base can be adjusted to bring the reading as close as possible to 0.
This procedure is repeated, usually two to five times, until the measurement of the tilt in both directions is within requirements.

\begin{figure}[htbp]
	\centering
	\includegraphics[keepaspectratio, width=\textwidth]{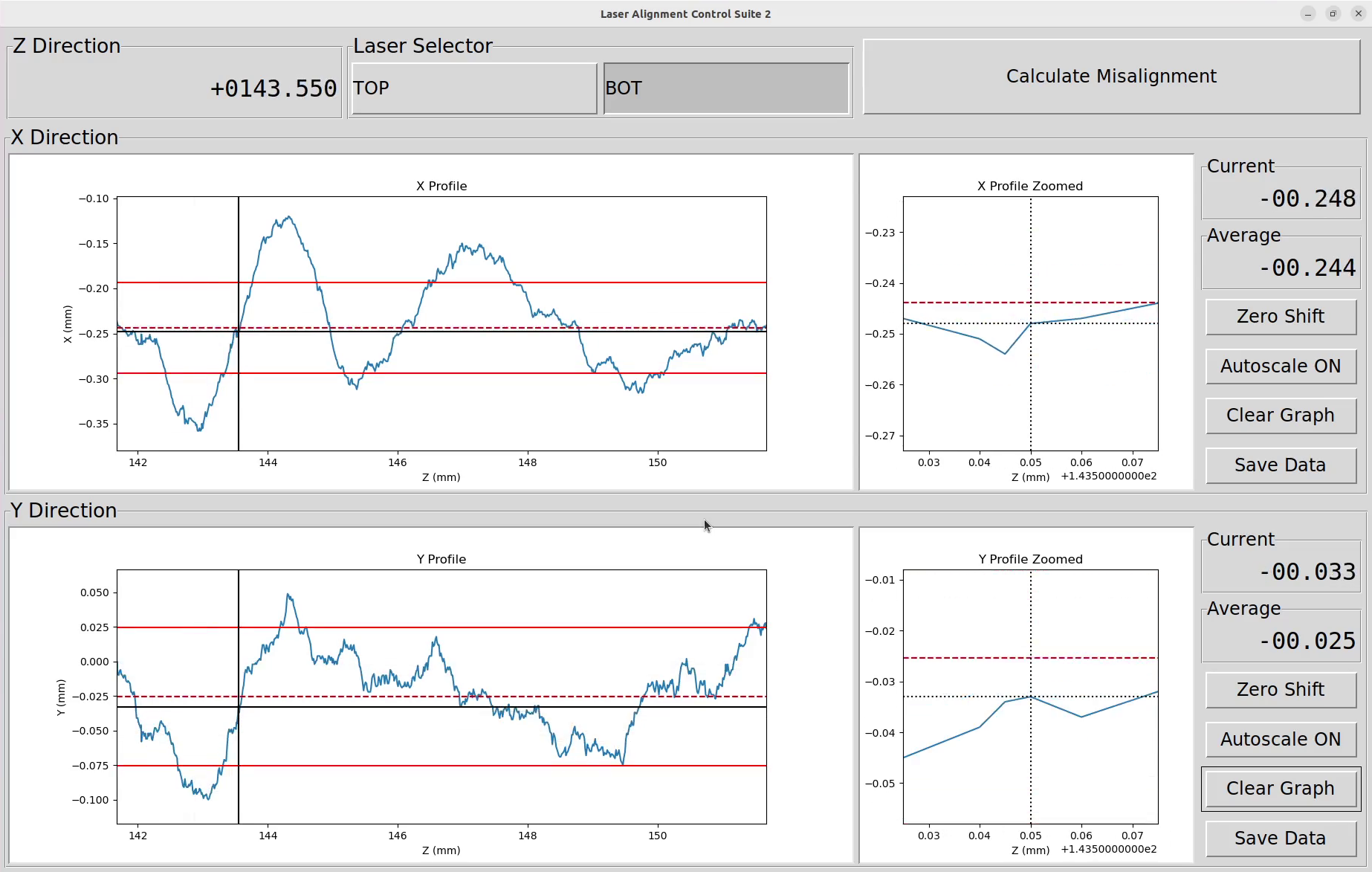}
	\caption[Final version of alignment interface]{Screenshot of the final version of the alignment interface. A function to automatically scale the plots was added to display effectively both small scans and the full profile of the mandrels. Zoomed-in views facilitate precise trolley positioning for zeroing and performing the adjustments.}
	\label{screen2}
\end{figure}

\begin{figure}[htbp]
	\centering
	\includegraphics[keepaspectratio, width=\textwidth]{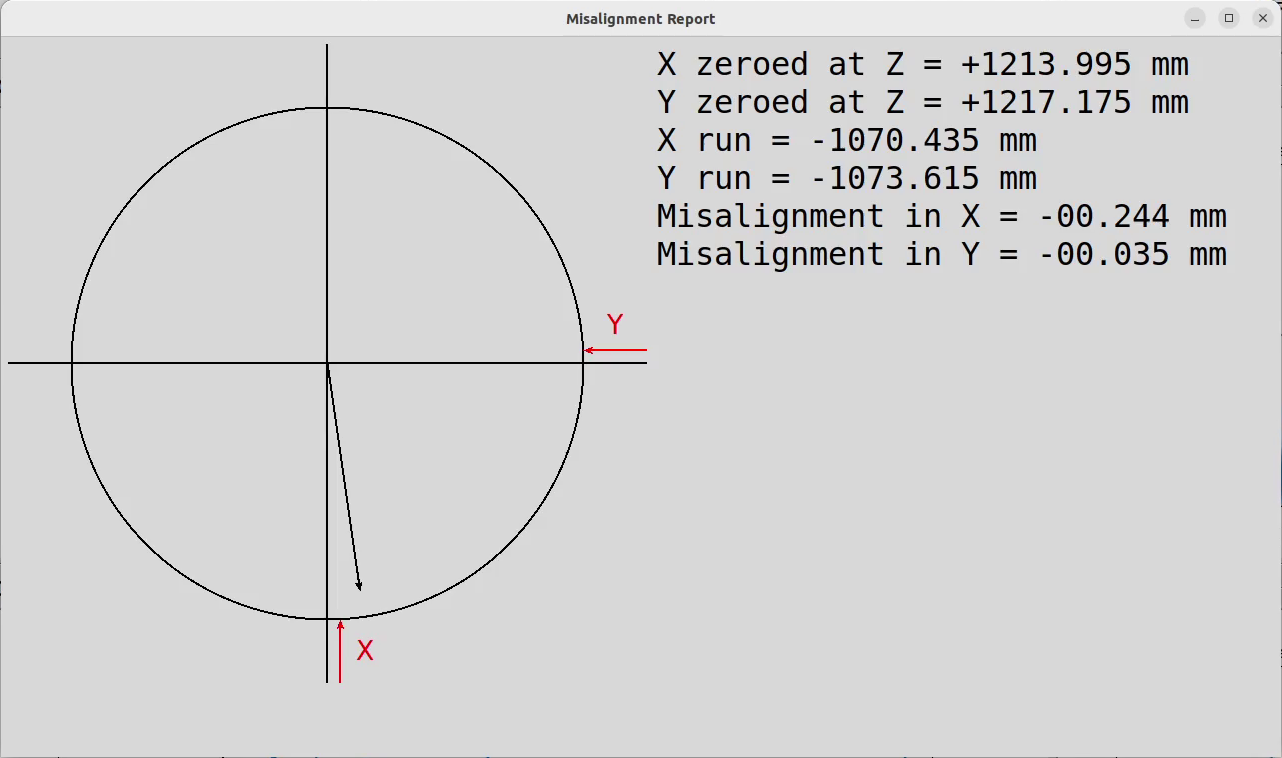}
	\caption[Misalignment report window]{Screenshot of the misalignment report window of the interface.}
	\label{reportwin}
\end{figure}

The final version of the interface, a screenshot of which can be found in figure$\,$\ref{screen2}, was reworked for this new procedure. Secondary, zoomed-in profile monitors were added to aid in the manual positioning of the trolley onto the average values, together with a function to display correction parameters and tilt direction, in figure$\,$\ref{reportwin}.

\begin{figure}[htbp]
	\centering
	\includegraphics[keepaspectratio, width=\textwidth]{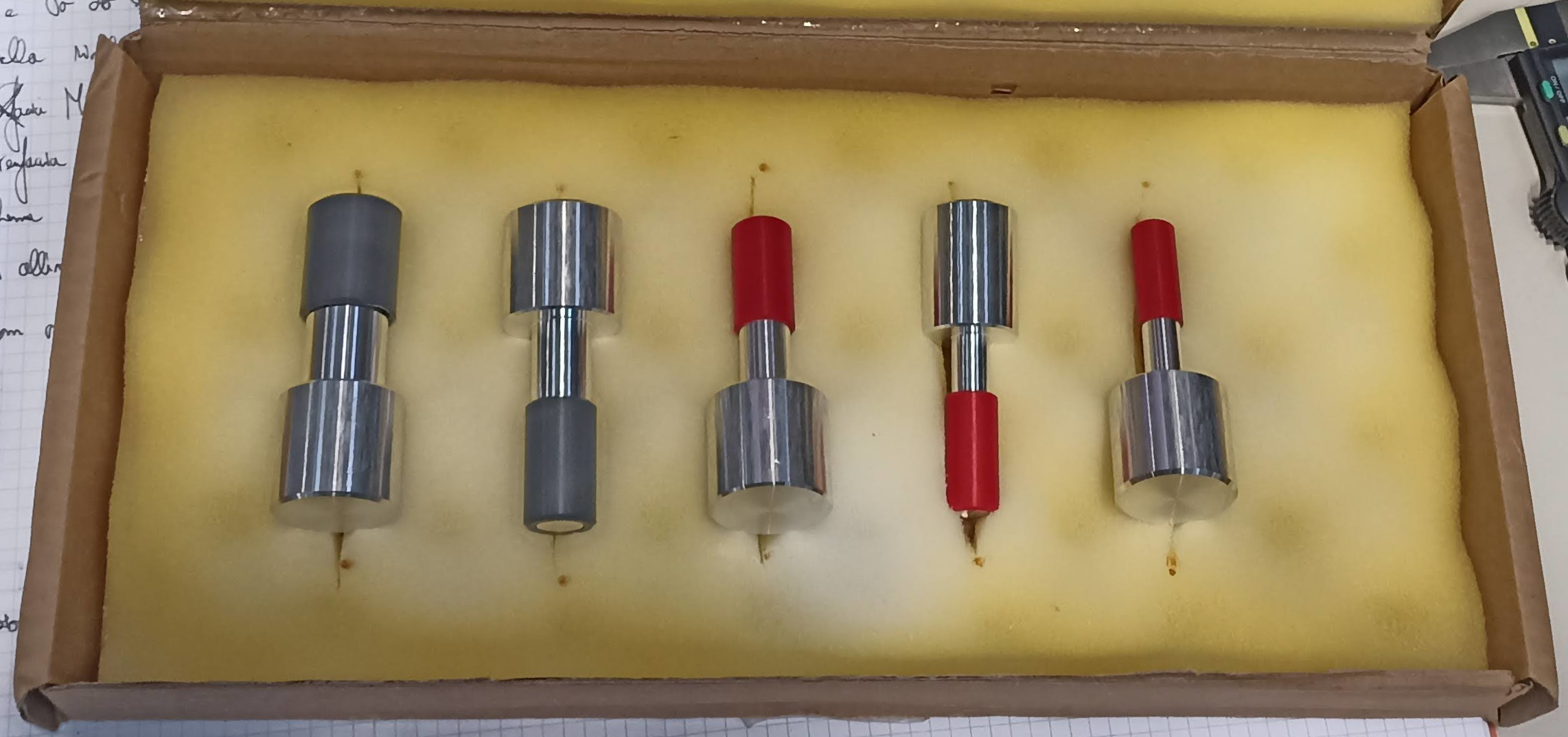}
	\caption[Custom gauge pin set]{Custom set of precision gauge pins, precisely turned to match the nominal clearance between the trolley's top flange and each mandrel's top end.}
	\label{pins}
\end{figure}

\begin{figure}[htbp]
	\centering
	\includegraphics[keepaspectratio, width=\textwidth]{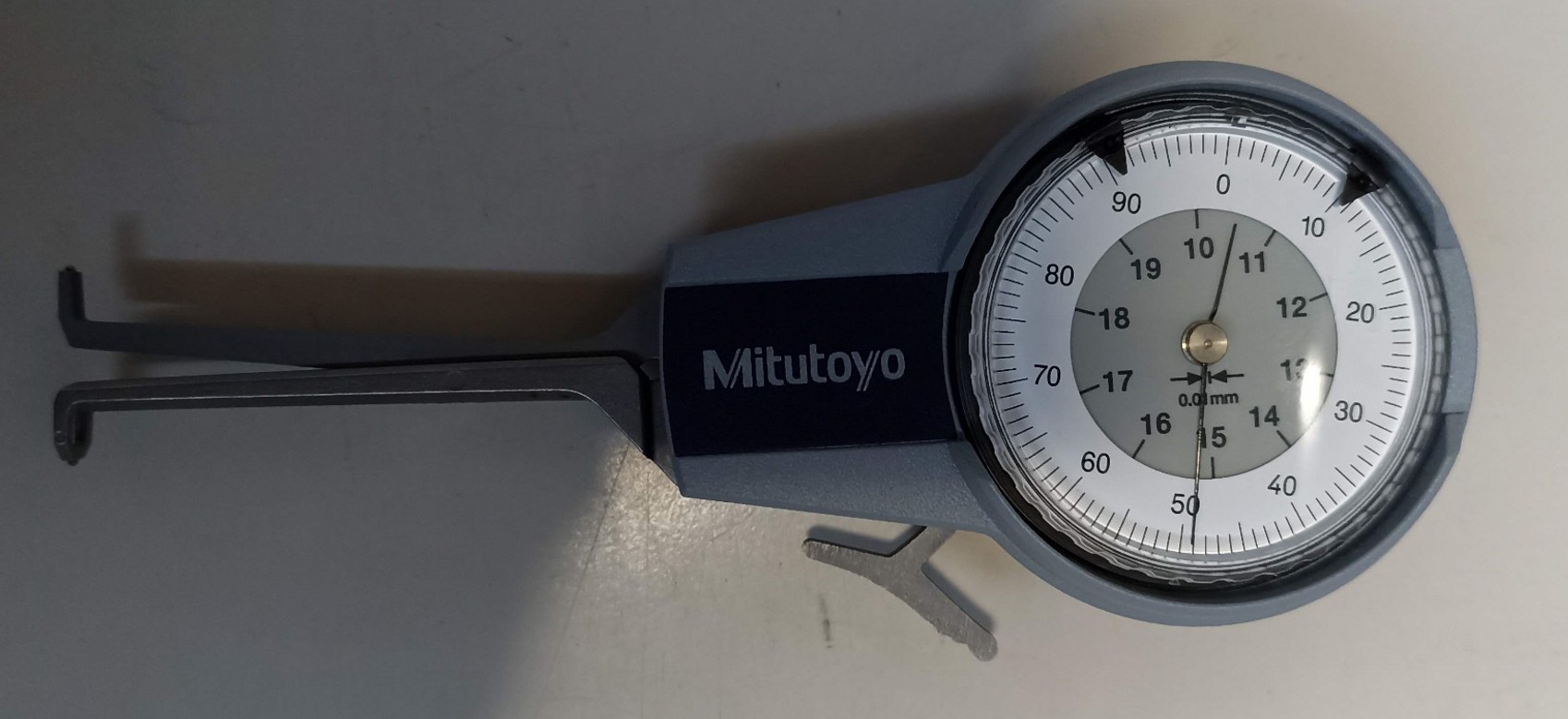}
	\caption[Internal dial gauge]{The internal dial gauge procured for improving the centering procedure.}
	\label{internal_gauge}
\end{figure}

The centering procedure also underwent extensive review, with the intention of improving upon the inaccuracies incurred when measuring with a caliper.
A set of precisely turned pin gauges, closely matching the nominal clearance between the flange and the top of each mandrel, was produced.
Sliding these pins, in figure$\,$\ref{pins}, between the two cylindrical surfaces informs a first round of corrections performed using the control wheels at the base of the machine. Quantitative feedback from using an internal dial measurement gauge, in figure$\,$\ref{internal_gauge}, can then help perform even finer tuning.

\FloatBarrier

\subsubsection{Towards an Iterative Alignment Procedure}
While conducting the tests necessary to the development and validation of the alignment system, two more issues arose:
\begin{itemize}
\item The adjustments of tilt and centering are coupled: correcting the tilt moves the top of the mandrel, affecting its centering, and moving the base to center the mandrel was observed to have an effect on its tilt.
\item Removing a mandrel from the machine and then placing it back onto it has as sizable impact on both centering and tilt.
\end{itemize}
This second problem could be mitigated by manufacturing a new pin for the machine's chuck to hold onto and by taking particular care when positioning the mandrels.
A series of good practices were adopted, such as: cleaning the base and the three-spoke flanges before lowering the mandrels, opening and closing the chuck's jaws several times after making contact with the machine's base, and avoiding to overtighten the chuck.
This allowed to reduce the error accrued from about 100$\,\upmu$m/m to maximum 50$\,\upmu$m/m.
To cope with the remaining error, the effective requirement for tilt alignment had to be shifted to 50$\,\upmu$m/m, which is still within reach thanks to the systems accuracy when making static relative measurements.

\begin{figure}[htbp]
	\centering
	\includegraphics[keepaspectratio, width=\textwidth]{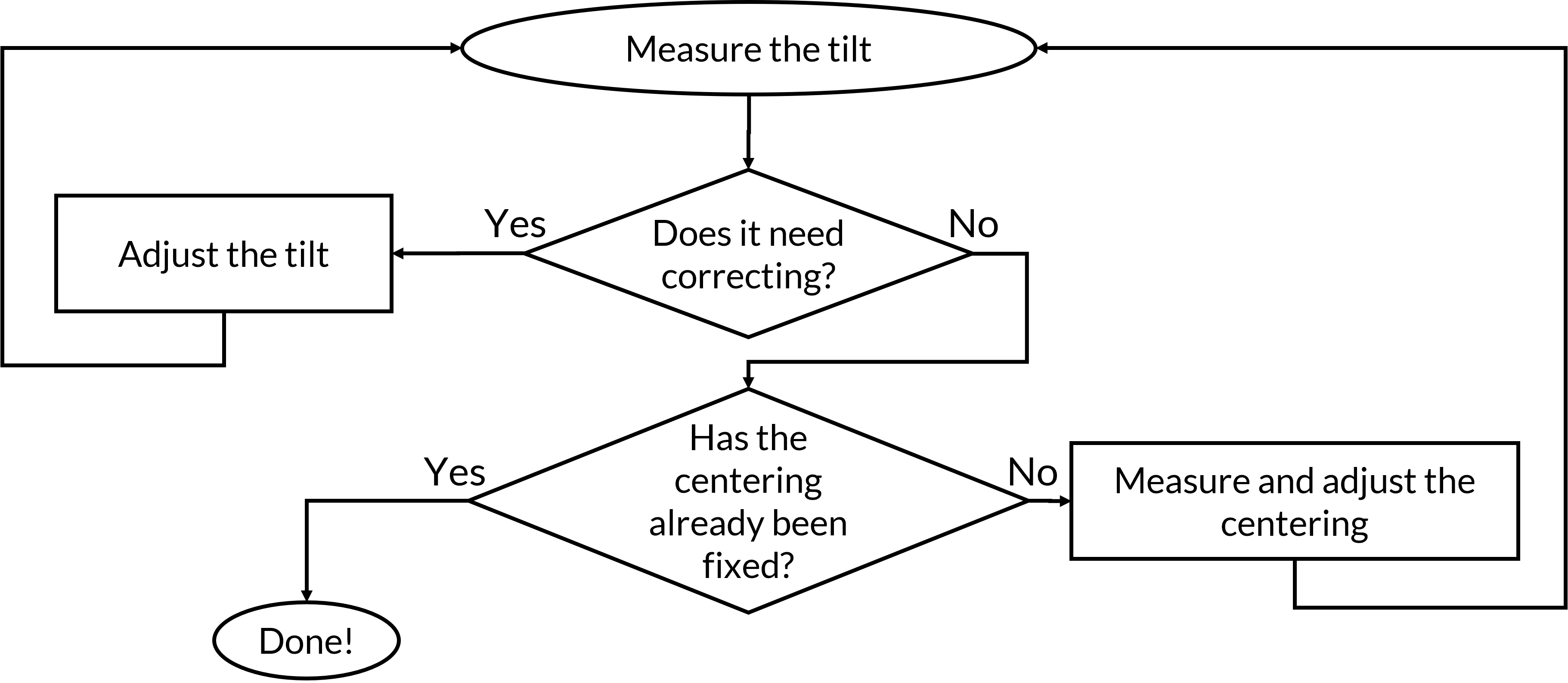}
	\caption[Iterative alignment procedure flowchart]{Flowchart summarizing the iterative alignment procedure.}
	\label{flowchart}
\end{figure}

The correlation of tilt and centering required yet another change of approach with respect to what had been done in the past.
Tilt and centering have to be progressively adjusted and re-checked until both can be brought simultaneously within tolerance, as shown in the flow diagram of figure$\,$\ref{flowchart}. Such iterative procedure requires the trolley to move up and down along its full run many times, increasing the overall duration of a full alignment.

The system and the new procedure underwent a full test consisting in the alignment of all five mandrels of Layer 3.
The anode's mandrel was used as reference and aligned by adjusting the base of the machine.
The other mandrels were aligned by inserting shims underneath their  three-spoke flanges.
Once all mandrels had been thus aligned, each was placed back on the machine to check if the corrections applied still held true, despite the accrued positioning error.
All five mandrels remained well within requirements, and the shims were left in place under the assumption that differences in the mandrels orientation were due to their fabrication and assembly.

\FloatBarrier

\subsection{VIM Commissioning and Alignment in Beijing}
\begin{figure}[htbp]
	\centering	\includegraphics[keepaspectratio, width=.5\textwidth]{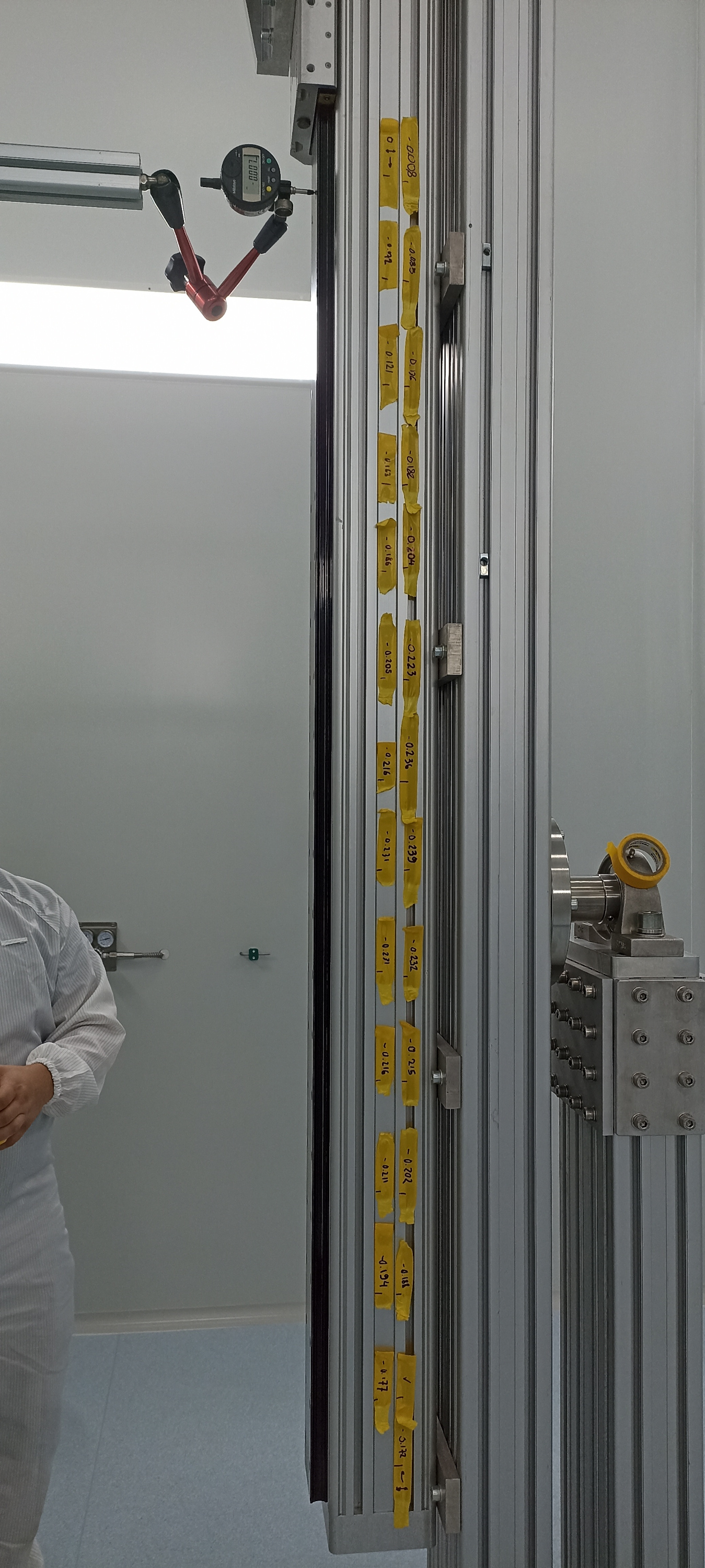}
	\caption[VIM commissioning checks]{Rails' parallelism checks performed during the commissioning of the VIM in Beijing.}
	\label{vimchecks}
\end{figure}

Following the construction of the electrodes, both the VIM and the mandrels were shipped to China by plane.
Upon arrival, the machine had to be unboxed, thoroughly cleaned, moved inside the new cleanroom, and bolted to the floor.
A full check-up of the machine was performed to ensure nothing had been damaged in the transport.
Dial gauges and the trolley's vertical movement were used to check the rails' parallelism, as shown in figure$\,$\ref{vimchecks}.
The results were better than those obtained from similar tests in Ferrara, likely due to the careful placement and the floor being more even.

The alignment of all five mandrels took three working days overall.
Tilt and centering of each mold were checked before making any adjustment.
After that, the anode's mandrel was again taken as reference and re-aligned using the base's adjustments.
Most of the other mandrels did not require any change in the shims' disposition.
Tilt values in both directions before and after the adjustments are reported for each mandrel in table$\,$\ref{beijingalign}.

\begin{table}[htbp]
\centering
\begin{tabular}{lrrrr}
\multicolumn{5}{c}{\textbf{Alignment in Beijing}}                                                       \\\midrule
\multirow{2}{*}{\textbf{Mandrel}}        & \multicolumn{2}{c}{\textbf{Before}}           & \multicolumn{2}{c}{\textbf{After}}            \\\cmidrule(l){2-3}\cmidrule(l){4-5}
 & \multicolumn{1}{c}{X} & \multicolumn{1}{c}{Y} & \multicolumn{1}{c}{X} & \multicolumn{1}{c}{Y} \\\midrule
Anode       & -0.080                & 0.030                 & -0.004                & 0.037                 \\
GEM 3      & -0.177                & 0.073                 & -0.027                & -0.038                \\
GEM 2      & -0.161                & 0.016                 & -0.087                & 0.022                 \\
GEM 1      & -0.177                & 0.057                 & -0.054                & -0.070                \\
Cathode       & -0.088                & 0.151                 & 0.070                 & -0.007        \\  
\end{tabular}
\caption[Alignment results]{Tilt measured on all five mandrels of Layer 3, before and after the adjustments.}
\label{beijingalign}
\end{table}

Switching to a contactless system, which can remain in place and functional during the assembly, proved to have other benefits.
The bottom laser sensors were used to re-center the mandrels during the assembly sequence.
This was achieved by zeroing them on the top end of the mandrel, after extracting an electrode, and then equalizing the distance measured from the next one before beginning the insertion.
The top sensors were used to keep track of the detector's position during the assembly and to check for deformation when inserting the pins to grab the electrodes.
Although some movement could be observed during the rotation, the detector always remained within 120$\,\upmu$m from its original position both when upright and upside down.

The efficacy of the system was finally proven by the unprecedented ease that characterized the insertion of all electrodes.
Despite having reduced the clearance on the radius from a few millimeters to 0.3$\,$mm for the entire length of the detector, no adjustments had to be made on the fly during assembly, and a clearance of at least 125$\,\upmu$m could still be measured between the rings at both ends.

\FloatBarrier

\section{Grids and Vertical Assembly of the Detector}
The design of the grids was slightly altered from the one adopted in the making of the mock-up.
The rod's extremities were chamfered and the diameter of the two outermost rings was slightly reduced to better guide the passage of the fiberglass rings in case of misalignment.
Differently from what was done for the KLOE2-IT, it was decided to introduce an additional grid in the induction gap, bringing the total number of spacers in the detector to four.
Grid parts were manufactured at CERN EST-DEM Workshop but BIEGLO's Dexnyl\textsuperscript{\textcopyright} PEEK film was used in place of VICTREX\textsuperscript{\textregistered} APTIV\textsuperscript{\textregistered} 1000, since it has an opaque face to improve gluing.
All parts had to be thoroughly cleaned with acetone, to remove both shavings and the adhesive that keeps the film in place as it is machined.
The last of the cleaning cycles was performed inside the cleanroom, just before assembly.

The mandrels housing the electrodes were also used as supports to ship the grids.
All electrodes were wrapped first in a Nylon sheet, then in peel-ply textile.
Each grid was then positioned on the corresponding mandrel and a first bag was built around it.
Nitrogen was flowed into the bag to remove as much humidity as possible before sealing it.
A second bag, built around the first, provided additional protection and a sacrificial layer that could be removed before entering the cleanroom to save some cleaning time.
Upon arrival, the grids were inspected and cleaned again, before they could be glued to the electrodes.

\begin{figure}[htbp]
	\centering
	\includegraphics[keepaspectratio, width=.75\textwidth]{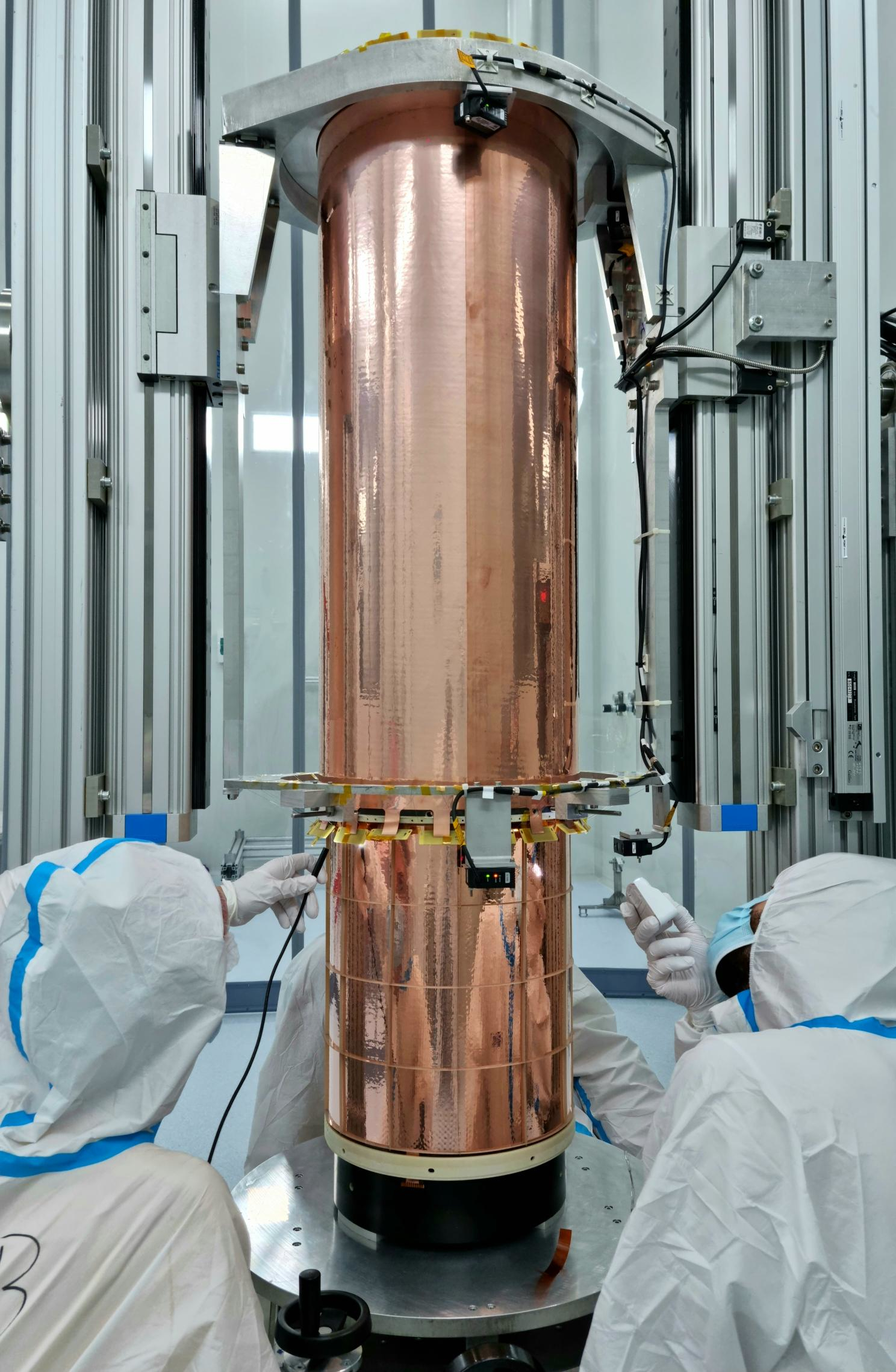}
	\caption[Cathode's insertion]{Photo taken during the insertion of the cathode into the other assembled electrodes.}
	\label{insertion}
\end{figure}

During assembly, the grids reduce the radial clearance form 2 and 5$\,$mm to 0.3$\,$mm over most of the run of the trolley.
Figure$\,$\ref{insertion} shows a snapshot of the cathode's insertion.
The electrodes are monitored from all angles, and any sign of movement of the grid, of the detector, or of the electrodes' surface prompts a temporary stop.
During a stop, the clearance is checked with endoscopes and lenses, while the grid's stiffness to a light touch is used as an indicator of friction.
Most of the movement observed was due to the grid, which is open on one side, not staying perfectly in place and adapting at the passage of the fiberglass ring.
No signs of compression or stress transferred to the electrodes were observed during the entire assembly.

After being sealed on both sides the detector was removed from the machine on July 10th 2023.
Despite the complexity involved in building a detector with spacer grids, the experience accrued in building the mock-up, the extensive quality control campaign, the numerous tests performed, and the improvements to the alignment system led to one of the smoothest constructions in the brief history of the CGEM-IT.
    		\chapter{Commissioning of the CGEM-IT}
\label{commissioning}

\section{Validation of a CGEM Layer}
After their construction, all CGEM layers are subjected to a series of tests meant to assess the success of the sealing and the absence of internal damage.
These are performed before the installation of the on-detector electronics and do not require the detector to be filled with the operating gas mixture.
The validation can be divided in two parts: the gas leakage test and the successive electrical tests.

\subsection{Gas Leakage Test}
\begin{figure}[htbp]
	\centering
	\includegraphics[keepaspectratio, width=\textwidth]{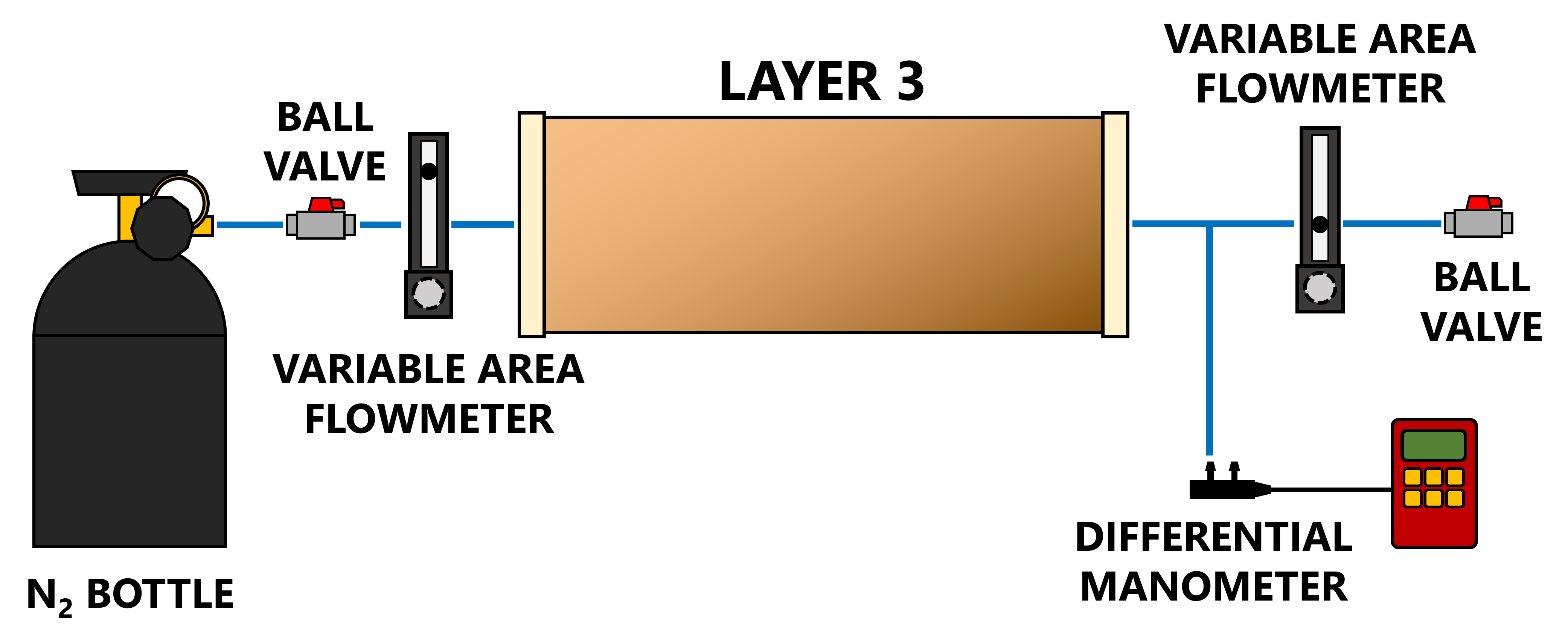}
	\caption[Leakage test setup]{Gas circuit used for measuring gas leakage rate.}
	\label{gasleakage}
\end{figure}

The detector's leakage rate is measured in two distinct ways, both exploiting the gas circuit in figure$\,$\ref{gasleakage}.
High purity nitrogen from a gas bottle is fed to the detector through a ball valve and a variable area flowmeter.
A differential manometer, connected downstream with respect to the detector, provides continuous pressure monitoring.
A second flowmeter and ball valve set is used to control the gas outlet.
Pin valves integrated into the inlets of both flowmeters can be used to regulate the gas flow passing through the instruments.

A first, static measurement of the gas leakage rate is performed by closing the outlet, filling the detector until its internal pressure reaches 20$\,$mbar, closing the inlet, and recording pressure variation over time.
The upstream flowmeter is used to fill the detector in a slow and controlled manner, while the ball valve allows to shut the incoming flow at the reaching of the target pressure.
For the measurement to be accurate, all fittings and pipe connections must be checked thoroughly, so to isolate contributions from the detector.

The second way to measure the leakage rate more closely resembles the normal operating condition of the detector.
In this case, both the flowmeter and the valve located downstream are fully open.
Upstream, the ball valve is fully open as well, while the flowmeter is used to set the input flow to the detector.
The difference in the flow measured by the two instruments provides an indication of the leakage rate at the input pressure and flow values.
All considerations on the gas tightness of the piping still hold true, with the flowmeters' calibration also affecting the accuracy of the results.

The leakage rate of Layer 3 was initially found to be above acceptable values.
Leaks were located using leak-finder fluid and then sealed with epoxy by mounting the detector on a rotating support.
The rate after the intervention was about 1$\,$l/h, depending on incoming flow and upstream pressure.
With the internal gas volume of Layer 3 being 10.4$\,$l, such value falls well within the acceptable limit of 20\% of the layer volume per hour set by the collaboration, allowing the detector to be safely operated inside BESIII.

\subsection{Electrical Tests}
The validation of the layer proceeds with a series of electrical tests, which begin with systematic measurements performed with a multimeter.
Since no high voltages are involved, these can be performed while flowing any inert filtered gas, including dry air, into the detector.
The first quantity measured is the resistance between each micro-sector and both its macro-sector and neighboring micro-sectors.
As different sectors correspond to separate HV channels, all values must be larger than the measurement range of any conventional multimeter.
A null or measurable resistance is indicative of a short circuit, usually due to some contaminant on the foil.

The capacitance between each micro-sector and the corresponding macro-sector is measured next.
Values differing considerably from the nominal capacitance of 3.4$\,$nF hint at some problem with the foil, with contamination being again one of the most likely causes.
A null value in this case may also signify a disconnected sector, which can usually be recovered by resoldering the connector to the HV tail.

\begin{figure}[htbp]
	\centering
	\includegraphics[keepaspectratio, width=\textwidth]{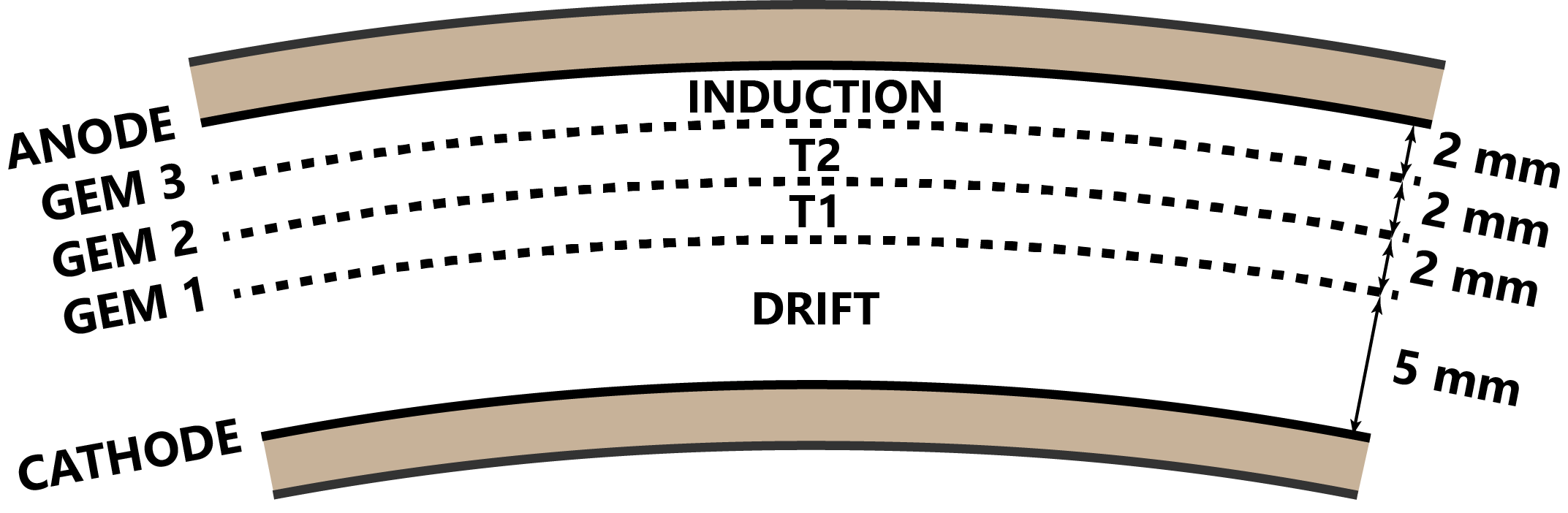}
	\caption[Naming convention of electrodes and fields]{Schematic drawing showing the arrangement of fields inside the detector, the electrodes involved in their establishment, and the size of the gap they inhabit.}
	\label{field_scheme}
\end{figure}

Once these preliminary measurements are completed, the layer can be cabled and connected to the HV power distribution system.
As of this moment, current flowing in each mesh of the voltage divider can be monitored through a dedicated LabVIEW interface.
Each layer has 7 cascading voltage levels, named: induction, GEM 3 (G3), transfer 2 (T2), GEM 2 (G2), transfer 1 (T1), GEM 1 (G1), and drift.
These correspond to the voltages applied to neighboring electrodes to operate the detector.
The schematic in figure$\,$\ref{field_scheme} summarizes the arrangement of the fields inside the detector, the electrodes involved in their generation, and the size of the gaps they occupy.
The anodes are capacitively coupled with grounded copper surfaces on the outer shell of each layer and all levels are conventionally negative.

All of the initial HV validation is performed in pure nitrogen to mitigate the damaging effects of a potential discharge.
Its first step entails raising all levels to 10$\,$V.
Already at these low voltages, the sensitive amperometers of the power distribution system started registering abnormal current absorption.
As the construction ended in July, and despite the tests being conducted inside the cleanroom, rampant leakage currents could be observed due to the high humidity of Beijing's summer weather.

Many precautions were adopted in order to limit the effects of humidity, such as: raising the entire power distribution system above floor level, pointing fans and heating lamps at the endcaps, insulating the HV connections with Kapton tape, and flowing nitrogen inside the patch panels.
These temporary fixes allowed to continue the validation and by the end of September Layer 3 had been powered on at nominal HV settings.
This means it had also cleared all previous steps of the HV validation without issues.
The order of the steps in the procedure aims to spot problems early on and to minimize discharge propagation.
The steps of the HV validation are as follows:
\begin{enumerate}
\item Individual power on of each GEM and each field up to nominal values.
\item Simultaneous power on of all fields (Induction, T2, T1, Drift) at both sub-nominal and nominal values.
\item Simultaneous power on of all GEMs (G3, G2, G1) at both sub-nominal and nominal values.
\item Power on of the whole detector at both sub-nominal and nominal values.
\end{enumerate}
Nominal and sub-nominal HV settings for all three layers of the CGEM-IT are provided for reference in table$\,$\ref{hvset}.
The outcome of the validation was ultimately positive, except for a single micro-sector that appeared to be shorted with its macro-sector.

\begin{table}[htbp]
\centering
\begin{tabular}{@{}lrrrr@{}}
\multicolumn{5}{c}{\textbf{CGEM-IT's HV settings in} $\mathrm{\mathbf{{Ar:iC_4H_{10}\,\,(90:10)}}}$}                                                                                                                                          \\ \midrule
\multirow{2}{*}{\textbf{Setting}} & \multicolumn{2}{c}{\textbf{Sub-nominal}} & \multicolumn{2}{c}{\textbf{Nominal}}\\
\cmidrule(l){2-3}\cmidrule(l){4-5}  
& \multicolumn{1}{r}{\textbf{$\Delta$V (V)}} & \multicolumn{1}{r}{\textbf{E (kV/cm)}} & \multicolumn{1}{r}{\textbf{$\Delta$V (V)}} & \multicolumn{1}{r}{\textbf{E (kV/cm)}} \\
\midrule
Induction                         & 700                                 & 3.5                                    & 900                                 & 4.5\\
G3                                & 200                                 & 40                                     & 275                                 & 55 \\
T2                                & 500                                 & 2.5                                    & 600                                 & 3  \\
G2                                & 200                                 & 40                                     & 280                                 & 56 \\
T1                                & 500                                 & 2.5                                    & 600                                 & 3  \\
G1                                & 200                                 & 40                                     & 280                                 & 56 \\
Drift                             & 650                                 & 1.3                                    & 750                                 & 1.5                                   
\end{tabular}
\caption[CGEM-IT's HV settings]{Sub-nominal and nominal HV settings of the three layers of the CGEM-IT}
\label{hvset}
\end{table}

\FloatBarrier

\section{The Recovery of Pathological HV Sectors}
An electrical cleaning procedure can be attempted to recover pathological sectors, which are responsible for abnormally high currents on the detector's electrodes.
To do so, the detector must be flushed with pure nitrogen, replacing many times its internal gas volume, so that as much residual humidity as possible can be extracted.
A Megger\textsuperscript{\textregistered} MIT420 insulation tester is then used to apply 500$\,$V between the micro-sector under test and its macro-sector.
A healthy sector should display a resistance above 100$\,$G$\Omega$.
Any reading below that is either indication of a problem with the foil or due to external humidity causing leakage on the HV contacts.
Upon the tester ramping up the voltage, an electric arc could be triggered, vaporizing the contaminant.
After the discharge, the nominal resistance between the micro and maro-sector should be restored.
If this does not happen, a more drastic approach is necessary.

\begin{figure}[htbp]
	\centering
	\includegraphics[keepaspectratio, width=.7\textwidth]{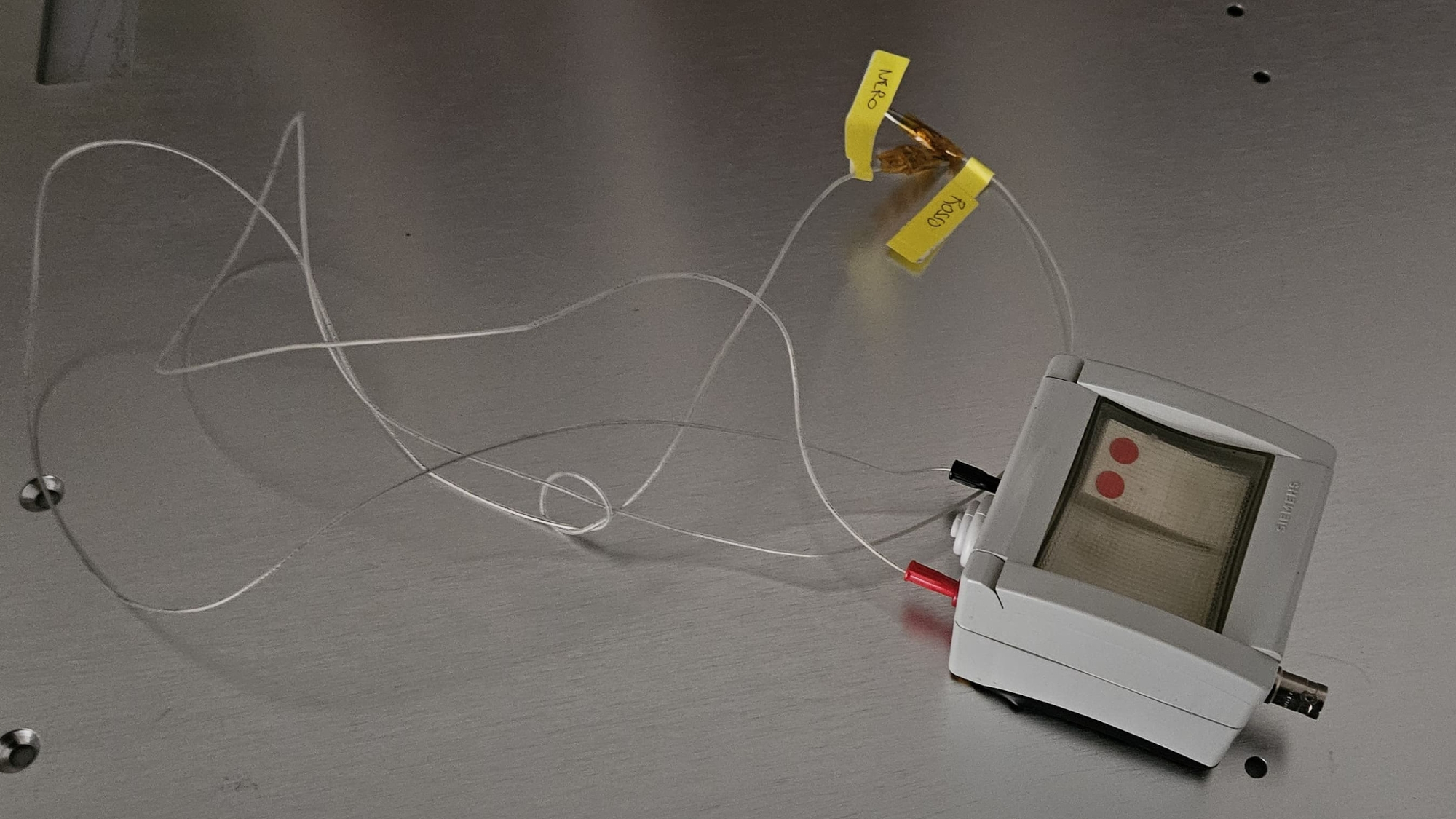}
	\caption[Device for electrical GEM cleaning]{Repurposed domestic power switch used for the electrical cleaning of pathological GEM sectors}
	\label{zapper}
\end{figure}

An alternative procedure was suggested by the manufacturer of the foils$\,$\cite{zprcom}, and closely resembles one of the steps used in the recovery of contaminated GEMs at CERN EST-DEM workshop.
A domestic power switch, in figure$\,$\ref{zapper}, was fitted with HV connectors so that it could be wired to a CAEN N1470 power supply and to the foil.
A voltage of 600$\,$V, a current limit of 5$\,\upmu$A, and the shortest possible trip time of 0.1$\,$s are set on the power supply.
The HV channel is ramped up to 600V and the switch is flipped to quickly apply the voltage to the GEM sector.
The sudden change in potential is more likely to generate an arc, and therefore has proven itself more effective for cleaning larger or more conductive debris.
When the arc is achieved the power supply trips and everything can be reset safely.
The procedure can be repeated many times until the power supply stops tripping.
This may indicate that the obstruction has been removed or that the hole has been disabled and therefore stopped hindering the sector's functionality.

This second procedure allowed the recovery of the shorted micro-sector identified during the validation, together with several others that showed abnormal electrical behavior afterwards.
Moving or handling the detector always carries the risk of dislodging or perturbing contaminants and therefore triggering electrical problems in the foils.
The aforementioned procedures were proven successful in treating all electrical problems affecting single GEM foils.
More complex issues involving electrodes in separate foils, like those observed during the power on of the first Layer 3, cannot be addressed in the same way and remain, as of now, without solution.

\section{Assembly of the Three Layers}
\begin{figure}[htbp]
	\centering
	\includegraphics[keepaspectratio, width=\textwidth]{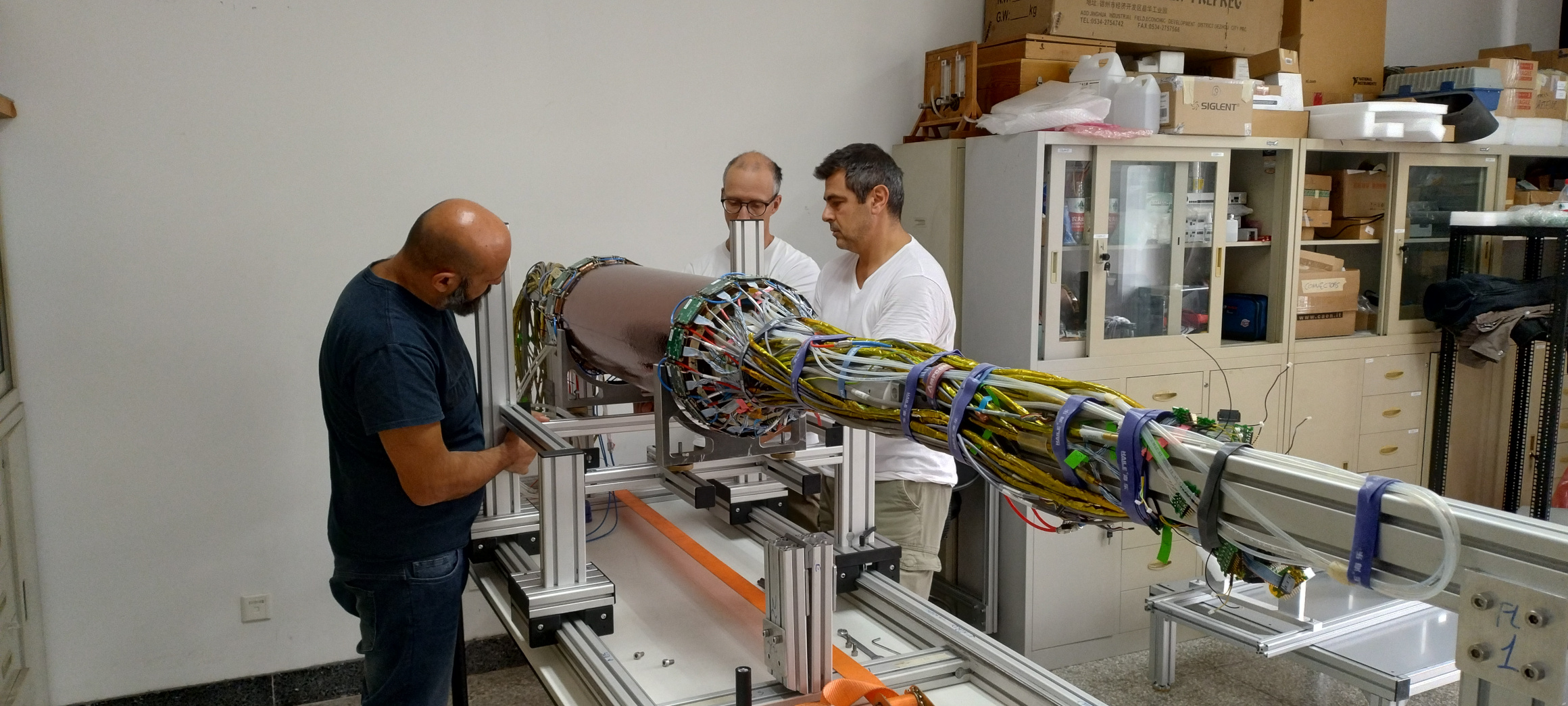}
	\caption[Layer assembly machine]{Custom machine used for the assembly of the three layers}
	\label{layerassymachine}
\end{figure}

Once Layer 3 cleared the validation, the three layers had to be assembled together so that the CGEM-IT as a whole could start collecting cosmic ray data.
A dedicated machine, in figure$\,$\ref{layerassymachine}, is used to assemble the layers horizontally, from the innermost to the outermost.
While on the machine, the detector is supported by the insertion trolley.
This cylindrical support is also designed to install the detector inside BESIII and will be discussed thoroughly in chapter$\,$\ref{installation}.
The insertion trolley is supported by a frame built out of item\textsuperscript{\textregistered} extruded aluminum profiles.
The struts supporting the main horizontal beam can be folded to allow the passage of the layers.
As they are sled in position, the layers are supported by cradles mounted on yet another trolley, moving on rails built into the machine's base.

\begin{figure}[htbp]
	\centering
	\begin{subfigure}{\textwidth}
		\centering
		\includegraphics[keepaspectratio, width=\textwidth]{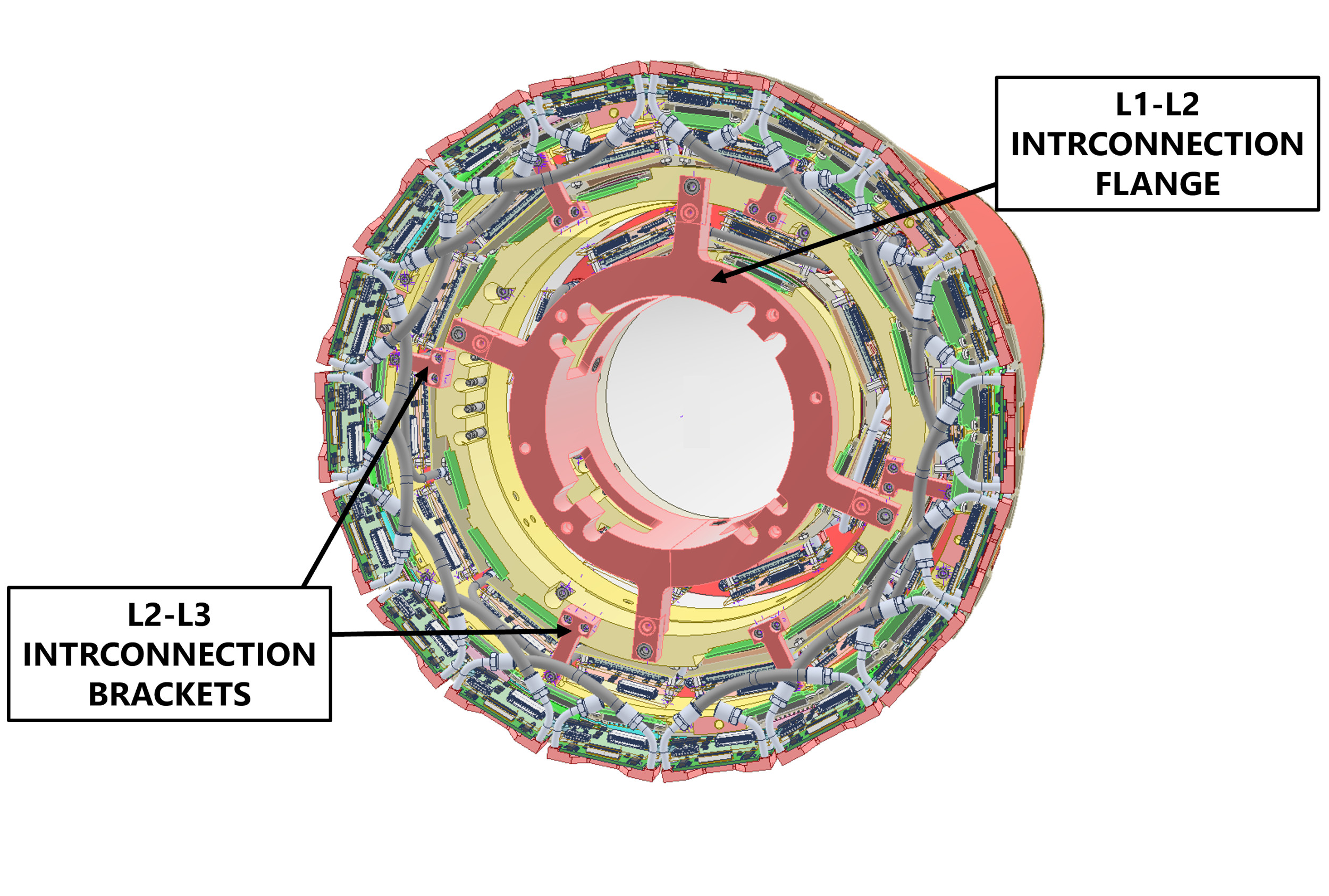}
		\caption[]{The flange connecting Layer 2 to Layer 1 is segmented to allow the passage of the former around the latter during assembly. The number of interconnection brackets differ between the two sides.}
		\label{intercon_flanges}
	\end{subfigure}
	\begin{subfigure}{\textwidth}
		\centering
		\includegraphics[keepaspectratio, width=\linewidth]{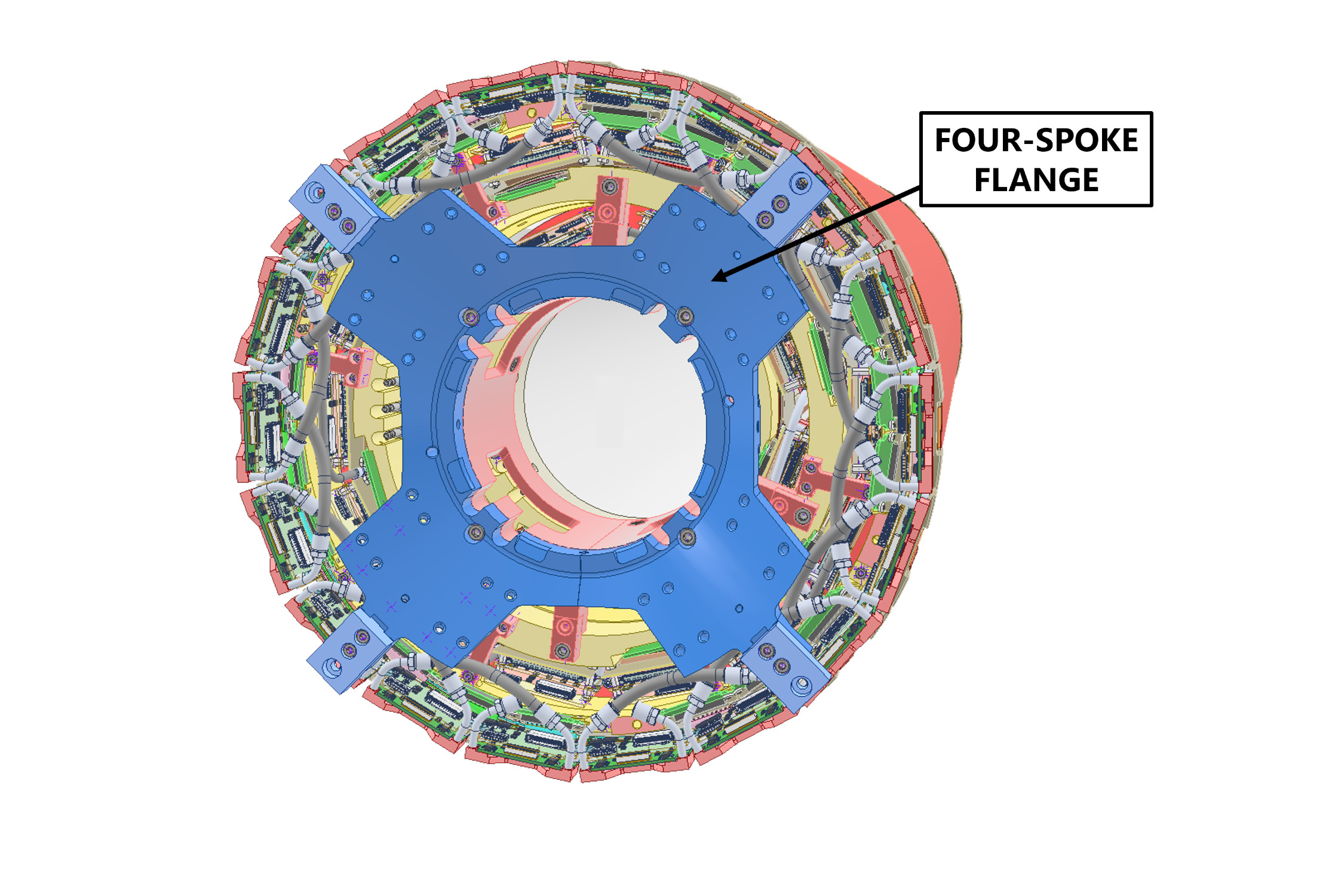}
		\caption[]{The four brackets attached to the four-spoke flange were changed for the installation. The design shown here is the one used to support the detector during the cosmic ray data taking.}
		\label{fourspoke}
	\end{subfigure}
	\caption[CGEM-IT interconnection scheme]{CAD drawing of the eastern endcap of the CGEM-IT; parts connecting different layers are colored in red, while those connecting CGEM-IT and MDC are colored in blue.}
\end{figure}

Layers 1 and 2 are assembled together with their front-end boards installed and fully cabled.
Layer 3's electronics are shaped and oriented differently, and so they have to be mounted afterwards.
Layer 2 is connected to Layer 1 by means of two large fiberglass flanges, while Layer 3 is fixed to Layer 2 by a set of interconnection brackets of the same material, as shown in figure$\,$\ref{intercon_flanges}.
As the detector is very sensitive to compression and stretching, the screws holding the layers together are fully tightened only on one side.
The clearance between the layers is large enough that the alignment can be performed on the fly, using adjustments built into the cradles.
Once all three layers have been connected, and after mounting the on-detector electronics of Layer 3, the four-spoke flanges for connecting the CGEM-IT to the outer MDC, in figure$\,$\ref{fourspoke}, can be installed at each side of the detector.

\begin{figure}[htbp]
	
\end{figure}

\section{A Testing Station for the Acquisition of Cosmic Ray Data}
The machine used for assembling the layers was deemed unsuitable to house the CGEM-IT during the cosmic ray data taking.
The aluminum profile running through the middle of the detector had in fact been observed to be the source of multiple scattering events, and therefore degrade the quality of the data for assessing the detector's performance.
Moreover, the machine's size made it more difficult to encase it in an enclosure for better environmental control.

\begin{figure}[htbp]
	\centering
	\includegraphics[keepaspectratio, width=.6\textwidth]{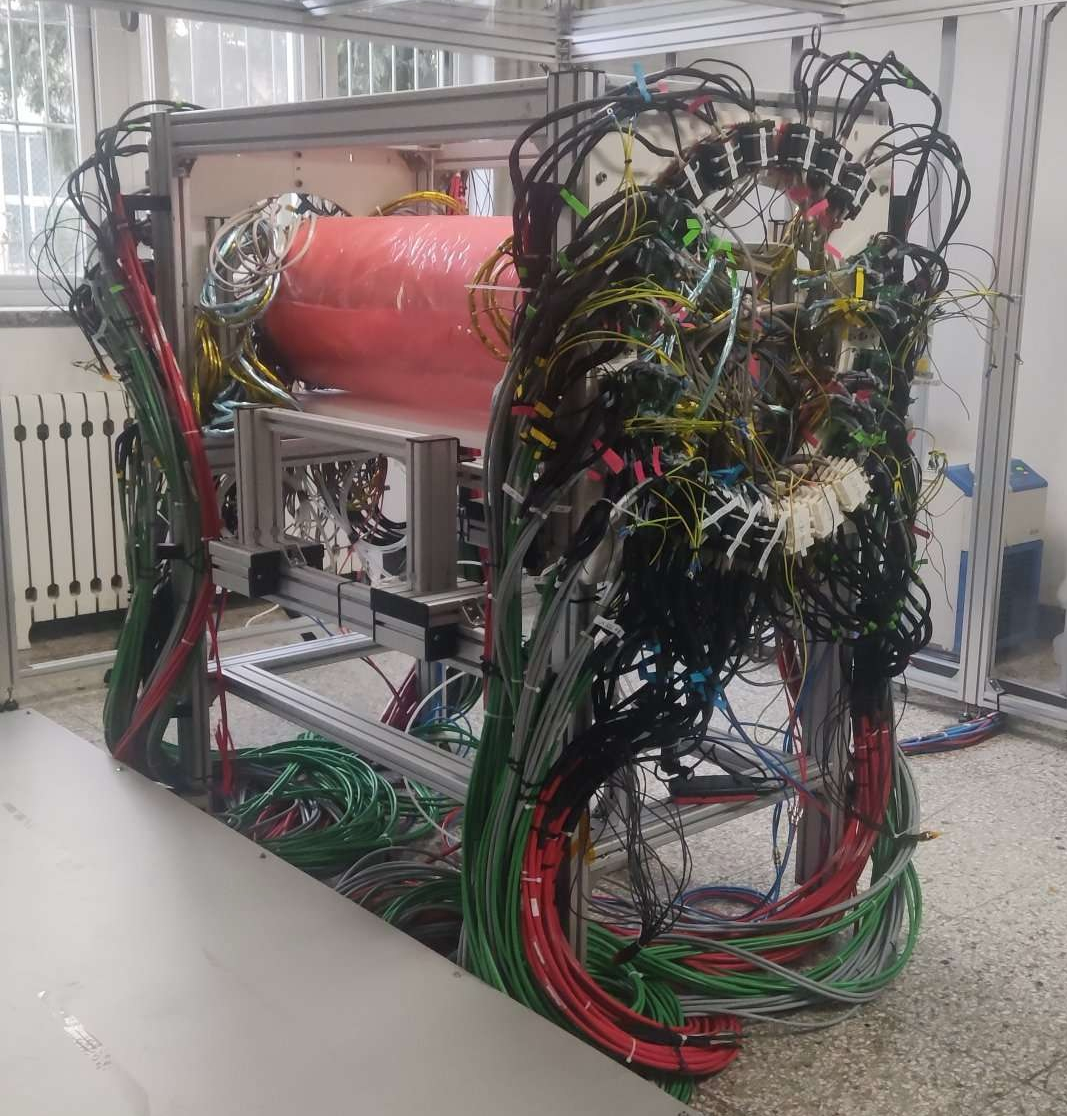}
	\caption[Testing station for the cosmic ray data taking]{The CGEM-IT fully cabled inside the testing station for the acquisition of cosmic ray data.}
	\label{telescope}
\end{figure}

\begin{figure}[htbp]
	\centering
	\includegraphics[keepaspectratio, width=.8\textwidth]{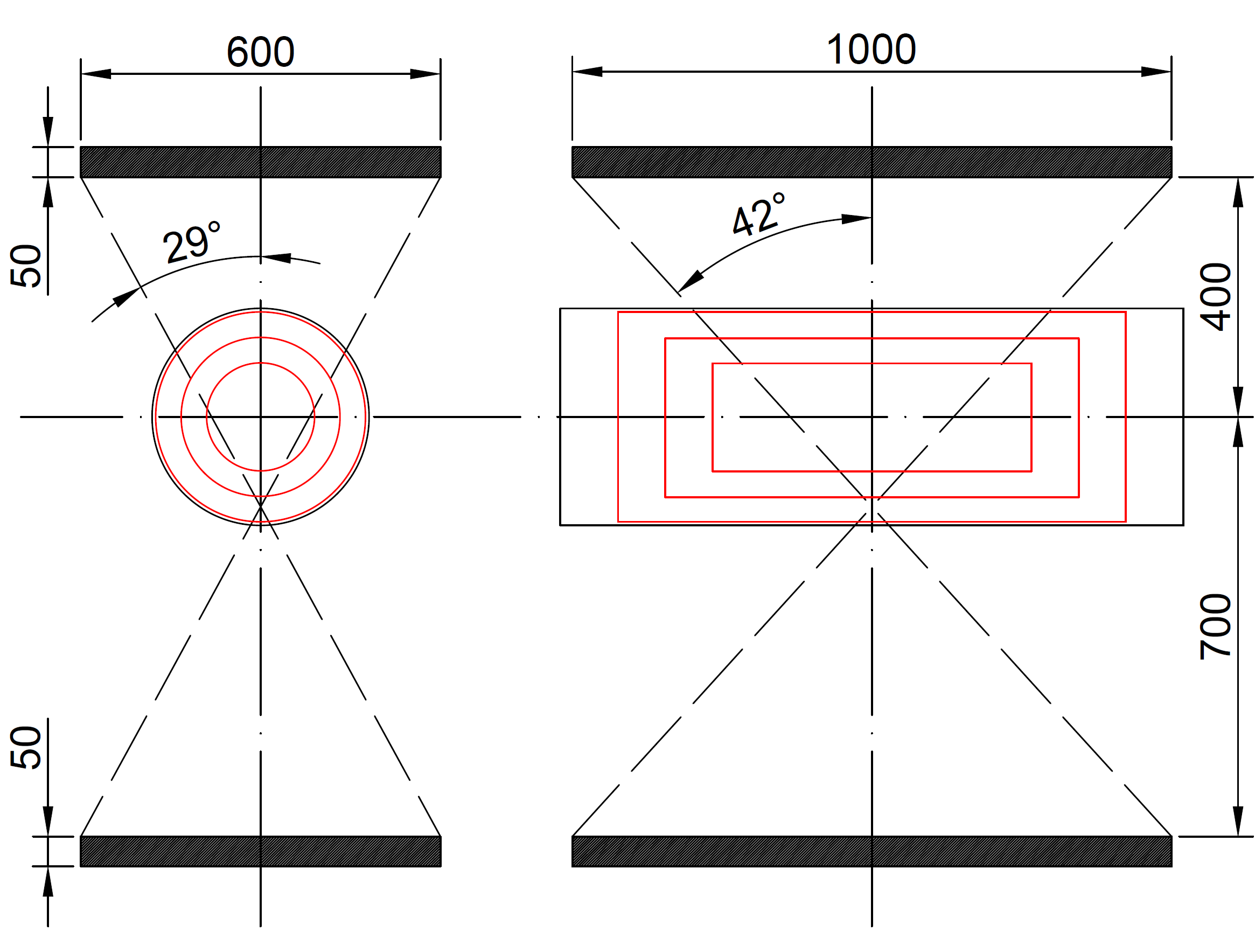}
	\caption[Relevant dimensions of the cosmic telescope setup]{Relevant geometries of the testing station for the determination of the solid angle acceptance. All units are in mm. The shapes in red represent the active surfaces of the three anodes in the system, while the black rectangles are the plastic scintillators. The maximum angles of acceptance in the two directions, assuming perfect alignment, are indicated in the two views.}
	\label{geometry}
\end{figure}

The machine's trolley was therefore refashioned to hold the detector from the mounting points on the four-spoke flanges, meant for fixing the CGEM-IT to the MDC.
The test station's outer framing was made by resizing and stacking the pre-existing supports for the individual layers.
Two modular panels were designed and machined from PVC to house the patch cards connecting short and long haul cables.
The patch card's mounting points were made detachable to ease maintenance and debugging of the towers.
A transparent PMMA slab provided the station with a see-through roof, capable of supporting the heavy scintillator of the trigger system.
The end result, in figure$\,$\ref{telescope}, was a robust structure on wheels capable of housing the CGEM-IT, two $600\times1000\times50\,\mathrm{mm^3}$ scintillators read by PMTs, all the patch cards, and cable holders to secure the terminal part of long haul cables.
The geometry of the testing station determines the solid angle acceptance for the cosmic ray data taking. The relative positions of the scintillators and the detector are reported in the drawing of Figure$\,$\ref{geometry}.

\section{Operation of the CGEM-IT}
The purpose of the data taking was to assess the detector's resolution, efficiency, and stability before its installation in BESIII.
To do so, all the systems necessary for safe and prolonged operation of the CGEM-IT were deployed together for the first time.
Aside from the final HV distribution system and readout chain already described in chapter$\,$\ref{intro}, the detector was provided with temporary, standalone cooling and gas systems.
To ensure safe operation during the prolonged stability studies, an interlock based on a Controllino PLC was installed and integrated with all ancillary services.
A complete description of the CGEM-IT's interlock can be found in reference$\,$\cite{tesimatias}.

\begin{figure}[htbp]
	\centering
	\includegraphics[keepaspectratio, width=\textwidth]{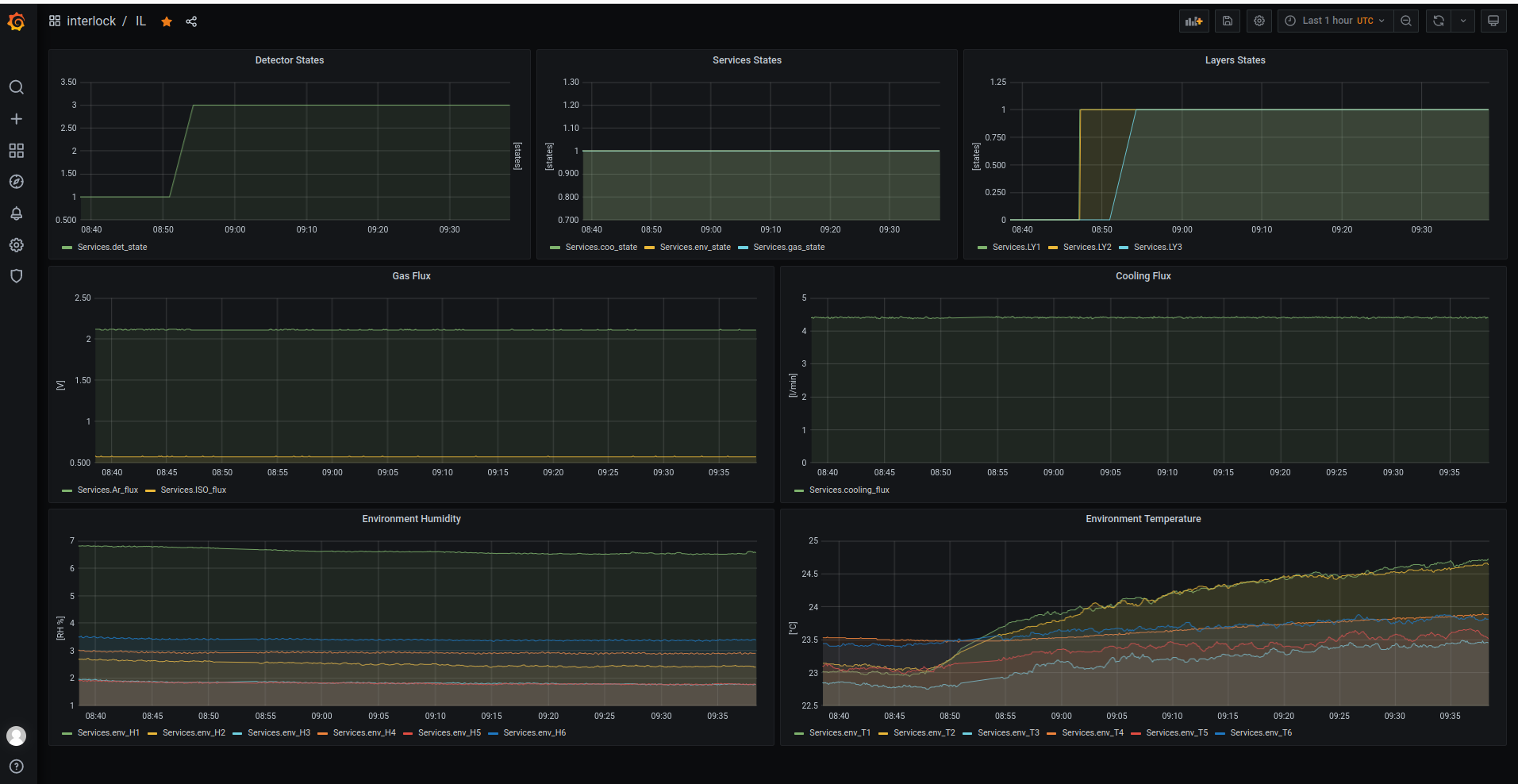}
	\caption[Interlock dashboard]{Online dashboard of the interlock system showing relevant environmental parameters and the status of ancillary services.}
	\label{grafana}
\end{figure}

The control of the detector HV settings and the monitoring of the currents pulled by its electrodes was handled by a LabVIEW interface residing on a dedicated computer.
Configuration and diagnosis of the electronics, together with the management of the data acquisition, was provided by the GUFI (Graphical User Fronted Interface) suite$\,$\cite{Bortone:2021vok}.
An online dashboard based on the Grafana$\,$\cite{Grafana} framework, allowed access to environmental parameters and status information from all ancillary systems, as gathered by the interlock and its network of sensors.
A screenshot of such dashboard is provided in figure$\,$\ref{grafana}.
A fast online data analysis, performed by CIVETTA (Complete Interactive VErsatile Test Tool Analysis)$\,$\cite{Bortone:2021vok}, could also be run and uploaded on a similar dashboard to check the quality of the data as they were being collected.

\subsection{Results of the Cosmic Ray Data Taking}
The analysis of data collected by the CGEM-IT is performed using CGEMBOSS (CGEM BESIII Online Software System), a collection of alignment, reconstruction, and tracking algorithms developed to integrate with the existing BESIII software framework.

\begin{figure}[htbp]
	\centering
	
	\begin{subfigure}[t]{.49\textwidth}
		\centering
		\includegraphics[keepaspectratio, width=\linewidth]{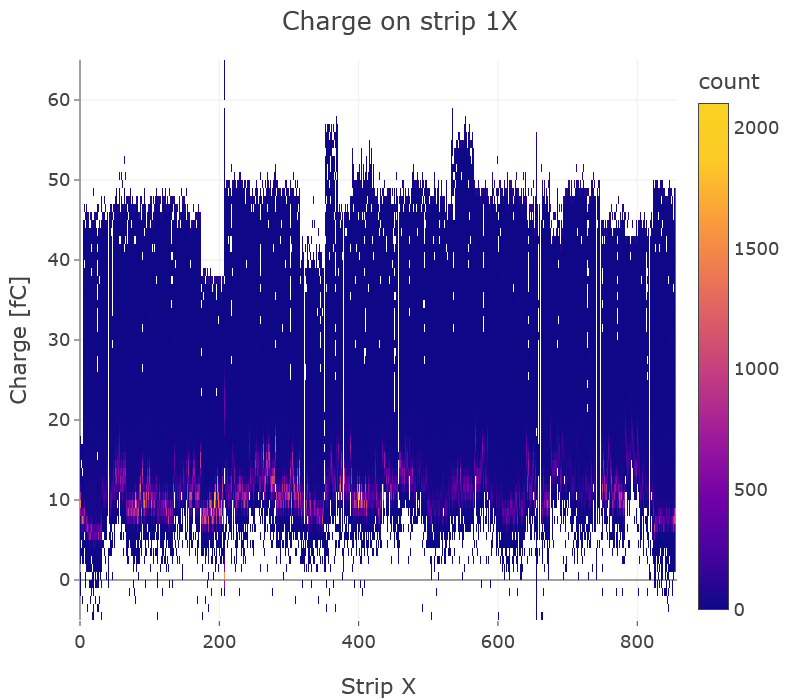}
	\end{subfigure}
	\hfill
	\begin{subfigure}[t]{.49\textwidth}
		\centering
		\includegraphics[keepaspectratio, width=\linewidth]{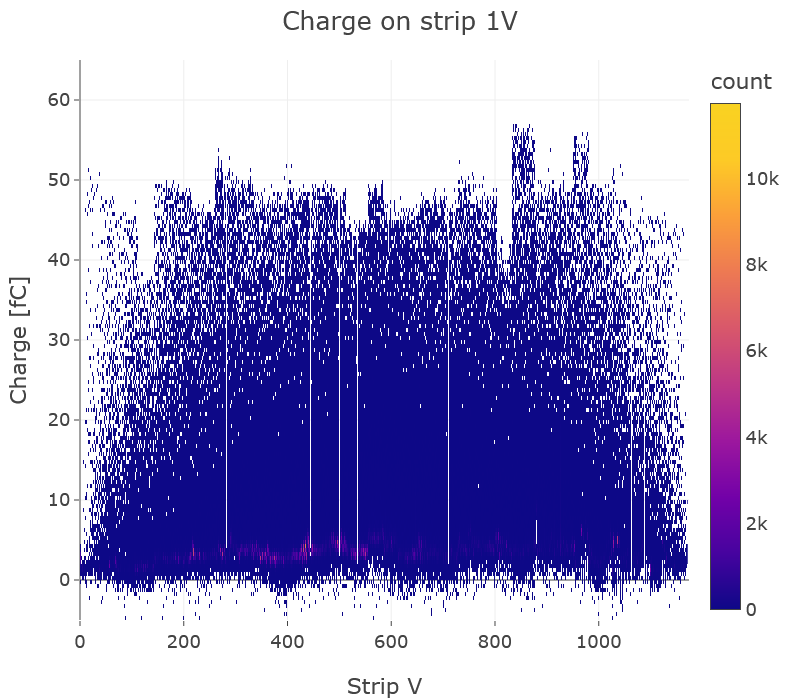}
	\end{subfigure}
	
	\begin{subfigure}[t]{.49\textwidth}
		\centering
		\includegraphics[keepaspectratio, width=\linewidth]{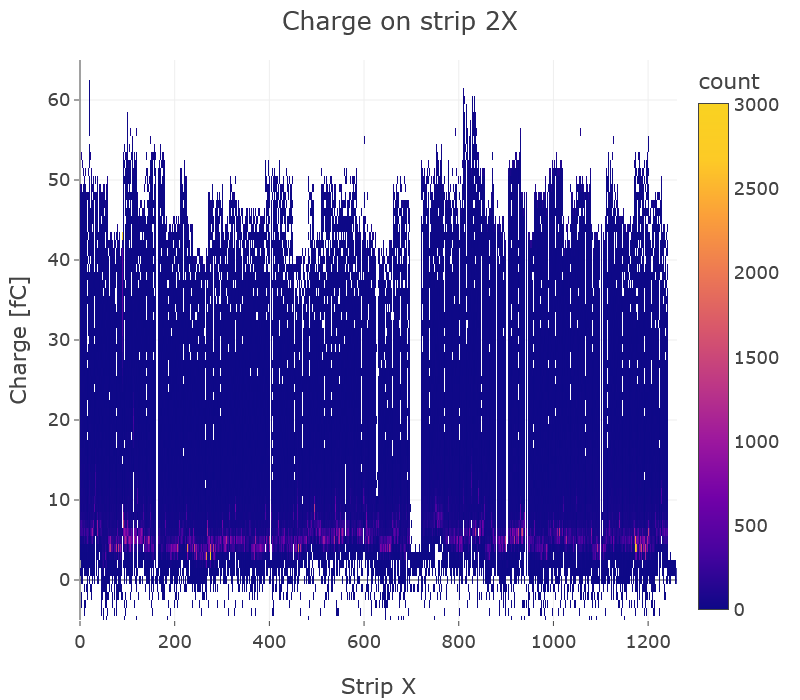}
	\end{subfigure}
	\hfill
	\begin{subfigure}[t]{.49\textwidth}
		\centering
		\includegraphics[keepaspectratio, width=\linewidth]{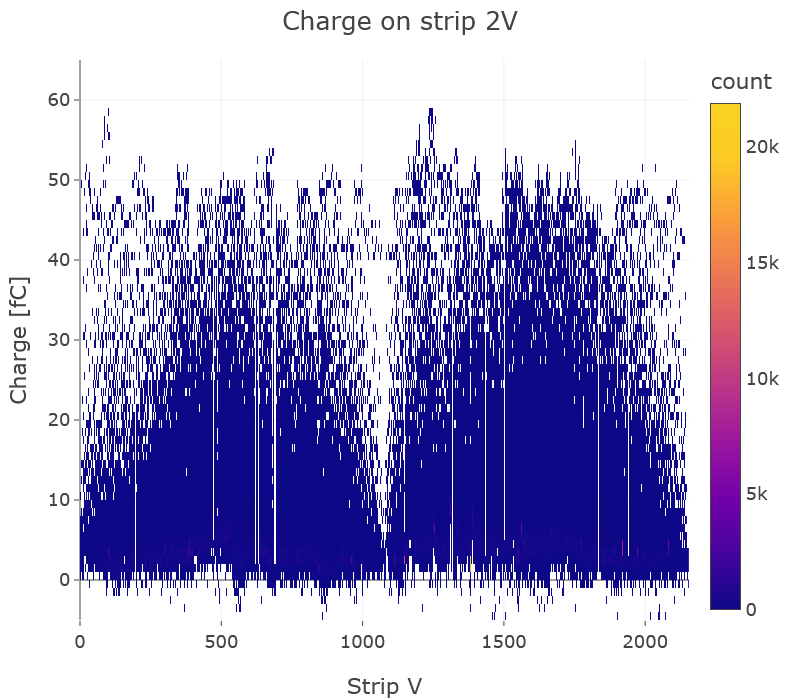}
	\end{subfigure}

	\begin{subfigure}[t]{.49\textwidth}
		\centering
		\includegraphics[keepaspectratio, width=\linewidth]{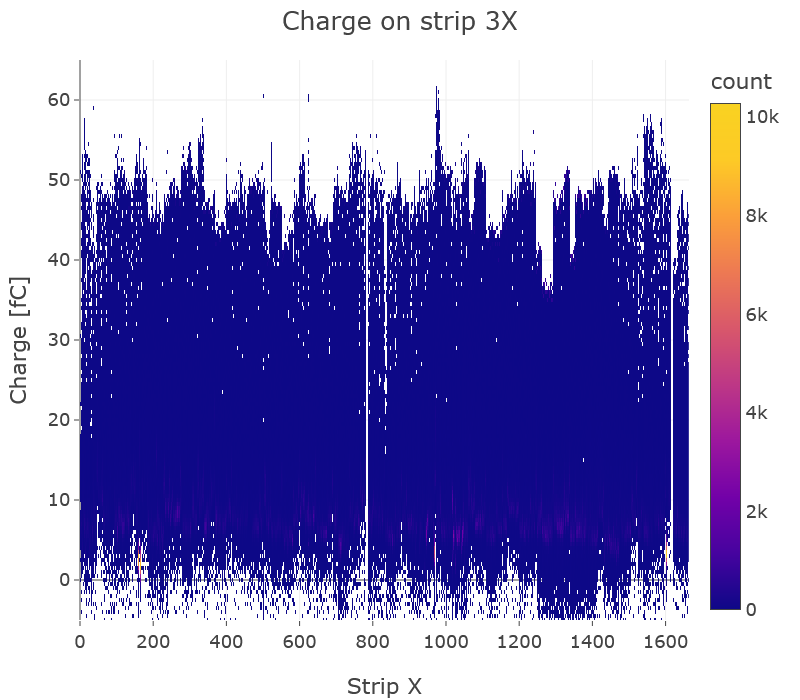}
	\end{subfigure}
	\hfill
	\begin{subfigure}[t]{.49\textwidth}
		\centering
		\includegraphics[keepaspectratio, width=\linewidth]{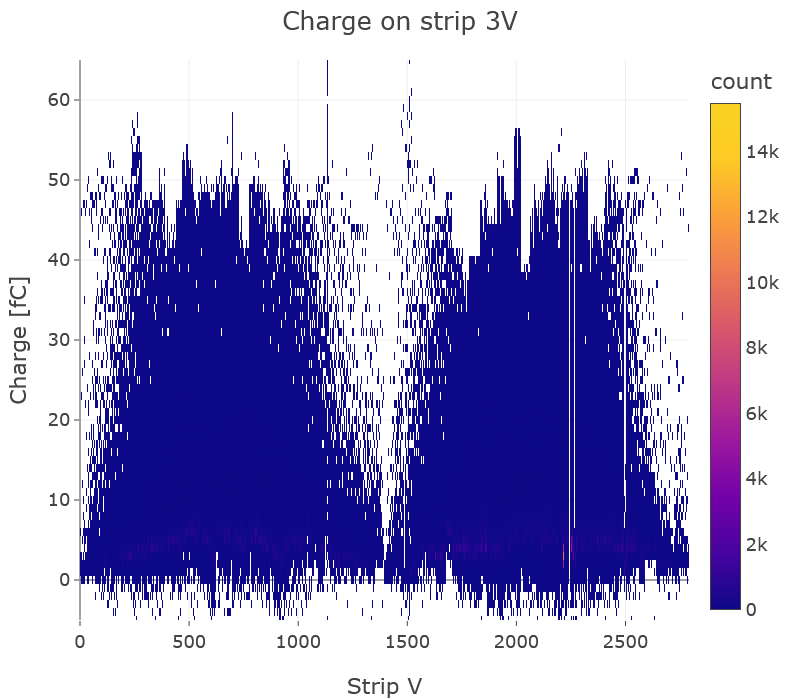}
	\end{subfigure}
	
	\caption[Hit charge per view]{Charge of all hits collected during a typical run for both views of all three layers.}
	\label{qvsstrip}
\end{figure}

Figure$\,$\ref{qvsstrip} shows the charge of the hits collected by both views of the three layers in a typical run.
At this level, no distinction is made between noise and signal hits, those belonging to clusters selected for tracking further into the analysis.
For stereo views, the shape of the distributions is dominated by the different length of the V strips.
As noise accounts for most of the plot entries, the smaller capacitance of the shorter strips reduces the probability of high charge noise hits.
A lesser contribution also comes from the geometry of the test station and the position of the strips in the detector, with those located near its sides less likely to acquire good tracks from cosmic rays.
Differences in the saturation levels of the TIGER chips are responsible for the discontinuity of the maximum charges measured, which give the plots their banded appearance.
The negative charge values are the result of the electronics, which were designed for amplified signals from gaseous detectors, misinterpreting the irregular signal shape of noise hits.

These plots are particularly helpful in diagnosing and locating issues such as: broken channels, disconnected HV sectors, miscablings, and software mismappings.
Broken channels output no data and therefore appear as white vertical bands.
These are mostly due to infant mortality in the on-detector electronics.
They cannot be repaired and thus require the replacement of the affected frontend board, making it wothwile only when a large number of problematic channels are located on the same board.
A damaged connection to the anode's strips is instead responsible for the wider bands observed on layer 2.
Disconnected HV sectors can be localized in the X view plots, whose strips run parallel to them, and appear as bands where the high charge region is sparsely populated.
The lower gain makes high charge hits rarer, but the electronics still registers a regular baseline.
Miscablings and software mismappings have similar phenomenologies: the application of wrong calibration parameters prevents the affected chips from measuring the charge correctly, generating discontinuities in the distributions and erratic values.
Both can be addressed either by swapping the affected cables at either end of the electronic chain or by modifying the software mapping.

\begin{figure}[htbp]
	\centering
	
	\begin{subfigure}[t]{.49\textwidth}
		\centering
		\includegraphics[keepaspectratio, width=\linewidth]{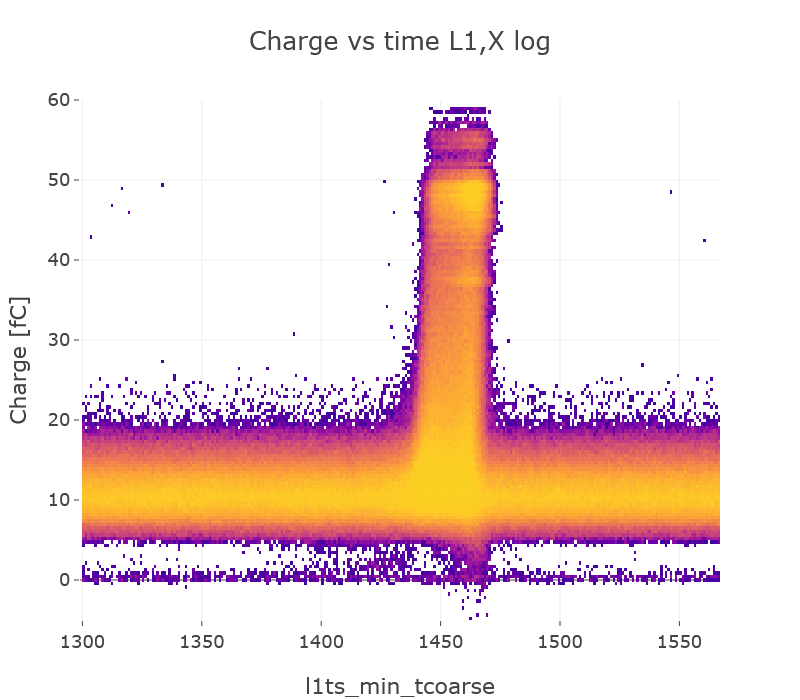}
	\end{subfigure}
	\hfill
	\begin{subfigure}[t]{.49\textwidth}
		\centering
		\includegraphics[keepaspectratio, width=\linewidth]{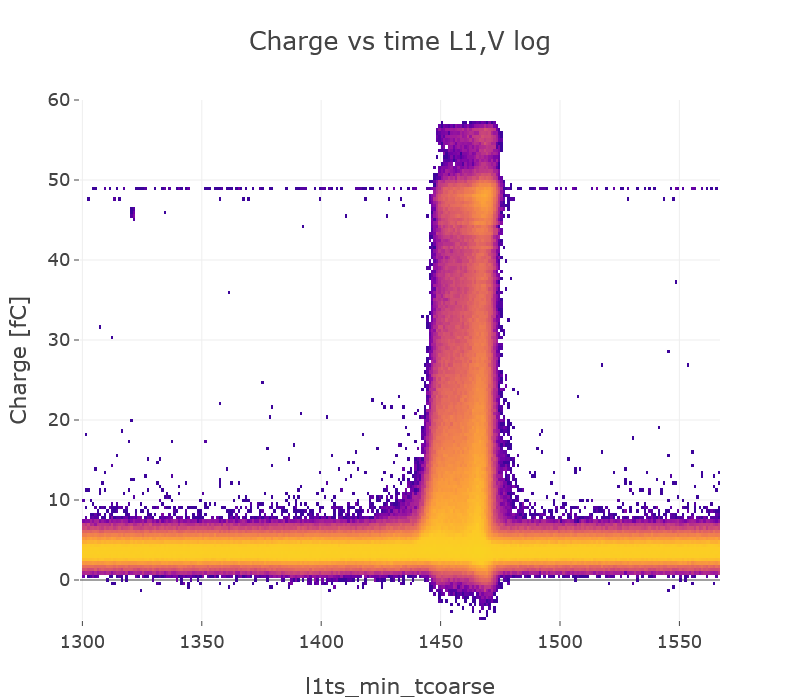}
	\end{subfigure}
	
	\begin{subfigure}[t]{.49\textwidth}
		\centering
		\includegraphics[keepaspectratio, width=\linewidth]{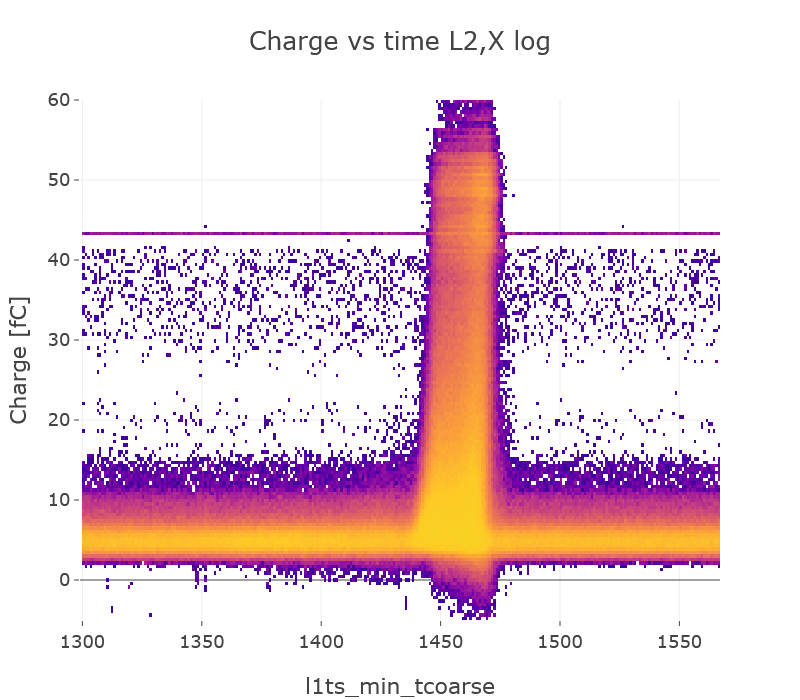}
	\end{subfigure}
	\hfill
	\begin{subfigure}[t]{.49\textwidth}
		\centering
		\includegraphics[keepaspectratio, width=\linewidth]{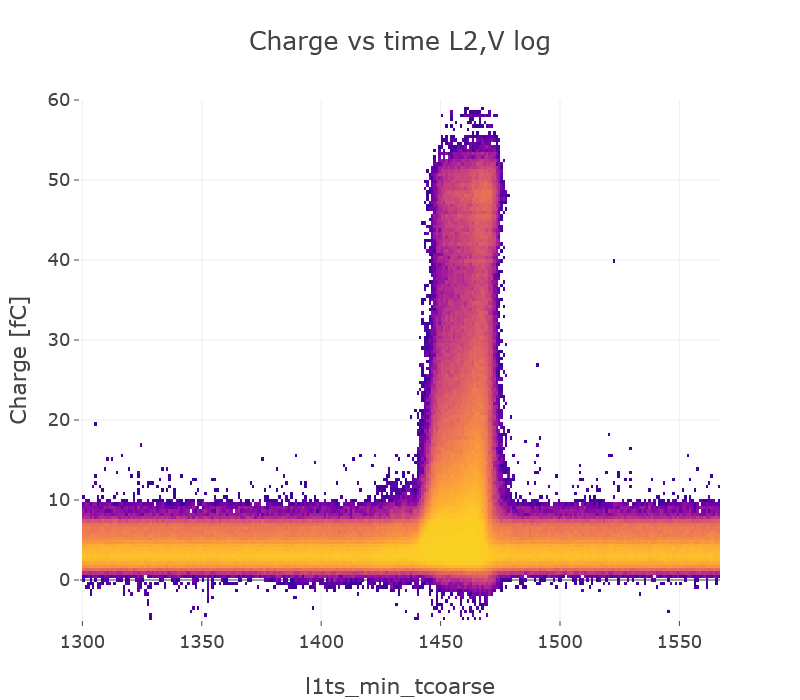}
	\end{subfigure}

	\begin{subfigure}[t]{.49\textwidth}
		\centering
		\includegraphics[keepaspectratio, width=\linewidth]{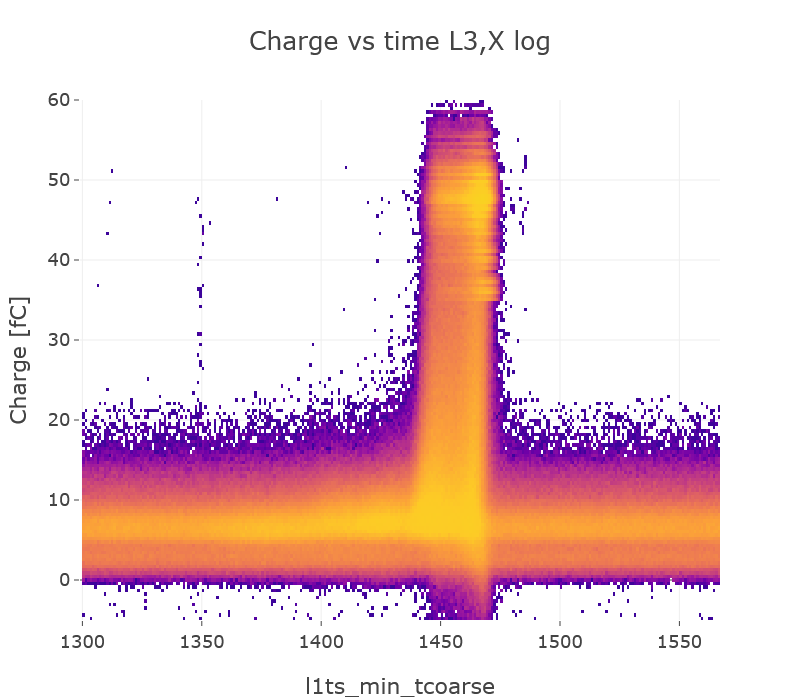}
	\end{subfigure}
	\hfill
	\begin{subfigure}[t]{.49\textwidth}
		\centering
		\includegraphics[keepaspectratio, width=\linewidth]{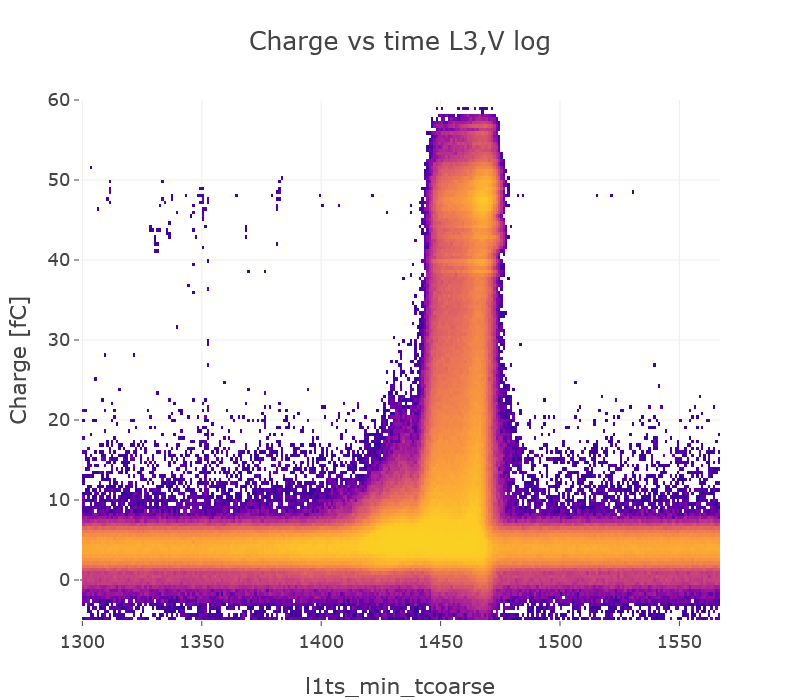}
	\end{subfigure}
	
	\caption[Hit charge in time]{Two-dimensional histograms displaying the time-charge distribution of hits. Signal concentrates in a narrow band, while noise is uniformly distributed}
	\label{qvst}
\end{figure}

Figure$\,$\ref{qvst} shows that most of the high charge hits belong to on-time signal in a narrow band, while noise is uniformly distributed in time.
The longer and the wider the strips are, the noisier they should be, due to their larger capacitance.
This is mostly true aside for the X strips of Layer 1, which display the highest noise level.
This is likely due to the presence of a carbon fiber layer just below Layer 1's anode, which may be the source of parasitic capacitive coupling.
Layer 1's V strips, facing the gas volume and therefore further from the composite skin, are less affected.

\begin{figure}[htbp]
	\centering
	\begin{subfigure}[t]{.49\textwidth}
		\centering
		\includegraphics[keepaspectratio, width=\textwidth]{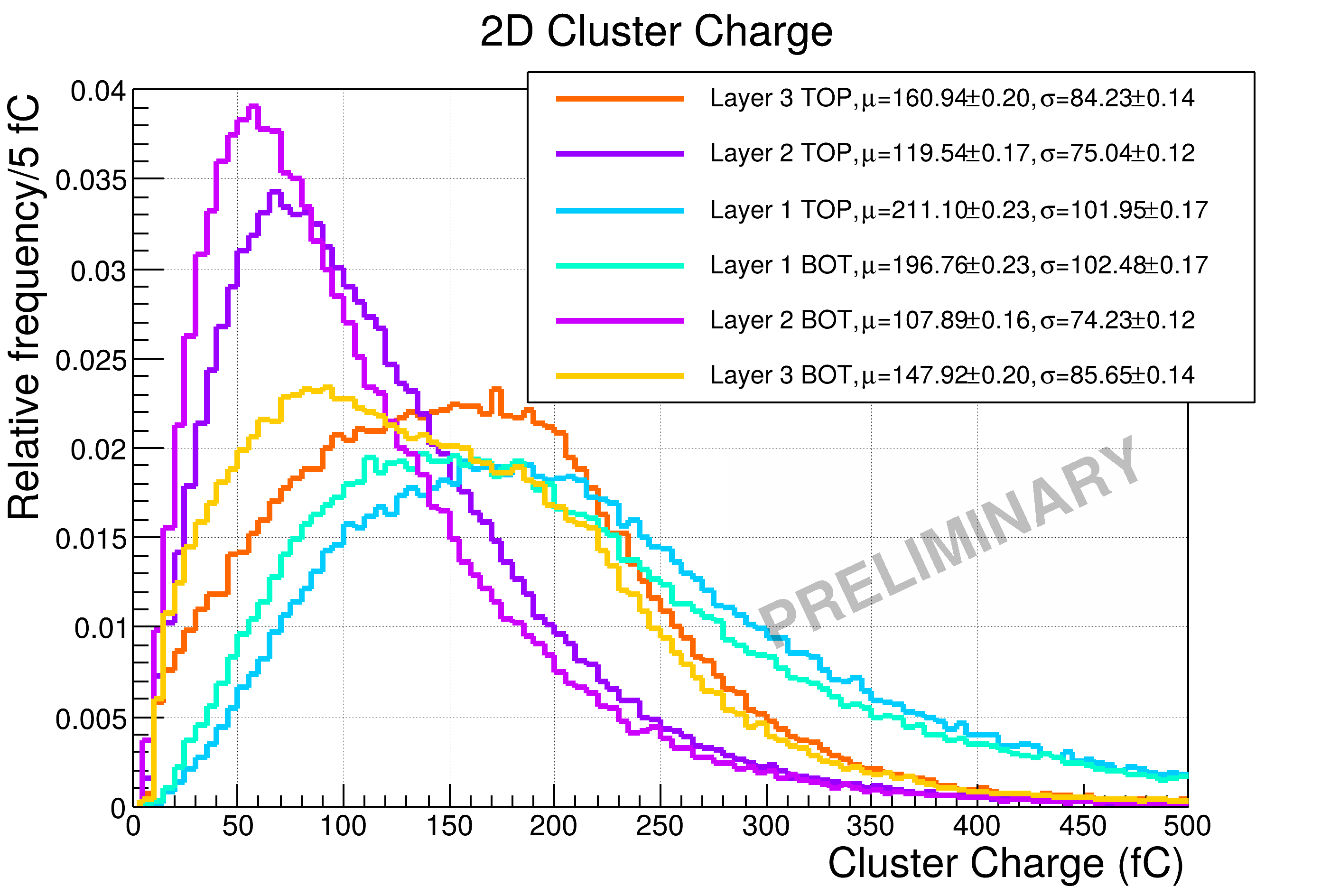}
		\caption[2D cluster charge]{Integrated charge of two-dimensional signal clusters.}
	\end{subfigure}
	\hfill
	\begin{subfigure}[t]{.49\textwidth}
		\centering
		\includegraphics[keepaspectratio, width=\textwidth]{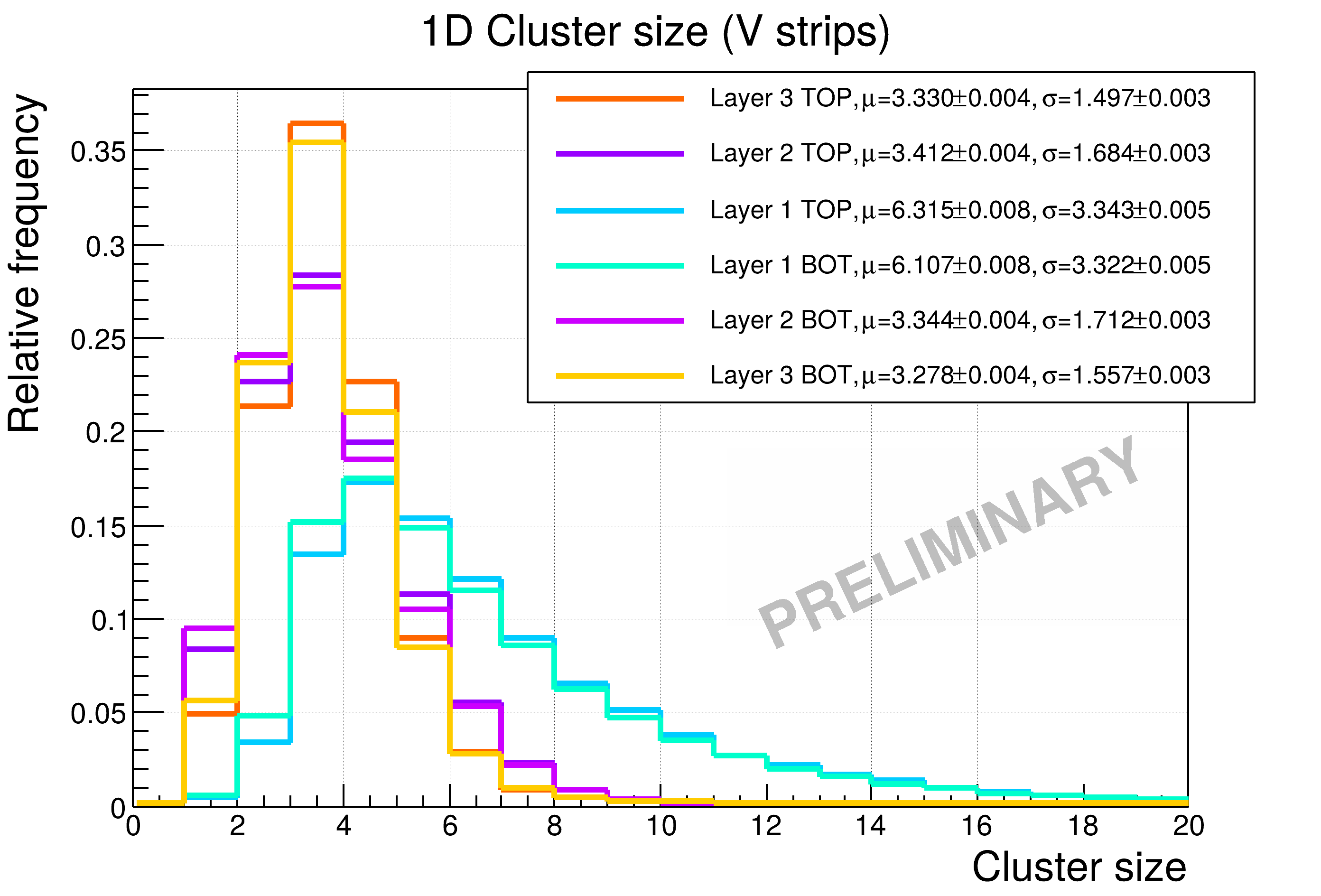}
		\caption[1D cluster size in the stereo view]{Distribution of the hit multiplicity in the stereo view for clusters selected as signal.}
	\end{subfigure}
	\caption[Signal cluster variables]{Distributions of relevant variables for signal clusters, highlighting some differences in the behavior of the three layers$\,$\cite{Gramigna:2024gyu}.}
	\label{sizecharge}
\end{figure}

\begin{figure}[htbp]
	\centering
	\includegraphics[keepaspectratio, width=\textwidth]{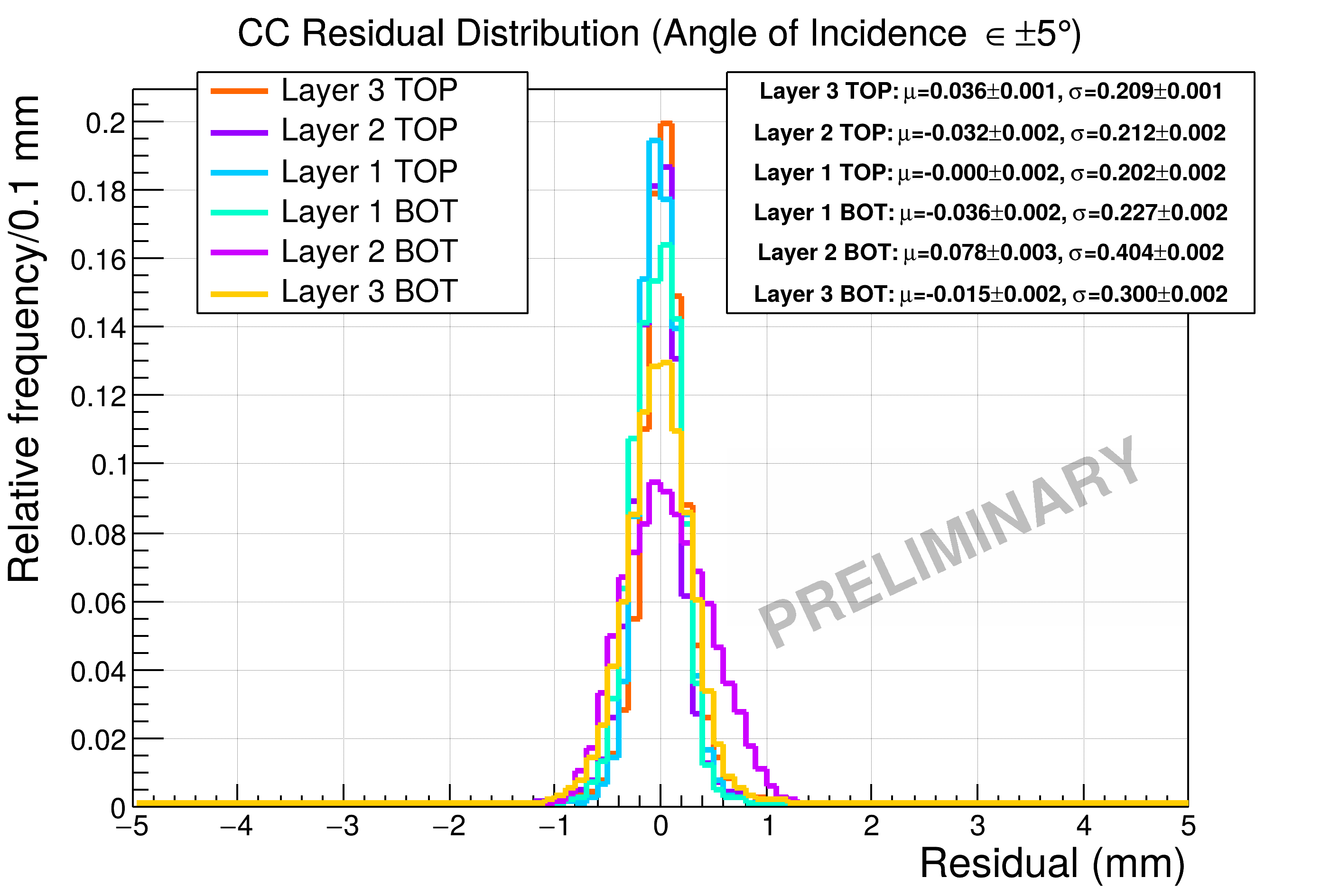}
	\caption[CC residual distribution]{Distribution of the residuals calculated from positions reconstructed with the CC algorithm. Only tracks with angles of incidence within 5$^\circ$ from the perpendicular are included in the study.}
	\label{residual}
\end{figure}

In the analysis of cosmic ray data, each layer is divided into halves along the horizontal plane, with each half-cylinder treated as an independent detection surface.
For layers 2 and 3, this means treating separately the two foils that make up the anodic circuit.
Size and charge of the signal clusters, the ones selected to fit the tracks, are shown in the plots of figure$\,$\ref{sizecharge}.
Layer 1 collects more charge and has larger clusters.
The reasons for this are still under investigation, with some hypotheses being: the greater rate at which its gas volume was renewed; the anode capacitively coupling with the carbon fiber layer just below it leading to more noise hits being grouped into the signal clusters; less defocusing due to its smaller radius of curvature.
The size of the clusters in layers 2 and 3 appears similar but the charge collected by Layer 2 is substantially lower.
This is due to the fact that Layer 2, the oldest and most fragile layer, was being operated at lower transfer fields to safeguard its electrodes.
It was later powered on at the nominal 3$\,$kV/cm and displayed no electrical instability but, due to the beginning of the detector's installation, no data was taken in this configuration.
An asymmetry can be observed in the charges collected by the top and bottom halves of layers 2 and 3. At the time of writing, the reason for this discrepancy, mirrored also in the residuals distribution of figure$\,$\ref{residual}, is not well understood.
The leading hypothesis being leaks located in the bottom halves of the two layers, possibly in correspondence of the gas outlets, injecting oxygen and nitrogen into the system and thus lowering the gain.
Tests have been performed at higher flows, leading to higher internal pressure and more volumes per hour of gas being changed.
Raising the flow increases the gain, partially reducing the discrepancy.
The gas distribution system employed in the cosmic ray data taking was limited in maximum flow, so the discrepancy could never be fully resolved in this way.
More tests are planned for when the final gas distribution system is going to be implemented after the installation.

Figure$\,$\ref{residual} reports the residuals' distribution for tracks at small angles of incidence. Residuals are the distances between the position of the particle reconstructed by the surface under study and the expected one, obtained tracking the particle with all other detection surfaces in the system and calculating the intercept.
The position of the clusters is here reconstructed using the Charge Centroid (CC) algorithm, which performs better when the tracks are almost perpendicular to the surface. 
The $\upmu$TPC reconstruction algorithm, meant to improve performance for angled tracks, is still being fine-tuned and does not yet reach its target performance.
The standard deviation of the residuals distribution for most half layers is about 200$\,\upmu$m, with the aforementioned bottom halves of layers 2 and 3 at 400$\,\upmu$m and 300$\,\upmu$m, respectively.
Since these values still include a contribution due to the tracking system, the actual spatial resolution is expected to be even better.

\begin{figure}[htbp]
	\centering
	\begin{subfigure}[t]{.49\textwidth}
		\centering
		\includegraphics[keepaspectratio, width=\textwidth]{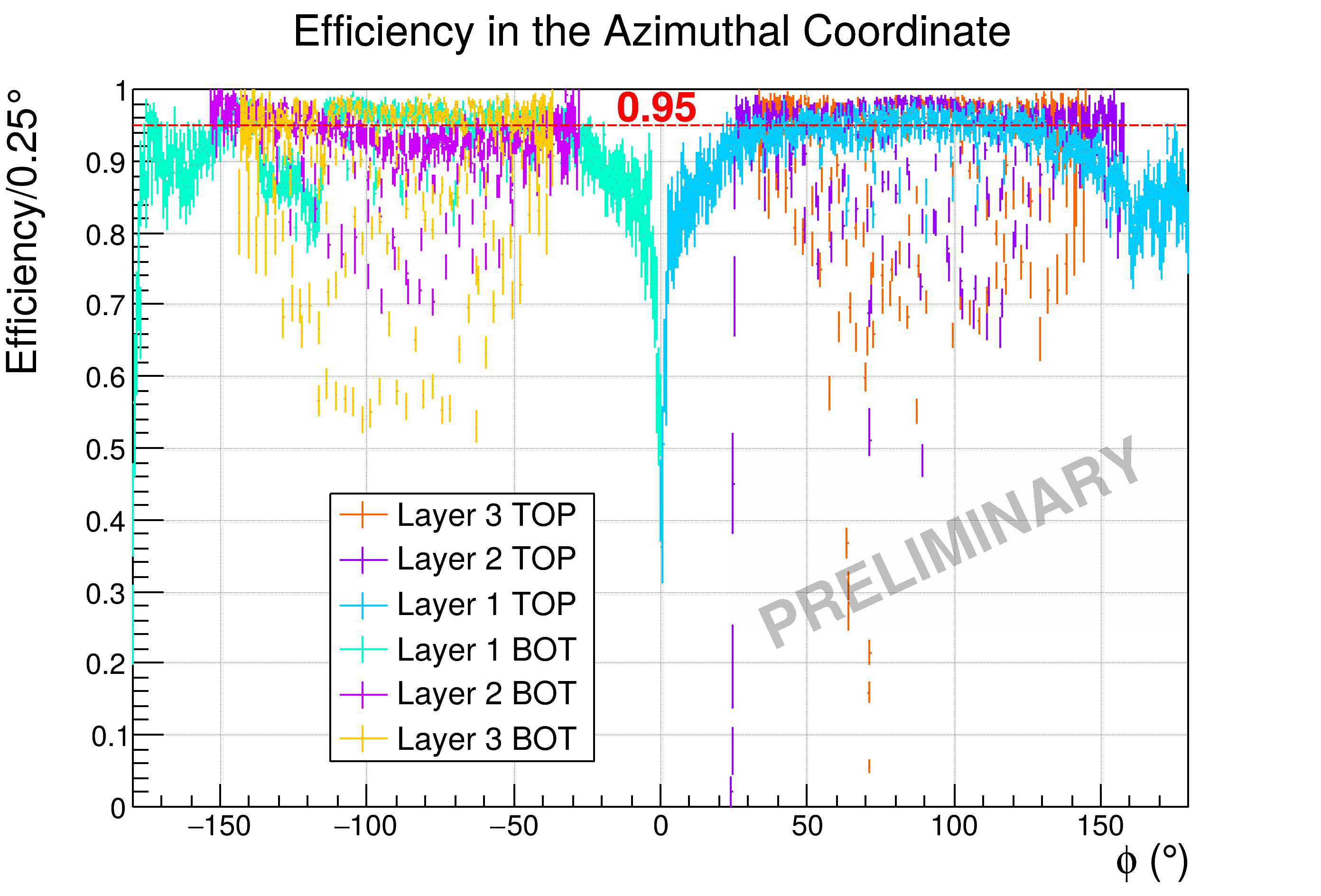}
		\caption[Efficiency in the azimuthal coordinate]{Efficiency in the azimuthal coordinate. The origin is located on the horizontal plane passing through the overlaps of the foils and corresponds with the southern direction for the nominal position of the detector in BESIII. }
	\end{subfigure}
	\hfill
	\begin{subfigure}[t]{.49\textwidth}
		\centering
		\includegraphics[keepaspectratio, width=\textwidth]{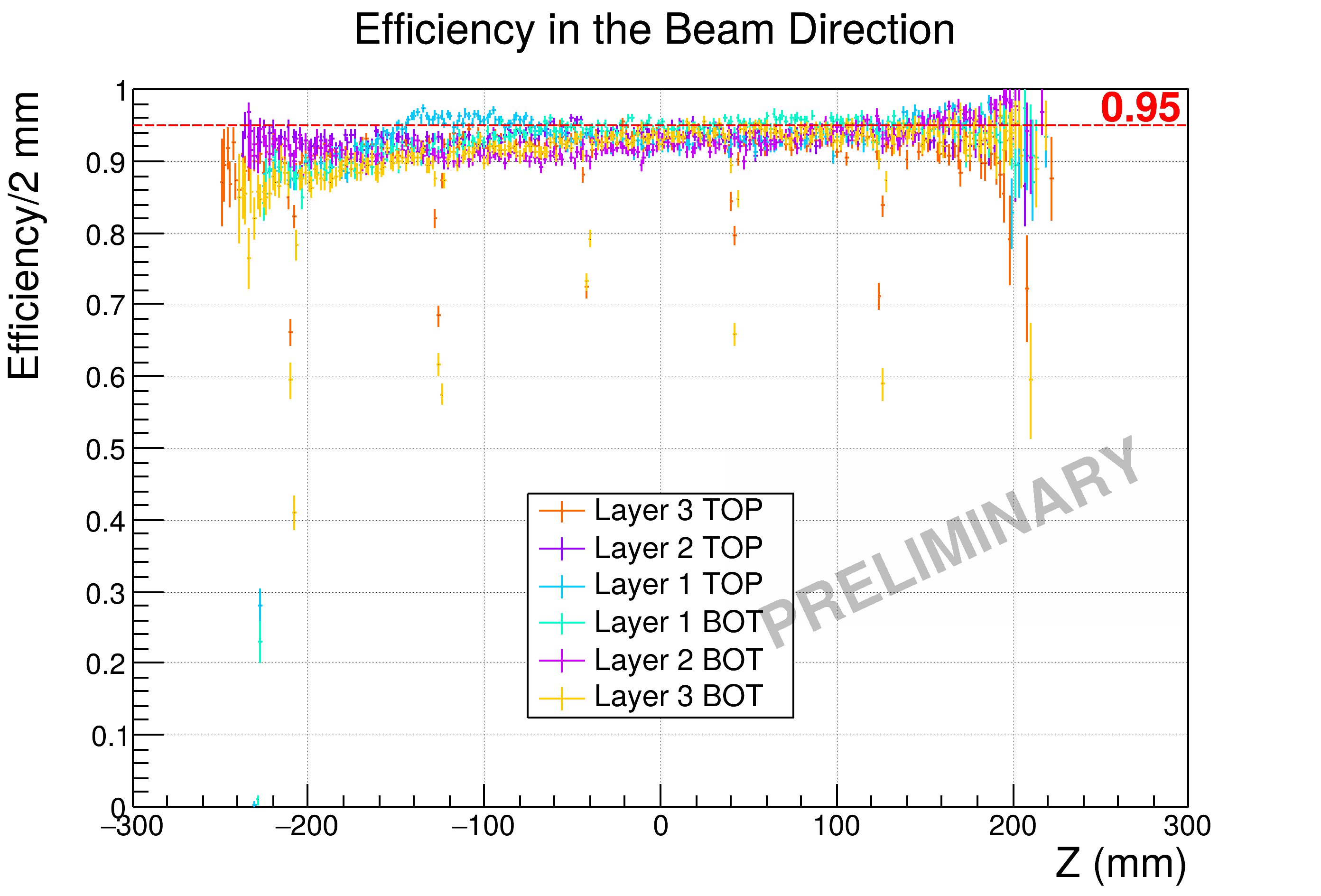}
		\caption[Layer 3 efficiency in the beam direction]{Efficiency in the beam direction. The origin is located at the center of the detector. The evenly spaced dips are due to the presence of the grids' rings.}
	\end{subfigure}
	\caption[Efficiency]{Efficiency in the Beam direction and in the azimuthal coordinate. Tracks are reconstructed using the CC algorithm to determine the clusters position.}
	\label{efficiency}
\end{figure}

Figure$\,$\ref{efficiency} shows the efficiency of the detector both in the beam direction and in the azimuthal coordinate.
In the latter, aside from localized drops due to the insulating areas separating the micro-sectors and the regions near the overlaps where the angle of incidence becomes too large for CC reconstruction to be effective, efficiency is greater than 95\% for all three layers.
Along the beam direction the efficiency appears lower but this is due to the aforementioned inefficient areas due to the sectorization being included in the average when integrating over the azimuthal angle.
The evenly spaced dips in efficiency that can bee seen in this direction are due to the rings of the PEEK spacer grids, which are aligned and therefore generate dead regions.

\begin{figure}[htbp]
	\centering
	\includegraphics[keepaspectratio, width=\textwidth]{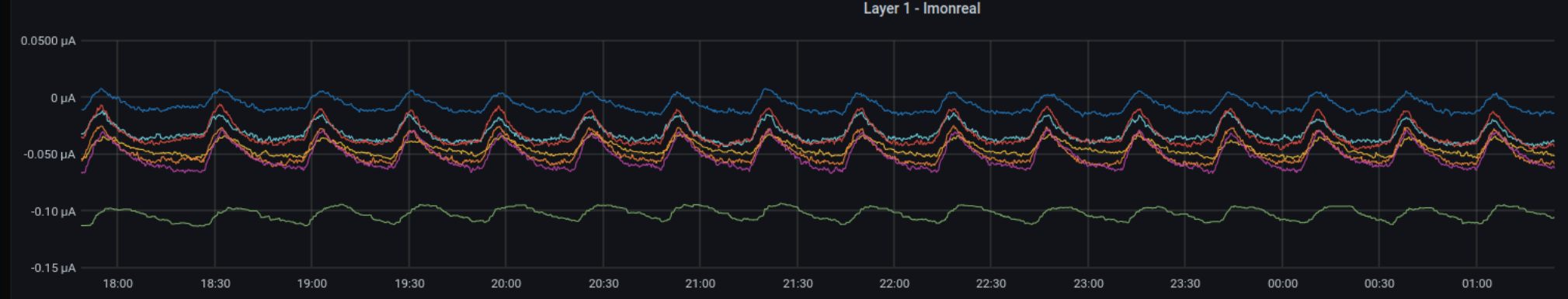}
	\includegraphics[keepaspectratio, width=\textwidth]{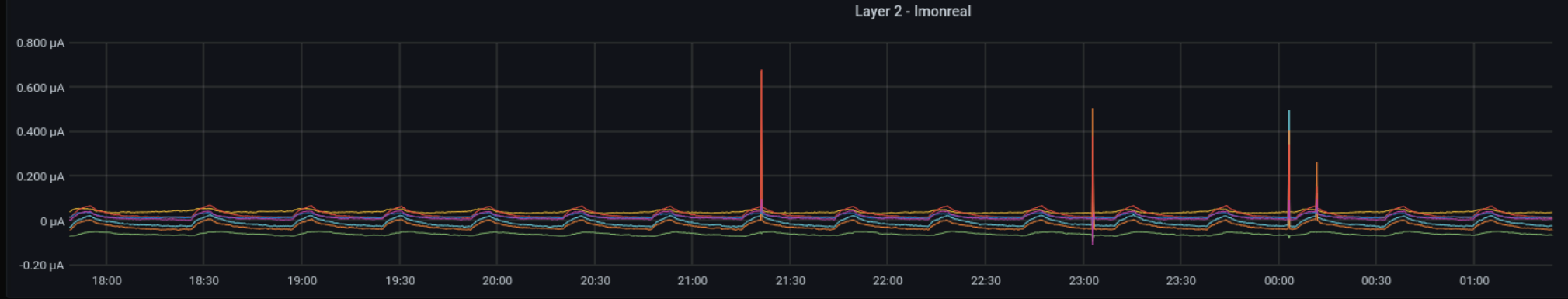}
	\includegraphics[keepaspectratio, width=\textwidth]{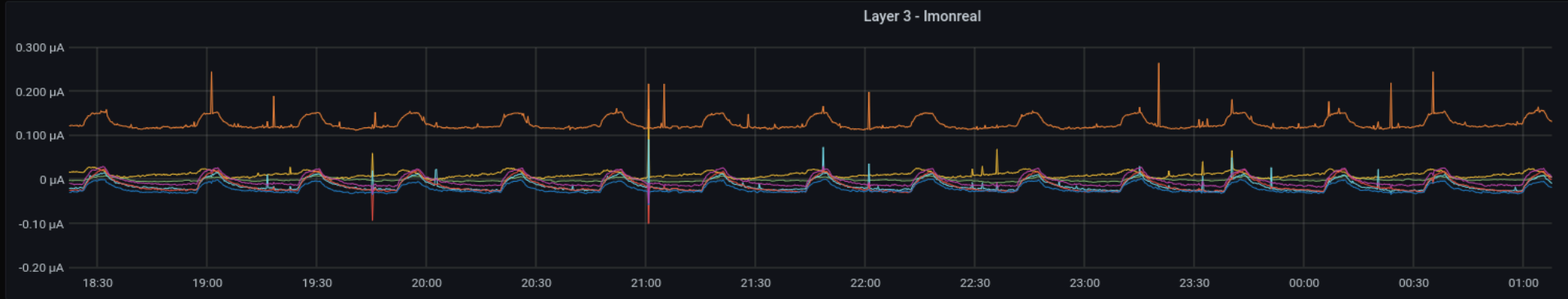}
	\caption[Detector Stability]{Operating currents for the three layers during a seven hours long data taking run.}
	\label{stability}
\end{figure}

The performance reached by the CGEM-IT in the cosmic ray data taking meets the requirements for the upgrade.
Over a period of about three months, the detector was operated on a daily basis to collect large data samples in different operating conditions.
The HV stability of the detector was confirmed, with only a single trip due to a particularly violent discharge occurring within this timespan.
After an initial adjustment period, in part coinciding with the newest layer's conditioning, the detector settled into a stable behavior, occasionally perturbed by small discharges.
Figure$\,$\ref{stability} captures a snapshot of the detector's operating currents during a seven hours long data taking run. All three layers are sensitive to ambient humidity, with the leakage currents of the detector fluctuating in accordance with the cycle of the dehumidifier present in the enclosure. Layer 2 experiences the largest discharges among the three, reaching up to a few hundreds nA.
Occasionally a discharge on a single foil propagates to the whole electrode stack and, if large enough, it can disrupt the electronics. Such events were rare and in the longest tests data could still be collected for several days uninterruptedly.
The issue was ultimately solved by splitting its power distribution among two patch panels.
The GEMs of Layer 3 discharge more frequently, although at lower amplitude.
The nature of the phenomenon is still unknown but it is possible that the electrical conditioning of the relatively new Layer 3 may still be incomplete, with residual contamination or malformed holes being the cause of the discharges.

An internal review committee was called to evaluate the project's readiness for installation.
The project underwent four reviews, encompassing all critical aspects: performance, stability, software, and integration.
Each time a favorable judgment was expressed alongside recommendations to improve upon the presented results.
While the initial concerns regarding the detector's efficiency have been addressed, both further developement of the $\upmu$TPC algorithms and improvements in the reconstruction of multi-track events were advised.
In the last review meeting, on June 29\textsuperscript{th} 2024, the CGEM-IT received the final approval to replace the inner MDC of the BESIII experiment.
    		\chapter{Installation of the CGEM-IT}
\label{installation}
The definition of the procedures for the detector's installation through dedicated, large-scale tests was an integral part of the project's review.
Due to the limited time available, most of this work had to be done in parallel to the detector's commissioning and cosmic ray data taking.
I was tasked with coordinating the work for the installation and the team charged with it.
Two main tasks were addressed: how to safely insert the detector into BESIII and how to cable it in the cramped spaces of the MDC's endcaps. 

\section{Insertion}
\begin{figure}[htbp]
	\centering
	\includegraphics[keepaspectratio, width=\textwidth]{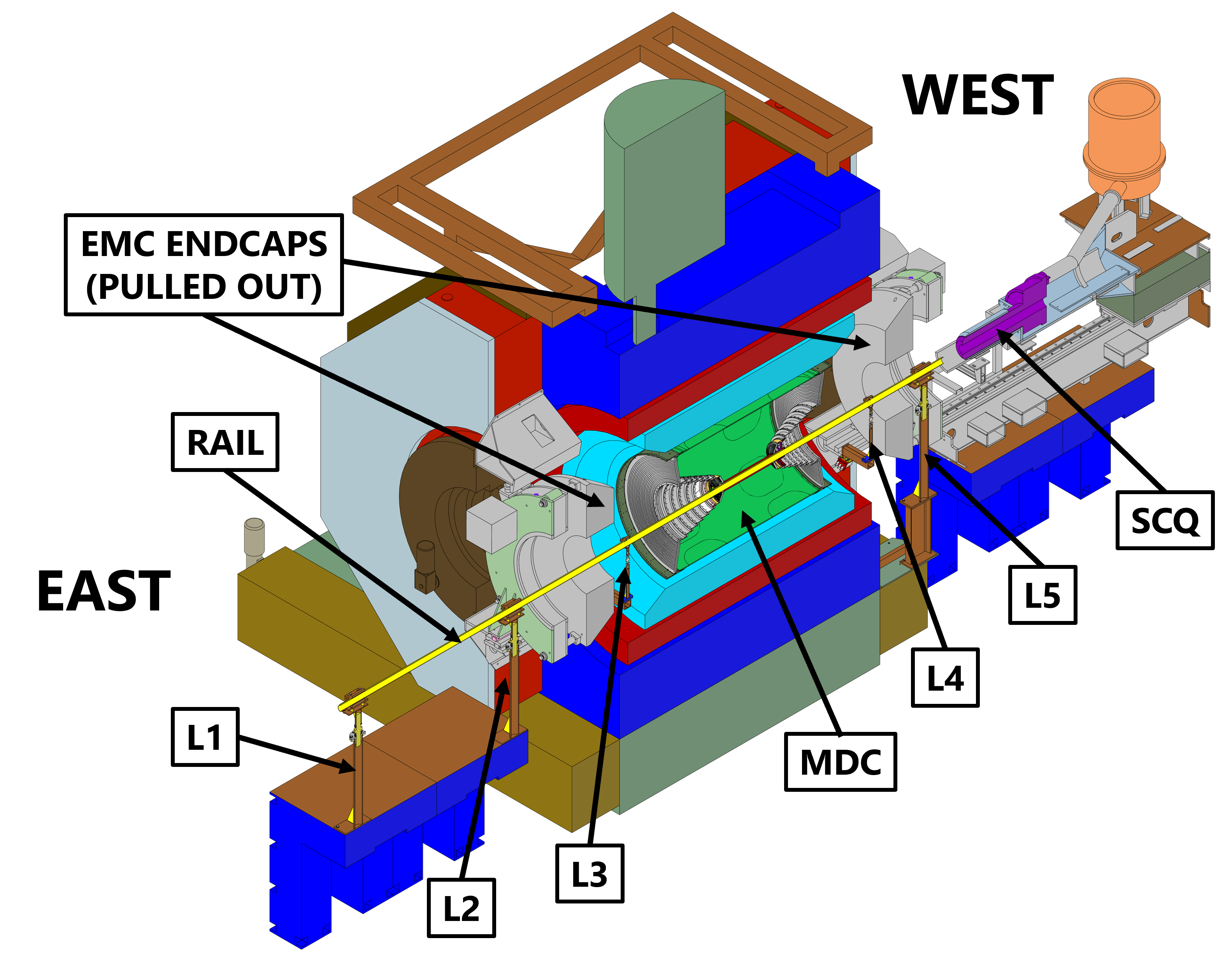}
	\caption[Installation setup as of 2017 documentation]{The installation setup according to the 2017 documentation.}
	\label{setup_2017}
\end{figure}

A preliminary study of the detector's insertion had already been conducted in 2017.
The general idea was to move the detector into place using a trolley, which would slide on a tubular rail mounted on adjustable legs, as of figure$\,$\ref{setup_2017}.

Since the removal of the inner MDC and the insertion of the CGEM-IT had similar needs, the rail and its legs were shared, while custom trolleys and fixtures were designed separately for the two detectors.
All the installation tooling had already been manufactured: the about 10$\,$m long rail, built in two parts to be welded together; all support legs, larger ones to be mounted outside the spectrometer and smaller ones to be fixed to the EMC endcaps; the extractor for the inner MDC; the framing to reinforce the outer chamber; the CGEM-IT insertion trolley, built into the horizontal assembly machine.

Only the main steps of the procedure had been drafted, with many finer details still to be investigated and defined.
The east interaction magnet had to be removed, to make space for the majority of the installation tooling, while the west one could just be retracted, to allow the positioning of two out of the five support legs.
A custom extractor had been designed to pull the inner chamber from the east side while guaranteeing the structural integrity of the outer chamber.
Once the inner chamber had been removed, a thin carbon fiber cylindrical shell was used to seal the gas volume of the outer one.
The CGEM-IT could then be moved in place on its insertion trolley and fixed to the MDC's endcaps.
With the detector secured, the insertion trolley could be pulled out and the rail dismantled to facilitate the cabling of the detector.

Neither the procedures nor the tooling had ever been validated, so dedicated tests were scheduled for both the removal of the inner MDC and the insertion of the CGEM-IT.
The former was conducted by a team of researchers and technicians from IHEP, while the latter was the responsibility of the CGEM-IT's working group.

\subsection{First Insertion Test}
Due to the tight schedule, it was decided that the first test would not have tried to replicate exactly the installation setup, with its 10$\,$m long rail and five legs.
The smaller setup already available at IHEP and used for validating the tooling for the extraction of the inner MDC, consisting of a 6$\,$m long section of rail supported by only three legs, was repurposed and used instead.
This allowed to collect feedback earlier and to have enough time to implement the necessary changes before a second, more representative test.

A seven people insertion team, supported by colleagues from IHEP, was deployed for the week dedicated to the test, the objectives of which were:
\begin{itemize}
\item Define a step-by-step procedure for inserting the CGEM-IT into BESIII, mechanically connect it to the rest of the spectrometer, and remove the trolley used to slide the detector in position.
\item Assess the risks involved in the aforementioned operations both for the new tracker and for the other subsystems of the experiment.
\item Evaluate readiness and suitability of the existing tooling and assess the need for additional instruments.
\end{itemize}

\subsubsection{Setup and tooling}
\begin{figure}[htbp]
	\centering
	\includegraphics[keepaspectratio, width=\textwidth]{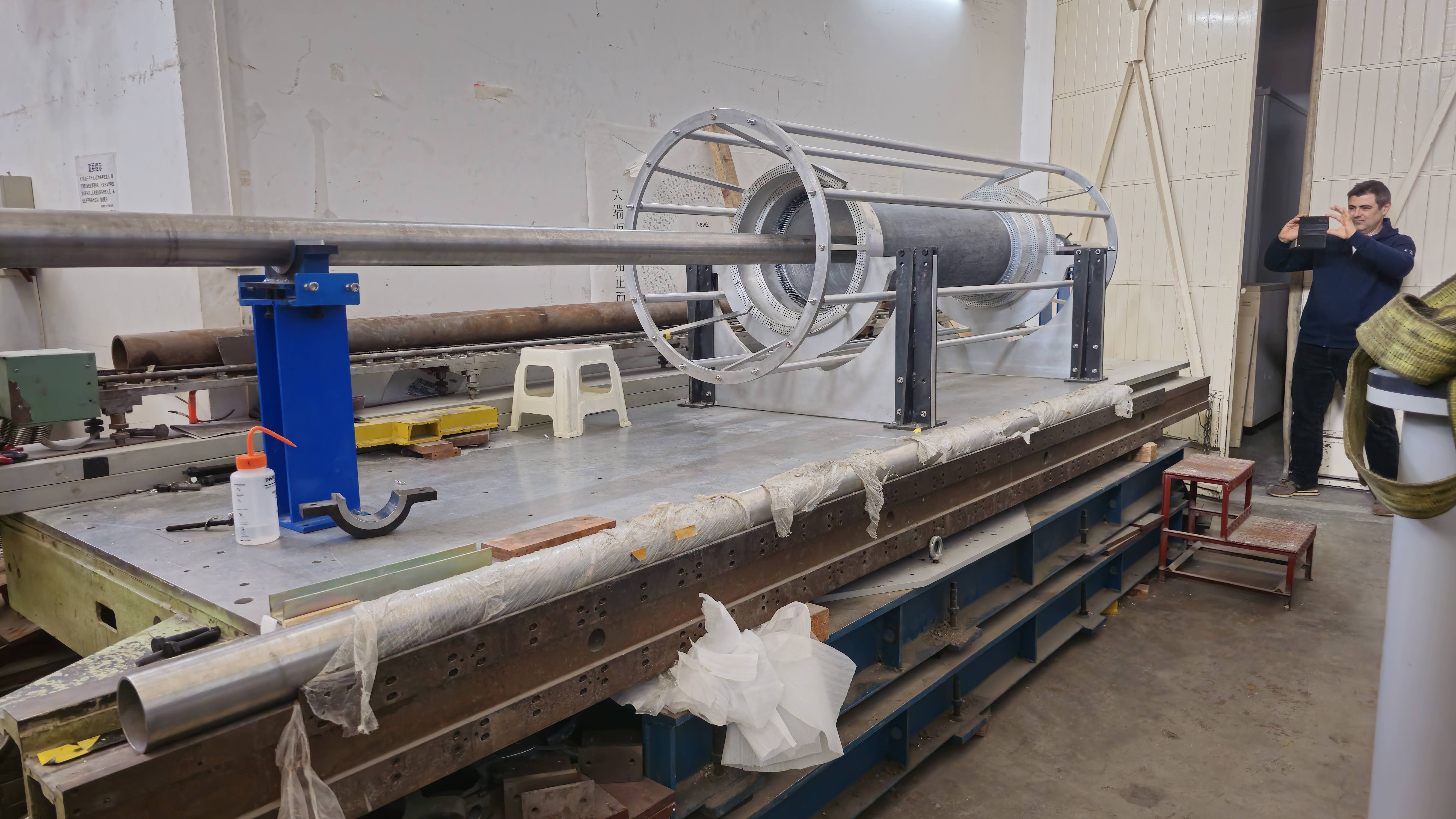}
	\caption[First insertion test's setup]{The setup used in the first insertion test}
	\label{firstsetup}
\end{figure}

The setup used for studying the extraction of the inner drift chamber, in figure$\,$\ref{firstsetup}, consisted of a mock-up of the outer drift chamber, a section of the rail, and three legs.
The outer MDC's mock-up can house the separation cylinder and thus replicates the exact dimensions of the cavity.
Aluminum frames at both sides roughly mimic the shape and size of the stepped endplates, and exact replicas of the MDC's flanges provide mounting points for the detector.
The longest of the two sections of the rail runs through the mock-up supported by three legs, two located on the east side and one on the west side.
These legs were not the ones manufactured for the installation but smaller analogs with similar adjustment mechanisms.

\begin{figure}[htbp]
	\centering
	\includegraphics[keepaspectratio, width=\textwidth]{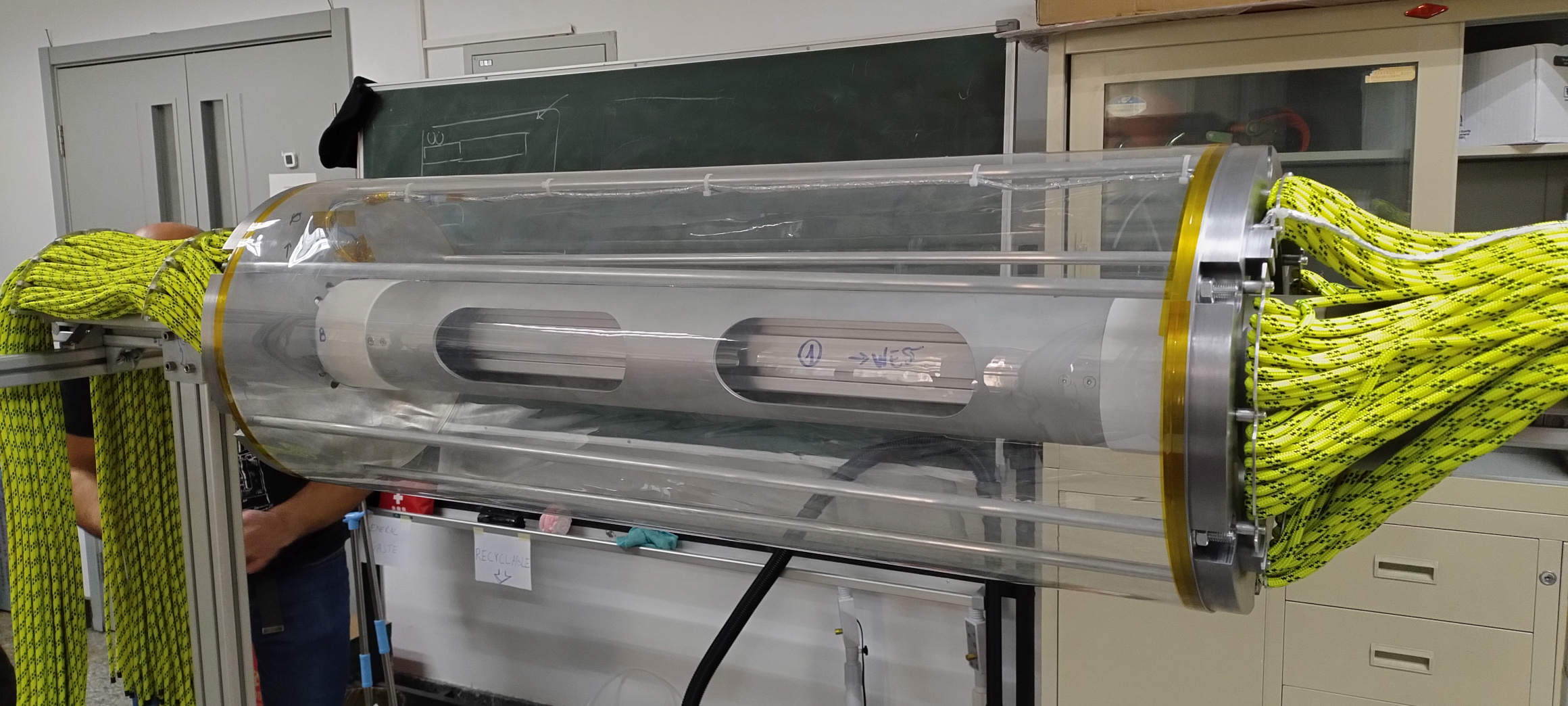}
	\caption[CGEM-IT mock-up in the first insertion test]{Mock-up of the CGEM-IT used in the first insertion test resting on the first version of the insertion trolley.}
	\label{mkp1}
\end{figure}

A mock-up of the CGEM-IT, in figure$\,$\ref{mkp1}, was manufactured for the test.
Its design was meant to provide slightly pejorative conditions for what concerns all the critical aspects for the insertion: diameter, weight, and clutter of the cables.
Steel flanges at the extremities replicate the mounting points for anchoring the detector and its maximum outer diameter.
The combined weight of the flanges, approximately 50$\,$kg, matches a conservative estimate for the detector and all its short haul cables.
Rope segments secured to the flanges simulate the occupancy of the cables.
Their length is equal to the longest cables, their diameter matches the cable's thickest sections, and their overall number is slightly in excess.
Spacers are used to replicate the overall length of the detector, by maintaining the correct distance between the flanges.
These were kept intentionally feeble to make the mock-up fragile against bending and torsion.
A Mylar sheet wrapped around the flanges and pulled taut with Kapton tape provides a reference surface to check for contacts with the cavity and other signs of damage.
The same MEMS accelerometers used in the drop test, mounted on the inside of each flange, continuously monitored inertial stresses during the test.

The mock-up was built on top of the actual insertion trolley, which had been freed from the assembly machine after the detector was moved to the testing station for collecting cosmic ray data.
This first version of the trolley consists of three sections of aluminum pipe joined by two shorter Teflon parts.
The aluminum sections have cutouts to reduce the overall weight of the trolley and, consequently, the deflection of the rail.
The two Teflon junctions, together with inserts of the same material on the external aluminum sections, ensure the trolley can slide smoothly on the rail and favor its extraction once the detector has been fixed to the outer chamber.
Crown-shaped cable holders allow to neatly arrange the cables on the exterior of the lateral pipe sections.
A makeshift telescopic extension had to be devised to keep the cables protruding from the front of the trolley stretched.
The ones in the back could simply be strapped to the rail with Velcro tape.

\subsubsection{Phases of the Test}
A step by step procedure of all the necessary operations was written for the test.
This first draft would then become the basis for the procedures followed in the second insertion test and during the installation of the detector, the last one provided in full in appendix$\,$\ref{insertproc}.
In this first version, the test was divided in six main phases:

\paragraph{Transport}
This phase was not intended as a test of the final transport solution but as a way to gather accelerometric data while carrying the mock-up by hand with a stretcher.
Long horizontal handles were fashioned from aluminum profiles so that the weight of the trolley and the mock-up could be shared among four people.
The accelerations measured were very low, and transport by hand was later implemented in the final transport solution as way to traverse uneven terrain without subjecting the detector to stresses.

\paragraph{Rail Coupling and Transfer of the Trolley onto the Rail}
In the 2017 documentation, how to transfer the trolley onto the rail had not been investigated.
This part of the procedure was therefore devised from scratch, with limited time to refine the solutions adopted.
A coupling implement was manufactured to join the rail and the aluminum profile supporting the trolley.
Makeshift adjustable legs were built using aluminum profiles to align the two before joining them.

The support legs closest to the joint must be removed to allow the passage of the trolley, so the joint must be able to sustain the full weight of the detector and it must be rigid enough to limit deflection at the maximum span.
On its profile side, the early version of the coupling implement relied on butt fasteners, while, on the rail side, it had an aluminum cylinder tightly fitting the rail's inner diameter.
Lacking the time and tools for machining the rail, no bolting was foreseen between rail and cylinder.
This greatly increased the deflection at the passage of the trolley, to the point that part of its weight had to be relieved manually.

Each makeshift leg had four adjustable feet for tilt and vertical adjustment.
Horizontal alignment was addressed by moving legs and trolley together.
Never intended as a proposal for the final solution, the improvised legs stressed the need for fine adjustments and sufficient degrees of freedom.
The difficulties incurred when traversing the joint also revealed a critical fragility in the design of the insertion trolley.
Bending was observed at the Teflon junctions, and these stresses could be transferred to the detector.

\paragraph{Travel towards the MDC}
A single person pushes the trolley, and legs may have to be removed and reinstalled along the way to allow its passage.
The main risks involved are those tied to the removal of the heavy support legs, which sometimes needed to be done in close proximity to the trolley and the detector.

A laser level and reference markings on the rail were used to measure deflection during this phase.
The deflection observed was larger than what the calculations predicted, the most likely cause being the poor rigidity of the makeshift legs.
No precautions were adopted at this time since this was pejorative with respect to the real case and therefore it did not invalidate the test.

\paragraph{Insertion}
The insertion is the most critical part of the installation.
Nominally, a radial clearance of 1.8$\,$mm separates the cavity from the detector.
The clearance during the test was reduced to about 1.5$\,$mm, by the thickness of the Mylar surface and the Kapton tape.
No aids had been foreseen to guide the insertion.
To evaluate the clearance, strong spotlights were pointed at the two endcaps, to be alternatively obscured so that their diffused light could be observed from the opposite side.

\begin{figure}[htbp]
	\centering
	\includegraphics[keepaspectratio, width=\textwidth]{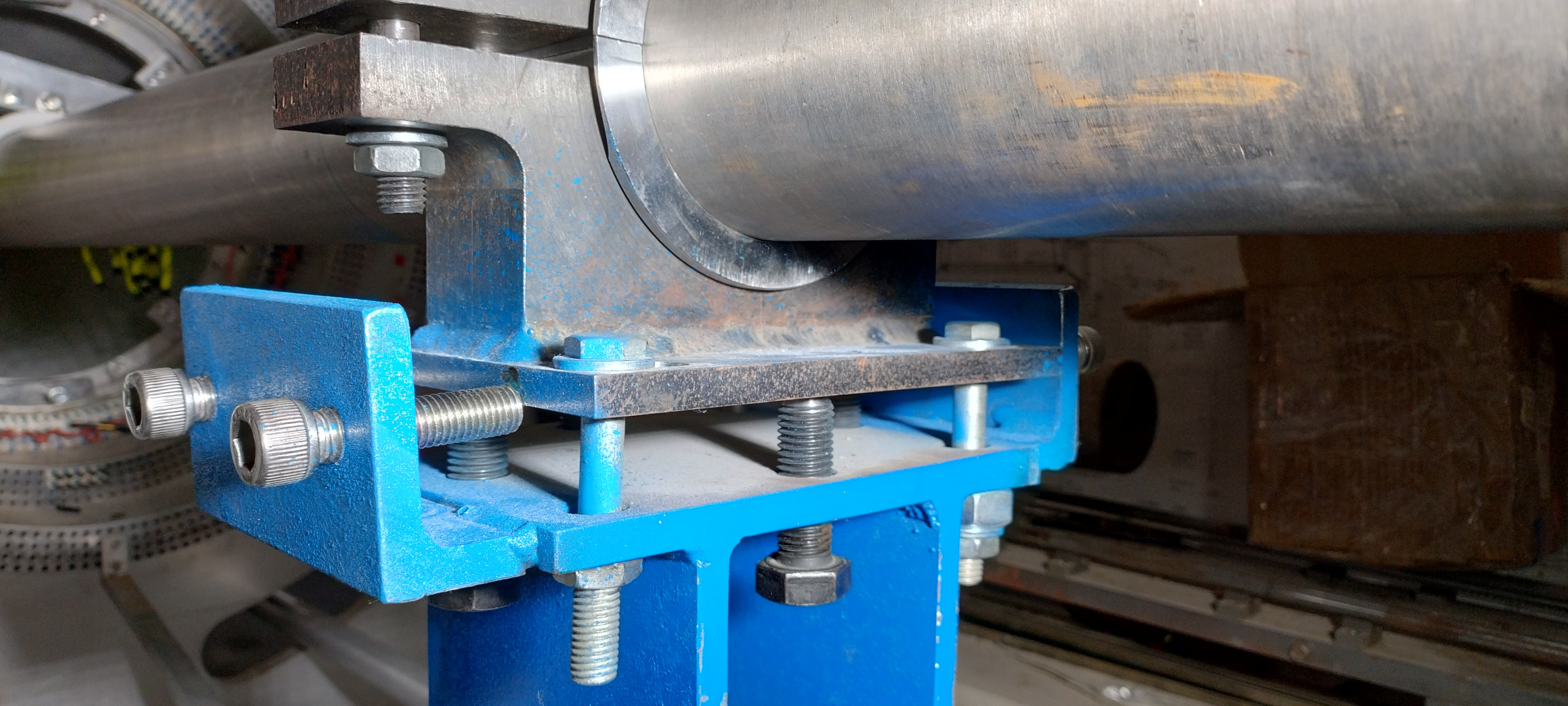}
	\caption[Makeshift legs adjustment mechanism]{Adjustment mechanism of the makeshift support legs used for the first insertion test.}
	\label{adjmech}
\end{figure}

The limited clearance and the large deflection of the rail required many adjustments to be performed at the rail support legs, from which the second critical issue with the tooling arose.
Figure$\,$\ref{adjmech}, shows the mechanism used for adjusting the makeshift legs, in which vertical and horizontal movements are unlocked by loosening the same nuts.
When trying to correct the alignment of the rail in the horizontal plane, unlocking the mechanism would release tension applied to the base and let the clamp tip.
Small displacements at the legs are amplified near the cavity due to the long lever arm, impacting the alignment in the vertical direction.
This effect could be in part counteracted by pre-loading the vertical pusher screws, with often unpredictable results.

Despite the unreliable adjustments and the difficulty involved in evaluating clearance, the insertion was completed without damage to the reference surface or the separation cylinder.
The large deflection of the rail did not make the insertion impossible, but due to the shorter distance of the legs with respect to the real case this was not enough to ensure its feasibility.

\begin{figure}[htbp]
	\centering
	\begin{subfigure}{\textwidth}
		\centering
		\includegraphics[keepaspectratio, width=\textwidth]{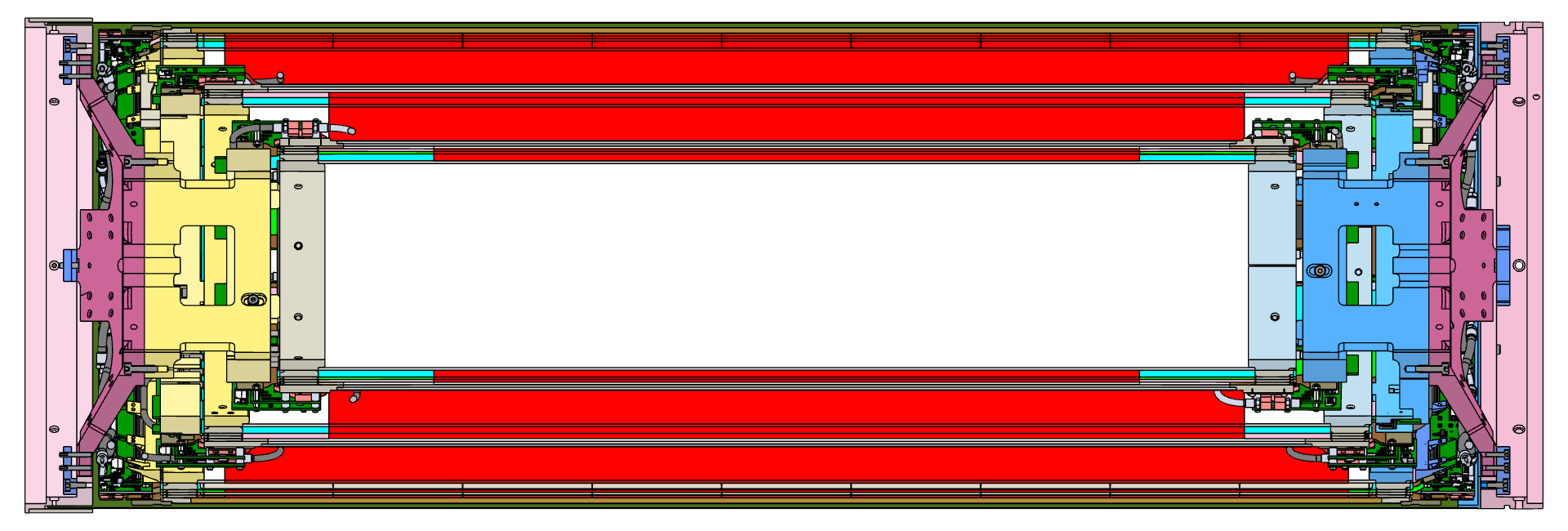}
		\caption[]{Lateral cross-section.}
		\label{anchoring}
	\end{subfigure}
	
	\vspace{.5cm}	
	
	\begin{subfigure}{\textwidth}
		\centering
		\includegraphics[keepaspectratio, width=\textwidth]{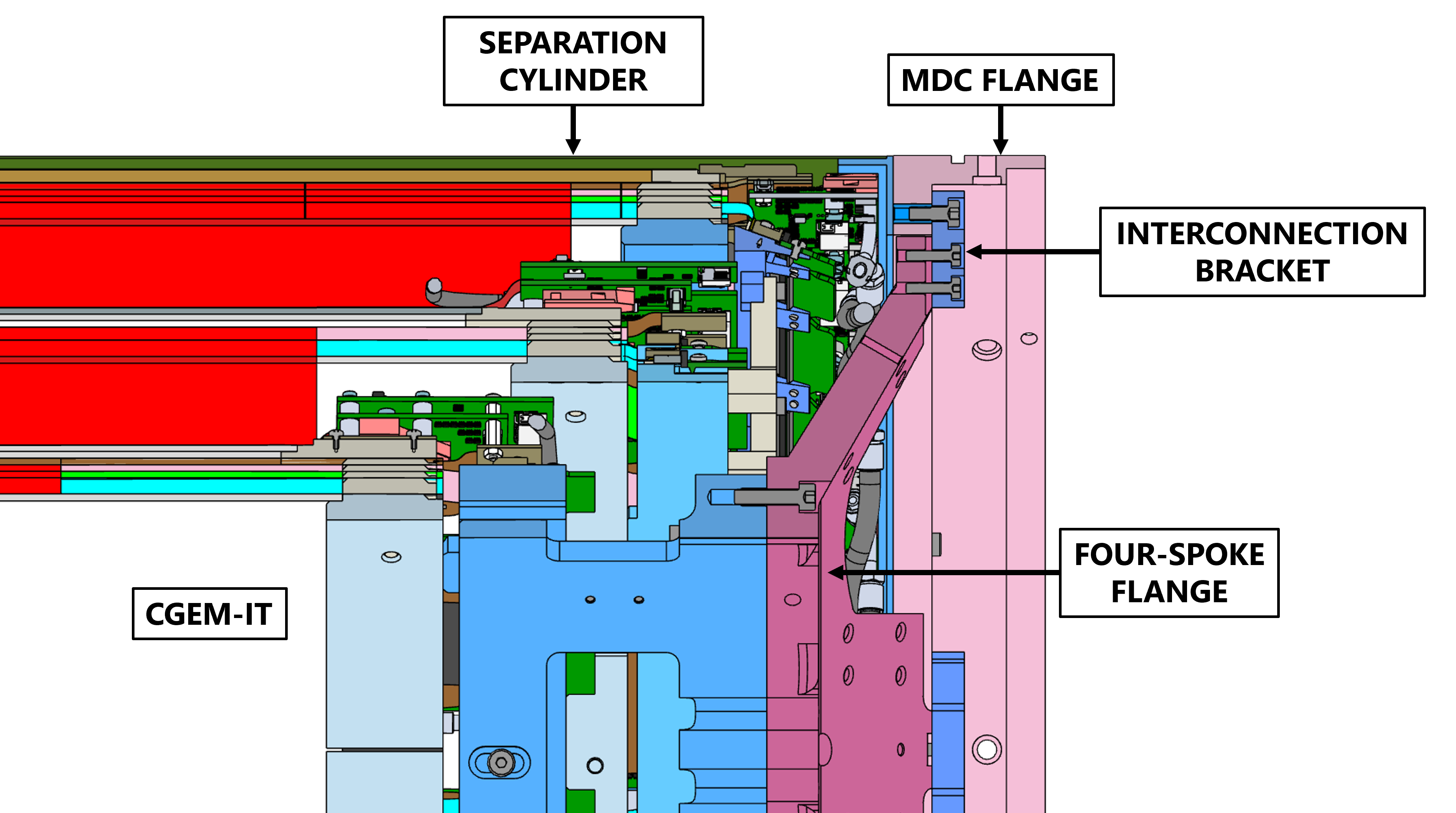}
		\caption[]{Detail of the west side fixing solution.}
	\end{subfigure}
	\caption[]{Cross-sectional views of the CGEM-IT in the cavity. The section is angled 45$^\circ$ so that it runs through the middle of the spokes and interconnection brackets.}
	\label{detail_anchoring}
\end{figure}

\paragraph{Anchoring}
The CGEM-IT's outer diameter is too large for it to pass through the central hole of the MDC's flanges.
The east flange is supposed to be removed together with the inner MDC and later replaced with a newly manufactured version after the insertion of the new inner tracker, while the west one remains in place.
Two sets of four interconnection brackets are used to fix the detector to these flanges, as shown in figure $\,$\ref{anchoring}.
The west side of the spectrometer provides the longitudinal reference, with mating surfaces and fully tightened screws, while fixing on the east side is instead intended to be compliant, with a large backlash and pins instead of screws.
This serves two purposes: on the one hand, it helps to cope with the uncertainties in the length of the tracker and in the distance between the flanges; on the other, it safeguards the detector from accidental compression.

During the first insertion test, all interconnection brackets shared the same design, which made necessary to use shims to compensate for the difference in the size of the two mock-ups.
After the mock-up of the CGEM-IT was fully inserted, a copy of the eastern flange was installed on the mock-up of the MDC.
The two flanges were not perfectly parallel so additional shims had to be used to compensate.
Many delicate adjustments, performed with the mock-up fully inserted into the cavity were necessary, which reinforced the need for precise and reliable control over the rail's position.

\paragraph{Extraction of the Trolley}
Before the trolley can be extracted, the weight of the detector must be transferred to the structure of the outer MDC and the cables must be freed and moved out of the way.
The former can be achieved by lowering the rail until some clearance appears all around the trolley, while the latter requires to remove the cables from their holders and fan them around the cone.
All protruding attachments on the west side of the trolley, like the cable holders, have to be removed as well, so that they do not interfere while the trolley is being pulled out from the east.
If the weight of the detector is fully resting on the structure, the trolley should offer almost no resistance as it is being extracted.
The first insertion test was considered concluded once the trolley had been fully extracted from the mock-up.

\subsubsection{Results}
Despite the difficulties encountered, the first test was successful.
The mock-up of the CGEM-IT had been inserted into the mock-up of the MDC, anchored to it, and the trolley had been extracted without indication that the detector might have incurred any substantial damage during the operations.
The test, although, highlighted many sources of risk to be addressed and several faults in the design of the installation tooling.
The logistics and the coordination of the operations also required a review.
Losing track of the procedure led to skipping important steps and communication between teams stationed at the opposite sides of the spectrometer could be improved.
A report of all the changes and improvements to the procedure, the tooling, and the logistics was compiled and presented at the first review meeting.

\subsection{Overhaul of the Tooling}
Having a first test as quickly as possible, even forgoing fidelity, proved to be strategically correct.
This in fact granted a time window to redesign the unsatisfactory tooling and manufacture it before the second test.
The main issues to address were the trolley lacking rigidity and the horizontal adjustment of the rail support legs generating unwanted vertical movement.
The design and the production of the new legs and trolley were tackled in parallel.
The production of the legs' components was outsourced to several Chinese companies thanks to the help of IHEP.
In Italy, where outsourcing was not possible due to time constraints, the parts of the trolley were produced internally, splitting the burden between the workshops of the INFN sections of Ferrara and Turin.

\subsubsection{Insertion Trolley}
\begin{figure}[htbp]
	\centering
	\begin{subfigure}{\textwidth}
		\centering
		\includegraphics[keepaspectratio, width=\textwidth]{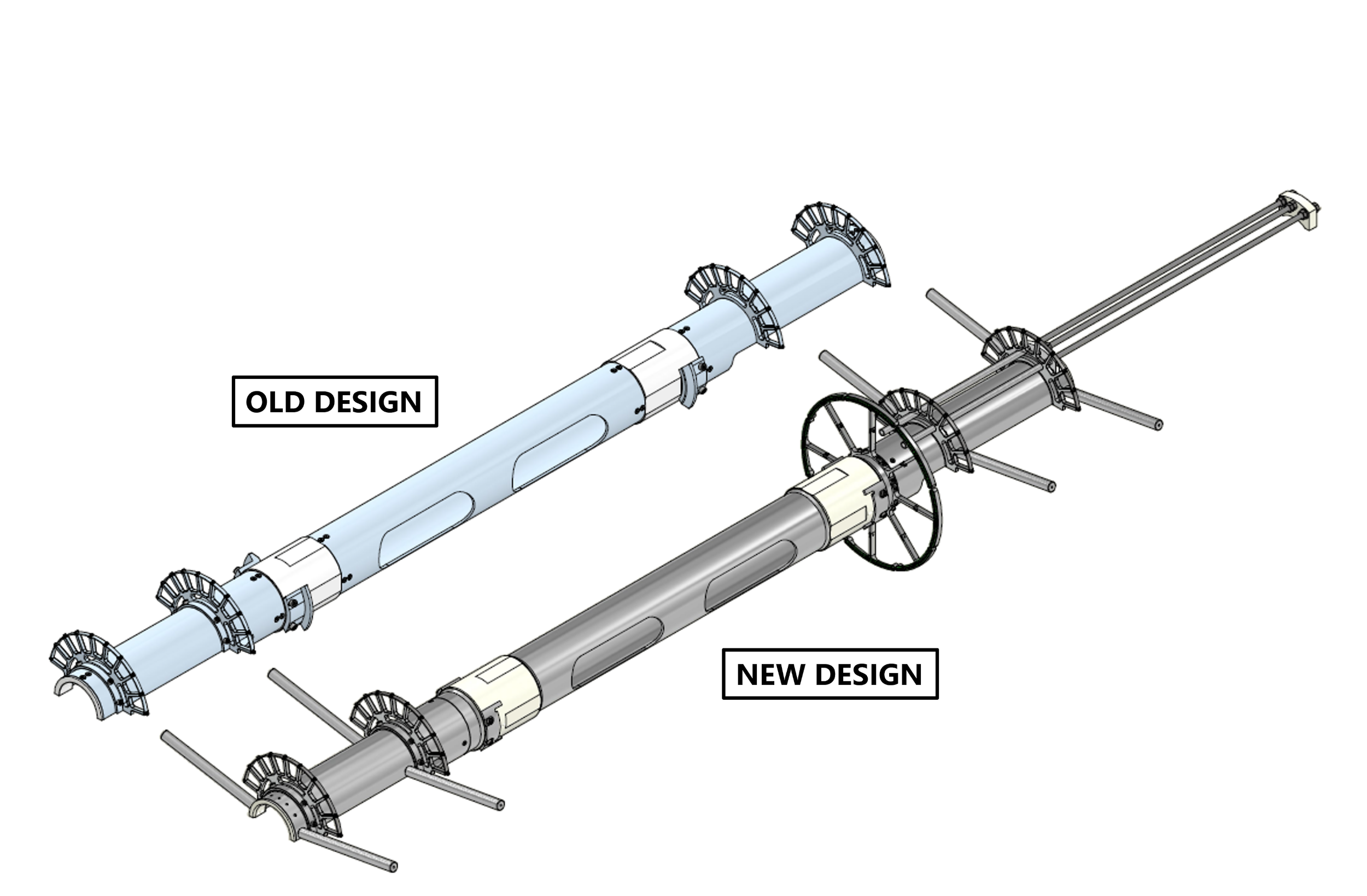}
		\caption[]{3D models of the two trolleys side by side.}
	\end{subfigure}
	
	\vspace{.5cm}
	
	\begin{subfigure}{\textwidth}
		\centering
		\includegraphics[keepaspectratio, width=\textwidth]{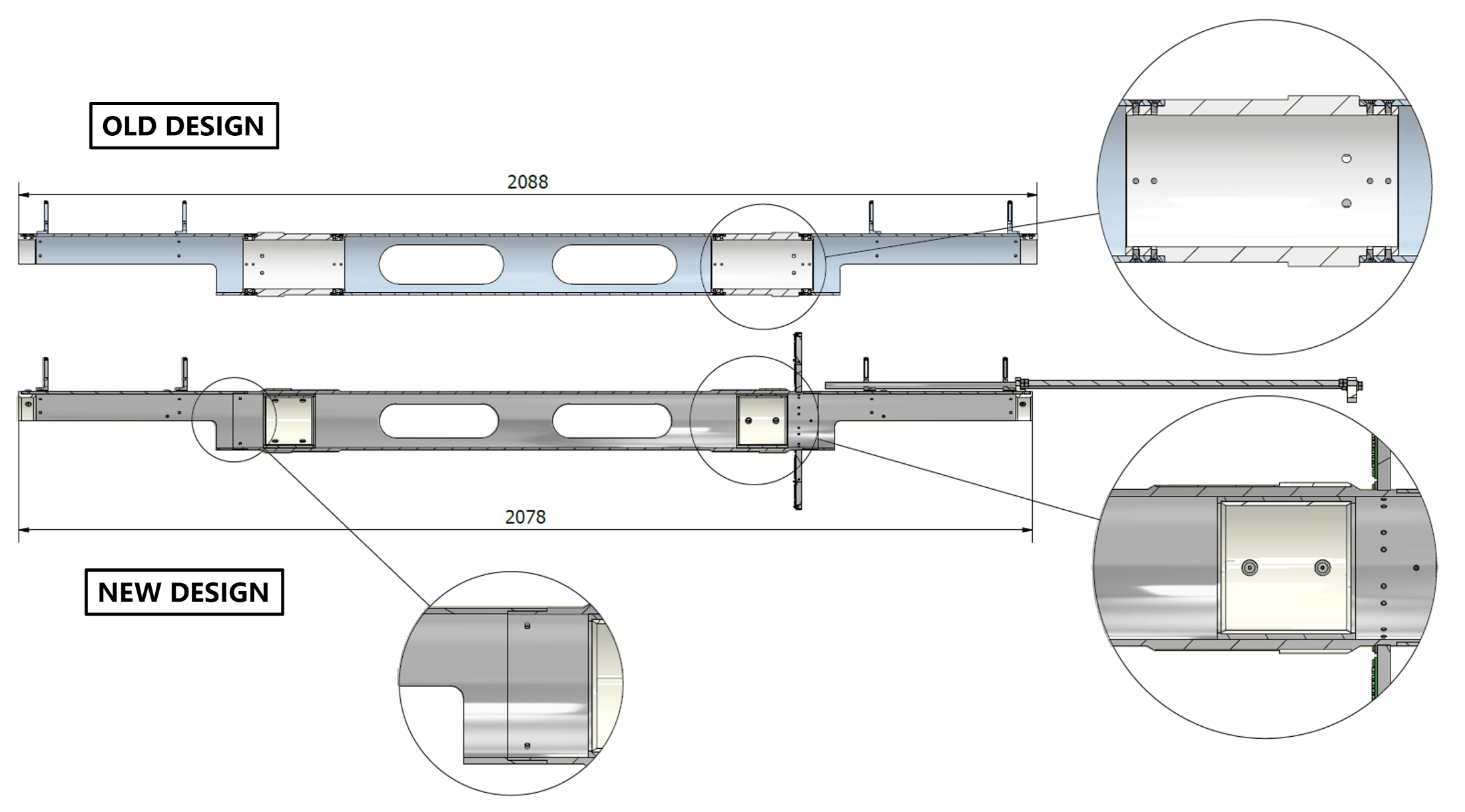}
		\caption[]{Cross sectional view of both designs}
	\end{subfigure}
	\caption[Comparison between old and new trolley design]{Comparison between the two designs of the insertion trolley.}
	\label{trolley_comparison}
\end{figure}

Figure$\,$\ref{trolley_comparison} shows a comparison between the new and the old design of the insertion trolley.
The trolley was redesigned with the intention of addressing the critical issues identified in the test without changing what worked of the original idea.
To facilitate its transport and match the working area of the available machines, the main body of the new trolley would also consist of separate sections of pipe to be assembled on site.
The Teflon junctions were removed and the central section was lengthened, with the detector now resting entirely on a single, solid part and therefore protected from bending.
To regain an internal sliding surface, aluminum bushings with Teflon inserts were manufactured and fixed to the central pipe with radial screws.
The outer Teflon surfaces, the only parts contacting the detector, were achieved by inlaying Teflon directly into the central pipe.
Bevels were machined on all Teflon parts to facilitate overcoming the rail's joints and extracting the trolley.
The crown-shaped cable holders of the previous design were retained, and the telescopic extension devised for the first insertion test was improved and fully integrated.

\begin{figure}[htbp]
	\centering
	\includegraphics[keepaspectratio, width=\textwidth]{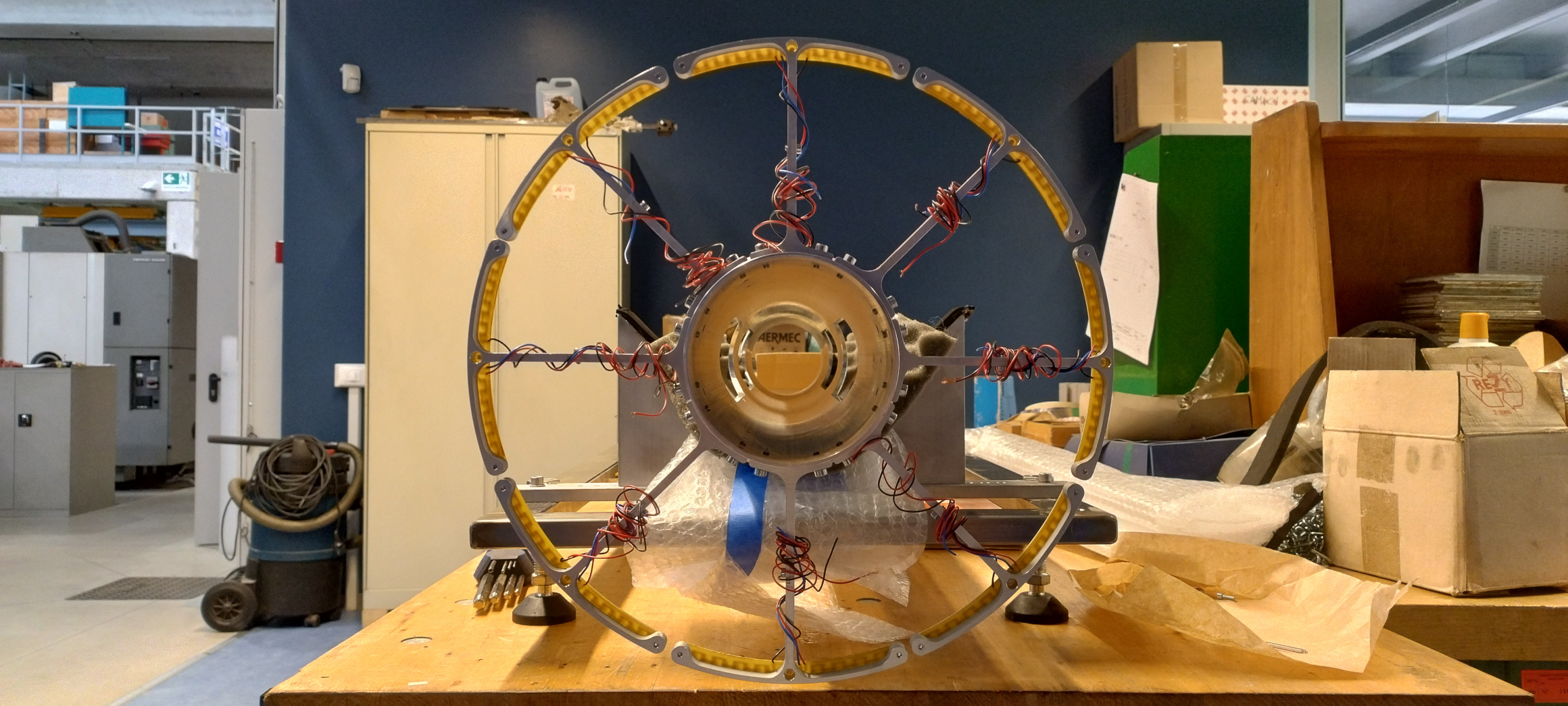}
	\caption[New trolley's guard ring]{The new trolley's guard ring, fully assembled on the central pipe for performing the quality control measurements.}
	\label{guard_ring}
\end{figure}

The remaking of the trolley also provided an opportunity to address the difficulties encountered in evaluating the clearance with respect to the cavity.
A reference guard ring, in figure$\,$\ref{guard_ring}, was designed to serve several purposes at the same time.
Positioned in front of the detector, it would be the first thing to contact the walls of the cavity in case of misalignment, therefore shielding the delicate electronics of Layer 3 from the impact.
Equipped with an array of led diodes, it would also provide a distributed source of light to help evaluate clearance at both sides without having to rely on the external spotlights.
The team operating on the west side could also benefit from the regular shape of the ring as a reference to determine the detector's position.
Finally, it could provide mounting points for the installation of an array of simple contact sensors to identify the contact's direction in case the optical evaluation would have still proven difficult.

Eight T-shaped brackets make up the full ring.
Each bracket is divided in two, to allow disassembling the top half at the reaching of the west flange.
Mating surfaces ensure correct positioning, and a single screw holds the two halves together.
The nominal diameter of the ring formed by the brackets is 362$\,$mm, 0.6$\,$mm larger than the detector's.
Kapton and copper tape, the former used for insulation and the latter as a sensitive surface, reduce the clearance with respect to the cavity even further.
A groove on the trolley ensures correct positioning and alignment of all brackets.
Since the clearance between the detector and the trolley is 0.3$\,$mm, the two are not coaxial.
Shims can be positioned below each bracket to compensate for this effect and guarantee that the guard ring can always effectively shield the detector.
Simple PCBs mounted to the top half of each bracket carry high luminosity LEDs, for illumination, and a red LED, for signaling contacts.
The electrical circuit constituting the rest of the high-sensitivity contact sensor is also housed on the same PCB.

The sensor works as a mechanical switch.
Copper tape on top of the brackets is wired to the circuit and kept at low voltage.
The aluminum foil on the inner surface of the separation cylinder is grounded.
When a contact occurs, a transistor flips and lets the current through the LED diode.
The circuit was made particularly sensitive so that it could spot contacts early on, before any damage to the cylinder or the array could occur.

\subsubsection{Rail Support Legs}
The first version of the rail support legs had horizontal adjustment mechanisms similar to those used in the test.
Pusher screws are used to shift the block supporting the clamp, and a series of nuts can be tightened to lock it in place.
When the mechanism is free, nothing opposes the overturning moment and the rail is free to sag further.
Another issue was identified while reviewing the legs' design after the test.
The vertical adjustment, based on an M80x2 shaft with both left and right-handed threading, could be subject to a lot of friction.
This may in turn apply torque to the rail, making it unintentionally flex in the horizontal plane.
The new design of the legs had to address these issues while also ensuring the feasibility of the production within a limited time window.
Because of this, both the recovery of components from the previous design and the use of commercially available parts was strongly advised.

\begin{figure}[htbp]
	\centering
	\includegraphics[keepaspectratio, width=\textwidth]{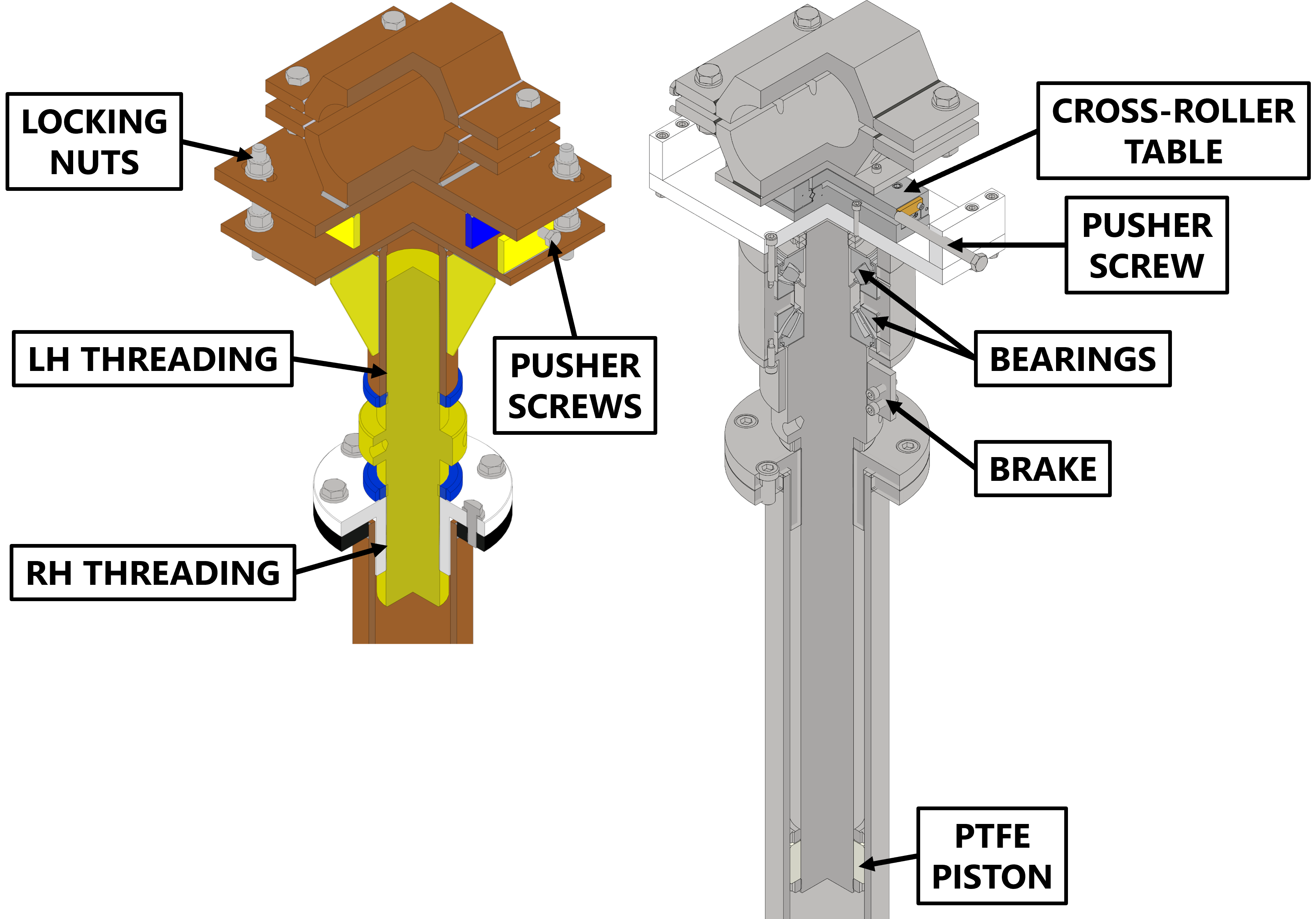}
		\caption[Comparison between rail support legs designs]{Comparison between the old leg's design, on the left, and the new one, on the right.}
	\label{leg_comparison}
\end{figure}

Th new and the old design of the rail support legs are compared in figure $\,$\ref{leg_comparison}.
The large welded assembly, constituting the foot of the leg, was reused, while the heads were entirely redesigned.
The M80x2 shaft controlling the legs' height was lengthened, and a Teflon piston was added at its bottom to increase its stability.
Two large SKF 29414E thrust bearings decouple the shaft from the upper part of the head housing the horizontal adjustment mechanism.
The conical spherical roller bearings also allow the head to tilt slightly so that the clamps can seamlessly adapt to the rail, without the need for additional mechanisms.
Horizontal adjustment is achieved by means of a cross-roller table operated with pusher screws.
Two strong springs keep the mechanism loaded so that a single screw can be used for control while a second, opposing one can lock the movement.

A U-shaped interface was designed to allow the support legs to house the aluminum profile supporting the trolley.
In this way, the same legs, already equipped with all the adjustments necessary, could also be used for transferring the trolley onto the rail.
A rolling pin integrated into the interface provided the last degree of freedom necessary to align the trolley with the rail, allowing the profile to be tilted without bending.

\subsection{Second Insertion Test}
\begin{figure}[htbp]
	\centering
	\includegraphics[keepaspectratio, angle=90, width=.8\textwidth]{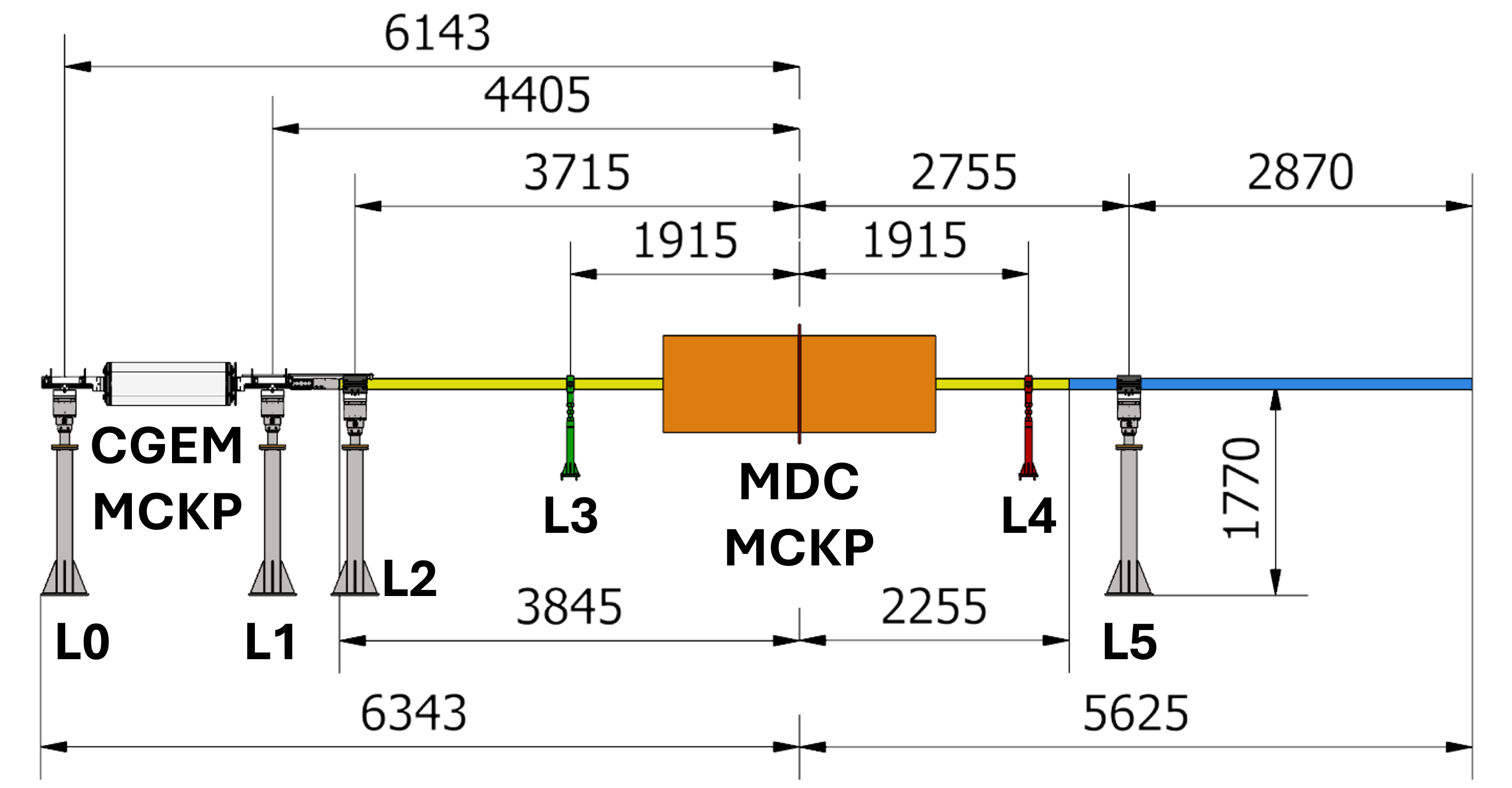}
		\caption[Setup for the second insertion test]{Drawing depicting the setup used for the second insertion test. All units are in mm.}
	\label{2nd_setup}
\end{figure}

The second insertion test shared the objectives of the first, while also aiming to prove the feasibility of the insertion in a realistic setting.
The setup, in figure$\,$\ref{2nd_setup}, recreated in 1:1 scale the relative position of legs and cavity as they would be inside BESIII's experiment hall.
After the extraction of the MDC, the rail would have to be cut to make space for the tooling needed to transfer the trolley onto it.
To replicate this condition, the rail was shifted westward and its protruding length was counterbalanced to prevent it from affecting the statics of the problem.
The mock-up of the MDC had to be raised from the platform, so that the longer final legs could be used instead of the makeshift ones employed in the previous test.

The coordination of the test was also changed to simulate the difficulty of going and conveying information from one side of the spectrometer to the other.
At the beginning of each operation, the people taking part in the test were assigned to either the east or the west team.
Each team had a leader carrying the team's radio.
People were prevented from crossing over to the other side and communication between the teams was restricted to the use of the radios.
As the operations' expert, I was located at the side dealing with the most critical operations, and I was able to cross to the other side if required.
A person dedicated solely to calling out the procedure, requesting feedback from the teams, and checking that the order of the operations was respected carried the fourth and last radio.

\begin{figure}[htbp]
	\centering
	\includegraphics[keepaspectratio, width=\textwidth]{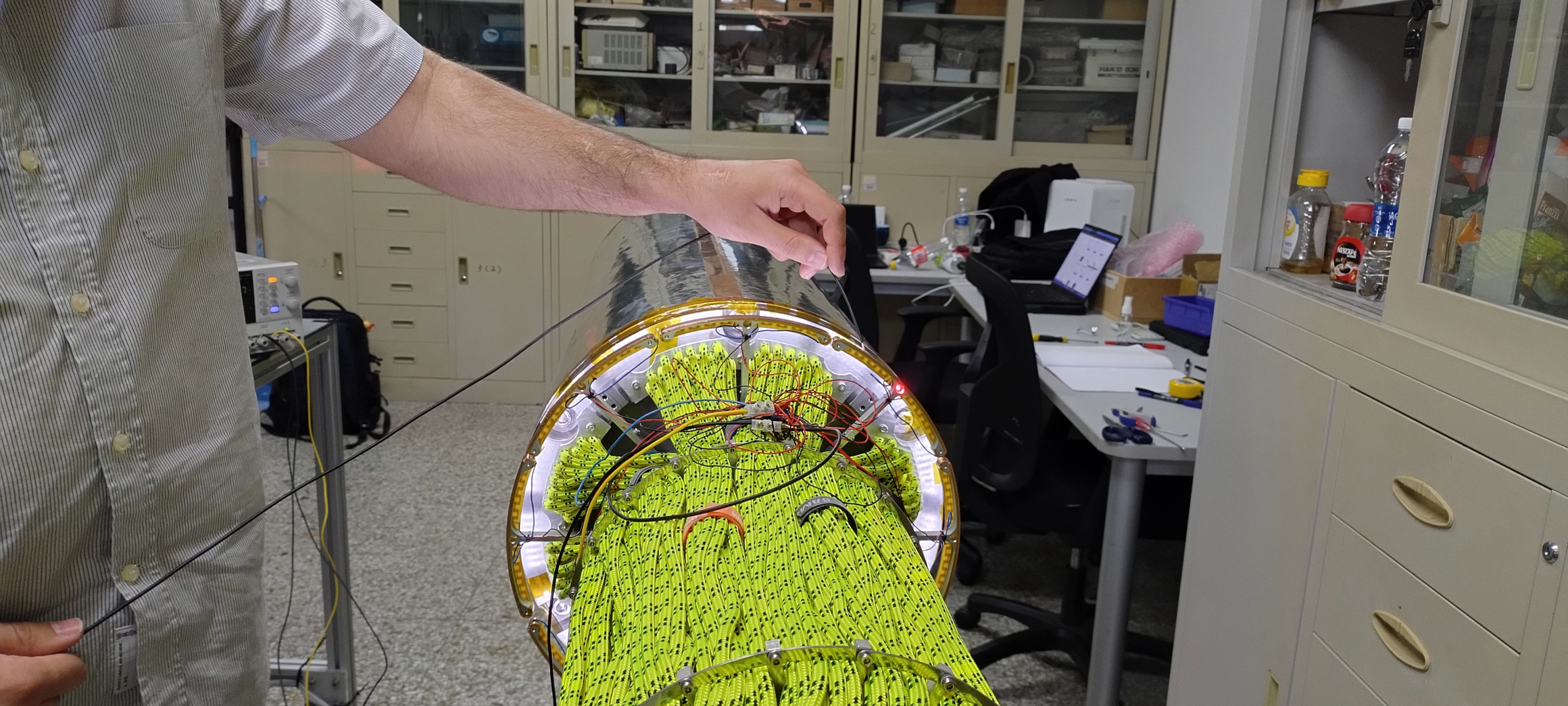}
		\caption[Test of the contact sensor array]{Photograph of the contact sensor array being tested. Individual sensitive surfaces are grounded to check the connections.}
	\label{sensor_test}
\end{figure}

\begin{figure}[htbp]
	\centering
	\includegraphics[keepaspectratio, width=\textwidth]{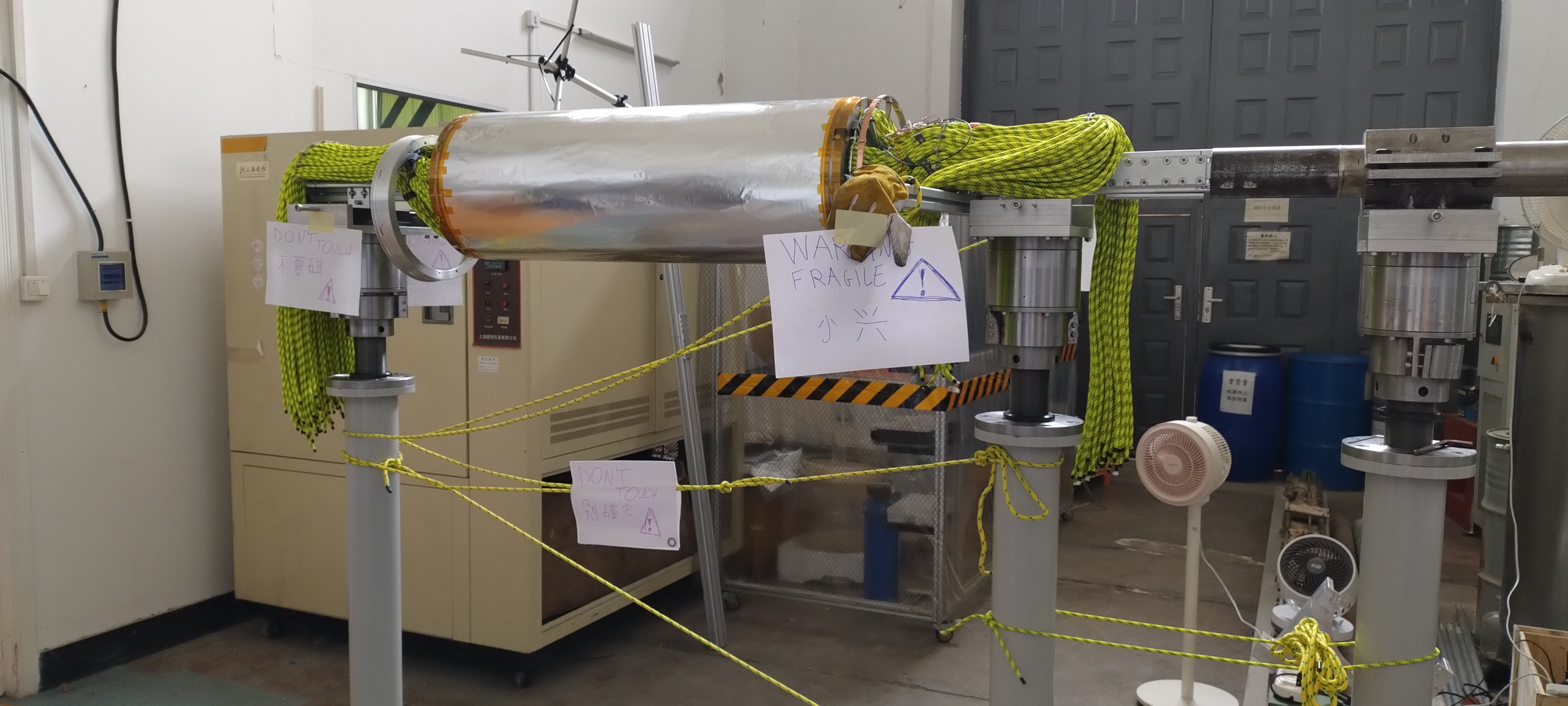}
		\caption[Trolley support coupled to the rail]{The aluminum profile supporting the trolley, fully aligned and coupled with the rail.}
	\label{coupling}
\end{figure}

Before the test, the mock-up was reassembled and given an outer conductive surface made of aluminum coated Mylar.
This was connected to one of the spare circuits of the contact sensor array, to work as a collision alarm.
The detector's Faraday cage would later be used in the same way during the installation.
Figure$\,$\ref{sensor_test} shows these systems being tested in the laboratory before the mock-up was moved to the testing site.
Dedicated tests had been scheduled for validating the final transport solution, so the second insertion test did not encompass this phase.
The coupling between aluminum profile and rail was completely redesigned, as of figure$\,$\ref{coupling}.
Bolted plates on the sides and on the bottom fixed the joint on the profile side.
Slots for radial screws were machined on a pipe to be welded onto the rail, so that the other side of the joint could be bolted as well.
The increased rigidity of both the joint and the legs in the second test greatly reduced the deflection.
The only issues in crossing the joint were due to the uneven surface resulting from welding the slotted extension.

\begin{figure}[htbp]
	\centering
	\includegraphics[keepaspectratio, width=\textwidth]{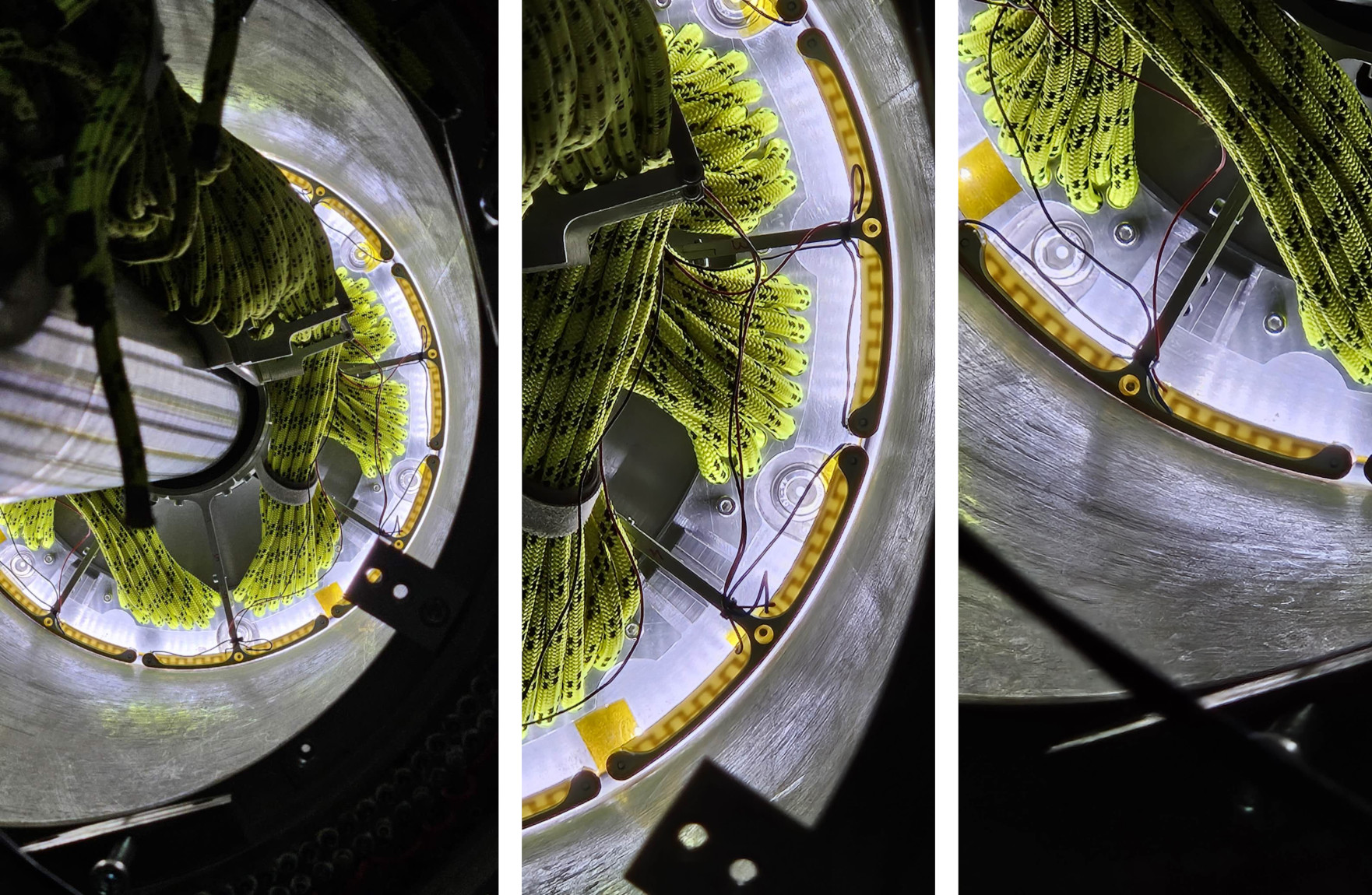}
		\caption[Clearance evaluation in the second test]{Photos, taken during the insertion phase of the second test, showing how the guard ring helps evaluate clearance and position of the detector with respect to the cavity.}
	\label{ring_on}
\end{figure}

The clearance during the second insertion test was reduced by the presence of the ring, which although also provided a much better reference to evaluate it, as shown by the photographs in figure$\,$\ref{ring_on}.
This, combined with the improvements to the legs' adjustment mechanisms, greatly simplified and expedited the insertion procedure.
The detector was fully inserted without ever triggering the contact sensors or the alarm on the outer surface, since optical evaluation proved sufficient.
Anchoring was again affected by imperfections in the geometry of the MDC's mock-up, probably aggravated since it was moved to a greater height.
As in the previous test, these were addressed by adding shims below the interconnection brackets to compensate for the deformation.

A new part was added to the procedure to place the mock-up back on the trolley at the end of the test and safely extract it from the cavity.
Its outer surface was then inspected, comparing any scratch with a photographic survey taken before the insertion.
No new scratches or signs of damage could be found.
The success of the second test confirmed the feasibility of the installation and the validity of the upgrades to the tooling.
A few adjustments were made to the procedure, and the implemented coordination hierarchy ensured it was rigorously followed during the entire test.

\section{Cabling}
The CGEM-IT has separate cables for data, low voltage, and high voltage, all of which are divided into Long Haul (LH) and Short Haul (SH) cables.
LH cables are sturdier and cover most of the travel towards the detector.
SH cables are shorter and more flexible, to facilitate routing them inside the MDC's cone-shaped endcap.
Data Low Voltage Patch Cards (DLVPCs) connect the respective LH and SH cables.
HV LH cables bifurcate at the end: one side can be directly connected to the SH, while the other carries the ground reference.
Low profile PEEK boxes insulate the HV connections, while up to seven ground references belonging to the same layer can be gathered by a single ground patch card (GNDPC).
The DLVPCs, The PEEK boxes, and the GNDPCs can be stacked, atop the MDC's preamplifier boards, in towers fixed to the endcaps.

When the CGEM-IT is fully assembled, FEBs and HV connections of the innermost layers are inaccessible.
Moreover, once fully inserted into the cavity Layer 3 becomes unreachable as well.
Because of this, the detector must be inserted with all the SH cables already attached.
3D printed cable holders, to be mounted on the four-spoke flanges, were designed to fix the cables as close to the connectors as possible, in order to prevent them from being accidentally disconnected during the operations.
Their shape is meant to keep the cables parallel and flat against the cone of the MDC as they depart from the detector, not to interfere with the superconducting quadrupole.

A certain number of preamplifier boards would have also been removed alongside the inner MDC, freeing the space for installing the CGEM-IT patch cards.
The number and position of the boards to be removed was irregular and differed for the two sides.
The distribution of the CGEM-IT cables, in table $\,$\ref{cable_distro},is also slightly asymmetric, as the cathodes are all powered from the west side.

\begin{table}[htbp]
\centering
\begin{tabular}{lrr}
\textbf{Cable} & \textbf{West} & \textbf{East}\\
\midrule
Data & 40 & 40 \\
LV & 40 & 40 \\
HV & 41 & 36
\end{tabular}
\caption[CGEM-IT cables distribution]{Distribution of the cables among the two sides of the CGEM-IT}
\label{cable_distro}
\end{table}

As the possible combinations are many, but not necessarily all feasible, it was necessary to find a way to study and validate the cabling scheme beforehand.
Addressing the problem using a CAD software, with nobody in the group having experience of it and the 3D model of the detector lacking the necessary references, would have required a long time.
A more direct approach of building a mock-up of the CGEM-IT and MDC endcap to phisically check the validity of the cabling schemes was pursued instead.

\subsection{The mock-up for the Cabling Studies}
\begin{figure}[htbp]
	\centering
	\includegraphics[keepaspectratio, width=.7\textwidth]{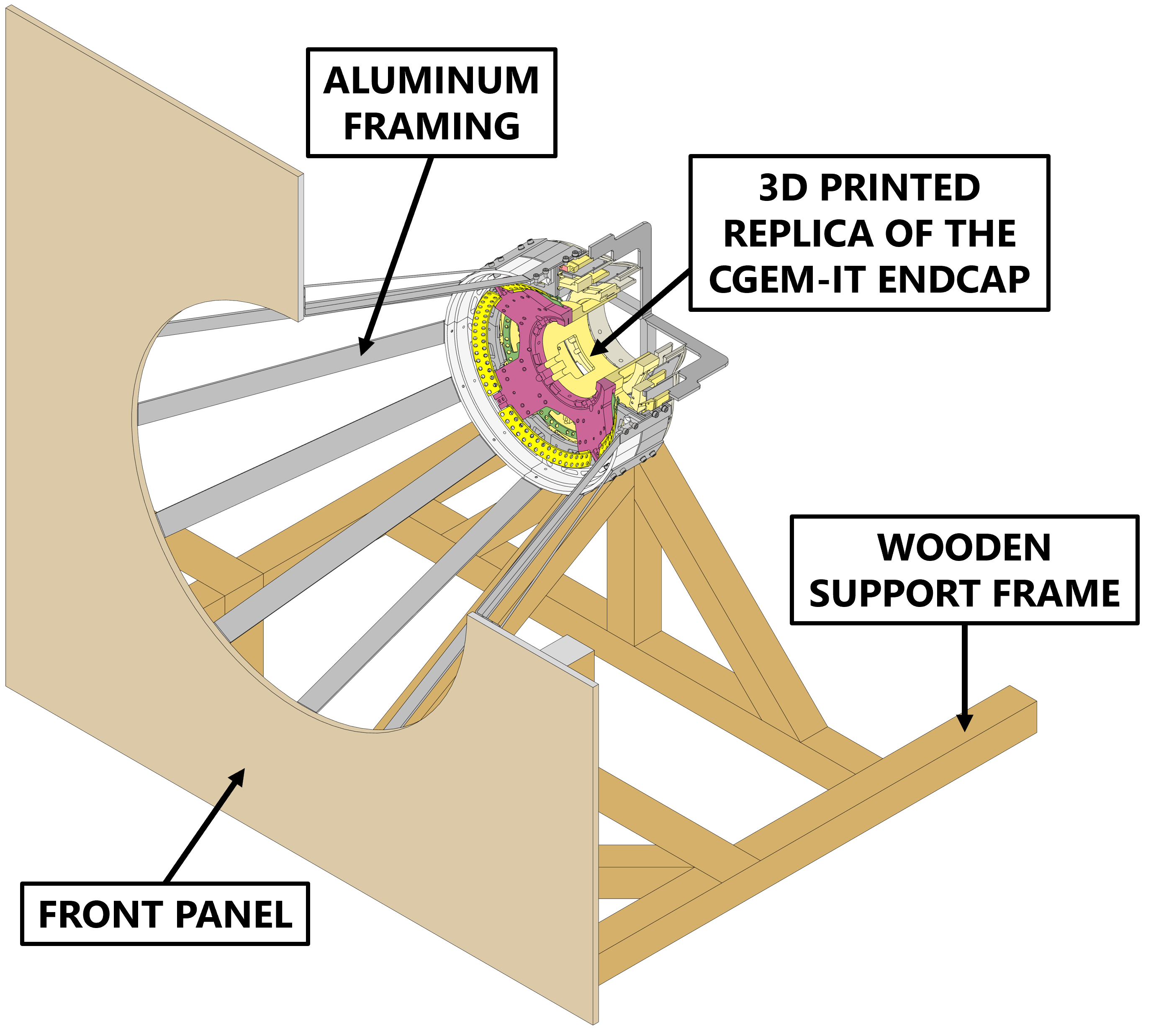}
	\caption[Mock-up for the cabling studies]{3D model of the mock-up used to study the cabling shcemes.}
	\label{cabling_mockup}
\end{figure}

Figure$\,$\ref{cabling_mockup} shows the CAD model of the mock-up constructed for studying the cable schemes.
The design in this case was intended to be inexpensive and quick to build.
The CGEM-IT's endcap and MDC's flange were replicated by gluing together several 3D printed parts, some of which were salvaged from previous setups.
Rope segments of different colors, lengths, and sections, glued to the endcap, simulate the different types of SH cables.
These analogs are labeled according to their position, and match the corresponding cables on the detector.
A wooden panel with a large central hole provides a support to staple full-scale drawings depicting the position of the removed preamplifier boards.
A conical aluminum frame connects the panel and the 3D printed endcap, replicating the shape of the MDC's cone, while a wooden one is used to support everything and provide a stable base.

Tentative cabling schematics were first prepared according to a set of criteria and then tested on the mockup.
The criteria adopted were derived from the experience accrued in cabling the detector for the cosmic ray data taking and from environmental constraints.
In order of priority these were:
\begin{enumerate}
\item DLVPCs and HV connections must be kept on separate towers.
\item HV connections of separate layers should be kept on separate towers, in order to to minimize the number of GNDPCs necessary.
\item Cables cannot cross near the cable holders, not to interfere with the interaction magnet.
\item The number of crossed cables should be minimized far from the patch cards, due to the limited clearance between the shielding plates.
\end{enumerate}

\begin{figure}[htbp]
	\centering
	\begin{subfigure}[t]{.49\textwidth}
		\centering
		\includegraphics[keepaspectratio, width=\linewidth]{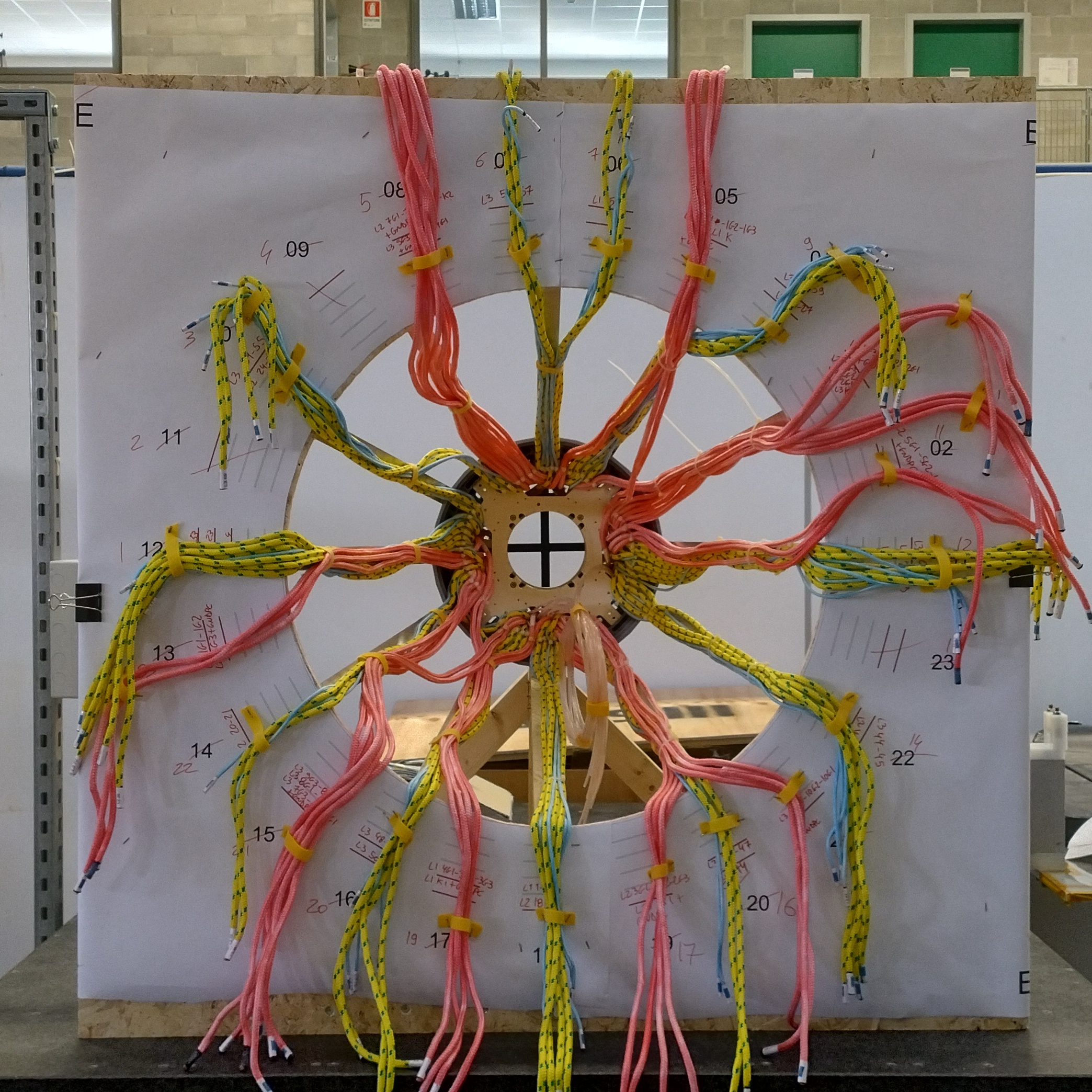}
		\caption[]{West side.}
	\end{subfigure}
\hfill
	\begin{subfigure}[t]{.49\textwidth}
		\centering
		\includegraphics[keepaspectratio, width=\linewidth]{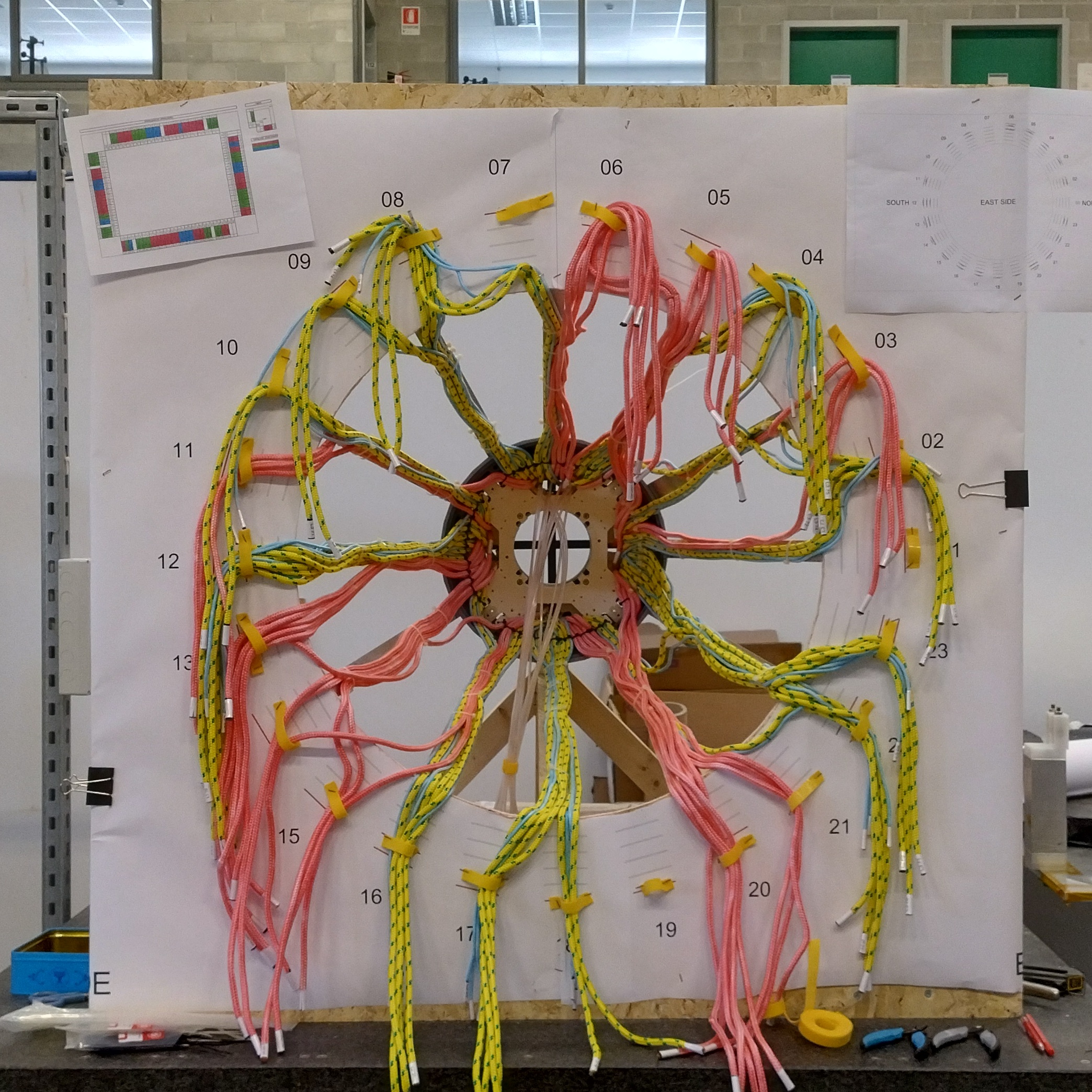}
		\caption[]{East side.}
	\end{subfigure}
	\caption[Test of the cabling schemes]{Results of the test of the cabling schemes for both sides of the detector.}
	\label{cabling_test}
\end{figure}

\begin{figure}[htbp]
	\centering
	\includegraphics[keepaspectratio, angle=90, width=\textwidth]{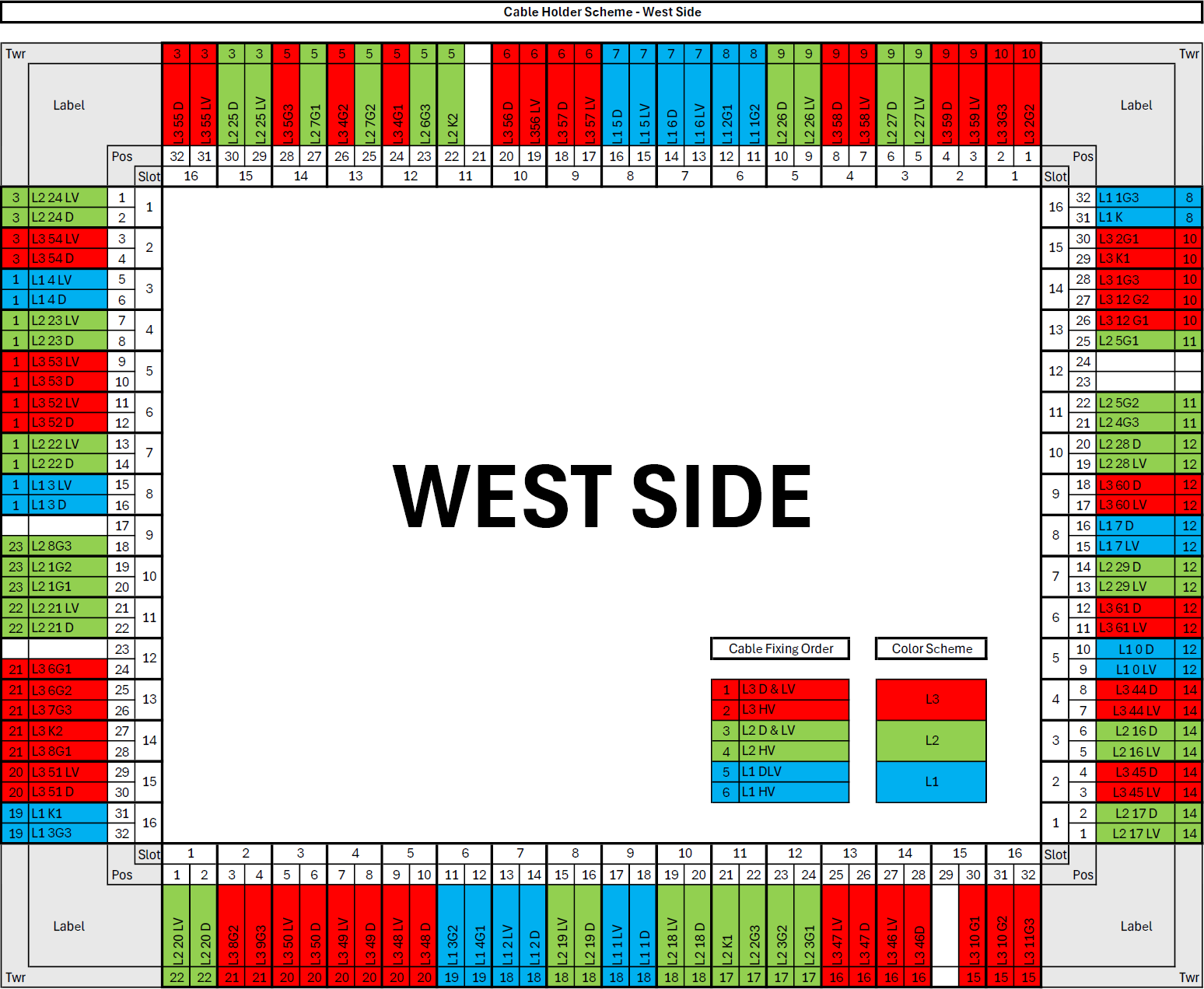}
	\caption[Cable holder schematic example]{An example of the schematics used to inform the fixing of the cables onto their holders. This schematic contains a complete set of the cabling information and can therefore be used as a map to locate issues when debugging the connections.}
	\label{holder_scheme}
\end{figure}

\begin{figure}[htbp]
	\centering
	\includegraphics[keepaspectratio, angle=90, width=\textwidth]{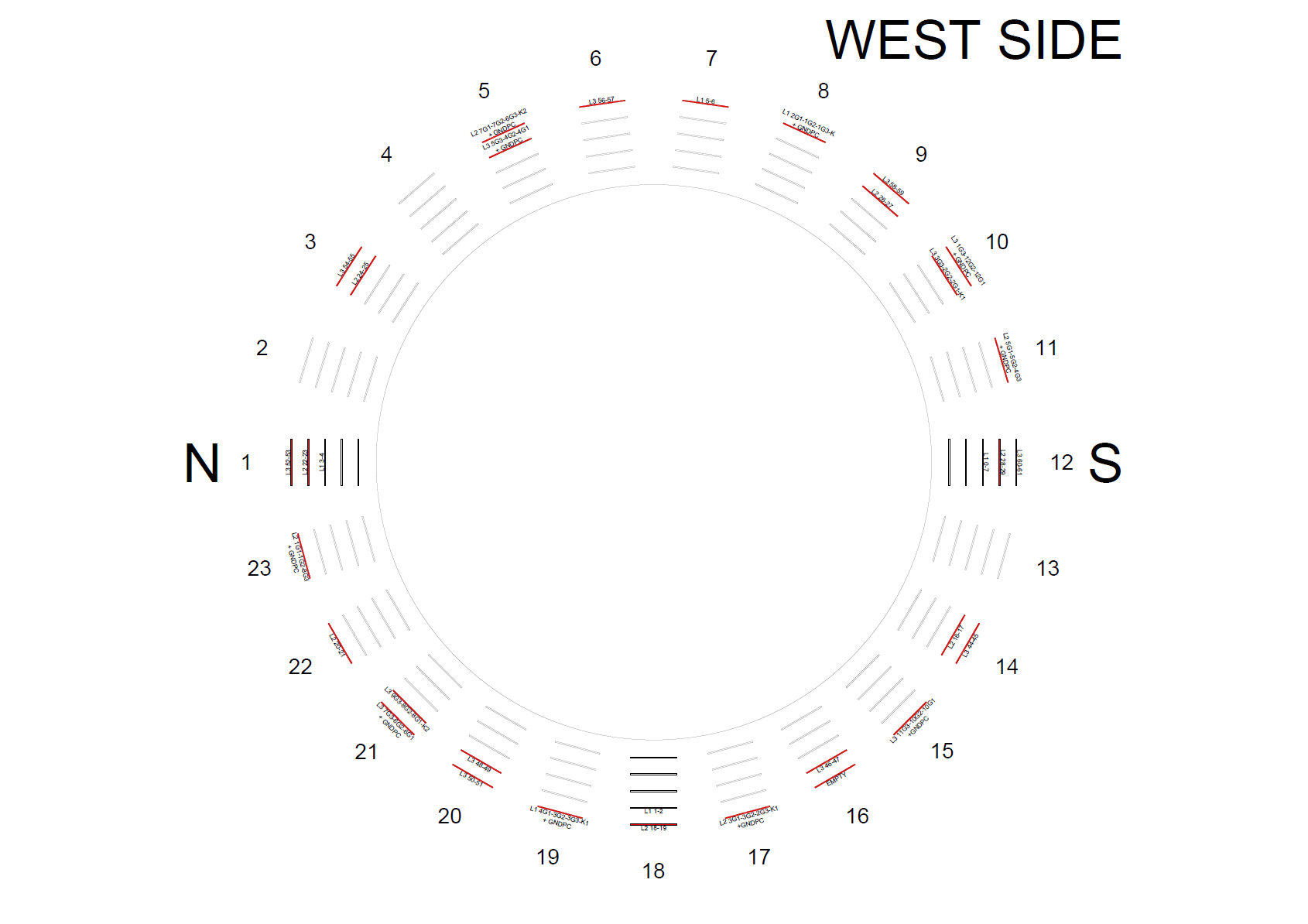}
	\caption[Patch card placement schematic example]{Schematic used to easily locate the tower of destination of SH cables. These schematics, divided in half and printed in A0 format, were stapled to the front panel of the mock-up to represent the location of the slots where patch cards could be installed on the apparatus}
	\label{tower_scheme}
\end{figure}

The last two criteria required to arrange and fix the cables on their holders according to the position of the towers of of arrival.
The mock-up could be converted to represent either side by flipping the 3D printed endcap 180° and adding or removing a few cables.
Several iterations were necessary before satisfactory solutions could be found for both sides, and figure$\,$\ref{cabling_test} showed the result of their testing conducted on the mockup.
The end product of the tests was the validation of cabling schematics as those shown in figure$\,$\ref{holder_scheme} and \ref{tower_scheme}.
These were designed to collect in a concise way all the information necessary to fix the cables and route them to the patch cards.

\FloatBarrier

\section{The Installation up to Now}
\begin{figure}[htbp]
	\centering
	\includegraphics[keepaspectratio, angle=270, width=7cm]{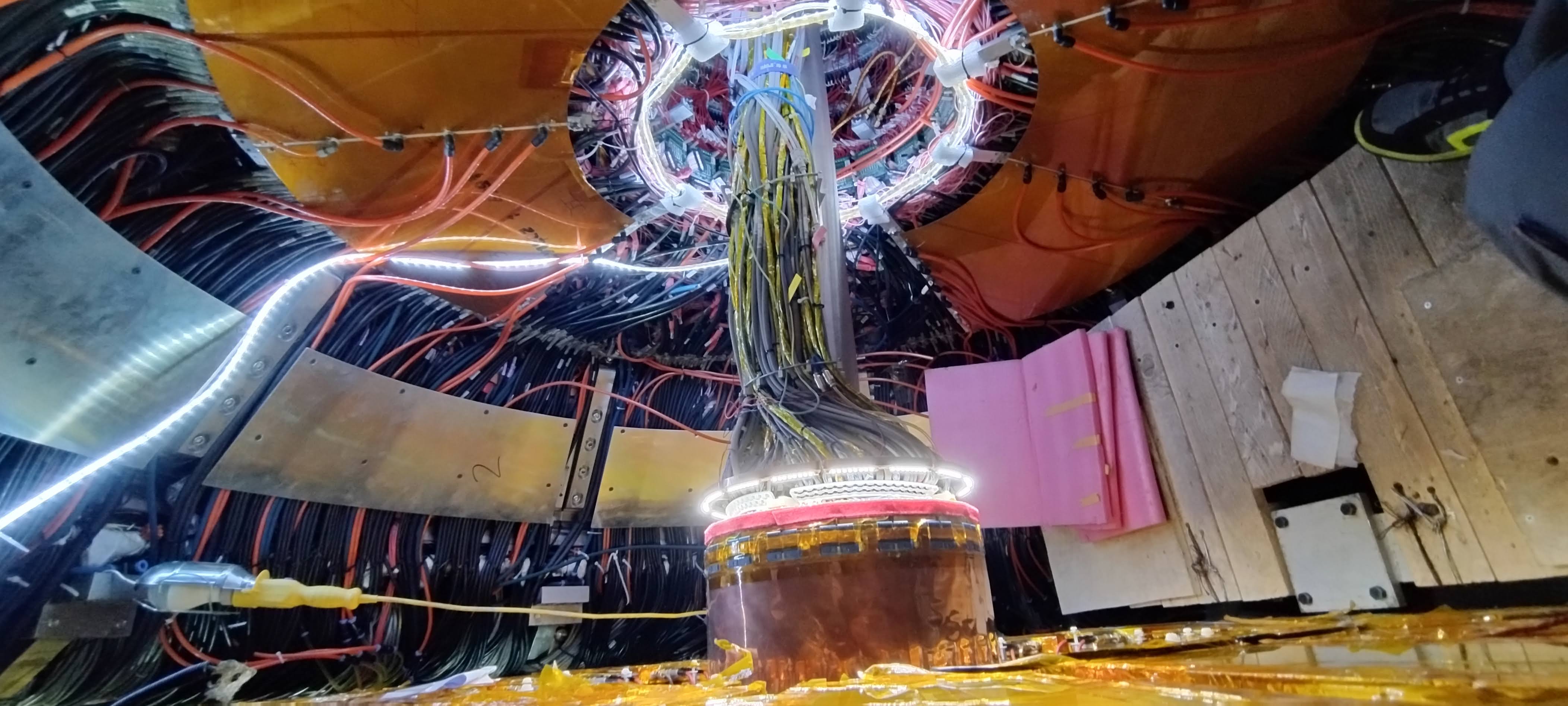}
	\caption[Cavity between the endcap of the EMC and the endplates of the MDC]{The narrow cavity separating the endcaps of the EMC and the endplates of the MDC, where most of the operations took place. The photograph was taken as the detector approached the MDC at the beginning of the insertion.}
	\label{office}
\end{figure}

The time window allocated for the installation of the CGEM-IT, inclusive of a 15 days validation with cosmic ray data, was of 45 days.
All the preparatory work was meant to ensure the safety of the detector during the operations and to optimize the workflow within this time frame.
The insertion of the detector and its cabling required about 10 days each, with the remaining time dedicated to grounding, debug, optimization, and installation of ancillary systems.
The cramped working environments were the cause of most of the difficulties encountered in the installation.
A large part of the operations had to be performed in the narrow cavity between the endcaps of the EMC and the endplates of the MDC, shown in figure$\,$\ref{office}.
Maximum three but more often only two people could occupy this space and work on the detector at the same time.
The occupancy of the preamplifier boards was also underrepresented in the mock-up used in the two insertion tests.
This eventuality had been considered and specialized tools had been purchased to cope with the narrower spaces.
The contact sensor array was never triggered as optical evaluation of the clearance proved sufficient to prevent collisions.
As in the two tests, shims had to be added to the west side while anchoring the detector.
Aside from this, no major issues arose during the operations, and only minor deviations from the procedure were rendered necessary, mostly by the different conditions offered by the environment.

\begin{figure}[htbp]
	\centering
	\includegraphics[keepaspectratio, width=\textwidth]{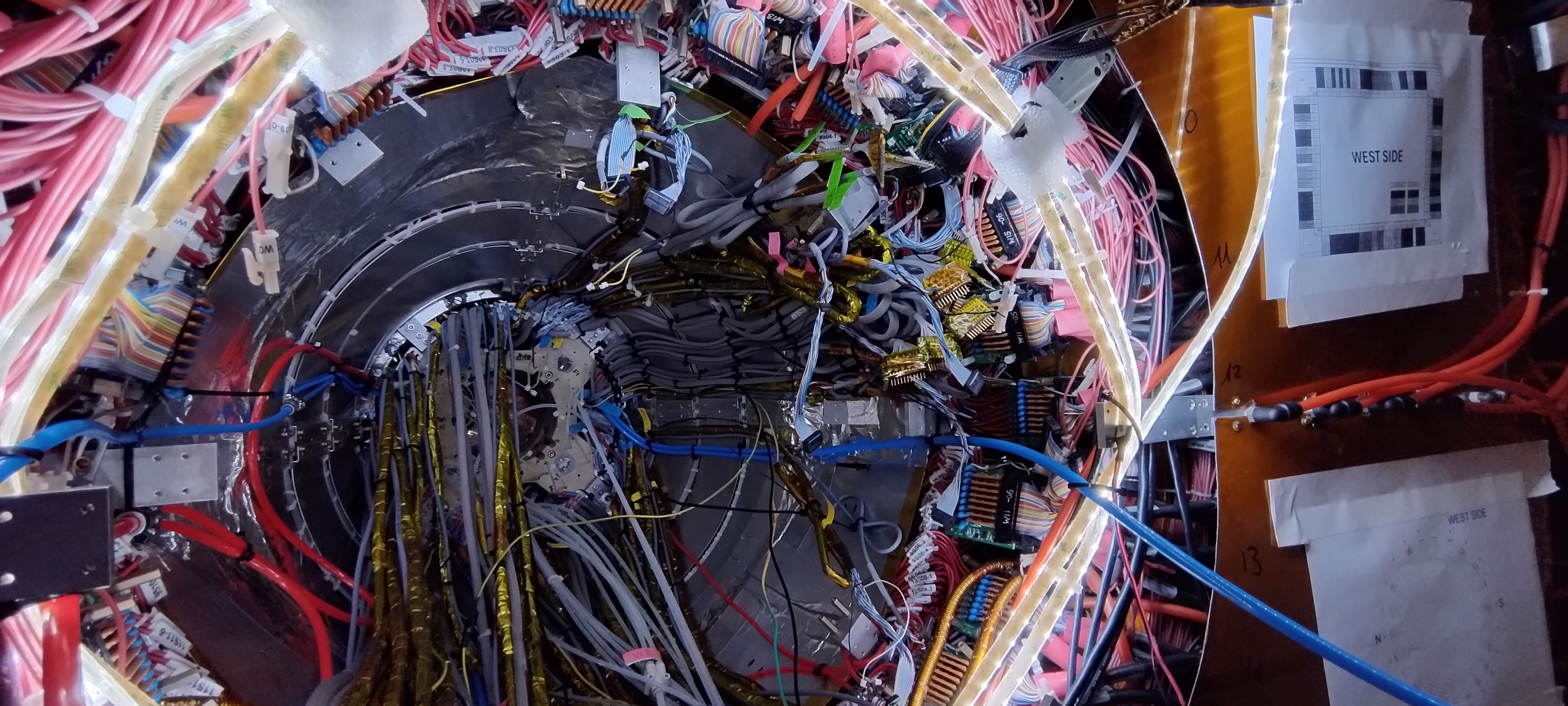}
	\caption[Routing of the cables inside the MDC's cone]{Routing of the cable withing the MDC's cone. Cables are kept parallel and flat against the shielding closer to the inner tracker, where the clearance is narrower, while excess length is collected near the destination towers.}
	\label{cabling_progress}
\end{figure}

\begin{figure}[htbp]
	\centering
	\includegraphics[keepaspectratio, width=\textwidth]{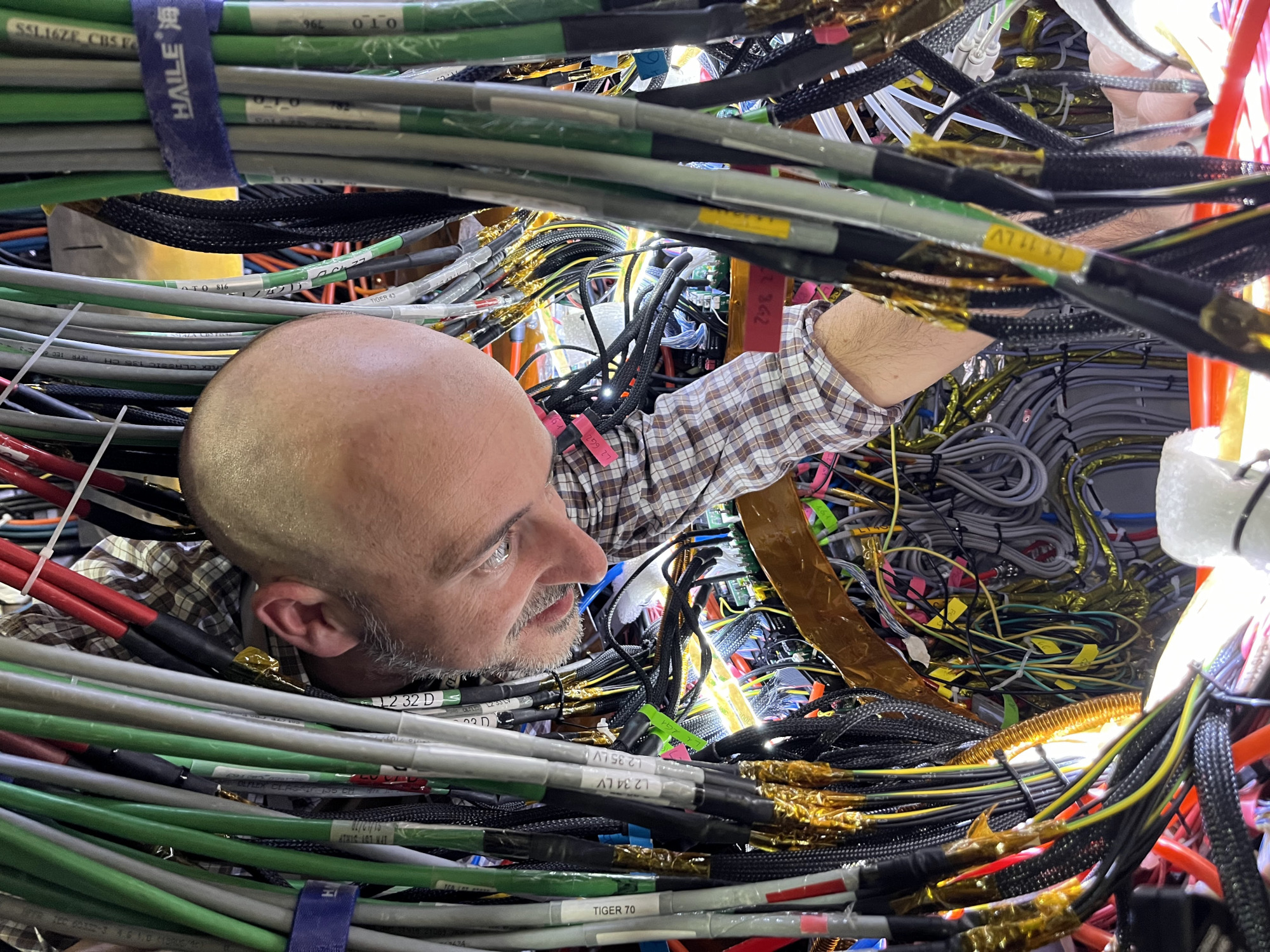}
	\caption[Difficulty in operating inside the MDC's cone after cabling]{Photograph depicting the difficulty of conducting operations inside the cone as the long haul cables progressively restricted the available space.}
	\label{gigi}
\end{figure}

The routing of the cables within the shielded cone of the MDC was by far the most time consuming operation.
The schematics proved invaluable in minimizing improvisation and mistakes that could have led to long delays.
The extra length of the cables was collected near their destination towers, as shown in figure$\,$\ref{cabling_progress}, where more space is available below the EM shielding.
The conduits where the MDC's cables reside could not be accessed, so the cables of the CGEM-IT had to exit the sprectrometer along the beam pipe.
A lot of planning was involved in making sure that the cables wouldn't get entangled, but their presence still made accessing the endcaps increasingly difficult as more and more cables were added, as shown in figure$\,$\ref{gigi}.

The detector was powered on at nominal HV settings, for the first time inside BESIII, on the 19th of October 2024.
All on and off-detector electronics has since been tested, showing noise levels comparable with those observed in the laboratory during the detector's validation with cosmic ray data.
At the time of writing, the installation of the detector is still ongoing, and now focused on grounding improvements and validation through the acquisition of cosmic ray events.
Once the operational status of all systems will be confirmed, the reassembly of the spectrometer and of the beam pipe will commence, to allow the commissioning of the detector with the particle beams provided by BEPCII.
The BESIII data taking is scheduled to restart in January 2025, while six months to a year of commissioning with collision data is estimated before the CGEM-IT can begin to operate at full efficiency.
    		\chapter*{Conclusions}
\addcontentsline{toc}{chapter}{Conclusions}
BESIII is a general purpose high energy physics experiment pursuing a diversified physics program centered around charmonium and $\tau$ physics.
Its inner tracker, the innermost part of a large multilayer drift chamber, has been suffering a decrease in performance that may start to hinder the data-taking in the face of planned upgrades to the BEPCII collider.
The Italian component of the collaboration proposed, developed, and it is now installing a replacement based on cylindrical GEM technology.
The CGEM-IT is a three-layer, triple-GEM detector with analog readout aiming to match the drift chamber's spatial resolution in the radial azimuthal direction and to improve it by at least a factor 2 in the beam one.
The aging resistant MPGD technology should allow BESIII to continue its data taking, scheduled to end in the year 2030, despite the increasing luminosity of the machine.

The first version of the largest layer failed to power on when the internal floating electrodes collapsed due to buckling-induced deformation.
The review of the mechanical design that ensued led to the addition of PEEK spacer grids, to constrain the amplitude of the buckling lobes within tolerance.
The solution was tested by letting a mock-up of layer 3 take controlled falls at increasing inertial loads and then comparing snapshots of its internal structure collected through X-ray CT scans.
The test confirmed the validity of the solution and set a new limit of 7.5$\,$g for the critical load of a new version of Layer 3 reinforced with PEEK spacers.

The new limit was not deemed enough to ensure safe transport of the whole detector by airplane, so a split construction was pursued instead.
The electrodes were built in Italy, shipped to the People's Republic of China while protected by their mandrels, and finally assembled on-site.
One of the main obstacles to overcome was how to align the assembly machine, which required dial gauges to slide on the surface of the mandrels.
The solution was to upgrade the VIM with a new contactless alignment system based on laser triangulation sensors.
The new system, and the procedure developed alongside it, ensured a very good alignment, leading to an unprecedentedly smooth assembly despite the reduction in clearance due to the grids.

With the completion of Layer 3, the CGEM-IT entered its pre-com\-mis\-sion\-ing phase.
All systems necessary to acquire data with the three layers were deployed together for the first time in a dedicated cosmic ray telescope setup.
All layers operated stably, despite some discrepancies between their individual behaviors.
The detector managed to achieve the target performance in terms of stability, resolution, and efficiency, and was therefore approved for installation by the internal review committee.

A series of tests were conducted for defining the procedures to safely insert the detector in the spectrometer.
The results of these tests led to redesigning and manufacturing anew most of the installation tooling.
The rigidity of the insertion trolley was increased and a guard ring was added in front of the detector to facilitate evaluating clearance with respect to the cavity.
The legs supporting the rail used to slide the CGEM-IT in place were also redesigned; their adjustment mechanisms were refined to ensure reliable and smooth control over the alignment of the rail.
A thorough study of the cabling scheme was also deemed necessary in order to complete the installation within the allotted time frame.
The problem was tackled by building a 1:1 scale mock-up of the CGEM-IT's endcap inserted into the MDC and by routing cable analogs made of rope to chose among several cabling options.
All the preparatory work done helped ensure a smooth insertion and time-efficient cabling during the actual installation of the detector, which is still ongoing at the time of writing.
The CGEM-IT was powered on at nominal HV settings inside BESIII for the first time on the 19\textsuperscript{th} of October 2024.
No electrical issues indicative of internal damage have manifested and all the detector electronics appear to be operating nominally.
    
 	\appendix
 		\chapter{GEM Foils HV Test Procedure}
\label{hvprocedure}
\section{Insertion of the GEM Foil into The Test Box}
\paragraph{Initial Condition}
The foil to be tested is inside the storage package, the package is resting on a table.
The HV test box is empty and closed.
The nitrogen flow to the box is closed.
\paragraph{Procedure}
\begin{itemize}
\item Open all the latches and remove all the clamps keeping the HV test box closed.
\item Lift the test box lid and set it aside.
Ideally, the lid should be placed upside down on a table, with the inner surface facing the ceiling. If placed vertically, with one side touching the floor, remember to clean it thoroughly with isopropyl alcohol before closing the box again.
\item Open the nitrogen flow to the ionizing gun.
To do so:
	\paragraph{Initial Condition} The nitrogen tank's valve is open.
	The line connecting the tank and the box's pressure regulator is open.
	The box's pressure regulator is set to 0.5$\,$bar or less (on a closed pipe).
	The valve to the ionizing gun and the valve to the box are closed.
	The pipes connecting the pressure regulator to the box/gun fork are under pressure.
	\begin{itemize}
	\item Open the valve to the ionizing gun at the fork.
	\item Turn on the ionizing gun's power supply.
	\item point the gun towards the ceiling or towards the floor, away from surfaces that may have collected dust, and let the gas flow for a few seconds to clean pipe and nozzle.
	\end{itemize}
\item Clean the inside of the box first with isopropyl alcohol and then by blowing it with nitrogen.
Always blow at low pressure ($< 1\,$bar), at a grazing angle with respect to the surface to clean, and moving outwards from the center. Do not start blowing while pointing at the surface to clean. Point the gun in the air and then move towards the surface to clean when the flow is low and stable while keeping the trigger pulled.
\item Open the GEM package.
\item Position the tray near the GEM package, so that moving a GEM foil on top of it would not require taking a step. Have a person keep the tray in place if there is no space for setting it down next to the package.
\item Free the GEM foil from any protective cover: Nylon film, thin plastic sheets, etc.
\item Transfer the GEM foil on the tray.
This is done by two people, each positioned at one of the shorter edges of the foil, holding the GEM by its corners.
The movement must be performed without taking any steps to minimize the risk of relaxing the foil and generating creases on its surface.
The TOP face of the GEM, where the micro-sectors are, should be facing upward.
If not, flip the GEM foil.
\item Move the tray next to the open test box.
Have a person keep it in place.
Make sure that the GAS OUT side of the foil, the one with the smaller gas holes, faces the GAS OUT side of the test box.
\item Transfer the GEM foil inside the box.
This is done by two people positioned at the shorter sides of the foil.
The movement must be performed without taking a step.
The active area of the GEM should rest at the center of the box, on top of the Mylar foil covering its bottom.
The two shorter edges of the GEM foil rest on top of the black rubber shims at each end of the box.
The HV tails protrude from the black rubber shims to allow for the placement of the green HV connectors.
\item Check that the GEM is in the correct position: the TOP face, where the micro-sectors are, should be facing upward; the GAS OUT side of the GEM, where the small gas holes are, should face the GAS OUT side of the box; the HV tails should be protruding from the black rubber shims at each side.
\item Cover the active area with a rectangular sheet of vacuum bag material.
\item Connect the HV tails to the green HV connectors.
To do so:
\begin{itemize}
\item Unscrew the two Nylon screws so that the slit in the connector becomes large enough for the stiffener of the HV tail to fit.
\item Carefully insert the tail and the stiffener inside the connector.
Once in the correct position, the stiffener should protrude slightly (about 0.1$\,$mm) from the connector.
Have someone press on the tail and on the edge of the foil to keep them in place from this moment up to the taping down of the connectors.
Make sure that the ground side of the connector (it should be written on the connector) matches the macro-sector track on the HV tail.
In case the identification of the ground side of the connector results difficult, remember that the lone wire on the connector corresponds to micro-sector number 10.
\item Tighten the Nylon screws to to secure the HV tail inside the connector.
\item Clamp the connector with a small black document clamp to further secure the contact between the connector and the HV tail.
\item Tape the connector and the cables in place using Kapton tape so that they do not put strain on the HV tail when released.
\item Let go of the HV tail and of the edge of the foil. Check that none of them are under stress due to the position of the connector.
\item Repeat these operations to connect all the remaining connectors to the respective  HV tails.
\end{itemize}
\item Check the connection between the female banana sockets on the outside of the box and the HV tracks, where they fan out towards the active area, using the tester's beeping mode.
To check the macro-sector connection, slide some copper-clad Kapton or some copper tape underneath the foil, where the macro-sector track is located, and press down on it to ensure electrical contact before using the tester.
\item If any sector results disconnected, check or reattach the green HV connector and repeat the beeping test.
\item Check for short-circuits between neighboring micro-sectors, using the tester beeping function, by plugging the probes in the female banana connectors outside the box two by two.
\item Measure the capacitance between each micro-sector and its macro-sector.
The capacitance for the Layer 3 GEMs should be around 3.4$\,$nF and remain relatively stable during the measurement.
\item Remove the rectangular vacuum bag sheet covering the active area.
\item Blow the active area with nitrogen using the ionizing gun. Always blow at very low pressure ($\leq\,0.5\,$bar on a closed pipe), at a grazing angle with respect to the surface to clean, and moving outwards from the center.
Do not start blowing while pointing at the surface to clean to avoid the first few moments of high speed flow.
Point the gun in the air and then move towards the surface to clean, while keeping the trigger pulled, after the flow is low and stable.
\item Clean the inner surface of the test box lid with isopropyl alcohol and by blowing it with nitrogen.
\item Close the test box with its lid.
\item Check that the O-ring is inside the groove all along the box perimeter and that no wires are being pinched.
\item Close the lid's latches and place the clamps around the box.
Start with the longer sides, then close the shorter ones and keep the corners last.
Place the PTFE shims under the clamps' bolts to preserve the box's surface.
\item Measure once again the capacitance between each micro-sector and its macro-sector.
The capacitance should be around 3.4$\,$nF and remain relatively stable during the measurement.
\item Close the flow to the ionizing gun and start the flushing the box with nitrogen.
To do so:
	\paragraph{Initial Condition} The nitrogen tank's valve is open.
	The line connecting the tank and the box's pressure regulator is open.
	The box's pressure regulator is set to 0.5$\,$bar or less (on a closed pipe).
	The valve to the ionizing gun is open, the valve to the box is closed.
	The pipes connecting the pressure regulator to the box/gun fork are under pressure.
	The flow-meter is completely open, both exhaust lines are connected.
	\begin{itemize}
	\item Turn off the ionizing gun's power supply.
	\item Close the valve to the ionizing gun.
	\item Disconnect the exhaust line.
	\item Disconnect the pipe connecting the box to the fork at the fork side.
	\item Open the valve to the box. The nitrogen will flow, the pressure will fall rapidly and the flow will quickly become stable.
	\item While the nitrogen is flowing connect the box at the fork. This is to avoid ram pressure issues when opening the valve with the box already connected.
	\end{itemize}
\end{itemize}

\section{HV Test}
\paragraph{Initial Condition} The foil to be tested is inside the box.
The box is closed, latches and clamps are in position.
Nitrogen is flowing inside the box.
The computer, the amperometer and the HV power supply are off.
The power supply interlock switch is in the closed position.
\paragraph{Procedure}
\begin{itemize}
\item Turn on the computer, the amperometer, and the HV power supply.
\item Check that the interlock switch of the power supply is closed.
If not switch it to closed.
\item On the computer open the VI for the test.
\item Type the name of the test inside the text box in the top left corner of the dashboard.
The naming convention is:\\ <<name\_of\_the\_foil>>\_m<<macrosector\_number>>.\\
Where <<name\_of\_the\_foil>> is g3a, g2b, g1c, etc., while the macrosectors are numbered starting from 0, left to right while looking at the foil from the GAS IN side.
Some examples of what should be written in the text box are: g3a\_m0, g2c\_m3, g1b\_m5, etc.
\item Start the VI.
If errors occur check:
\begin{itemize}
\item That the power supply is on and connected to the computer via Ethernet.
\item That the amperometer is on and connected to the computer (the connector at the amperometer side is pretty loose).
\item That the Arduino inside the computer is on, lights should be visible through the computer frontal grate, and connected to the computer via USB.
\item That the COM ports used for communicating with the amperometer and the Arduino in the VI are COM1 and COM15, respectively.
\end{itemize}
If none of these solve the issue, power cycle everything. If it still doesn't work, call an expert.
\item Wait for the humidity to fall below 5\%, the red light next to the humidity value should turn off as well.
\item Begin the actual HV test:
\paragraph{Initial Condition} Humidity is below 5.0\%, the humidity alarm light is off.
The current for all channels is around 1.5-2.0$\,$nA.
The voltage for all channels is 0$\,$V.
The HV power supply interlock is closed.
\paragraph{First Power On} Set the voltage to 10 and press enter on the keyboard or click the up arrow on the dashboard control, in this case it will show 11 but it will enter 10 anyway.
Check that all channels behave in the same way, pulling approximately the same amount of current until the new voltage is established.
After reaching the desired voltage the current should fall back to 1.5-2.0$\,$nA.
If everything is stable proceed to the next phase.
If one or more channels do not fall back to the background current level of 1.5-2.0 nA evaluate if to proceed with the test to try and burn up the contaminant or to open the box and try cleaning it with nitrogen.
\paragraph{Ramp Up}
Increase the voltage progressively according to the following steps:

100$\,$V, 200$\,$V, 300$\,$V, 350$\,$V, 400$\,$V, 450$\,$V, 475$\,$V, 500$\,$V, 525$\,$V, 550$\,$V, 560$\,$V, 570$\,$V, 580$\,$V, 590$\,$V, 600$\,$V.

After each step wait 15-30 seconds, or the time necessary to update the logbook, and record the channel behavior during this time.
The possibilities are:
\begin{itemize}
\item \textbf{Stable behavior} After the ramp the current falls back to the background level.
\item \textbf{Instability} The current fluctuates randomly.
A ramp or a discharge have been observed triggering some instability in the past.
After some instability arises wait for 1-2 mins of stable behavior before raising the voltage again.
Log channel and approximate duration of the unstable behavior.
\item \textbf{Discharge} The current spikes and falls back to the background level.
Log channel and maximum current reached.
After a discharge occurs wait for 1-2 mins of stable behavior before raising the voltage again.
If during this time another discharge occurs on the same channel, turn off the light and check for the position of the discharges through the box window to check if the discharges always happen at the same location.
\item \textbf{Repeated discharges} Multiple discharges occurring at the same location.
Repeated discharges may be due to contaminants too big to burn away.
It may be necessary to open the box and try to remove the contaminant by blowing the affected area with nitrogen.
\item \textbf{Current absorption} After a ramp, a discharge, or a period of instability, the current does not fall back to the background level of 1.5-2.0$\,$nA but either stabilizes itself at a different level or keeps increasing slowly.
Log the behavior and decide if to try and burn the contaminant by raising the voltage or if to open the box and flush with nitrogen.
\end{itemize}
\paragraph{600V Conditioning}
After reaching 600V wait for 5 minutes before starting the Long Term Test (LTT). During this time log anomalous behavior.
\paragraph{600V Long Term Test}
The long term test lasts 30 minutes. During this time log any anomalous behavior in the channels under test alongside the time at which discharges occur. More than two discharges from the same channel during the 30 minutes disqualify the foil for construction. The foil can be flushed and tested again or simply discarded.
\paragraph{Scoring}
A scoring system may be implemented for choosing the best candidates for construction.
\begin{itemize}
\item \textbf{Minor Mechanical Defects (1 point per defect)} Small localized mechanical imperfection that surely will not affect performance.
\item \textbf{Moderate Mechanical Defects (5 points per defect)} Pronounced and localized or diffused mechanical defects that may affect performance but will not prevent operation of the detector.
\item \textbf{Major Mechanical Defects (immediate exclusion)} Pro\-nounced and localized or diffused mechanical defects that may prevent the detector from functioning altogether.
\item \textbf{Minor Discharge (1 point per discharge)} Discharges that reach a maximum current lower than 1$\,\upmu$A.
\item \textbf{Moderate Discharge (2 point per discharge)} Discharges that reach a maximum current between 1$\,\upmu$A and 5$\,\upmu$A.
\item \textbf{Major Discharge (3 point per discharge)} Discharges that reach a maximum current above 5$\,\upmu$A.
\item \textbf{Resolved Instability (1 point per occurrence)} Any instance of instability, as long as it was resolved.
\item \textbf{Unresolved Instability (immediate exclusion)} Unresolved, recurring or long lasting instability.
\item \textbf{Resolved Current Absorption (1 point per occurrence)} Any instance of current absorption, as long as it was resolved either electrically or mechanically.
\item \textbf{Unresolved Current Absorption (immediate exclusion)} Unresolved, recurring or long lasting current absorption.
\item \textbf{Resolved Repeated Discharges (score each discharge independently)} Score each charge according to its peak current.
\item \textbf{Unresolved Repeated Discharges (immediate exclusion)} Unresolved, recurring or long lasting repeated discharges.
\end{itemize}
The two lowest scoring candidates are chosen as the best candidates for construction. More than two discharges recorded on the same channel during the long term test disqualify the foil regardless of its score.
\end{itemize}
 		\chapter{CGEM-IT Insertion Procedure}
\label{insertproc}
\paragraph{DISCLAIMER}
The format of the procedure was subject to optimization during course of the various insertion tests.
Because of this it is here presented in full and in the original formatting.
Some of the names used to describe parts or components have although been changed to uniform nomenclature with the main body of the thesis.
Heading hierarchy as also been altered to fit within this appendix.

\section{Introduction}
\subsection{Premise}
This document aims to provide a guide for all the operations leading to the installation of the CGEM-IT in BESIII. These were defined and practiced in two tests, which took place in early February and early July 2024 at the Institute of High Energy Physics in Beijing.

This procedure is written for the people who will be involved in the installation of the CGEM-IT and who are therefore already familiar with the adopted terminology. Previous knowledge of many aspects of BESIII and the CGEM-IT is assumed.

A risk assessment report, highlighting the main risks the detector will be exposed to during operations, is included at the end of this document.

\subsection{Initial Condition}
\begin{figure}[htbp]
\captionsetup{singlelinecheck=off}
\centering
\includegraphics[keepaspectratio, width=\textwidth]{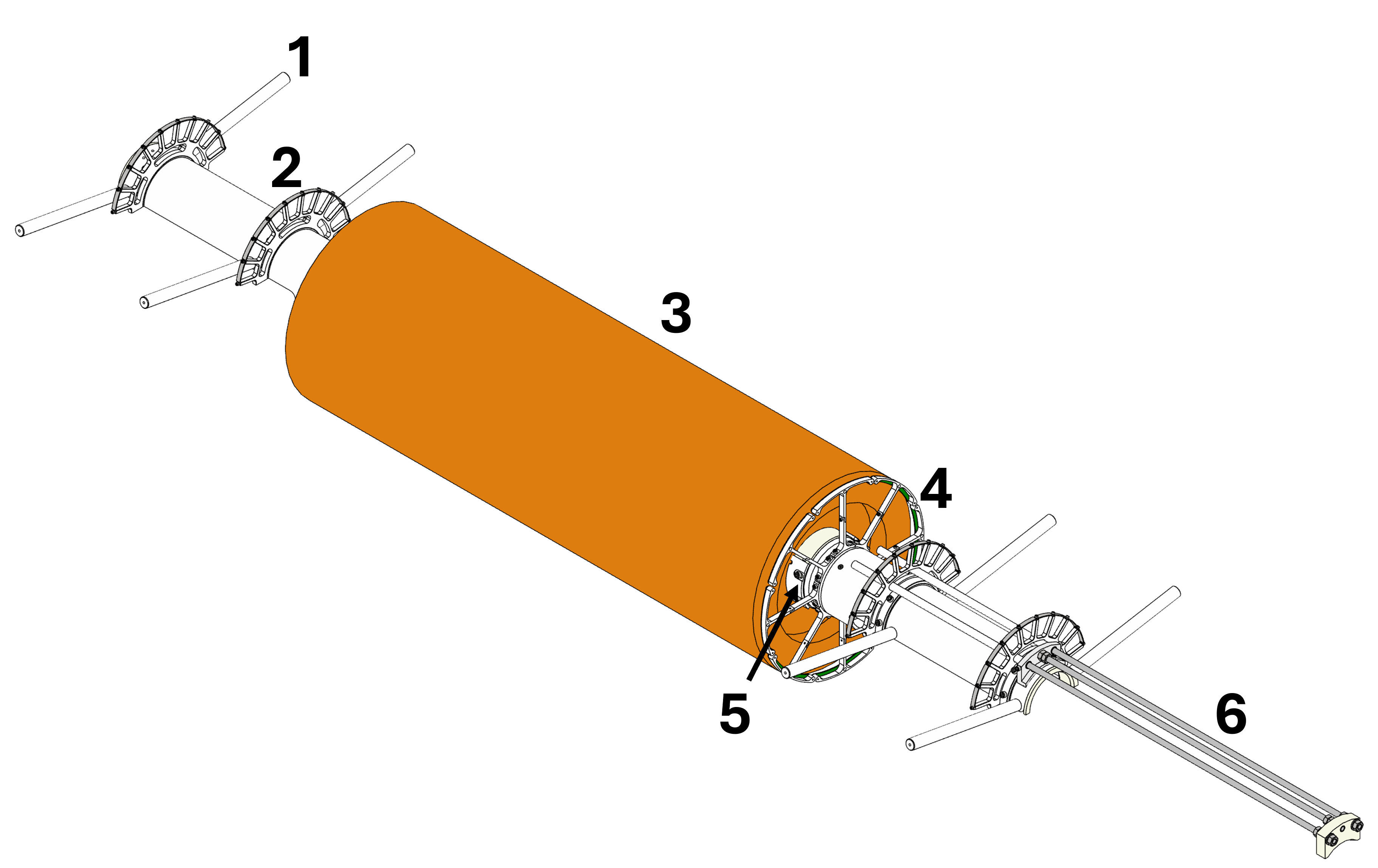}
\caption[Insertion trolley]{Volume occupied by the CGEM-IT on the insertion trolley. The numbers refer to different parts of the trolley:
\begin{enumerate}
\item Handles
\item Cable holders
\item Volume occupied by the detector
\item Guard ring and sensor array
\item Detector stoppers
\item Telescopic cable holder
\end{enumerate}
}
\label{det_on_trolley}
\end{figure}

The CGEM-IT is resting on the insertion trolley as of fig.$\,$\ref{det_on_trolley}. Four aluminum stoppers, henceforth referred to as detector stoppers, prevent the detector from sliding on the trolley. Two of these are shaped so to also prevent rotation with respect to the trolley. The telescopic extension is initially missing and meant to be installed at a later time. The trolley's handles are only meant to help controlling the trolley once it is on the rail and so they are not installed. The sensor array is installed and the power and ground cables coming from the circuits are connected together in bundles. One of the sensor circuits is connected to the Faraday cage, to act as a contact alarm.

The trolley is resting on an extruded aluminum profile, supported by the stretcher used for transporting the detector to the experiment hall. Another set of stoppers, called trolley stoppers, prevent the trolley from sliding along the profile and from rotating freely.

\begin{figure}[htbp]
\centering
\includegraphics[keepaspectratio, angle=90, width=.8\textwidth]{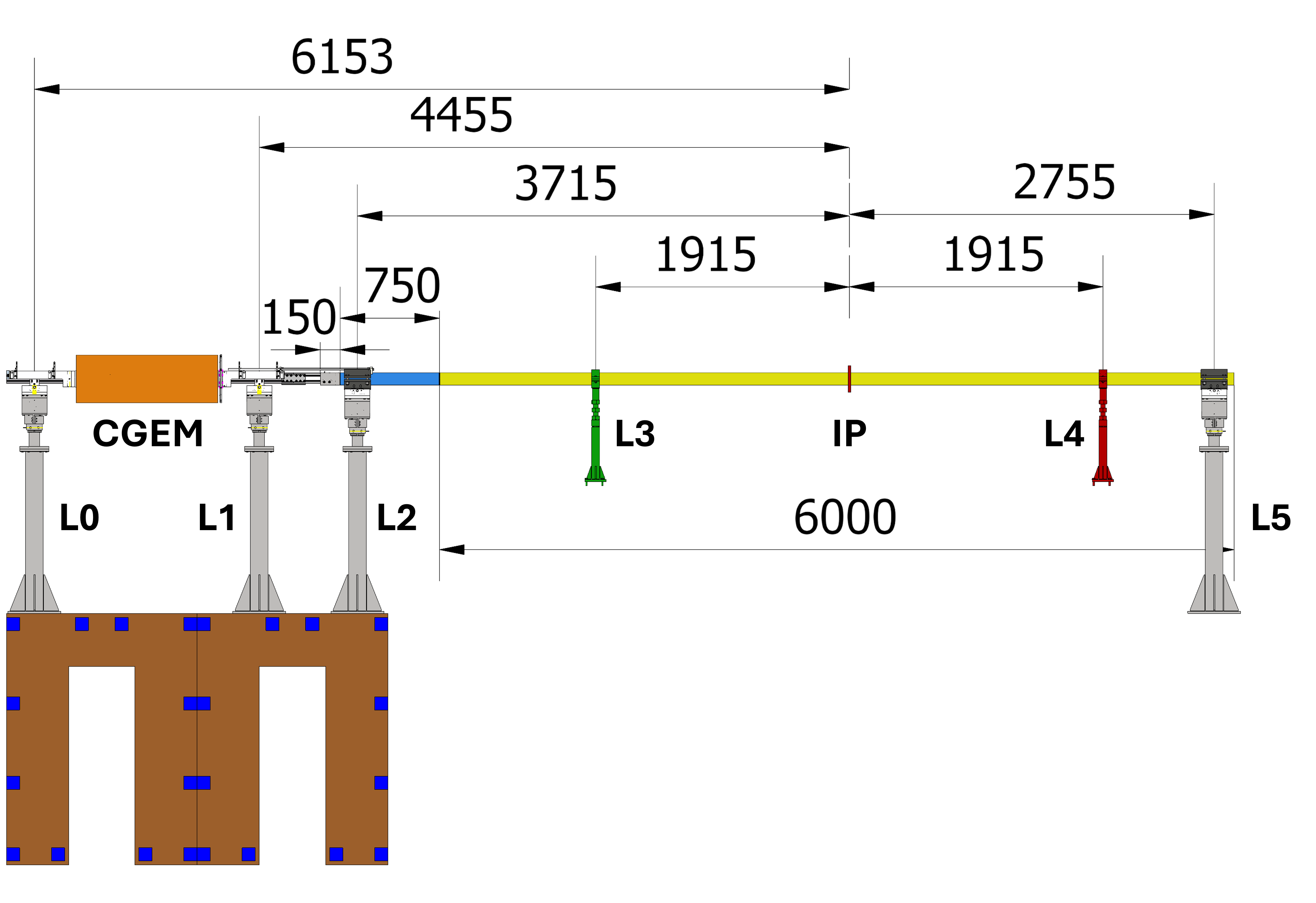}
\caption[Insertion equipment layout]{Layout of the insertion equipment in the BESIII experiment hall. All dimensions are in millimeters.}
\label{setup}
\end{figure}

The trolley is located near the spectrometer's east side, in a position that can be reached by the overhead crane. Suitable anchor points are available on the stretcher to facilitate lifting the detector. The rail and its support legs, as well as the legs used to place the trolley onto the rail, are installed as of fig.$\,$\ref{setup}. The rail is horizontally centered within the cavity.

\FloatBarrier

\subsection{Logistics}
Traveling from one side of the detector to the other is going to be time consuming. Operations that require acting at both sides of the spectrometer will be conducted by two quasi-stationary teams communicating via radio. The composition of each team may vary during the operations.

Each team will be led by a team leader, carrying the team's radio. Two procedure's overseers, one per each side, will call out the order of operations, keep track of progress, and facilitate coordination between the two teams. The operations' expert will always be part of the team handling the most critical operations. Each team can request help or directions from the procedure's overseers or the operations' expert at any moment.

\subsubsection{Radio Communication Protocol}
\begin{itemize}
\item Radios are in the hands of each team leader, of the procedure's overseer, and of the operations' expert.
\item Before initiating any radio communication the speaker should identify themselves by name or team of belonging.
\item The end of each communication should be followed by the word "over" ("passo" in Italian).
\item Radio chatter should be kept to a minimum and solely reserved for the exchange of relevant information.
\end{itemize}

\section{Preparation}

\subsection{Setup Preparation}
\begin{enumerate}[label=\labellazzo{{\arabic*}}{.\,\square}]
\item Install spotlights at each side of the spectrometer, pointing at the MDC.
\item Install led strips in the MDC cone at both sides of the spectrometer.
\item Position a laser level on the east side so that the horizontal line is as aligned as possible with the axis of the rail. Once positioned, the laser level should not be moved or touched until operations are completed.
\item Mark the position of the laser line in several places along the rail using a marker. The rail's deflection and/or position may later be evaluated by measuring the distance between the laser line and the markers using a caliper.
\item Check the assembly of each leg, tightening the top locknuts with bearing spanner keys if necessary. Make sure all the legs are oriented in the same way when reinstalling them.
\item Check vertical alignment and stability of each leg. Adjust if necessary by adding shims below the base.  Note down the position and overall thickness of the shims used for each leg.
\item Test mount the east MDC flange and measure relevant dimensions:

\begin{table}[h]
\centering
\begin{tabular}{|l|l|cc|}
\hline
\textbf{Parameter}               		& \textbf{Value} & \multicolumn{2}{c|}{\textbf{Passed}} \\ \hline
Distance between the two flanges 		&                & \multicolumn{1}{l|}{Y}   & N   \\ \hline
Flanges $\parallel$ w.r.t. each other   &                & \multicolumn{1}{l|}{Y}   & N   \\ \hline
Flanges $\perp$ w.r.t. cavity axis  	&                & \multicolumn{1}{l|}{Y}   & N   \\ \hline
Cavity diameter at $0 \times L$         		&                & \multicolumn{1}{l|}{Y}   & N   \\ \hline
Cavity diameter at $1/4 \times L$       		&                & \multicolumn{1}{l|}{Y}   & N   \\ \hline
Cavity diameter at $1/2 \times L$       		&                & \multicolumn{1}{l|}{Y}   & N   \\ \hline
Cavity diameter at $3/4 \times L$       		&                & \multicolumn{1}{l|}{Y}   & N   \\ \hline
Cavity diameter at $1 \times L$         		&                & \multicolumn{1}{l|}{Y}   & N   \\ \hline
\end{tabular}
\end{table}

\FloatBarrier
\item Horizontally center the rail within the cavity.
\end{enumerate}

\subsection{Fixing the Interconnection Brackets to the West MDC Flange}
\begin{figure}[htbp]
\centering
\includegraphics[keepaspectratio, width=\textwidth]{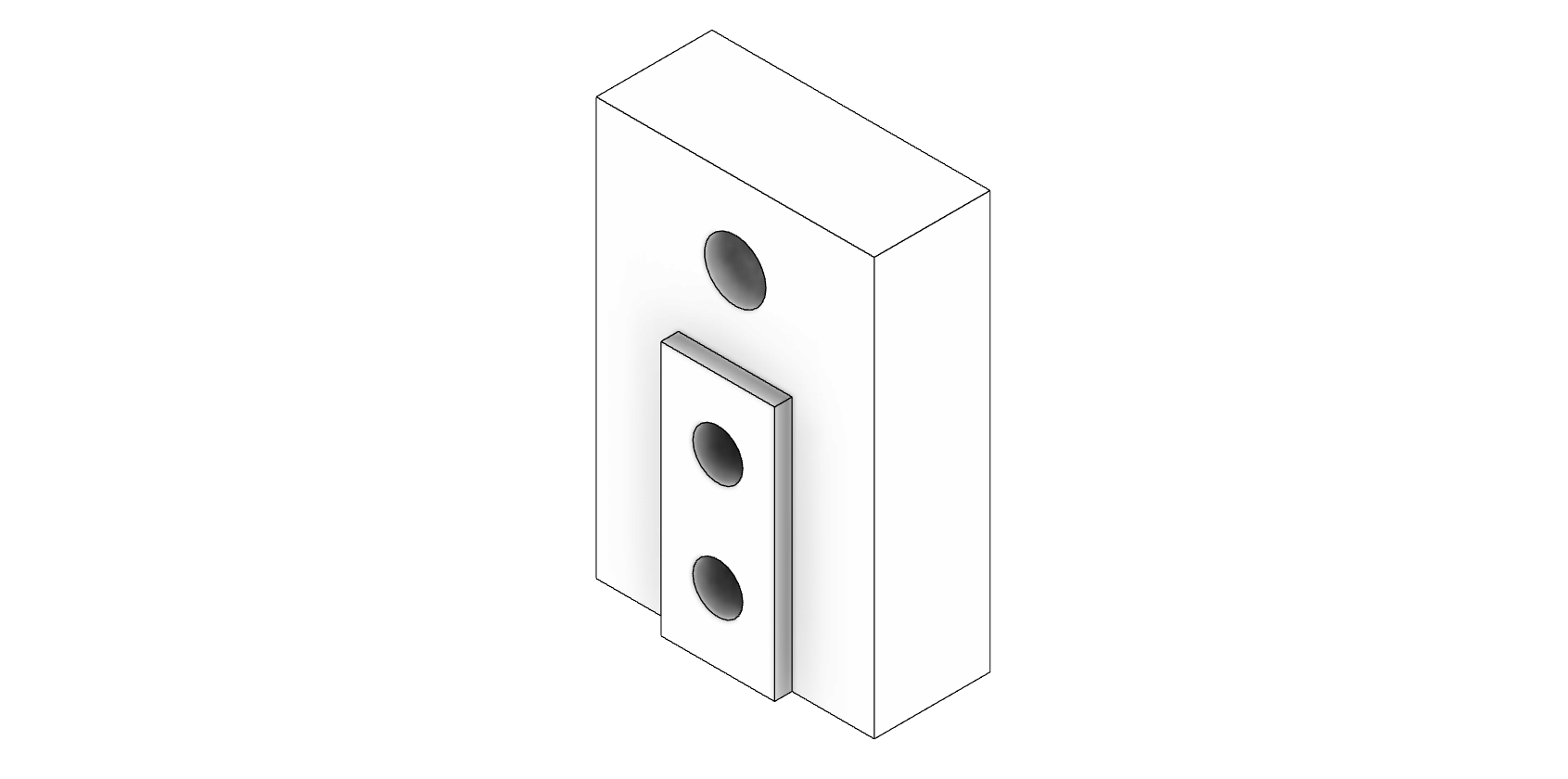}
\caption[Interconnection brackets]{Reference interconnection bracket. The other interconnection brackets lack the relief on the side facing the detector}
\label{interconnection_brackets}
\end{figure}

\begin{figure}[htbp]
\centering
\includegraphics[keepaspectratio, width=\textwidth]{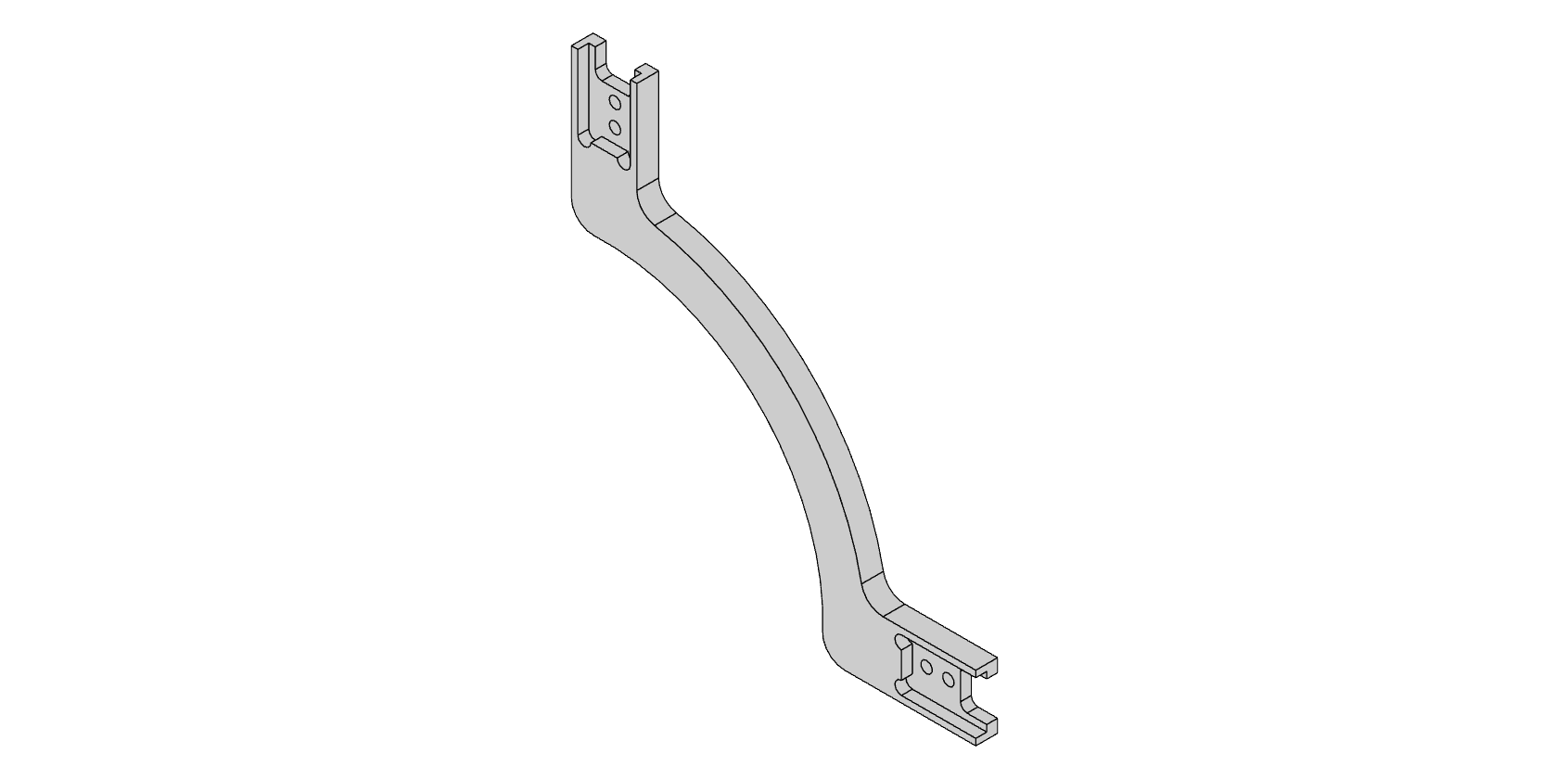}
\caption[Straight positioning tool]{Positioning tool.}
\label{straight_positioning_tool}
\end{figure}

\paragraph{Additional information:}
The reference interconnection bracket (fig.$\,$\ref{interconnection_brackets}) distinguishes itself from the other interconnection brackets due to the presence of a relief on the side facing the MDC.
\begin{enumerate}[label=\labellazzo{{\arabic*}}{.\,\square}]
\item Fix the reference interconnection bracket and a regular interconnection bracket to the positioning tool (fig. \ref{straight_positioning_tool}) using four M5x16 bolts.
\item With the help of the positioning tool, fix the two brackets to the west flange using two M5x16 bolts. Make sure that the reference bracket is located at the bottom right (south) corner.
\item Remove the positioning tool by unscrewing the four M5 bolts that connect it to the brackets.
\item Repeat the three previous points to fix the two remaining brackets to the west flange.
\end{enumerate}

\FloatBarrier

\subsection{Rail coupling}
\begin{figure}[htbp]
\centering
\includegraphics[keepaspectratio, width=\textwidth]{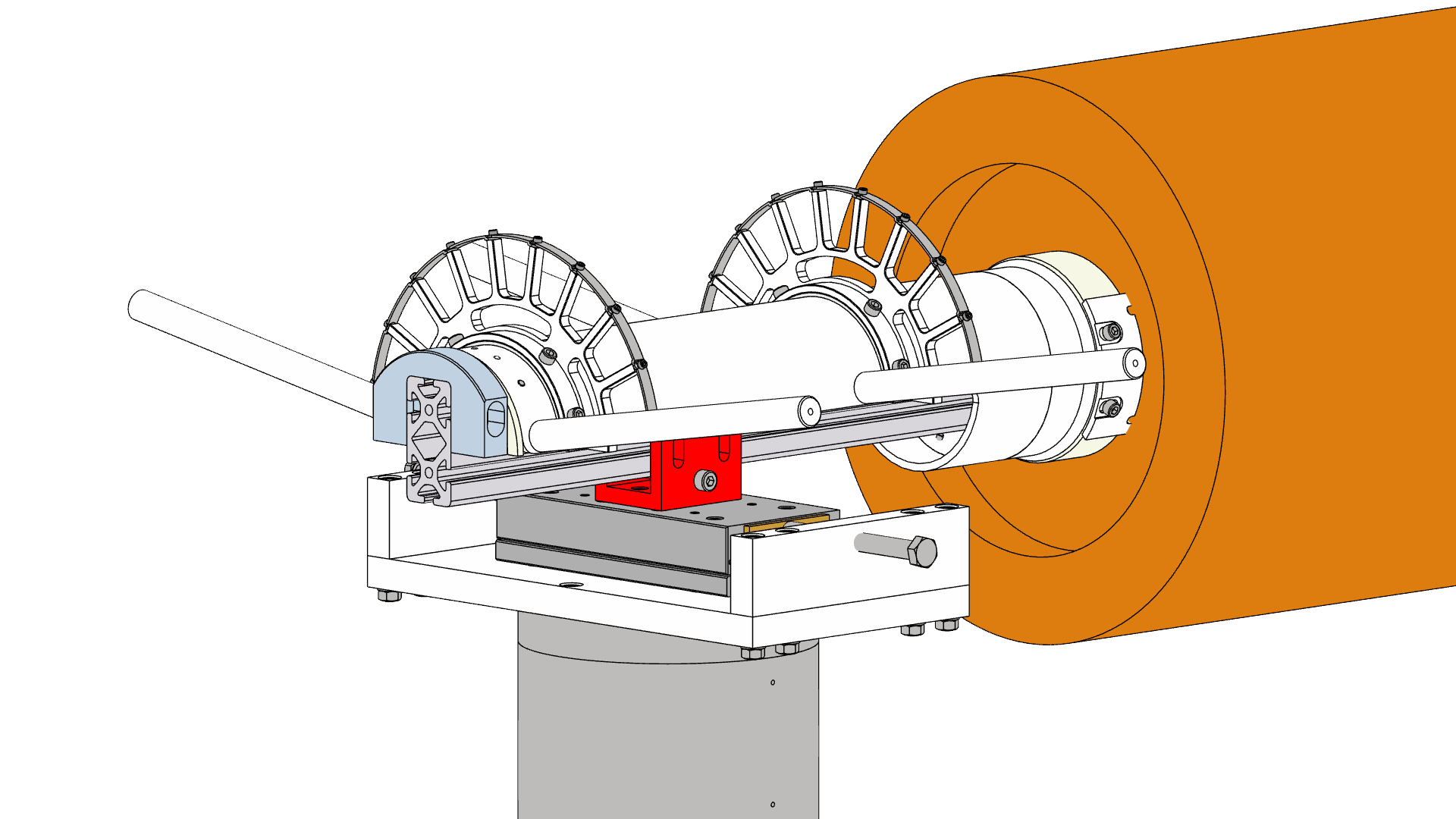}
\caption[U-bracket]{U-bracket.}
\label{c-bracket}
\end{figure}

\paragraph{Additional information:}
L0 and L1 mount the U-brackets (fig.$\,$\ref{c-bracket}) for housing the aluminum profile supporting the trolley, while L2 and L5 mount rail clamps.
All legs are assumed to be vertically aligned. The rail is assumed to be horizontally centered with respect to the cavity.

\subsection{Pre-alignment}
\begin{figure}[htbp]
\centering
\includegraphics[keepaspectratio, width=\textwidth]{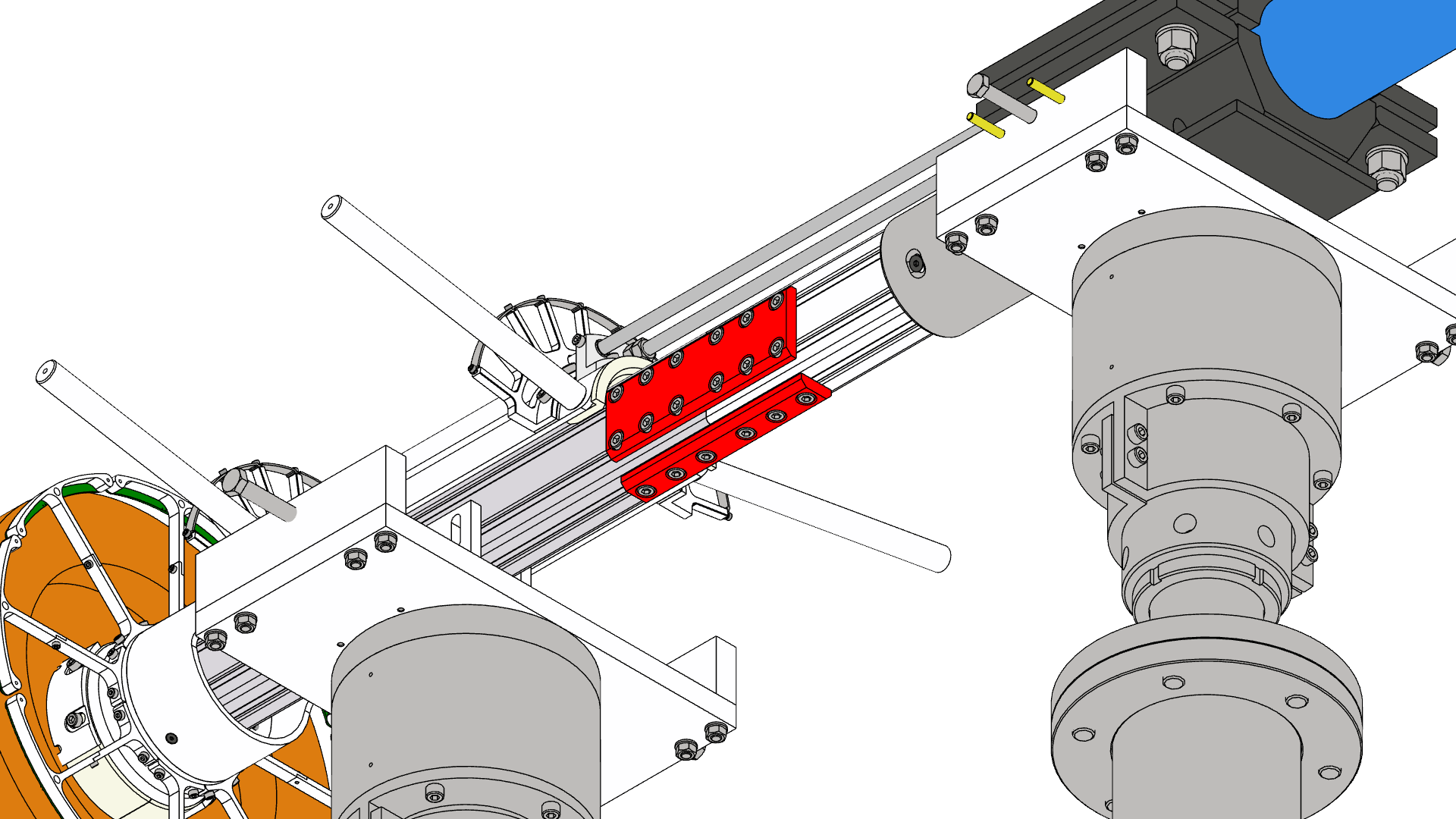}
\caption[Rail coupling plates]{Rail coupling plates.}
\label{plates}
\end{figure}

\begin{figure}[htbp]
\centering
\includegraphics[keepaspectratio, width=\textwidth]{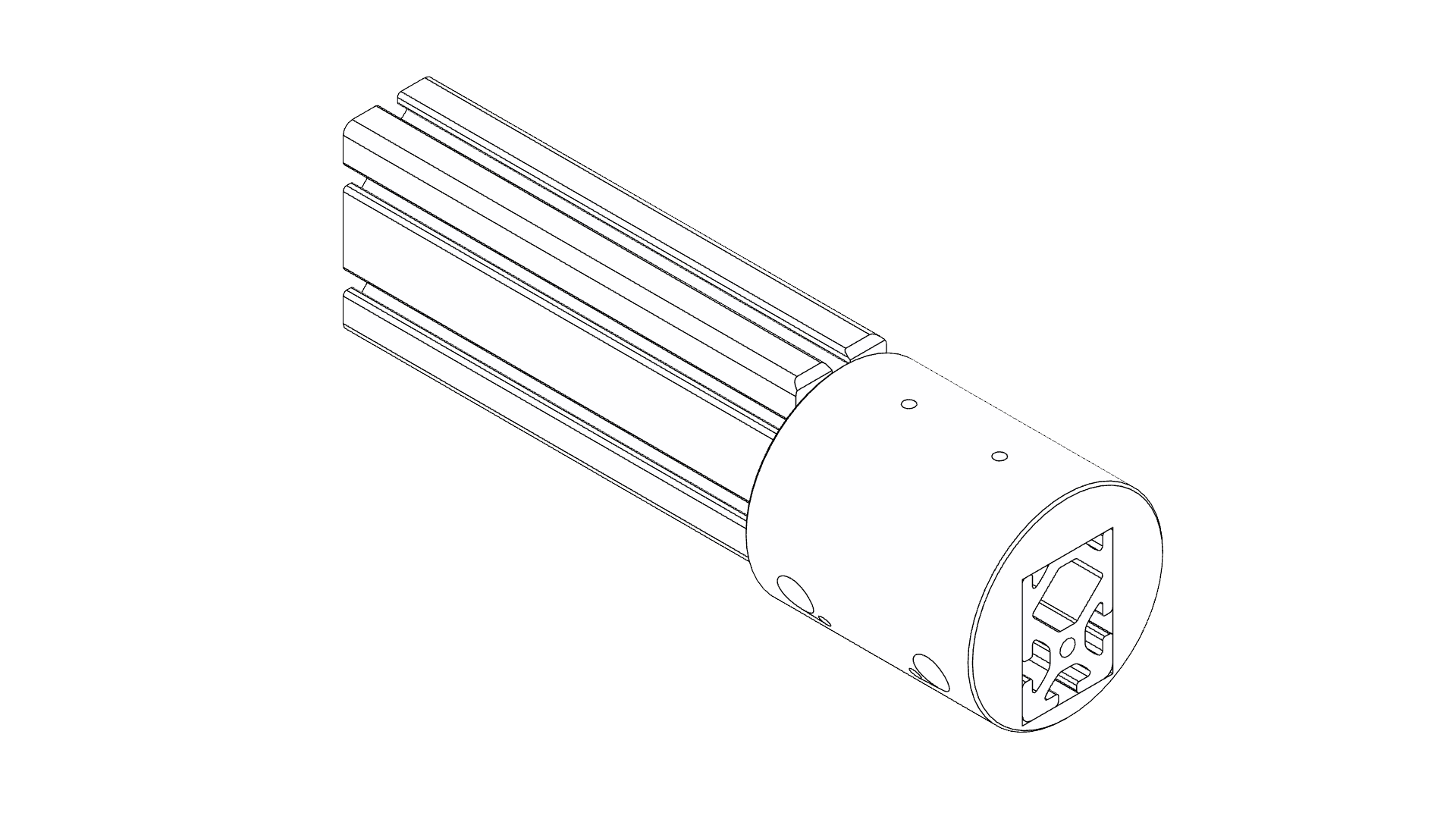}
\caption[Coupling implement]{Coupling implement.}
\label{coupling_implement}
\end{figure}

\begin{figure}[htbp]
\centering
\includegraphics[keepaspectratio, width=\textwidth]{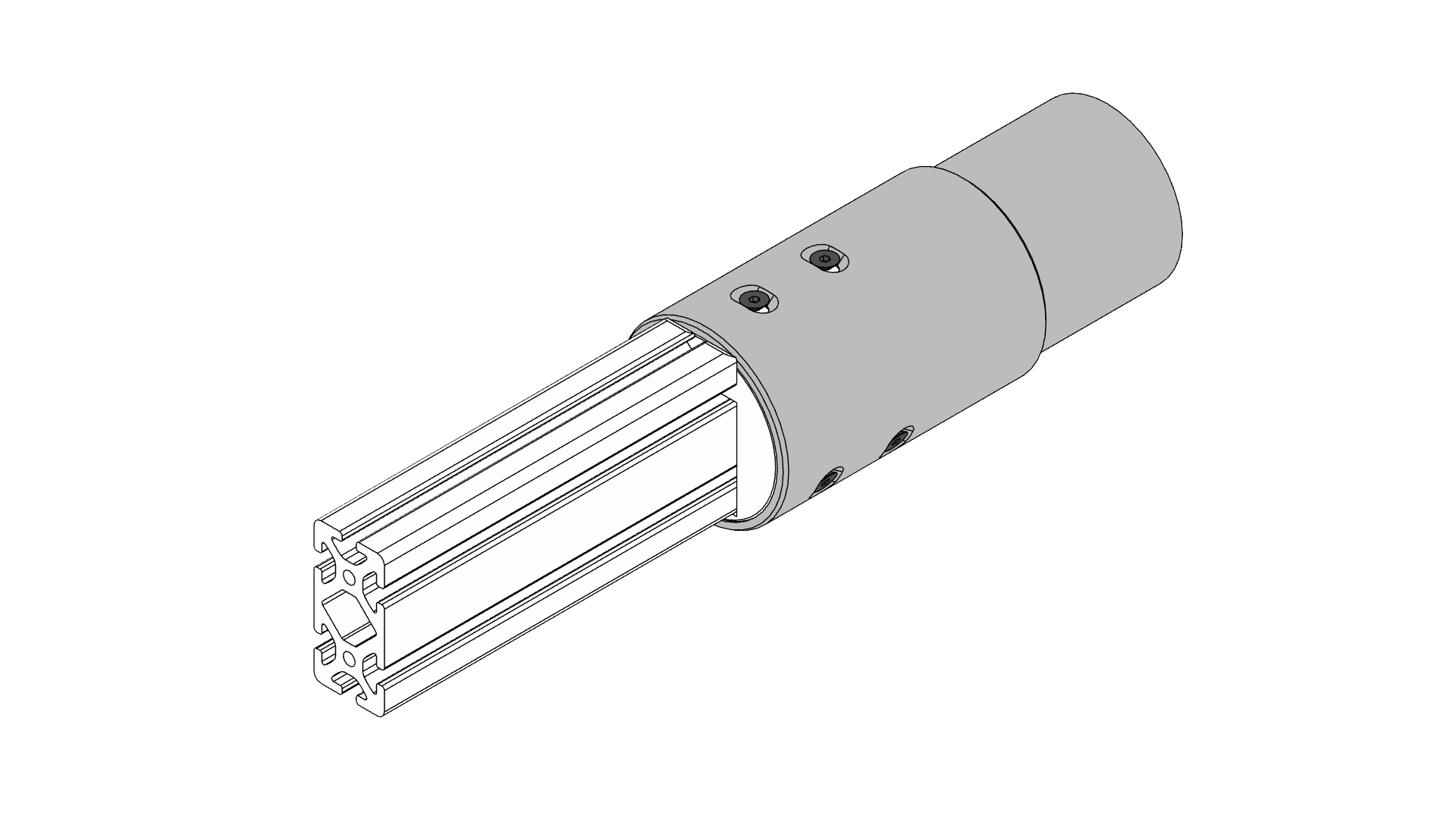}
\caption[Bolted coupling implement]{The coupling implement bolted to the slotted rail extension. The slotted rail extension is welded to the rail (not in picture).}
\label{bolted_coupling_implement}
\end{figure}

\paragraph{Additional Information:}
In this phase no team separation is necessary, all operations occur at the east side.
\begin{enumerate}[label=\labellazzo{{\arabic*}}{.\,\square}]
\item \textbf{Make sure that the 3 plates (fig.$\,$\ref{plates}) used for joining the coupling implement (fig.$\,$\ref{coupling_implement}) and the aluminum profile are inserted into the coupling implement.}
\item Insert the coupling implement into the rail and secure it in place with 6 M6x12 countersunk head screws as in fig.$\,$\ref{bolted_coupling_implement}. The screws may be left loose to allow the coupling implement to rotate as much as permitted by the slotted rail extension.
\item Position an 80x40 profile on L0 and L1.
\item Adjust the trolley support legs until the profile is aligned with the coupling implement.
\item If the profile cannot be aligned to the coupling implement due to a rotation of the coupling implement around the rail axis, rotate the coupling implement until alignment is achieved.
\item If necessary repeat the two previous steps until the profile and the coupling implement are aligned.
\item Tighten the 6 M6x12 countersunk head screws securing the coupling implement to the rail.
\item Lock the adjustment mechanisms of all legs.
\item Remove the 80x40 aluminum profile.
\end{enumerate}

\FloatBarrier

\subsection{Trolley Alignment}
\paragraph{Additional information:}
No team separation is necessary in this phase, all operations occur on the east side.
Part of the operations described in this section are strongly dependent on the design of the transport stretcher, which has not yet been finalized. Because of this, a few steps are kept generic to encompass as many cases as possible.

\begin{enumerate}[label=\labellazzo{{\arabic*}}{.\,\square}]
\item \textbf{Verify that all stoppers are in place:}
\begin{enumerate}[label=\labellazzo{{\arabic*}}{.\,\square}]
	\item Verify that all 4 detector stoppers are contacting the four-spoke flanges at both ends.
	\item Verify that the trolley stoppers are in position and tightened.
	\item Verify that the stoppers preventing the trolley's rotation are in position and tightened.
	\end{enumerate}	
\item \textbf{Position the east flange onto the trolley, between the two rear cable holders.}
\item Use the overhead crane to lift the trolley with its support profile and lower it onto the U-brackets of L0 and L1. Two people, stationed at L0 and L1 and wearing leather gloves, help guiding the profile into the U-brackets.
The stretcher should always be guided by hand when moved by crane.
\item If necessary, make sure that the profile supporting the trolley contacts the coupling implement by sliding it forward gently.
\item Use 8 T-slot nuts inserted into the profile's grooves and 8 M8 screws inserted into the U-bracket slots to secure the profile in place. The screws should be left loose to allow the profile to move according to the adjustments performed at the legs.
\item \textbf{Proceed with the steps between the horizontal lines only if the aluminum profile supporting the trolley requires further alignment.}\\\rule{10 cm}{1pt}
\item Unlock rotation and horizontal movement of L0 and L1.
\item Adjust the trolley support legs until the profile is aligned with the coupling implement.
\item If the profile cannot be aligned to the coupling implement due to a rotation of the coupling implement around the rail axis, loosen the 6 M6x12 countersunk head screws and rotate the coupling implement until alignment is achieved.
\item If necessary, repeat the two previous steps until alignment is achieved.
\item If necessary, tighten the 6 M6x12 countersunk head screws.
\item Lock rotation and horizontal movement of L0 and L1.\\
\rule{10 cm}{1pt}
\item Tighten the 8 M8 screws securing the profile to the U-brackets.
\item Disconnect the trolley from the overhead crane.
\item Dismantle any part of the stretcher that may have remained in place to facilitate lifting the trolley.
\end{enumerate}

\subsection{Coupling}
\paragraph{Additional Information:} Three plates for joining the profile and the coupling implement are inserted into the coupling implement.

\begin{enumerate}[label=\labellazzo{{\arabic*}}{.\,\square}]
\item If necessary, loosen all screws keeping the plates in place.
\item Slide the plates across the joint so that it is positioned in the middle of the plates as of fig.$\,$\ref{plates}.
\item Tighten the 6 screws on the bottom plate.
\item Tighten screws on the side plates (12 on each side).
\end{enumerate}

\subsection{Travel of the Trolley towards the MDC}
\begin{enumerate}[label=\labellazzo{{\arabic*}}{.\,\square}]
\item Remove L2:
	\begin{enumerate}[label=\labellazzo{{\arabic*}}{.\,\square}]
	\item Note down location and thickness of the shims used to align the leg vertically.
	\item Remove the top half of the clamp.
	\item \textbf{Unlock the vertical adjustment mechanism and release the rotation brake.}
	\item Fully Lower the leg. \textbf{Pay particular attention when the clamps detaches itself from the rail, as this could lead to unwanted jolting}. If necessary, keep the clamp steady by hand.
	\item Remove the 4 screws at the  base of the leg.
	\item Remove the leg.
	\end{enumerate}
\item Check the status of the joint.
\item Install the telescopic cable holder:
	\begin{enumerate}[label=\labellazzo{{\arabic*}}{.\,\square}]
		\item Clear access to the guides of the telescopic cable holder by moving the cables at the front of the trolley.
		\item Insert the two M10 threaded bars of the telescopic cable holder into the guides.
		\item Insert the spacer shaft that maintains the telescopic cable holder extended.
		\item collect the cables in two bundles and secure them to the telescopic cable holder using some Velcro tape.
	\end{enumerate}
\item Remove L1 while continuously checking the status of the joint:
	\begin{enumerate}[label=\labellazzo{{\arabic*}}{.\,\square}]
	\item Note down location and thickness of the shims used to align the leg vertically.
	\item Remove the 4 M8 screws securing the profile to the leg.
	\item \textbf{Unlock the vertical adjustment mechanism and release the rotation brake.}
	\item Fully Lower the leg while checking the status of the joint. \textbf{Pay particular attention when the U-bracket frees itself from the profile, as it could induce unwanted jolting.} Keep the head from rotating by hand if necessary.
	\item Remove the 4 screws at the  base of the leg.
	\item remove the leg.
	\end{enumerate}
\item Install handles on the trolley. Make sure the screws on the handles do not protrude more than 5 millimeters from the body of the handle before installation.
\item Remove all stoppers preventing the trolley from moving with respect to the profile:
	\begin{enumerate}[label=\labellazzo{{\arabic*}}{.\,\square}]
	\item Remove the front trolley stopper.
	\item Remove the rotation stoppers.
	\end{enumerate}
\item One person slowly pushes the trolley westward until the front of the trolley almost contacts L3. Two people, one at each side of the trolley, guides the cables around any obstacle while the trolley is moving. It may be necessary to slightly lift the front of the trolley when crossing the joint to ease its passage. When the telescopic cable holder almost reaches L3 either remove the spacer shaft to let it retract or remove the top part of L3's clamp to let it pass.
\item Reinstall L1:
	\begin{enumerate}[label=\labellazzo{{\arabic*}}{.\,\square}]
	\item Reposition the leg making sure to restore the shim configuration used to keep the leg vertical.
	\item Tighten the 4 screws at the base of the leg.
	\item Raise the leg and adjust the linear table to reinsert the profile into the U-bracket. \textbf{Pay particular attention when the U-bracket reaches the profile, keep the leg's head steady by head to prevent it from rotting and bumping into the profile.}
	\item Secure the profile to the leg using 4 M8x20 screws.
	\end{enumerate}
\item Remove L3 while continuously checking the status of the joint:
	\begin{enumerate}[label=\labellazzo{{\arabic*}}{.\,\square}]
	\item if necessary push/pull the trolley eastward to ensure safety of the detector during the operations.
	\item Remove the top half of the leg clamp, if not already absent.
	\item Fully lower the leg. \textbf{Pay particular attention when the clamps detaches itself from the rail as this could lead to unwanted jolting movement}. If necessary, keep the clamp steady by hand.
	\item Remove the 4 screws at the base of the leg.
	\item Remove the leg.
	\end{enumerate}
\item If it was removed, reinstall the spacer shaft of the telescopic cable holder.
\item Push the trolley westward until L2 can be reinstalled:
\item Reinstall L2:
	\begin{enumerate}[label=\labellazzo{{\arabic*}}{.\,\square}]
	\item Reposition the leg making sure to restore the shim configuration used to keep the leg vertical.
	\item Tighten the 4 screws at the base of the leg.
	\item Raise the leg and adjust the linear table until the bottom part of the clamp couples well with the rail. \textbf{Pay particular attention when the clamp reaches the rail and keep it steady by hand for preventing it from rotating and bumping into the rail.}
	\item Secure the rail in place using the top part of the clamp.
	\end{enumerate}
\item Decouple the aluminum profile from the coupling implement by loosening the screws securing the plates and sliding these all the way towards the coupling implement.
\item Remove the aluminum profile, loosening all the M8 screws securing it to the U-brackets.
\item Dismantle L1 to free some space around L2 for the next phases. L0 can remain in place or be dismantled at leisure.
\item Power and test the sensor system:
	\begin{enumerate}[label=\labellazzo{{\arabic*}}{.\,\square}]
	\item Position a DC power supply at the west side of the spectrometer.
	\item Run 2 cables through the cavity to connect the sensor system on the trolley to the power supply.
	\item Ground the inner aluminum surface of the separation cylinder to the power supply by means of a cable soldered on some copper tape on the west side, as close as possible to the edge of the separation cylinder.
	\item Turn on the power supply and set the channel to 24 V.
	\item Using a conductor to connect the inner surface of the separation cylinder to the sensors and the reference surface, check the functioning of all the sensors.
	\end{enumerate}
\item If mounted, remove the trolley's handles from the front of the trolley.
\end{enumerate}

\section{Insertion}
\paragraph{Additional information:}
The insertion requires good coordination between the teams stationed at each side of the detector. Only one person, and always the same person, moves the trolley. The clearance between the detector and the separation cylinder must be continuously monitored from both sides.

\begin{enumerate}[label=\labellazzo{{\arabic*}}{.\,\square}]
\item \textbf{E \& W: Check that the rail clamps on L2 and L5 are well tightened.}
\item \textbf{E: Remove both detector stoppers on the east side.}
\item \textbf{E:} Push the trolley until the guard ring is close to the east opening.
\item \textbf{E:} Remove the east flange from the trolley. Position it somewhere behind the trolley along the rail, in a place where it does not hinder operations.
\item \textbf{E \& W:} Adjust L2 and L5 to ensure the sensor ring can pass through the opening.
\item \textbf{E:} Push the trolley until the guard ring is fully inserted into the opening. Pay attention to the radio and stop pushing immediately in case a contact is signaled.\\
\textbf{W:} Monitor the clearance between the sensor array and the separation cylinder. Signal contacts and near contacts as soon as they occur.
\item \textbf{E \& W:} Adjust L2 and L5 to ensure the clearance between the sensor ring and the separation cylinder is homogeneous along the whole circumference.
\item \textbf{E:}Push the trolley until the west end of the detector is slightly inserted into the opening.\\
\textbf{W:} Monitor the sensor array to spot possible contacts. Signal contacts and near contacts as soon as they occur. Recover the cables powering the sensors as the trolley advances.
\item \textbf{E \& W:} Adjust L2 and L5 to ensure the clearance separating the detector and the separation cylinder is homogeneous along the whole circumference.
\item \textbf{E:} Slowly push the trolley deeper into the opening while continuously checking clearance between the detector and the separation cylinder.\\
\textbf{W:} Monitor the sensor array to spot possible contacts. Signal contacts and near contacts as soon as they occur. Recover the cables powering the sensors as the trolley advances.\\
\textbf{E \& W:} When necessary, Adjust L2 and L5 and ensure the clearance separating the detector and the separation cylinder is homogeneous along the whole circumference.
\item \textbf{E:} Stop when the space for disassembling the sensor array is sufficient.\\
\textbf{W:} Signal when it is time to pause the insertion for disassembling the sensor array.
\item \textbf{W:} Disassemble the top half of the sensor brackets:
	\begin{enumerate}[label=\labellazzo{{\arabic*}}{.\,\square}]
	\item Cut the two groups of wires just behind the solder that joins them together.
	\item Unscrew the M3 screw that joins the top and bottom half of the first bracket while keeping the top half of the bracket in your hand.
	\item Extract the top half of the bracket from the cavity and set it aside.
	\item Repeat the two previous steps until all the remaining 7 brackets have also been removed.
	\item Reconnect the reference surface contact alarm.
	\end{enumerate}
\end{enumerate}

\section{Anchoring}
\paragraph{Additional information:} This section is subject to changes as the anchoring is still being optimized. Information regarding the mounting of the east flange are incomplete at the moment and should be integrated at a later time.

\begin{enumerate}[label=\labellazzo{{\arabic*}}{.\,\square}]
\item Screw guiding pins into the innermost of the two holes at the lower right (south) and at the upper left (north) anchor points on the west side.
\item Either slowly push (from the east side) or pull (from the west side) the trolley towards the interconnection brackets mounted on the west flange. Clearance between the detector and the separation cylinder must be continuously monitored from both sides.
\item When necessary, slightly rotate the trolley to align the guiding pins with the holes on the interconnection brackets. Clearance between the detector and the separation cylinder should be continuously monitored from both sides.
\item Alternate between the two previous steps until at least one of the anchor points on the four-spoke flange contacts one of the interconnection brackets.
\item \textbf{Remove both detector stoppers on the west side.} Never rotate the trolley when both the detector stoppers on the west side and the guiding pins are constraining the rotation of the detector.
\item Check the coupling at the reference bracket and at the opposite bracket with an endoscope. If necessary, adjust the tilt of the rail operating L2 and L5 to obtain a good coupling for all brackets on the west side.
\item Check that all the holes on the interconnection brackets are aligned with those on the four-spoke flange. The reference bracket and the one opposite to it cannot be moved, the other two may be loosened and adjusted to match.
\item \textbf{\textit{Install the east flange.}}
\item Mount the (4) interconnection brackets on the east flange using M5x16 screws. Make sure that the threaded pins can be inserted into the holes on the interconnection brackets and screwed into those on the four spoke-flange.
\item If everything checks out, position and tighten all screws on the west side, removing the guiding pins where necessary.
\item Mount the threaded pins on the east side. Use the largest possible passing (non-interfering) pin to reduce backlash when the rail will be lowered.
\end{enumerate}

\section{Extraction of the Trolley and Cabling}
\subsection{Extraction of the Trolley}
\paragraph{The trolley is removed before cabling to minimize the risk of stressing the detector by leaning on the rail during the routing of the cables.}
\begin{enumerate}[label=\labellazzo{{\arabic*}}{.\,\square}]
\item Free the cables from the cable holders on both sides.
\item Temporarily spread and fix the cables in the cone at both sides so that they do not hinder the passage of the trolley.
\item Dismantle the telescopic cable holder extension.
\item \textbf{Remove the handles on the west side.}
\item \textbf{Remove the cable holders on the west side.}
\item \textbf{Remove all bottom halves of the sensor array brackets.}
\item For each leg write down the distance between the threaded flange and the brake flange, as well as the orientation of a single marked hole.
\item Unload the trolley by lowering the rail until the clearance between the trolley and the detector is homogeneous along the whole circumference on both sides.
\item Slowly extract the trolley from the detector. The difference in diameter between the Teflon and aluminum parts of the trolley could make the trolley latch onto the detector flanges during extraction. If necessary, adjust the rail legs to ease the passage of the Teflon parts.
\end{enumerate}

\subsection{Cabling}

Progressively free the cables previously affixed to the cone and route them towards the patch card mounting points. Refer to the attached schemes for cable routing and patch card placement. Avoid touching the rail while routing the cables.

\newpage

\section{Risk Assessment}
\subsection*{General Considerations}
\begin{table}[htbp]
\centering
\begin{tabular}{@{}m{.27\textwidth}m{.67\textwidth}@{}}
\textbf{Risk} & \textbf{Mitigation}\\
\midrule
Jolt the detector &
\vspace*{-\baselineskip}
\begin{itemize}[leftmargin=*, itemsep=0pt, parsep=0pt]
\item Always move the detector slowly
\item Slightly rotate the trolley before pushing/pulling to overcome static friction
\item Make sure joints/welds are level with the rail
\item Polish the rail
\end{itemize} \\
\midrule
Damage the legs &
Always make sure brakes and locking mechanisms are released before making adjustments
\end{tabular}
\end{table}
\FloatBarrier

\subsection*{Alignment and Coupling}
\begin{table}[htbp]
\centering
\begin{tabular}{@{}m{.27\textwidth}m{.67\textwidth}@{}}
\textbf{Risk} & \textbf{Mitigation}\\
\midrule
Dropping the detector while lifting it by crane &
\begin{itemize}[leftmargin=*, itemsep=0pt, parsep=0pt]
\item Double check anchor points and lifting ropes
\item Keep hands on the stretcher when moving it by crane
\end{itemize}\\
\midrule
Squeezing the detector between the lifting ropes &
\begin{itemize}[leftmargin=*, itemsep=0pt, parsep=0pt]
\item Use a lifting beam for connecting the stretcher to the overhead crane
\item Check the behavior of the lifting ropes when applying tension
\end{itemize}\\
\midrule
Bumping the detector against obstacles when moving it by crane &
\begin{itemize}[leftmargin=*, itemsep=0pt, parsep=0pt]
\item Clear the working area as much as possible before lifting the detector
\item Operate the crane at its slowest speed
\item Keep hands on the stretcher when moving it by crane
\end{itemize}
\\\midrule
Transfer sudden movements to the detector when lowering it onto the legs &
\begin{itemize}[leftmargin=*, itemsep=0pt, parsep=0pt]
\item Keep hands on the stretcher when moving it by crane
\item Operate the crane at its slowest speed
\item Guide the detector onto the U-brackets by hand
\end{itemize}
\end{tabular}
\end{table}
\FloatBarrier

\newpage

\subsection*{Travel along the Rail}
\begin{table}[htbp]
\centering
\begin{tabular}{@{}m{.27\textwidth}m{.67\textwidth}@{}}
\textbf{Risk} & \textbf{Mitigation}\\
\midrule
Jolt the detector when crossing the joint/welds &
\begin{itemize}[leftmargin=*, itemsep=0pt, parsep=0pt]
\item Evaluate interference and obstacles using the endoscope
\item Ease the passage of joints/welds by lifting the trolley
\item Ease the passage of joints/welds by temporarily supporting the rail by hand
\end{itemize} \\
\midrule
Cables getting snagged against obstacles &
\begin{itemize}[leftmargin=*, itemsep=0pt, parsep=0pt]
\item Check the cables before moving the trolley
\item Guide the cables around obstacles manually
\item Stretch the cables along the rail and fix them temporarily when necessary
\end{itemize}\\
\midrule
Jolt the detector when removing/reinstalling legs &
\begin{itemize}[leftmargin=*, itemsep=0pt, parsep=0pt]
\item prevent head rotation by hand when the coupling is about to become loose
\item Guide coupling elements in position by hand when raising the legs
\end{itemize}
\end{tabular}
\end{table}
\FloatBarrier

\newpage

\subsection*{Insertion}
\begin{table}[htbp]
\centering
\begin{tabular}{@{}m{.27\textwidth}m{.67\textwidth}@{}}
\textbf{Risk} & \textbf{Mitigation}\\
\midrule
Bumping the detector against the wall of the separation cylinder &
\begin{itemize}[leftmargin=*, itemsep=0pt, parsep=0pt]
\item Move the trolley slowly
\item Maintain coordination and communication between the two teams
\item Always keep the sensors in check
\item Check clearance on both sides
\item Push/pull the detector while keeping the force applied as aligned as possible with the rail to avoid unwanted movement
\item Coordinate leg adjustments and move only a single leg at a time by small amounts
\item Monitor sensor array and clearance during leg adjustment
\end{itemize}
\end{tabular}
\end{table}
\FloatBarrier

\subsection*{Anchoring}
\begin{table}[htbp]
\centering
\begin{tabular}{@{}m{.27\textwidth}m{.67\textwidth}@{}}
\textbf{Risk} & \textbf{Mitigation}\\
\midrule
Bumping the detector against the wall of the separation cylinder &
Check the same item in the risk assessment table for the insertion \\
\midrule
Compressing and/or stretching the detector &
This should be handled at design level, but:
\begin{itemize}[leftmargin=*, itemsep=0pt, parsep=0pt]
\item Move the trolley slowly
\item Make sure the detector stoppers on the east side have been removed before beginning the insertion
\item Always check coupling/interference before tightening screws on the west side
\end{itemize}\\
\midrule
Bending the detector &
\textbf{NEVER} adjust the rail when the west side screws have been tightened, unless the east side has been fixed as well
\end{tabular}
\end{table}
\FloatBarrier

\newpage
\subsection*{Extraction of the Trolley}
\begin{table}[htbp]
\centering
\begin{tabular}{@{}m{.27\textwidth}m{.67\textwidth}@{}}
\textbf{Risk} & \textbf{Mitigation}\\
\midrule
Bending the detector &
This should be accounted for at design level, although:
\begin{itemize}[leftmargin=*, itemsep=0pt, parsep=0pt]
\item Remove as much backlash as possible with the use of properly sized threaded pins before lowering the rail
\item Lower the rail at both sides simultaneously to maintain inclination
\end{itemize}\\
\midrule
Bump the trolley into the detector &
\begin{itemize}[leftmargin=*, itemsep=0pt, parsep=0pt]
\item Check clearance between the trolley and the four-spoke flanges using feeler gauges before extracting the trolley
\item Extract the trolley slowly
\item Pay particular attention when reaching the points where the outer diameter of the trolley varies
\end{itemize}
\end{tabular}
\end{table}
\FloatBarrier

\subsection*{Cabling}
\begin{table}[htbp]
\centering
\begin{tabular}{@{}m{.27\textwidth}m{.67\textwidth}@{}}
\textbf{Risk} & \textbf{Mitigation}\\
\midrule
Pulling on the cables &
This should be handled by the cable holders mounted on the four-spoke flange, but:
\begin{itemize}[leftmargin=*, itemsep=0pt, parsep=0pt]
\item Handle the cables gently
\item Provide a second fixing point for the cables within the cone as soon as possible and as near as possible to the detector
\end{itemize}\\
\midrule
Applying force to the detector through the rail &
\begin{itemize}[leftmargin=*, itemsep=0pt, parsep=0pt]
\item Extract the trolley before cabling
\item Clear the rail of any accessories that may increase occupancy (cable holders, handles, etc.)
\item Avoid touching the rail during operations
\end{itemize}\\
\midrule
\end{tabular}
\end{table}
\FloatBarrier
 	\pagestyle{empty}
 	\pagenumbering{gobble}
 	\bibliographystyle{ieeetr}
 		\bibliography{biblio}
\end{document}